\documentclass[floatfix,aps,preprint,showpacs,superscriptaddress,amsmath,amssymb,longbibliography]{revtex4-2}

\usepackage[utf8]{inputenc}
\usepackage{pgfplots}
\usepackage{subcaption} 
\usepackage{pgfplotstable}
\usepackage{multirow}
\usepackage{booktabs}
\usepackage{colortbl}
\usepackage{color}
\usepackage{array}
\usepackage{dcolumn}
\usepackage{physics}
\usepackage{amsmath}
\usepackage{amssymb}
\usepackage{dsfont}
\usepackage{float}
\usepackage{mhchem}
\usepackage{placeins}
\usepackage{rotating}
\usepackage[unicode=true,
 bookmarks=true,bookmarksnumbered=true,bookmarksopen=true,bookmarksopenlevel=2,
 breaklinks=false,pdfborder={0 0 1},backref=false,colorlinks=true]
 {hyperref}
\usepackage{cleveref}
\crefname{section}{Sec.}{Secs.}
\Crefname{section}{Section}{Sections}
\crefname{subsection}{subsection}{subsections}
\Crefname{subsection}{Subsection}{Subsections}
\crefname{subsubsection}{subsection}{subsections}
\Crefname{subsubsection}{Subsection}{Subsections}
\crefname{equation}{Eq.}{Eqs.}
\Crefname{equation}{Equation}{Equations}
\crefname{table}{Tab.}{Tabs.}
\Crefname{table}{Table}{Tables}
\crefname{figure}{Fig.}{Figs.}
\Crefname{figure}{Figure}{Figures}

\usepackage{rotating}
\usepackage[margin=2.5cm, portrait]{geometry}

\usepackage{tabularx}

\usepackage{diagbox}        % provides \diagbox  :contentReference[oaicite:0]{index=0}

\usepackage{mathtools}      % gives \mathmakebox
\usepackage{calc}           % gives \widthof (used inside the macro below)  :contentReference[oaicite:1]{index=1}

\newcommand{\mathtextover}[3][l]{%
  \mathmakebox[\widthof{\(#3\)}][#1]{#2}%
}

\usepackage{tikz}
\usetikzlibrary{tikzmark,shapes.arrows,shapes,arrows,shadows,patterns,pgfplots.groupplots,decorations.markings,chains,calc,positioning}

\pgfkeys{/pgf/number format/.cd,
fixed,
fixed zerofill,
dec sep = {.}}

\usepackage[autostyle=true]{csquotes}
\usepackage{times}
\usepackage{amsmath}
\usepackage{mathtools}
\usepackage{textcomp}
\usepackage{graphicx}
\usepackage{siunitx}
\usepackage{ulem}
\usepackage{physics}
\usepackage{appendix}

\usepackage{ulem}

\hypersetup{linkcolor=blue, citecolor=blue, urlcolor=blue, filecolor=blue, pdfpagelayout=OneColumn,
  pdfnewwindow=true, pdfstartview=XYZ, plainpages=false}

\usepackage[all]{hypcap}
\usepackage{color}

\newcommand{\vecJ}{\vec{J}}

\newcommand{\beq}{\begin{equation}}
\newcommand{\eeq}[1]{\label{#1} \end{equation}}
\newcommand{\beqar}{\begin{equation}\begin{aligned}}
\newcommand{\eeqar}[1]{\label{#1} \end{aligned}\end{equation}}

\newcommand{\Threej}[6]{ \begin{pmatrix}
    #1 & #2 & #3 \\
    #4 & #5 & #6
   \end{pmatrix}}

 \newcommand{\Sixj}[6]{ \begin{Bmatrix}
    #1 & #2 & #3 \\
    #4 & #5 & #6
   \end{Bmatrix}}

\pgfkeys{/pgf/number format/.cd,
fixed,
fixed zerofill,
dec sep = {.}}

\makeatletter
\def\mathcolor#1#{\@mathcolor{#1}}%
\def\@mathcolor#1#2#3{%
  \protect\leavevmode%
  \begingroup\color#1{#2}#3\endgroup%
}%
\newcommand{\msout}[1]{\text{\sout{\ensuremath{#1}}}}%

\newcommand{\sam}[2]{%
\ifmmode%
  {#2}%\msout{#1}\mathcolor{red}{#2}%
\else%
  {#2}%\sout{#1}\textcolor{red}{#2}%
\fi}
\newcommand{\samO}[2]{%
\ifmmode%
  \msout{#1}\mathcolor{magenta}{#2}%
\else%
  \sout{#1}\textcolor{magenta}{#2}%
\fi}

\newcommand{\cyr}[2]{%
\ifmmode%
  \msout{#1}\mathcolor{red}{#2}%
\else%
  \sout{#1}\textcolor{red}{#2}%
\fi}

\newcommand{\gab}[2]{%
\ifmmode%
  \msout{#1}\mathcolor{green}{#2}%
\else%
  \sout{#1}\textcolor{green}{#2}%
\fi}

\makeatother

\makeatletter

\newcommand\Autoref[1]{\@first@ref#1,@}
\def\@throw@dot#1.#2@{#1}% discard everything after the dot
\def\@set@refname#1{%    % set \@refname to autoefname+s using \getrefbykeydefault
    \edef\@tmp{\getrefbykeydefault{#1}{anchor}{}}%
    \xdef\@tmp{\expandafter\@throw@dot\@tmp.@}%
    \ltx@IfUndefined{\@tmp autorefnameplural}%
         {\def\@refname{\@nameuse{\@tmp autorefname}s}}%
         {\def\@refname{\@nameuse{\@tmp autorefnameplural}}}%
}
\def\@first@ref#1,#2{%
  \ifx#2@\autoref{#1}\let\@nextref\@gobble% only one ref, revert to normal \autoref
  \else%
    \@set@refname{#1}%  set \@refname to autoref name
    \@refname~\ref{#1}% add autoefname and first reference
    \let\@nextref\@next@ref% push processing to \@next@ref
  \fi%
  \@nextref#2%
}
\def\@next@ref#1,#2{%
   \ifx#2@ and~\ref{#1}\let\@nextref\@gobble% at end: print and+\ref and stop
   \else, \ref{#1}% print  ,+\ref and continue
   \fi%
   \@nextref#2%
}

\makeatother

\begin{document}

\def\be{\begin{equation}}
\def\ee{\end{equation}}

\def\myvec#1{{\bf #1}}
\def\Esw{\myvec{E}_{sw}}
\def\Hsw{\myvec{H}_{sw}}

\title{The hyperfine interaction as a probe of the microscopic structure of the atomic nucleus}

\affiliation{Center for Quantum Nanoscience, Institute for Basic Science, Seoul 03760, Republic of Korea}
\affiliation{Universit\'e de Strasbourg, CNRS, Institut de Physique et Chimie des Mat\'eriaux de Strasbourg, UMR 7504, F-67000 Strasbourg, France}

\author{D. Jankovi\'c}
\thanks{These two authors contributed equally}
\affiliation{Center for Quantum Nanoscience, Institute for Basic Science, Seoul 03760, Republic of Korea}
\affiliation{Universit\'e de Strasbourg, CNRS, Institut de Physique et Chimie des Mat\'eriaux de Strasbourg, UMR 7504, F-67000 Strasbourg, France}
\affiliation{Ewha Womans University, Seoul 03760, Republic of Korea}

\author{J. -G. Hartmann}
\thanks{These two authors contributed equally}
\affiliation{Universit\'e de Strasbourg, CNRS, Institut de Physique et Chimie des Mat\'eriaux de Strasbourg, UMR 7504, F-67000 Strasbourg, France}
\affiliation{Universit\'e de Strasbourg, CNRS, Institut de Chimie, UMR 7177 F-67000 Strasbourg, France}

\author{J. Bartel}
\affiliation{Universit\'e de Strasbourg, CNRS, Institut pluridisciplinaire Hubert Curien, UMR 7178, F-67000 Strasbourg, France}

\author{H. Molique}
\affiliation{Universit\'e de Strasbourg, CNRS, Institut pluridisciplinaire Hubert Curien, UMR 7178, F-67000 Strasbourg, France}

\author{L. Bonneau}
\affiliation{LP2i Bordeaux, UMR5797, University of Bordeaux, CNRS, F-33170 Gradignan, France}

\author{P. -A. Hervieux}
\email{hervieux@unistra.fr}
\affiliation{Universit\'e de Strasbourg, CNRS, Institut de Physique et Chimie des Mat\'eriaux de Strasbourg, UMR 7504, F-67000 Strasbourg, France}

\date{\today}

\begin{abstract}

The study of highly charged electronic and muonic hydrogen-like ions, provides an intriguing way to probe the internal structure of their atomic nuclei. In this work, we use nuclear structure calculations to accurately calculate the hyperfine splitting of electronic and muonic hydrogen-like ions, focusing in particular on the incorporation of finite-volume corrections, such as Bohr-Weisskopf and Breit-Rosenthal, due to the penetration of the electron and muon wavefunction into the nuclear electric charge and magnetic dipole densities. These corrections are essential for refining our understanding of the nuclear magnetic dipole and electric quadrupole moments. Our simulations use a Skyrme-Hartree-Fock-BCS model known for its effectiveness in modeling well-deformed nuclei such as ${}^{159}\mathrm{Tb}^{64+}$ and ${}^{165}\mathrm{Ho}^{66+}$, with particular emphasis on ${}^{161,163}\mathrm{Dy}^{65+}$ isotopes. It can also be generalised to multi-electron ions by studying the hyperfine anomaly between two isotopes.

\end{abstract}

\maketitle

\tableofcontents

\section{Introduction}\label{Introduction}

Hyperfine interactions serve as a fundamental bridge between atomic and nuclear physics, connecting the electromagnetic properties of the nucleus with the atomic energy levels of the electrons. In a hyperfine transition, the magnetic dipole and electric quadrupole moments of the nucleus interact with the electromagnetic field of the electron cloud, producing small shifts or splittings in atomic energy levels. This phenomenon effectively links two regimes of vastly different scales and characters. The ability of hyperfine measurements to probe nuclear properties has long been recognised as an important avenue for cross-disciplinary insights. Moreover, hyperfine interactions play a crucial role in many physical and chemical processes, with applications ranging from metrology to condensed matter physics and quantum computing on molecular complexes..  

In metrology, the definition of the SI second is based on a hyperfine transition ($F=4\leftrightarrow 3$) in ground-state $^{133}$Cs \cite{ESSEN1955}. More recently, in ${}^{229}$Th nuclear clocks, the hyperfine coupling mediates electronic-bridge and internal-conversion pathways that connect optical electronic excitations to the $\sim8$ eV nuclear isomer, critically shaping excitation rates, linewidths, and readout \cite{PeikOkhapkin2015,Seiferle2019,Campbell2011,Morgan2025,Karpeshin1998,Tkalya2016}. Another example in the field of solid-state physics is ferromagnetic nuclear resonance (FNR), a technique that provides unique insights into ferromagnetic films, multilayers, and nanocomposites. Its spectra reveal the distribution of hyperfine fields, offering valuable information about chemical configurations, site symmetries, phases, structures, and defects \cite{Meny2017}. Finally, even in emerging areas like quantum information, hyperfine couplings between electron and nuclear spins are exploited for qubit storage and control \cite{Thiele2014,Godfrin2017,Yu2025}, reflecting the same fundamental interaction at work in disparate contexts. These examples highlight how hyperfine spectroscopy provides a common thread linking atomic-scale measurements to nuclear-scale phenomena, albeit one that demands both high precision and careful disentangling of nuclear and electronic contributions.

Likewise, muonic atoms -- where an electron is replaced by a heavier muon -- dramatically magnify hyperfine effects, allowing nuclear charge and magnetisation distributions to be extracted with high sensitivity~\cite{Pohl2010,Antognini2013}. 

One fundamental challenge in hyperfine physics is
therefore the enormous disparity in spatial scale between the nucleus
and the atom. The atomic electron cloud extends over tens of thousands
of fermis, while the nucleus occupies only a few fermis. As a
consequence, the overlap of the electronic wavefunction with the tiny
nucleus is {extremely small}. Bridging this gap with reliable theory is difficult: atomic physicists must incorporate nuclear-structure corrections, and nuclear physicists must deliver those corrections in a form useful for atomic calculations.

Traditionally, simple nuclear models have been used to approximate the nucleus in hyperfine calculations. In the most naive picture, one treats the nucleus as a point-like source of charge and magnetism. This approximation often suffices for order-of-magnitude estimates. More refined descriptions introduce spherical or Fermi-like charge distributions and uniformly magnetised spheres, to capture finite-size corrections -- known respectively as the Breit-Rosenthal (BR) and Bohr-Weisskopf (BW) effects~\cite{BohrWeisskopf1950,RosenthalBreit1932,Karpeshin2015}. Yet such parametrisations come with limitations: they may capture the gross nuclear radius but not detailed variations in deformation or the distinct contributions of orbital and spin currents to the magnetic dipole moment. In heavy or deformed nuclei, and in particular muonic ions, these simplifications can introduce significant uncertainties.

These deviations manifest empirically as hyperfine
anomalies: departures from the expected scaling of hyperfine constants
with nuclear moments across isotopes. Such anomalies encode
nuclear-structure effects that are not captured by point-like or
simplistic models and can therefore serve as indirect
diagnostics{. Very few} studies have investigated
{this matter}~\cite{Roberts2021}. Crucially, hyperfine
anomalies can be exploited in ways that negate {\it electronic}
uncertainties in such measurements. By taking specific combinations or differences of
hyperfine measurements, physicists can design observables that are
largely independent of the atomic wavefunction {in the region of the nucleus}, thereby isolating nuclear structural effects. A prominent example is the ``specific difference'' between hyperfine splittings in hydrogen-like (H-like) and lithium-like (Li-like) heavy ions, designed to eliminate electronic-structure uncertainties~\cite{Volotka2012}.

Motivated by these challenges, in the present study we go beyond traditional
models by implementing a microscopic nuclear-structure framework based
on the Skyrme-Hartree-Fock (Skyrme-HF) model, including Bardeen-Cooper-Schrieffer
(BCS) pairing correlations~\cite{Bender2003}. We shall refer to this framework as Skyrme-HF-BCS, or equivalently the abbreviated ``HFBCS''. This method is well-suited for describing deformed and open-shell nuclei {close to the valley of $\beta$-stability}, and provides spatially-resolved charge and current densities directly from nuclear many-body {structure}. An important benefit of using Skyrme-HF-BCS densities is the ability to {\it dissect the origin of the nuclear magnetisation}: we can compute separate contributions to the hyperfine constant coming from the motion of charged nucleons (orbital magnetism) versus their intrinsic spin alignment (spin magnetism). This opens a new window onto the magnetic hyperfine constant $a$: rather than obtaining only an empirical $a$-value, we can quantify how much of it is due to, say, a polarised looping current from a core proton as opposed to a net spin alignment of an {unpaired} neutron. Similarly, our microscopic approach naturally yields the nuclear electric quadrupole tensor, allowing us to compute the electric quadrupole hyperfine constant $b$ without assuming a point-like quadrupole moment. We can thereby study how nuclear deformations (deviations from sphericity) influence $b$, an effect usually collapsed into a single number $Q$ (the electric quadrupole moment) in conventional treatments. 

The rare-earths are home to some of the largest observed nuclear deformations; dysprosium, in particular, has a rich isotopic spectrum (36 isotopes, 7 of which are stable) and contains some of the most prolate known nuclei. In addition, the dysprosium atom plays a major role in quantum computing because it carries a partially filled $f$-electron shell -- characteristic of the lanthanide family -- and can be chemically protected by suitable ligands. Molecular complexes such as TbPc$_2$ exhibit hyperfine levels that can be selectively addressed using radio-frequency pulses of different energies. This addressability is enabled by the hyperfine constant $b$, which is directly linked to nuclear deformation. The tuning of this parameter for optimisation -- such as by substituting isotopes within the molecule, producing different isotopologues -- lies at the core of the quantum-computing strategy developed over several years by Ruben’s and Wernsdorfer's groups at KIT \cite{Moreno2018}. Indeed, it is the non-equidistant hyperfine energy levels arising from nuclear deformation that are essential for implementing quantum algorithms in these molecular complexes (for example, Grover’s quantum search algorithm on TbPc$_2$ \cite{Godfrin2017}). Consequently, microscopic modeling of the hyperfine interaction in dysprosium and its isotopes remains a central challenge in this field.

By validating our calculations against the known magnetic dipole moments and
electric quadrupole moments of dysprosium, we can verify that the Skyrme-HF-BCS
model reproduces empirical nuclear observables before predictions are
made for hyperfine constants. {In} addition to our primary focus on the even-proton isotopes $^{161}$Dy and $^{163}$Dy, we also compute and analyse nuclear densities and currents for neighbouring odd-proton isotopes $^{159}$Tb and $^{165}$Ho. These nuclei provide valuable comparative baselines for exploring systematic trends and the interplay of nuclear structure and magnetic observables across isotopic and isotonic chains. The key novelty of this microscopic approach compared with phenomenological models is that it naturally incorporates deformation, shell structure, core polarisation and the interplay of proton and neutron contributions in both charge and magnetisation densities, and yields observables compatible with measurements without requiring free parameters from experiments.

On the atomic side, we focus on hyperfine interactions in highly charged hydrogen-like ions of the element dysprosium, specifically the aforementioned isotopes: $^{161}$Dy and $^{163}$Dy. In such ions, with only a single electron bound to a high-$Z$ nucleus, theoretical calculations are comparatively clean: any electron-electron correlation is absent, and the one-electron Dirac equation can be solved to very high precision for a given nuclear potential. At the same time, as already mentioned, the calculation of the hyperfine anomaly of these two isotopes in the single-electron case already provides insight into the multi-electronic case, as its calculation largely cancels electronic uncertainties, thus isolating nuclear effects. Moreover, the strong electric and magnetic fields near a bare $Z=66$ nucleus amplify relativistic effects, making this regime a demanding test of theory. We therefore employ a fully relativistic (Dirac) framework for the bound electron to properly account for high-$Z$ kinematics (e.g. the $1s$ electron in Dy$^{65+}$ moves at a significant fraction of the speed of light). The QED or radiative corrections (vacuum polarisation and self-energy) are then of the order of a few percent of the total energy of the hyperfine transition {\cite{Volotka2013}}. These corrections must be considered if the total energy of the hyperfine transition is to be accurately determined. Since the latter has already been the subject of extensive studies {\cite{Boucard1998, Boucard2000, Jentschura2011, Shabaev2001, Artemyev2001, Oreshkina2022}}, in this work we only focus on the corrections due to the effect of the finite nucleus on the classical electromagnetic potentials. Our work coupled with QED corrections can produce theoretical predictions comparable with modern high-precision experiments, enabling sensitive tests of nuclear structure or potential physics beyond the Standard Model.

Importantly, we extend this analysis to muonic hydrogen-like dysprosium ions. Since the muon’s Bohr radius is much smaller, its wavefunction overlaps more strongly with the nucleus, thereby amplifying the sensitivity to nuclear structure by several orders of magnitude compared with the electronic case. Consequently, differences between nuclear-model approximations lead to measurable differences in Bohr-Weisskopf and Breit-Rosenthal corrections, hyperfine anomalies, and quadrupole shifts. We show that, unlike electronic H-like dysprosium -- where such differences may be masked by electronic uncertainties -- muonic systems allow these distinctions to emerge clearly, and are reachable with current experimental capabilities. Muonic atoms, therefore, stand out as promising benchmarks for refining microscopic nuclear models. Moreover, the close proximity of the muon to the nucleus inverts the hierarchy of the dipolar and quadrupolar shifts, resulting in much larger quadrupolar effects~{\cite{Persson2025}} and the ensuing ``hyperfine quadrupole anomaly'', which are also subjects of this work.  

In essence, the isotopes investigated in this study allow us to demonstrate how a detailed nuclear-structure model can improve the interpretation of hyperfine data for mid-$Z$ nuclei, which are heavy enough to demand relativistic treatment yet light enough that laser spectroscopy of H-like ions may be feasible, with modern high-intensity beams even providing the relevant infrastructure for muonic matter~\cite{PSISMS}.

The remainder of this article is organised as follows. 
In \cref{Nuclear description} we present the nuclear-structure framework used in this work: the Skyrme effective interaction, the Skyrme-Hartree-Fock equations for spherical and axially deformed ground states including BCS pairing correlations, the treatment of odd nuclei, and the extraction of the nuclear charge densities and magnetic dipole moments. \Cref{atomic-structure} is devoted to the atomic-structure framework of this work: in \cref{Atomic description} we summarise the atomic description of highly charged hydrogen-like ions and the relativistic one-electron wavefunctions used in the calculations. 
\Cref{Multipole} is devoted to the multipole expansion of the electric and magnetic potentials generated by the microscopic nuclear charge and current distributions; we discuss in particular the electric monopole and quadrupole contributions, and the magnetic potential.
In \cref{sec Hyperfine interactions} we define the hyperfine interaction and introduce the magnetic dipole and electric quadrupole hyperfine-structure constants that will be our main observables.
\Cref{sec Isotope Shift} discusses the isotope shift: we review the Breit-Rosenthal and Bohr-Weisskopf finite-size corrections, introduce a hierarchical scheme for composing hyperfine anomalies across increasingly realistic nuclear charge and current models, going from point-like to more involved models, and define the magnetic-dipole and electric-quadrupole contributions to the anomaly.
Our numerical results and their discussion are presented in \cref{Results and discussions}. We first analyse the nucleon density distributions and nuclear currents obtained from the Skyrme-HF-BCS calculations for $^{159}$Tb, $^{161}$Dy, $^{163}$Dy, and $^{165}$Ho, and then combine the nuclear and atomic ingredients to evaluate charge and magnetic corrections to the hyperfine constants, the resulting hyperfine anomaly, and the correction to the electric-quadrupole term for the ${}^{161}$Dy$^{65+}$ and ${}^{163}$Dy$^{65+}$ electronic and muonic ions.
Finally, \cref{Conclusions} contains our conclusions and perspectives.

% {\color{red} Add examples for which the HF interaction plays a crucial role: Qdits, electron conversion in Th etc...

% Add a text explaining the role of HF in the context of quantum computing with lanthanides. "In chemistry, isotopologues are molecules that differ only in their isotopic composition. They have the same chemical formula and bonding arrangement of atoms, but at least one atom has a different number of neutrons than the parent."

% Add simple model for describing the Nuclear magnetism: point-dipole, Fermi distribution (Indelicato and his PhD student)

% Muon facility in PSI. 
% }

\section{Nuclear description}\label{Nuclear description}
The general goal of low-energy nuclear theory is to obtain a precise
description of the structure of nuclei and their elementary excitation
and decay modes. There are two fundamental difficulties in the
realisation of this project. First of all, since the nucleus is
generally composed of a rather large number of protons and neutrons,
it constitutes an $N$-body problem which (except for very small nuclei
at the very beginning of the nuclear table) cannot be solved exactly,
as is the case for the atomic problem. An additional problem arises
from the fact that, unlike the atomic problem (where the interaction
between the electrons is the exact Coulomb interaction), the
interaction between the nuclear constituents (protons and neutrons),
which is mainly due to the strong interaction, is known only very
approximately. Indeed, although the bare interaction between two
nucleons is rather well known, it is not so between two nucleons in
the nuclear environment, i.e. in the presence of all the other
$N-²2$ nucleons. It then becomes inevitable to use some
approximations. Moreover, a three-nucleon contribution is expected in
addition to the two-nucleon interaction. The Hartree-Fock (HF)
approach allows one to find a variational solution of the nuclear $N$-body problem by
requiring that the total nuclear energy of a Slater determinant
many-body wave function is minimal with respect to a simultaneous
variation of the wave functions of the $N$ particles, as is done in
atomic physics \cite{Friedrich2017}. As a result, there exists a mean
field potential $V$ to which all nucleons are subject and which is
generated by the interaction of all nucleons with each other. Such an
approximation is indeed justified, since the distance between two
nucleons inside the nucleus is large compared to the correlation
length (the so-called {\it healing distance}). It is therefore
possible to perform a HF calculation with a nucleon-nucleon
interaction, but this is an interaction in the nuclear medium. This
two-body interaction is therefore far from being the bare interaction
between two free nucleons, for the simple reason that it takes into
account the correlations between the nucleons, among which the {\it
  Pauli principle} plays an important role. One then speaks of an {\it
  effective interaction}, which can in principle be obtained from the
{\it bare} (or {\it free}) nucleon-nucleon interaction by the method
proposed by Brueckner \cite{Br55}, but which has to be corrected for
an important density dependence, as shown by Negele \cite{Ne70}. The
effective interaction obtained in this way is often referred to as the
Brueckner $G$ matrix derived from a given bare nucleon-nucleon
interaction. Such Brueckner-Hartree-Fock calculations do exist, though
mostly for infinite nuclear matter, or by considering the finite
nucleus as composed of pieces of nuclear matter at a given local
density (see e.g.\ \cite{Ne70,Ne82,DB70,CS72}), but they are generally
extremely time-consuming. It is therefore essentially out of the
question to perform such calculations on a large scale for finite
nuclei.

\subsection{The Skyrme effective interaction}
%verified

It is then necessary to resort to what is often called {\it
  phenomenological effective interaction}, which attempts to model the
interaction between the nucleons inside the nucleus in some analytical
form, taking into account the essential components of the bare
interaction. This interaction is of course finite, vanishing at
distances beyond about $2$ fm, with an attractive component between
$1$ and $1.5$ fm, and becoming strongly repulsive at short distances
below $\approx 0.7$ fm. One could imagine a linear combination of two
functions of Gaussian form, one attractive with a width of about $1.2$
fm, the other strongly repulsive with a width of about $0.7$ fm. As we
will see below, Gogny proposed such a finite-range effective
interaction \cite{Go75,DG80}, which, almost 50 years later, is still
one of the most successful phenomenological effective interactions. 
%verified

In light of the short-range nature of the nucleon-nucleon interaction,
a direct approach would be to parameterise it using a Dirac delta
distribution. This would effectively account for the short-range
character of the strong interaction in the extreme limit. Such
interactions include Moszkowski's modified delta force \cite{Mo70} and
the interaction originally proposed by Skyrme \cite{Sk59}. The latter
has been the subject of substantial research and has enjoyed
considerable success following the pioneering work of Vautherin and
Brink \cite{VB72, Va73}. In its original form \cite{VB72}, the nuclear
part of this interaction can be expressed in \sam{configuration}{coordinate} space as
follows 
%verified
\begin{equation}
{\displaystyle V_{\rm Sk} = V_{\rm Sk,2} +
  V_{\rm Sk,3} } \;,
\label{Eq01}
\end{equation}
i.e. as a sum of a two-body potential
\begin{equation}
\begin{aligned}
V_{\rm Sk,2}(\vec{R}_{1},\vec{R}_{2}) &=t_0 \!\left[\! 1 +
  x_0 P_{\sigma} \frac{}{}\!\!\right]\! \delta(\vec{R}_{12}) +
\frac{1}{2} t_1 \!\left[\frac{}{}\! \delta(\vec{R}_{12}) \, \vec k^2 +
  {\vec{k'}}^{2} \, \delta(\vec{R}_{12}) \right] \\ 
&+ t_2 \, \vec{k'} \cdot \delta(\vec{R}_{12}) \vec{k} + i W_0
(\sigma_i + \sigma_j) \cdot [\vec{k'}\times \delta(\vec{R}_{12})
  \vec{k}] \;, 
\label{Eq02}
\end{aligned}
\end{equation}
and a three-body term
\begin{equation}
V_{\rm Sk,3}(\vec{R}_{1},\vec{R}_{2},\vec{R}_{3}) = t_3
\delta(\vec{R}_{12}) \, \delta(\vec{R}_{23}) \;, 
\label{Eq03}
\end{equation}
where the coefficients $t_i$, $x_i$ and $W_0$ are defined as constant
parameters, which serve to define the Skyrme force and where
$\vec{R}_{12}$ in \cref{Eq02,Eq03} stands for the
relative distance $\vec{R}_{12} \!=\! \vec{R}_1 \!-\! \vec{R}_2$
between the nucleons, $\vec{k}$ for the operator $\left(\vec{\nabla}_1
- \vec{\nabla}_2 \right)/2i$ acting on its right and $\vec{k'}$ its
Hermitian conjugate, i.e. the operator $-\left(\vec{\nabla}_1 -
\vec{\nabla}_2 \right)/2i$ acting on its left. The operator
$P_{\sigma}$ is the spin-exchange operator and ${\sigma}_i, \;
i=1,2,3$ in \cref{Eq02} are the Pauli matrices. It can be shown
(see e.g. \cite{VB72}) that the first three terms of \cref{Eq02}
can be obtained, in the short-range limit, from a finite-range central
force of Gaussian type including exchange terms. The last term in \cref{Eq02}
is a zero-range two-body spin-orbit potential. Note that,
due to the presence of a $\delta$ distribution in the 
two-body potential, no other exchange operators, such as $P_r$
(position exchange) or $P_{\tau}$ (isospin exchange), are required in
$V_{\rm Sk,2}$.

Furthermore, as demonstrated in \cite{VB72}, when time-reversal
symmetry is respected, as is the case in the ground state of even-even
nuclei, the three-body potential in \cref{Eq03} is equivalent to a
density-dependent two-body interaction 
\begin{equation}
  V_{\rm Sk,3}(\vec{R}_{1},\vec{R}_{2},\rho) =
  \frac{1}{6} t_3 \left[ 1 + x_3 P_{\sigma}
  \frac{}{}\!\right] \delta(\vec{R}_{12}) \; \rho(\vec{\cal
  R}_{12})\;, 
\label{Eq04}
\end{equation}
where $\rho$ is the total nucleon density ($\rho \!=\! \rho_n \!+\!
\rho_p$, where $n$ is used for neutron and $p$ for proton) evaluated
at the center-of-mass position $\vec{\cal R}_{12} \!=\! (\vec{R}_1
\!+\! \vec{R}_2)/2$ of two interacting nucleons.
\Cref{Eq04} provides a straightforward  
representation of the aforementioned density dependence of the
effective nucleon-nucleon interaction. It
depicts the manner in which the interaction between two nucleons is
influenced by the presence of other nucleons within the nucleus. The
Skyrme interaction can therefore be understood as a kind of
phenomenological $G$ matrix, which %(see, for example \cite{La09}), which
takes into account the effect of short-range correlations. As
previously outlined, the linear dependence of the density-dependent
term in \cref{Eq04} is achieved with the use of a three-body
force. Because we want to include such a density dependence in a more
general way, we write this term with a density dependence of the form
$\rho^\alpha(\vec{\cal R}_{12})$, where the power $\alpha$ is a
parameter.

It should be noted that following this initial attempt, numerous
additional parametrisations of the phenomenological effective
nucleon-nucleon interaction of the Skyrme type have been proposed, as
has already been mentioned above. To this end, the form
of the interaction, as expressed in \cref{Eq02,Eq04}, has been generalised by incorporating exchange
parameters, designated as $x_j$, into the momentum-dependent terms,
which are characterised by strength parameters, denoted as $t_1$ and
$t_2$. This encompasses the density-dependent term with the corresponding
exchange parameter $x_3$. Furthermore, it would be advantageous to have
access to a density-dependent term that is not necessarily linear in
density. As a result, the correspondence of this term to the
three-body interaction of \cref{Eq03} is lost, but a more flexible
density dependence of the effective interaction is gained. As
previously discussed, this is the final objective, namely to bridge
the gap between a bare inter-nucleon interaction in the vacuum and an
effective interaction between nucleons inside the nucleus. One thus
writes
%verified
\begin{equation}\begin{aligned}
V_{\rm Sk}(\vec R_1,\vec R_2) &=t_0 (1 + x_0 P_{\sigma})
\delta(\vec{R}_{12}) + \frac{t_1}{2} (1 + x_1 P_{\sigma})
\big[\delta(\vec{R}_{12}) \, \vec k^2 + \vec {k'}^{2} \,
  \delta(\vec{R}_{12})\big]  \\
&+ t_2 (1 + x_0 P_{\sigma}) \, \vec{k'} \cdot
\delta(\vec{R}_{12}) \vec{k} + \frac{1}{6} t_3 (1 + x_3 P_{\sigma})
\delta(\vec{R}_{12}) \, \rho^{\alpha}(\vec{\cal R}_{12})  \\
&+ i W_0
(\vec \sigma_1 + \vec \sigma_2) \cdot \vec{k'} \times \delta(\vec{R}_{12})
  \vec{k}] \;. 
\label{Eq05}
\end{aligned}\end{equation}
If we consider a nucleus whose ground state is described by a Slater
determinant $|\Phi\rangle$ of single-particle states
$|\varphi_{k}\rangle$, we can write down
the expectation value $E$ of the Hamiltonian operator ${}^{\rm nuc}H$ of
the nucleus
\begin{equation}
{}^{\rm nuc}H = T_{\rm intr} + V_{\rm Sk} + V_{\rm Coul}\,,
\label{Eq06}
\end{equation}
where $T_{\rm intr}$ is the intrinsic kinetic energy,
$V_{\rm Sk}$ the density-dependent Skyrme interaction and
$V_{\rm Coul}$ the Coulomb interaction, as the integral 
\begin{equation}
E = \langle \Phi | {}^{\rm nuc}H | \Phi \rangle = \int \mathcal H(\vec{R})\,
d^3R 
\label{eqE}
\end{equation}
of a Hamiltonian density $\mathcal H(\vec{R})$ given by~\cite{EBG75,GQ80,HHB12} 
\begin{align}
  \mathcal H(\vec{R}) & =
  \Big(1-\frac 1 A\Big)\,\frac{\hbar^2}{2m} \, \tau(\vec{R})
  + B_{1} \, \rho^2(\vec{R}) + B_{10}\,\vec s^{\,2}(\vec R)
  + B_{3} \, \Big(\rho(\vec{R}) \, \tau(\vec{R}) - {\vec j}^{\,2}(\vec
  R)\Big) \notag \\
  & \phantom{=}
  + B_{5} \, \rho(\vec{R}) \, \Delta \rho(\vec{R})
  + B_{18} \, \vec s(\vec R) \cdot \Delta \vec s(\vec R)
  + \rho^\alpha(\vec R) \, \Big(B_{7} \, \rho^2(\vec R) + B_{12} \,
  \vec s^{\,2}(\vec R)\Big) \notag \\
  & \phantom{=} 
  + B_{9} \, \rho (\vec R) \, \vec{\nabla} \cdot \vecJ(\vec R)
  + B_{14} \Big( \sum_{\mu,\nu=x,y,z} J_{\mu\nu}^2(\vec R) -
  \vec s(\vec R) \cdot \vec T(\vec R) \Big)
\label{H_density}  \\ 
  & \phantom{=}
  + \sum_{q = n,p} \bigg[ B_{2} \, \rho_q^2(\vec R)+ B_{11}\,\vec s_q^{\,2}(\vec R)
    + B_{4} \, \Big(\rho_q(\vec R)\, \tau_q(\vec R) -
    \vec j_q^{\,2}(\vec R)\Big)  \notag \\ 
  & \phantom{= + \sum_q \Big[}
      + B_{6} \, \rho_q(\vec R) \, \Delta\rho_q(\vec R)
      + B_{19} \, \vec s_q(\vec R) \cdot \Delta \vec s_q(\vec R)
    + \rho^\alpha(\vec R) \Big(B_{8} \, \rho_q^2(\vec R) +
    B_{13}\,\vec s_q^{\,2}(\vec R)\Big)  \notag\\ 
  & \phantom{= + \sum_q \Big[}
      + B_{9} \, \rho_q \vec{\nabla} \cdot \vecJ_q
      + B_{15} \Big( \sum_{\mu,\nu=x,y,z} J_{\mu\nu,q}^2(\vec R) -
      \vec s_q(\vec R) \cdot \vec T_q(\vec R) \Big)
      \bigg] \;,\notag
\end{align}
where the $B_i$ coefficients are functions of the Skyrme-force
parameters $t_j$ and $x_j$. They have been introduced to simplify the
notation and are given in Appendix A of Ref. \cite{BMD15}. The functions of $\vec R$ appearing in the right member are called
local densities and are defined below. The charge index $q$ refers
to a neutron or proton contribution. Quantities without index $q$
correspond to the sum of neutron and proton contributions (e.g. $\rho
= \rho_n +\rho_p$, $\tau=\tau_n + \tau_p$, ${\vec s}={\vec s}_n +
{\vec s}_p$...). Because the proton mass $m_p$ and
the neutron mass $m_n$ are almost equal, one takes $m = (m_n+m_p)/2$
for neutrons and protons in the kinetic energy term. Their actual
difference is thus absorbed in the adjustment to experimental data
of the interaction parameters. Note that the factor $-\dfrac 1 A$
in the Hamiltonian density, \cref{H_density}, with $A=N+Z$ the
number of nucleons, reflects the one-body contribution to the
center-of-mass kinetic energy that has to be subtracted from the
total kinetic energy to obtain the intrinsic part only. The two-body
contribution to the center-of-mass kinetic energy is usually
omitted, as done here, and is thus absorbed in the fit of the
interaction parameters. \\

The various local densities for charge state $q$ appearing in 
the Hamiltonian density $\mathcal H({\vec R})$ are expressed in terms
of the matrix elements in the space-spin-isospin representation
$|\vec{R} \sigma q\rangle$ of the one-body density operator $\hat
\varrho$ associated with the Slater determinant $\Phi(\vec{R})$ 
\begin{equation}
  \varrho(\vec{R},\sigma,q;\vec{R}\,'\!,\sigma',q')
  = \langle \vec R\,'\sigma'q'|\hat\varrho|
  \vec{R}\sigma q\rangle
\label{Eq10} \;,
\end{equation}
where $\sigma=\pm1/2$ is the spin projection on the symmetry axis
The operator $\hat\varrho$ can be defined by its spectral
  decomposition
  \begin{equation}
    \hat\varrho = \sum_k v_k^2 \, |\varphi_k\rangle\langle\varphi_k| \;,
  \end{equation}
  where the eigenvalue $v_k^2$ represents the occupation probability of the
  single-particle state $|\varphi_k\rangle$. For a Slater determinant,
  one has $v_k^2 = 0$ if the state is empty and $v_k^2 = 1$ if it is
  occupied. Because single-particle states $|\varphi_k\rangle$ do not
  mix neutrons and protons, one has 
  \begin{equation}
    \varrho(\vec{R},\sigma,q;\vec{R}\,'\!,\sigma',q')
         = \delta_{q'q} \sum_k \! v_{k}^2
\, \varphi_{k}^{*}\!(\vec{R}\,'\!,\sigma',q)
\varphi_{k}(\vec{R},\sigma,q) \;.
  \end{equation}
The Hamiltonian density $\mathcal H(\vec R)$ defined by
  \cref{H_density} involves three local densities that are
invariant under time reversal, called ``time-even'' densities, and
three local densities which change sign under time reversal, called
``time-odd'' densities. The former densities are: 
\begin{itemize}
\item the nucleon matter density $\rho_q(\vec R)$
\begin{equation}
\label{eq_rho_q}
\rho_q(\vec R) = \sum_\sigma \varrho_q(\vec{R},\sigma,q;\vec{R},\sigma,q)
= \sum_{k,\sigma} \! v_{k}^2 \, |\varphi_{k}(\vec{R},\sigma,q)|^2 \;,
\end{equation}
\item the kinetic energy density
\begin{equation}
\tau_q(\vec R) = \sum_\sigma \!\vec{\nabla} \cdot\! \vec{\nabla}'
\varrho_q(\vec{R},\sigma,q;\vec{R}\,'\!\!,\sigma,q)|_{\vec{R}\,'=\vec{R}} 
=\! \sum_{k,\sigma} \! v_{k}^2 \, |\vec{\nabla} \varphi_{k}\!(\vec{R},\sigma,q)|^2
\;,
\end{equation}
where $\vec{\nabla}$ and $\vec{\nabla}'$ denote the
  gradient operators with respect to the coordinates $\vec R$ and
  $\vec R'$, respectively and,
\item the tensor density
  \begin{align}
    J_{q,\mu\nu}(\vec R) & = \frac 1{2i} \sum_{\sigma,\sigma'}
    \big(\partial_{\mu} - \partial'_{\mu}\big)
    \varrho_q(\vec{R},\sigma,q;\vec{R}\,',\sigma',q)|_{\vec{R}\,'=\vec{R}}
    \langle \sigma| \sigma_{\nu} |\sigma' \rangle  \\
    & = \frac{1}{2i}\sum_{k,\sigma,\sigma'}
    v_k^2\Big(\varphi_k^*(\vec R,\sigma,q)\big[\partial_{\mu}\varphi_k(\vec
    R,\sigma',q)\big] - \big[\partial_{\mu}\varphi_k^*(\vec
      R,\sigma,q)\big] \varphi_k(\vec R,\sigma',q)\Big)
    \langle\sigma|\sigma_{\nu}|\sigma'\rangle \;,
  \end{align}
  where $\mu$, $\nu$ are cartesian indices. From this density one
  deduces the so-called spin-orbit density $\vec{J}_q(\vec R)$, which
  is the antisymmetric part of $J_{q,\mu\nu}(\vec R)$ in the sense
  that its component $J_{q,\lambda}(\vec R)$ is defined by
  \begin{equation}
    J_{q,\lambda}(\vec R) = \sum_{\mu\nu}
    \varepsilon_{\lambda\mu\nu} J_{q,\mu\nu}(\vec R)
    \label{Eq13} \;,
  \end{equation}
  where $\varepsilon_{\lambda\mu\nu}$ is the Levi--Civita symbol.
\end{itemize}
%verified
The time-odd local densities are:
\begin{itemize}
  \item the spin density 
    \begin{equation}\begin{aligned}
\vec{s}_q(\vec R) &=\sum_{\sigma,\sigma\,'}
\varrho_q(\vec{R},\sigma,q;\vec{R},\sigma',q) \; \langle \sigma| {\vec
  \sigma} |\sigma' \rangle 
\\
&=\sum_{k,\sigma,\sigma'} \! v_{q,k}^2 \,
\varphi_{k}^{*}\!(\vec{R},\sigma,q) \varphi_{k}(\vec{R},\sigma',q)
\langle \sigma| {\vec \sigma}  |\sigma' \rangle \;, 
\label{Eq14}\end{aligned}\end{equation}
\item the current density
\begin{equation}\begin{aligned}
\vec j_q(\vec R) &=\frac{1}{2i} \sum_{\sigma}
\!\left[\left(\vec{\nabla} \!-\! \vec{\nabla}\,'\right) \varrho_q(\vec
  R,\sigma,q;\vec R\,',\sigma,q) \right]|_{\vec{R}\,'=\vec{R}} 
\\
&=\frac{1}{2i} \!\sum_{k,\sigma} v_{q,k}^2 \! \left[
  \varphi_{k}^{*}\!(\vec{R},\sigma,q) \vec{\nabla}
  \varphi_{k}\!(\vec{R},\sigma,q) - \varphi_{k}\!(\vec{R},\sigma,q)
  \vec{\nabla} \varphi_{k}^{*}\!(\vec{R},\sigma,q) \right] \;, 
\label{Eq15}
\end{aligned}\end{equation}
\item and the spin-kinetic density
\begin{equation}\begin{aligned}
\vec{T}_q(\vec R) &=\sum_{\sigma,\sigma\,'}
\left[\vec{\nabla} \cdot \vec{\nabla}\,' \varrho_q(\vec R,\sigma,q;\vec
  R\,',\sigma',q) \right]_{\vec{R}\,'=\vec{R}} \langle \sigma'| {\vec \sigma} |\sigma \rangle
\\
&=\sum_{k,\sigma,\sigma\,'} \! v_{q,k}^2 \! \left[ \vec{\nabla}
  \varphi_{k}^{*}\!(\vec{R},\sigma,q) \cdot 
\vec{\nabla} \varphi_{k}\!(\vec{R},\sigma,q) \right] \langle \sigma'|
    {\vec \sigma} |\sigma \rangle \; . 
\label{Eq16}
\end{aligned}\end{equation}
\end{itemize}
For Slater determinants $\Phi(\vec{R})$ that are
invariant under time-reversal symmetry, all time-odd
densities vanish. This is the case for the $0^+$ ground state of an
even-even nucleus. In contrast, these densities do not vanish in the
self-consistent solution describing an excited state of an even-even
nucleus, or any state of an odd-mass or doubly-odd nucleus.

As mentioned above, by introducing a density-dependent term in the
interaction, which is not necessarily linear in $\rho$, we have given
up the correspondence of such a term with a three-body force. However,
since the energy of the nucleus can still be written as a functional
of a certain number of local densities, and since this will continue
to be the case for all subsequent extensions of what we called at the
beginning the Skyrme phenomenological effective interaction, we will
speak in the following of an energy-density functional (EDF) of the
Skyrme type, the concept of an EDF being of
course well known from quantum chemistry \cite{TAK14} and materials Science
\cite{KS65} to investigate the electronic structure. 

The phenomenological effective interaction or EDF that we are
looking for should of course be applicable not only to a given nucleus
or group of nuclei in a given region of the
nuclid chart, but rather to the whole nuclear chart, with the possible
exception of very light nuclear systems, where the concept of
mean field loses its meaning and where {\it ab initio} methods would
be much better suited. With this exception, the basic ingredients of
an effective nucleon-nucleon interaction, which are rooted in the
strong interaction, obviously do not depend on the nuclear system
under study. It should then be expected that such a self-consistent
approach will be able to accurately describe the ground state
properties of a very large number of nuclei, but also to predict some
of their excitation and decay modes, such as their stability against
fission.

The first really very successful Skyrme-like effective interaction
that fulfilled these requirements, namely to adequately describe the
binding energies of nuclei along and near the $\beta$-stability line,
their ground-state radii and multipole moments, but also their low-energy
spectroscopy and decay probabilities, was the SIII parametrisation 
\cite{BFN75}. In the fitting procedure of this interaction, 
the authors also had in mind to reproduce at the same time some
nuclear-matter properties such as the energy per particle $E/A$, the
nuclear saturation density as it is realised in the center of heavy
nuclei, or the nuclear-matter incompressibility, which is essential
for the description of the nuclear isoscalar monopole resonance, the
so-called {\it breathing mode}. It should be noted that
for ground-state or spectroscopic properties the SIII parametrisation
is still one of the most successful Skyrme-type EDFs available today
and will be used in the nuclear magnetic-moment studies presented
here.

However, it quickly became apparent that essentially all of the
early-day EDFs failed to adequately account for the stability
against fission of heavy nuclei in the actinide region, and thus
produced fission barriers that were far too high. It was the Skyrme
SkM$^{*}$ parameterisation~\cite{BQB82} which showed that it was
indeed possible to obtain a fission barrier height compatible with the
experimental data, while retaining the ability to describe the binding
energies of nuclei along or near the $\beta$ stability line. It is
only when one moves away from the $\beta$ valley, i.e. when one
explores the regions closer to the neutron and proton drip lines, that
these early parameterisations begin to fail. 

However, the prospect of radioactive-beam facilities and the search
for {\it super-heavy elements} rapidly motivated the interest in using
these EDFs also for nuclei away from the $\beta$-stability line, which
requires studying more specifically the neutron-proton
part of the interaction. This requires an improvement of the available
parameterisations with respect to their isospin degrees of
freedom. Improving the effective interaction in this direction also
opens the way to studying nuclear matter under extreme conditions,
such as those encountered in astrophysical applications such as
calculations of the structure of neutron stars or their formation by
supernova explosions. Work in this direction has been carried out by
the Lyon group~\cite{CBH97,CBH98} and has led to EDFs such as SLy4 to
describe the spectroscopic properties of nuclei from the
$\beta-$stability line to the driplines of protons and
neutrons. Additional EDF parameterisations (SLy5, SLy6, SLy7) have
been proposed to be used when including a spin gradient term or/and
taking into account the two-body center-of-mass correction. 

Later attempts of improvement included a tensor part in the EDF
\cite{LBB07,BBD09,HHB12}, density-dependent generalisations of the
gradient ($t_1$ and $t_2$) terms \cite{GCP10}, or the introduction of
a true 3-body interaction \cite{SDM13}. One should also mention the
EDFs of the Bruxelles-Montreal collaboration, which try to include
terms simulating correlations beyond the mean field in order to obtain
a good reproduction of nuclear masses already at the level of the
Hartree-Fock or Hartree-Fock-Bogoliubov approach (see
\cite{GCP10,GCP13,PC22} and references therein).

In all these cases the Hamiltonian density $\mathcal H(\vec R)$ of
\cref{eqE} can always be written as a functional of the local
densities, so its space integral is an energy-density
functional. The concept of Density Functional Theory (DFT) has, for decades, had enormous success in atomic physics. In
atomic DFT, this EDF is used as a corrective
term to account for exchange effects and correlations, while the
dominant contribution comes from the Coulomb interaction between the
electrons and the nucleus and among the electrons themselves, and can in principle be
treated exactly. In contrast, in the nuclear case, the dominant
contribution comes from the interaction energy of the nucleons among
themselves, that one tries to put into some analytical form
because the nucleon-nucleon interaction is only known in some
approximate way.

Once the Hamiltonian density is known, it can be used to determine, as
a function of the force parameters, the expressions for the binding
energy per nucleon $B/A$ (where $B=-E >0$) in symmetric nuclear
matter, a hypothetical infinite system with $N = Z \to
\infty$, in the absence of the Coulomb field. From the second 
derivative of $B/A$ with respect to the Fermi momentum $k_f$,
the incompressibility of nuclear matter is obtained, as
explained, e.g., in the original article by Vautherin and Brink
\cite{VB72}. The case of nuclear matter is considered to be
approximately realised in heavy nuclei, and the nuclear
incompressibility is considered to be a quantity that determines the
energy of the isoscalar giant monopole resonance (the so-called
breathing mode). In addition, the energy per nucleon should be
stationary with respect to variations of the Fermi momentum, giving
the nuclear saturation condition $\partial(E/A)/\partial k_f = 0$. The
three equations that fix the energy per nucleon, the saturation
density and the nuclear incompressibility have traditionally been used
to get a first idea of the range of values that some of the Skryme
force parameters can take. In~\cite{BQB82} an attempt was made to
control the nuclear surface energy, which enters, together with the nuclear incompressibility, in a crucial way in the determination of the height
of the fission barriers in heavy (actinide) nuclei. 

\subsection{The Skyrme--Hartree--Fock equations}

Let us introduce the single-particle two-component Pauli
  spinor for a nucleon of charge state $q$
  $[\varphi_j](\vec R,q) = \begin{pmatrix}
    \varphi_j(\vec R,\sigma = 1/2,q) \\
    \varphi_j(\vec R,\sigma = -1/2,q)
  \end{pmatrix}$.
The Skyrme--Hartree--Fock equations can be obtained from the energy
density by requiring that the total energy in \cref{eqE} be
stationary with respect to the individual variations of the
single-particle states, under the constraint that these are normalised
\begin{equation}
\frac{\delta}{\delta [\varphi_{k}]^{\dagger}(\vec R,q)} \left[ E - \sum_k
  e^{(q)}_{k} \int [\varphi_{k}]^{\dagger}(\vec{R},q)\,
             [\varphi_k](\vec R,q)\, d^3R \right] = 0
\;, 
\label{Eq17}
\end{equation}
where the single-particle energies $e_k^{(q)}$ play the role of
Lagrange parameters. Using the explicit expression in \cref{H_density}
of the Hamiltonian density appearing in \cref{eqE} we find that
the single-particle spinors $[\varphi_{k}](\vec{R},q)$ must
satisfy the following set of equations 
\begin{equation}
  h_q[\varphi_k](\vec R,q) = e_k^{(q)}[\varphi_k](\vec R,q)
  \label{HFeq} \;,
\end{equation}
where the Hartree--Fock Hamiltonian, $h_q$, for charge state $q$ in coordinate
representation  takes the form~\cite{HHB12}
\begin{equation}
\begin{aligned}
h_q[\varphi_k](\vec R,q) = &
-\vec{\nabla} \cdot \Big[\Big(\frac{\hbar^2}{2m^*_{q}(\vec R)} +
\vec C_{q}(\vec R) \cdot \vec{\sigma} \Big)\vec{\nabla}
[\varphi_k](\vec R,q)\Big] \\
& + \Big(U_{q}(\vec R) +\delta_{q p}\,V_{\rm Coul}(\vec R)\Big)
[\varphi_k](\vec R,q) + i \, \vec{W}_{q}(\vec R) \cdot \Big(\vec{\sigma}
\times \vec{\nabla} [\varphi_k](\vec R,q) \Big)\\
& - i \sum_{\mu,\nu = x,y,z}\Big[ W_{q,\mu \nu}(\vec R) \sigma_{\nu} \partial_{\mu}
[\varphi_k](\vec R,q) + \partial_{\mu} \Big( W_{q,\mu  \nu}(\vec R)
\sigma_{\nu} [\varphi_k](\vec R) \Big) \Big]  
\\
& - i \, \vec A_{q}(\vec R) \cdot \vec{\nabla}[\varphi_k](\vec R,q)
+ \vec S_{q}(\vec R) \cdot \vec{\sigma}[\varphi_k](\vec R,q)  \:.\label{hHF}
\end{aligned}
\end{equation}
In \cref{hHF} the functions $m_q^*$, $U_q$, $V_{\rm Coul}$,
$\vec W_q$ and $W_q^{\mu\nu}$ are the effective-mass field, the
central-plus-density-dependent scalar field, the Coulomb field, the
spin-orbit vector field and the spin-current tensor field,
respectively. They are called ``time-even'' fields because they 
are invariant with respect to time reversal and are defined by~\cite{HHB12}
\begin{align}
U_q(\vec R) &= 2 \Big(B_1\,\rho(\vec R) + B_2\,\rho_q(\vec R)\Big) +
B_3\,\tau(\vec R) + B_4\,\tau_q(\vec R)
+ 2 \Big(B_5\,\Delta\rho(\vec R) + B_6\,\Delta\rho_{q}(\vec R)\Big) \notag \\ 
&\phantom{=} + \rho^{\alpha-1}(\vec R) \, \left[(2+\alpha)\,B_7\,\rho^2(\vec R)
+ B_8\,\alpha \,\Big(\rho_n^2(\vec R) + \rho_p^2(\vec R) + 2
\,\rho(\vec R)\,\rho_q(\vec R)\Big) \right] 
\label{Eq19}\\ 
&\phantom{=} + B_9 \, \Big(\vec{\nabla} \cdot \vecJ(\vec R) + \vec{\nabla} \cdot
\vecJ_q(\vec R)\Big) + \delta_{qp} V_{\rm Coul}(\vec R) \;,\notag
\\
 \frac{\hbar^2}{2 m_q^{*}(\vec{R})} & = \frac{\hbar^2}{2 m} +
 B_3\,\rho(\vec R) + B_4\,\rho_q (\vec R)
 \label{Eq23} \;,
 \\
\vec{W}_q(\vec{R}) & = B_9 \Big(\vec{\nabla} \rho(\vec R) +
\vec{\nabla} \rho_q(\vec R) \Big) \;, 
\label{Eq24}
\\
W_{q,\mu\nu}(\vec{R}) & = B_{14} J_{\mu\nu} + B_{15} J_{q,\mu\nu} \;.
\label{Eq25}
\end{align}
The Coulomb field concerning only protons is the sum of direct $V^{(\rm dir)}_{\rm
  Coul}$ and exchange $V^{(\rm exc)}_{\rm Coul}$ contributions 
\begin{equation}
V_{\rm Coul}(\vec R) = V^{(\rm dir)}_{\rm
  Coul}(\vec R) + V^{(\rm exc)}_{\rm Coul}(\vec R) \,.
\label{Eq20}
\end{equation}
While the direct Coulomb potential can be easily calculated as 
\begin{equation}
V^{(\rm dir)}_{\rm Coul}(\vec R) = e^2 \int
\frac{\rho_p(\vec R)}{|\vec R - {\vec R'}|} d^3R \; , 
\label{Eq21}
\end{equation}
one generally uses (to avoid very time-consuming calculations) the so-called Slater
approximation for the exchange contribution \cite{Sl51} 
\begin{equation}
V^{(\rm exc)}_{\rm Coul}(\vec R) = - e^2 \left(
\frac{3}{\pi} \rho_p \right)^{1/3} \;. 
\label{Eq22}\end{equation}
In contrast, the spin field $\vec S_{q}$, the current field $\vec A_{q}$ and
the spin-gradient field $\vec C_{q}$ appearing in \cref{hHF} change
their sign under time reversal and are called ``time-odd
fields''. Their expressions are~\cite{HHB12}
\begin{align}
\label{Sq}
\vec S_{q}(\vec R) & = 2\big(B_{10}+B_{12} \, \rho^{\alpha}(\vec R)\Big)
\vec s(\vec R) + 2\big(B_{11} + B_{13} \, \rho^{\alpha}(\vec R)\Big)
\vec s_{q}(\vec R) - B_9 \, \vec{\nabla} \times \Big(\vec j(\vec R) + \vec j_{q}(\vec
R)\Big)  \notag \\
& \phantom{=}
- B_{14} \, \vec T(\vec R) - B_{15} \, \vec T_{q}(\vec R) 
+ 2 \, \Big( B_{18} \, {\Delta} \vec s(\vec R) + B_{19} \, {\Delta} \vec
s_{q}(\vec R)\Big) \; , \\
\label{Aq}
\vec A_{q}(\vec R) & = -2\Big(B_{3} \, \vec j(\vec R) + B_{4} \,
\vec j_{q}(\vec R)\Big)
+ B_9 \, \vec{\nabla} \times \Big(\vec s(\vec R) + \vec s_{q}(\vec R)\Big) \; , \\
\label{Cq}
\vec C_{q}(\vec R) & = -B_{14} \, \vec s(\vec R) - B_{15} \, \vec
s_{q}(\vec R) \:,
\end{align}
where the vector Laplacian of $\vec s$ is related to the spin-kinetic
density through
\begin{align}
{\Delta} \vec s(\vec R) =  \, 2 \, \vec T(\vec R) + \sum_{k,q} v_k^2
\Big\{
\Big(\Delta [\varphi_k]^{\dagger}(\vec R,q)\Big)
\vec{\sigma}[\varphi_k](\vec R,q)
  + [\varphi_k]^{\dagger}(\vec R,q)\vec{\sigma}\Delta [\varphi_k](\vec R,q)
\Big\} \:.
\end{align}

The Skyrme-Hartree-Fock equations (\ref{HFeq}) are part of the self-consistent procedure and
have to be solved iteratively. Starting with some initial guess for
the fields entering this Schr\"odinger-type equation, one finds a
first solution for the wave functions $\varphi_{k}(\vec{R},\sigma,q)$ 
for neutrons and protons, which can then be used to determine the
densities \crefrange{Eq10}{Eq16} and through them the fields
\crefrange{Eq19}{Cq}, which enter the Skyrme-Hartree-Fock
equations (\ref{HFeq}). This procedure is iterated until convergence
of some observable quantities is obtained.

\subsection{Cases of spherical and axial symmetries of the nucleus
  ground-state shape}

The calculations described above can be carried out for
the ground state of nuclei close to magic (closed-shell) nuclei,
such as $^{16}$O, $^{40}$Ca, $^{48}$Ca, $^{90}$Zr, $^{132}$Sn,
$^{208}$Pb, by imposing spherical symmetry. These nuclei
are the ones mainly considered in the fitting procedures of many
parameterisations of the Skyrme EDF. They possess a spherical shape in
their ground state and a pure Slater-determinant wave function is a reasonably
good approximation to the exact ground-state wave function. In
practice the Skyrme-Hartree-Fock equations (\ref{HFeq}) can be  
solved in configuration space as described in Ref.~\cite{VB72}. The nucleon
wave functions $\varphi_{k}(\vec{R},\sigma,q)$ are then 
characterised by the principal quantum number $n$, the orbital angular
momentum $\ell$, the total angular momentum quantum number $j$ and the
magnetic quantum number $m_j$. From the definition of the local
densities, \crefrange{Eq10}{Eq16}, it follows that these depend
only on the radial variable $R$.

However, if the spherical symmetry is broken in the nuclear
ground-state shape, neither of these quantum numbers is conserved. If
one imposes axial symmetry around the $z$ axis, the projection
$\Omega$ onto the symmetry $z$ axis of the total angular momentum
$\vec{j}$ of a particle evolving in the mean field is conserved. If
left-right symmetry (reflection through the $xy$ plane) is additionally
imposed, the single-particle states $|\varphi_{k}\rangle$ are
eigenstates of the parity operator $\pi$, in addition to being
eigenstates of $j_z$. To solve the Skyrme--Hartree--Fock equations in
this case, one can perform an expansion of the wave functions
$\varphi_{k}(\vec{R},\sigma,q)$ in a suitable basis. Given the 
retained symmetries, an expansion in the basis of an axially deformed
harmonic oscillator (HO) is a reasonable choice~\cite{DPP69}, as 
in the pioneering work of Vautherin~\cite{Va73} to solve the Skyrme
Hartree-Fock equations for axially deformed nuclei.

Such a basis is necessarily truncated as only a finite number of
oscillator shells can be considered. The result of the converged
Hartree--Fock cycle will obviously depend on the chosen oscillator
parameters. Let us call these $N_{0}$, the number of main
oscillator shells included, and $\omega_z$ and $\omega_\perp$, the
oscillator frequencies in the symmetry and perpendicular directions
respectively, which will at the same time determine the nuclear
deformation. The optimal choice of this basis will then be the one
that minimises the total energy of a given system and thus at the same
time determines the nuclear ground state deformation. The number
$N_{0}$ of HO shells must, of course, be chosen large
enough so that the final result of the Skyrme--Hartree--Fock
calculation is approximately independent of this choice. For example,
a value of $N_{0}=14$, corresponding to 680 HO basis states, is enough
for rare-earth nuclei. Instead of $\omega_z$ and $\omega_\perp$ it is
often chosen to consider the following basis parameters~\cite{FQKV}
\begin{equation}
\eta = \frac{\omega_\perp}{\omega_z} \;\; \mbox{and} \;\; b =
\sqrt{\frac{m \omega_{0}}{\hbar}} \;\; \mbox{with} \;\; \omega_{0} =
\left(\omega_{\perp}^2 \omega_z \right)^{1/3} \;, 
\label{Eq26}
\end{equation}
or otherwise expressed
\begin{equation}
{\omega_z} = \omega_{0} \, \eta^{-2/3}
\;\;\;\; \mbox{and} \;\;\;\;
\omega_\perp = \omega_{0} \, \eta^{1/3} \;.
\label{Eq26b}
\end{equation}
Note that $\eta$ is denoted by $q$ in Ref. \cite{BMD15}. If the nuclear deformation does not exhibit axial symmetry, then
triaxial symmetry can be imposed and the Skyrme--Hartree--Fock
calculations are performed with this symmetry. For the rare earth isotopes (and
more particularly the lanthanide family, whose atoms have interesting
magnetic properties) we are interested in, almost all the
nuclei show some well-developed prolate deformation and axial symmetry
can safely be imposed.  
%
%%%%%%%%%%%%%%%%%%%%%%%%%%%%%%%%%%%%%%%%%%%%%%%%%
\begin{figure*}[ht!]
\begin{center}
\includegraphics[scale=0.75]{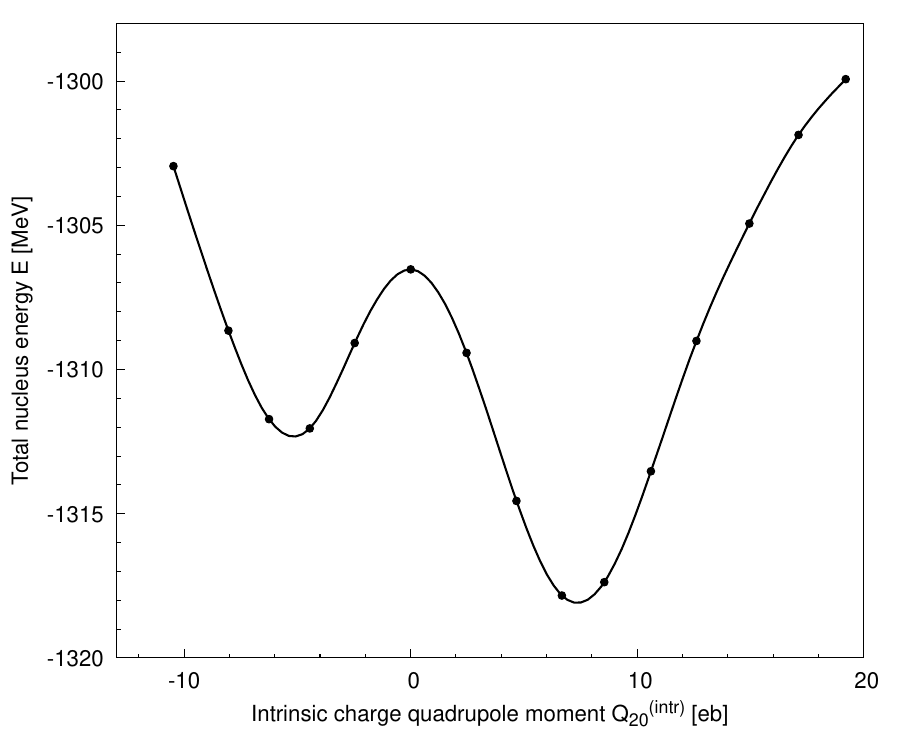}
\caption{Energy (in MeV) of the $^{162}$Dy nucleus as a function of
  the constrained expectation value of the intrinsic charge quadrupole
  moment $Q^{(\rm intr)}_{20}$ in units of $e$barn (see \cref{subsec_r_Q20} for its definition). The solid curve is an
  interpolation through the calculated values marked by filled circles.\label{Fig01}} 
\end{center}
\end{figure*}

%%%%%%%%%%%%%%%%%%%%%%%%%%%%%%%%%%%%%%%%%%%%%%%%

If one is only concerned with the ground-state properties of a given
nuclear system, the Hartree-Fock iterative cycle will automatically
converge to the minimum of the nuclear energy that is realised for
such a state (as long as the starting point of the iterative cycle is
reasonably chosen). If one is, however, interested in the nuclear
deformation energy, i.e.\ the nuclear energy as a function of
deformation, the variational calculation has to be carried out with a
constraint on some given multipole moment of the nucleon density. Such
constrained Hartree--Fock calculations are usually carried out by imposing
the mass quadrupole moment, but higher moments can of course also be
constrained. \Cref{Fig01} shows the energy of the nucleus
$^{162}$Dy as a function of the constrained expectation value of the
axial quadrupole moment operator
\begin{equation}
  \label{eq_Q20_op}
  Q_{20} = 2Z^2-(X^2+Y^2)\,.
\end{equation}
Note that this definition yields a value twice as large as that relying on the expression based on spherical harmonics $R^2Y_{20}(\Theta,\Phi)$, as introduced in Appendix \ref{Multipole}. This deformation-energy curve of $^{162}$Dy shows two local minima. The lower, ground state configuration, occurs for $Q_{20} > 0$ corresponding  a prolate shape, whereas the other configuration occurs for $Q_{20} < 0$ corresponding  an oblate isomeric state.

Once the deformation of a given nucleus is well established, the
resulting single-particle wave functions
$\varphi_{k}(\vec{R},\sigma,q)$ can be constructed and used to
determine all the local densities, \crefrange{Eq10}{Eq16}. Then one can compute any experimentally accessible
quantity such as the nuclear root mean-square (rms) radii
or multipole moments (see \cref{subsec_r_Q20,subsec_mu}).

\subsection{Including pairing correlations}

Pairing correlations are known to affect the nuclear structure in a
non-negligible way in all nuclei, with the very rare exception of the
regions in the immediate vicinity of closed shells (magic
numbers). At the mean-field level, such pairing
correlations should, in principle, be taken into account by performing
Hartree-Fock-Bogoliubov calculations in which both the mean field and
the pairing field are treated equally and determined simultaneously
from the same interaction. This is generally done for mean-field
calculations with the Gogny force, like in the D1S parameterisation, \cite{DG80} (see also
\cite{CPG15,GCV18} and references therein for more recent versions of
this nucleon-nucleon interaction, such as D1M* and D2). Skyrme EDFs,
on the other hand, have generally not been constructed to treat
pairing correlations with the same interaction, with the notable
exception of the SkP parameterisation of the Skyrme-type
EDF~\cite{DFT84}. Such correlations have often been taken into account
by performing BCS calculations {\it on top} of the Hartree-Fock iterative 
cycle using a schematic so-called seniority interaction (constant
pairing matrix elements between pairs of time-reversed states, see
below). Given a pairing strength $G_q$, one solves the BCS
equations to obtain the Fermi energy $\lambda_q$ and the gap parameter
$\Delta_q$ for each nucleon charge state ($q = \{n,p\}$), which then
determine the BCS occupation factor $v_k^2$ for a given
single-particle state $|\varphi_k\rangle$. In this framework, all
expressions of the local densities \crefrange{eq_rho_q}{Eq16}
remain unchanged and the nuclear energy $E$ receives an additional
contribution $E_{\rm pair}$ called ``pair condensation energy'' defined by
\begin{equation}
  E_{\rm pair} = \sum_q \sum_{\substack{k>0\\l >0}} (\langle \varphi_k\varphi_{\bar
    k}|V_{\rm pair}^{(q)}|\varphi_l\varphi_{\bar l}\rangle
  - \langle \varphi_k\varphi_{\bar
    k}|V_{\rm pair}^{(q)}|\varphi_{\bar l}\varphi_l\rangle) \, u_kv_ku_lv_l \;,
\end{equation}
where $u_k = \sqrt{1-v_k^2}$, $|\varphi_{\bar k}\rangle$ is the
canonically conjugate state of $|\varphi_{k}\rangle$ (which is
associacted with the same eigenvalue of $\widehat{\varrho}$ as is
$|\varphi_k\rangle$~\cite{Ring-Schuck}) and $V_{\rm pair}$ is the pairing interaction for
the nucleons of charge state $q$. The summation ``$k>0$'' means that
one takes only one member of a pair of canonically conjugate
states. For the above-mentioned seniority case one has $\langle
\varphi_k\varphi_{\bar k}|V_{\rm pair}^{(q)}|\varphi_l\varphi_{\bar
  l}\rangle -\langle \varphi_k\varphi_{\bar
      k}|V_{\rm pair}^{(q)}|\varphi_{\bar l}\varphi_l\rangle  = G_q$
($<0$ because $V_{\rm pair}$ is attractive).

The point now is to choose the pairing strength $G_q$ in such a way
that the experimental data for quantities such as proton and neutron
separation energies are reproduced as well as possible. It turns out
that such a choice can indeed be made by defining the pairing strength
as 
\begin{equation}
G_q = -\frac{G^{({0})}_q}{N_q + 11} \;, \;\;\; q=\{n,p\} \;,
\label{Eq27}
\end{equation}
with $N_n = N$, $N_p=Z$, and where the seniority force strengths $G^{(\rm
  0)}_n$ and $G^{(0)}_p$ can be chosen for a given mass region as
suggested in \cite{BMD15,NRL19}, where also some additional information about
the details of the seniority BCS pairing treatment used in our
approach can be found.

\subsection{Treatment of odd nuclei}
\label{subsec_odd_nuc}

At the heart of the pairing treatment described above is
the canonical conjugation as
a consequence of the twofold degeneracy of the eigenvalues $v_k^2$ varying between 0 and 1 of the one-body density operator $\hat \varrho$ (the eigenvalues
0 and 1 being highly degenerate). In conjunction with axial
symmetry, characterised by the commutation of $j_z$ and $\hat \varrho$, we can choose one of the two degenerate eigenstates of
$\hat \varrho$ as an eigenstate $|\varphi_k\rangle$ of $j_z$ with a
positive eigenvalue $\hbar\Omega_k$, while the other eigenstate of
$\hat \varrho$, denoted $|\varphi_{\bar k}\rangle$, is an eigenstate
of $j_z$ with the opposite eigenvalue $-\hbar\Omega_k$. According to
Herbut and Vuji$\rm\check{c}$i\'{c}~\cite{Herbut68}, there
exists an anti-unitary operator $P_a$, called pairing operator,
which maps $|\varphi_k\rangle$ to $|\varphi_{\bar k}\rangle$ and is
such that $P_a^2 = -1$. When the one-body density operator $\hat
\varrho_q$ commutes with the time-reversal operator $\mathcal T$,
the pairing operator coincides with $\mathcal T$. This happens only in the
self-consistent mean-field solution describing the ground state of an
even-even nucleus. In this case, one also has a twofold degeneracy
of the eigenvalues $e_k^{(q)}$ of the Hartree--Fock Hamiltonian
$h_q$, known as the Kramers degeneracy.

In contrast, when time-reversal symmetry is broken, the Kramer
degeneracy no longer holds and the pairing operator differs from the
time-reversal operator. This happens in an excited configuration of an
even-even nucleus or when the number of neutrons and/or protons is odd.
However, at least in axially symmetric Hartree--Fock--BCS
calculations, it turns out that for every single-particle state
$|\varphi_k\rangle$ with a good quantum number $\Omega_k > 0$
it is always possible to find a state
$|\varphi_{\widetilde{k}}\rangle$ of the same isospin (proton or
neutron), which we will call the {\it conjugate partner} of
$|\varphi_k\rangle$, with a good quantum number $-\Omega_k < 0$ and maximising
the overlap modulus $|\langle \varphi_{\widetilde k}|(\mathcal
T|\varphi_k\rangle)|$ with a value close to 1. The states
$|\varphi_k\rangle$ and $|\varphi_{\widetilde{k}}\rangle$ form what we
will call a {\it pseudo pair}~\cite{Pototzky10,BMD15}. This suggests
that the BCS approach presented above can be extended to odd-even or
odd-odd nuclei, provided that one works with {\it pseudo pairs}
instead of Cooper pairs, which involve time-reversed states. In a
dedicated study \cite{BMD15}, it has been shown that the polarisation effect of the
even-even core nucleus due to the presence of the unpaired nucleon, which
is responsible for the time-reversal symmetry breaking in the odd-mass
nucleus, is able to give a fair reproduction of the experimentally
measured magnetic dipole moments.

Let us next explain how to calculate properties of an odd-even nucleus using the
example of the dysprosium ($Z=66$) isotopes $^{161}$Dy and $^{163}$Dy,
which have an unpaired neutron hole ($^{161}$Dy) or neutron particle
($^{163}$Dy) with respect to the even-even $^{162}$Dy nucleus.  
For this investigation we use a standard Skyrme EDF without any tensor 
contribution, namely the SIII parametrisation~\cite{BFN75}, known for 
its good and well established spectroscopic properties. It is
implemented in what is called the {\it minimal scheme}.
In this scheme adapted to SIII, the
$B_{18}$, $B_{14}$, $B_{19}$ and $B_{15}$ terms
of the Hamiltonian density $\mathcal H(\vec R)$ in \cref{H_density}
are discarded, causing the spin-current tensor field $W_{q,\mu\nu}$ and the
spin-gradient field $\vec C_q$ defined by \cref{Eq25,Cq}, respectively, to contribute neither to the Hamiltonian density \cref{H_density} nor to the Hartree-Fock Hamiltonian in \cref{hHF}. This is consistent with the fact that these terms of the spin field $\vec S_q$
defined by \cref{Sq} have not entered the fitting process of this particular Skyrme EDF (see Ref.~\cite{BMD15} for details).

\begin{figure}[htb!]
    \centering
    \includegraphics[width=0.45\textwidth]{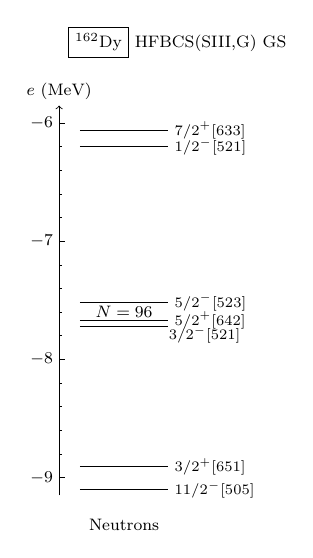}
    \includegraphics[width=0.395\textwidth]{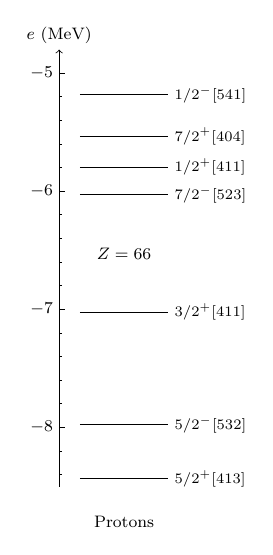}
    \caption{Single-particle spectra for the ground state of $^{162}$Dy near
      the Fermi level obtained from the HFBCS calculations. Each level
      is twice degenerate because of the time-reversal symmetry. They are both
      labeled by the same Nilsson quantum numbers, $\Omega^{\pi}[Nn_z\Lambda]$, representing the dominant contribution of either single-particle state. $\Omega > 0$ and $\Lambda$ are the projections onto the symmetry axis of the
      total and orbital angular momenta, respectively, $n_z$ is the number of oscillator quanta along that axis and $N$
      is the total number of oscillator quanta.
    \label{fig::sp_spectra_Dy162}}
\end{figure}

In a first step, we determine the ground state deformation of these
nuclei. Assuming that the two odd-mass dysprosium isotopes have the
same deformation as the common even-even $^{162}$Dy neighbour, one
varies the basis parameters $b$ and $q$ of \cref{Eq26} to
minimise the ground-state energy of $^{162}$Dy for a fixed value of
the number $N_{0}$ of HO major shells. Using $N_0=14$ turns out to
provide a large-enough HO basis (see subsection III.C above). One
finds that the optimal parameters are $b=0.49$ and $\eta=1.24$. Choosing
the pairing strengths $G^{(0)}_n = 16.2$ and $G^{(0)}_p = 15.1$ in 
\cref{Eq27} adjusted to give a good description of the odd-even
mass differences in the considered region of the nuclear
chart~\cite{NRL19}, the single-particle spectra of neutrons and
protons are then found as shown in \cref{fig::sp_spectra_Dy162}. 

%%%%%%%%%%%%%%%%%%%%%%%%%%%%%%%%%%%%%%%%%%%%%%%%%
%%%%%%%%%%%%%%%%%%%%%%%%%%%%%%%%%%%%%%%%%%%%%%%%%

\begin{figure}[h]
  \includegraphics[width=0.7\textwidth]{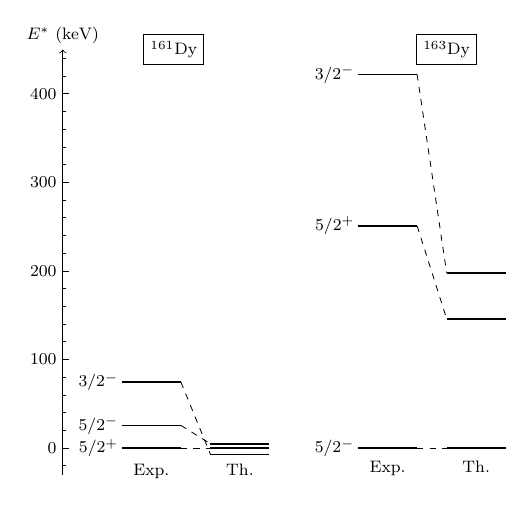}
  \caption{Excitation energies of the first three bandhead states in
    $^{161}$Dy and $^{163}$Dy isotopes. Experimental data (levels
    labelled ``Exp.'') are taken from \cite{Reich11_NDS112_Dy161} and
    \cite{Reich10_NDS111_Dy163}, respectively. The levels labelled
    ``Th.'' correspond to the results of the present
    Skyrme--HF--BCS calculations using the Skyrme SIII
    parametrisation and the seniority pairing strengths $G_n =
    -16.2/(11+N)$~MeV and $G_p=-15.1/(11+Z)$~MeV for neutrons and
    protons, respectively, with $N=96$ and $Z=66$ (see text for details).
  \label{fig_bandheads_Dy}}
\end{figure}

Experimentally, the odd dysprosium isotopes $^{161}$Dy and
$^{163}$Dy are found to have ground-state spin and parity 5/2$^{+}$
and 5/2$^{-}$, respectively~\cite{Reich11_NDS112_Dy161,Reich10_NDS111_Dy163}.
In the calculated single-particle neutron spectrum of the ground state
of $^{162}$Dy in \cref{fig::sp_spectra_Dy162}, where the label
$N=96$ corresponds to a filling of the lowest single-particle
twofold degenerate energy levels, we do find single-particle states
with these quantum numbers in the immediate vicinity of the neutron
Fermi level (corresponding to the last occupied state if pairing is
not taken into account). Since the single-particle states are expanded in the harmonic-oscillator basis, the displayed Nilsson quantum numbers $[Nn_z\Lambda]$ are not good quantum numbers but those of the dominant contribution to the given single-particle state. Encouraged by these results, we can then
carry out the self-consistent mean-field calculations for the two odd-mass dysprosium isotopes. To do so, we start the Hartree--Fock--BCS iterative process from the Hartree--Fock
Hamiltonian and local densities of the converged $^{162}$Dy ground-state solution,
and impose that the relevant single-particle state identified
above with quantum numbers $K$ and $\pi$ is now blocked with an
occupation probability equal to 1, while the {\it conjugate partner}
with quantum numbers $-K$ and $\pi$ is blocked with a vanishing
occupation probability. This is what we call mean-field
calculations with self-consistent blocking. This {\it pseudo pair} of
states therefore does not participate in pair excitations and this
quenches pairing correlations. The energy spectrum of the first three
bandhead states resulting from such calculations is displayed in
\cref{fig_bandheads_Dy} and compared with experimental spectra.
It can be considered as a good agreement with the experimental data to reproduce such a spectrum
within less than 200~keV, even much less in $^{161}$Dy.

\subsection{Calculation of the nuclear charge density}
\label{subsec_r_Q20}

As a result of Skyrme--Hartree--Fock--BCS
calculations, one obtains all the local densities,
defined in \crefrange{eq_rho_q}{Eq16},
from which we get information about
the structure of the odd-mass nucleus. In particular, we can
calculate observable quantities like the root-mean-square charge
radius and the electric quadrupole moment and compare them with
experimental data. \\

In a simple nonrelativistic approach, the nuclear charge
  density $\rho_{\rm c}(\vec R)$ can be obtained by folding the proton charge
  density $\rho_p(\vec{R})$, which corresponds to point-like
  particles, with a proton form factor $F(\vec r)$ that takes into
  account the charge distribution inside the proton generated by its
  quark substructure
  \begin{equation}
    \rho_{\rm c}(\vec R) = \int \rho_p(\vec R') \, F(\vec R-\vec R') \, d^3R' \,.
  \end{equation}
  The proton density $\rho_p(\vec R)$ is defined by
\cref{eq_rho_q} for the charge state $q = p$.
  The mean-square radius $R_{\rm c}^2$ of the charge density is given by
  \begin{equation}
    \label{eq_Rc2}
    R_{c}^2 = \dfrac{\displaystyle\int R^2\rho_{\rm c}(\vec R) \, d^3R}
    {\displaystyle\int \rho_{\rm c}(\vec R)\,d^3R} \,.
  \end{equation}
With a spherically symmetric Gaussian form factor $F(\vec R) =
(\sqrt{2\pi}\,d)^{-3}e^{-R^2/(2d^2)}$ normalised to unity, as in
Ref.~\cite{Gambhir90} for instance, the denominator of \cref{eq_Rc2} becomes
\begin{equation}
  \int \rho_{\rm c}(\vec R)\,d^3R = \int \rho_p(\vec R)\,d^3R = Z\,,
\end{equation}
and the root mean square radius $R_{\rm c}$ can be written in the form
\begin{equation}
R_{\rm c} = \sqrt{\langle R^2\rangle_p + r_p^2} \;,
\end{equation}
where $\langle R^2\rangle_p$ is the proton contribution to the
average value of $R^2$ considered as a one-body operator
\begin{equation}
  \langle R^2\rangle_p = \dfrac{\displaystyle\int R^2\rho_p(\vec R) \,
    d^3R}{\displaystyle\int \rho_p(\vec R)\,d^3R}
  = \dfrac 1 Z \int R^2\rho_p(\vec R) \, d^3R \;,
\end{equation}
and $r_p^2$ is the mean-square radius of the proton
\begin{equation}
  r_p^2=\int R^2 F(\vec R)\, d^3R = 3\,d^2 \,.
\end{equation}
The CODATA 2022 recommended value of $r_p$ is about $0.84$~fm~\cite{CODATA2022}.

The intrinsic charge quadrupole moment of the nucleus $Q_{20}^{({\rm
    intr})}$ is calculated as the proton contribution to the
expectation value of the
axial quadrupole operator $Q_{20}$, defined by
\cref{eq_Q20_op}, in the intrinsic nucleus state $\ket{\Phi}$. It thus takes the integral form:
\begin{equation}
  Q_{20}^{({\rm intr})} = e\int \rho_p(\vec R) \, \big[2\,Z^2 -
    (X^2+Y^2)\big] \, d^3R
%%   \sum_k \sum_{\sigma}  v_{q,k}^2 \int
%% \varphi_{p,k}^{*} (\vec{R'},\sigma') Q_{20}
%% \varphi_{p,k}(\vec{R},\sigma) d^3R\;, 
\label{Eq30}
\end{equation}
Indeed, using the above spherically symmetric proton charge form factor $F$,
one can show that the axial quadrupole moment of the charge density
$\rho_{\rm c}$ is equal to $Q_{20}^{({\rm intr})}$.

Because the intrinsic state $\ket{\Phi}$ and the corresponding
quadrupole moment $Q_{20}^{({\rm intr})}$ are defined in the
body-fixed frame, while the experimental quadrupole moment as
tabulated, e.g., by Stone \cite{St05} is measured in the laboratory
frame, one needs to perform a change of frame as explained in detail in
Appendix~\ref{App_Intrinsic vs. spectroscopic multipole moments}. In a
nucleus state of total angular momentum $I$ and projection $K$ on the
symmetry axis, the axial quadrupole moment in the laboratory frame, called
``spectroscopic'' and denoted by $Q_{20}^{({\rm spec})}$, is given by
\begin{equation}
Q_{20}^{({\rm spec})} = \frac{3K^2 - I(I+1)}{(I+1)(2I + 3)} \,
Q_{20}^{({\rm intr})} \;. 
\label{Eq29x}
\end{equation}
For the lowest-energy state of a rotational band, identified as the
bandhead state, one has $I = K$ (with rare exceptions when $K =
1/2$). In this case the spectroscopic quadrupole moment
$Q_{20}^{({\rm spec})}$ becomes
\begin{equation}
  Q_{20}^{({\rm spec})} = \dfrac{K(2K-1)}{(K+1)(2K+3)}\,Q_{20}^{({\rm intr})} \equiv {\cal I}^{E_2}_{{\rm intr/spec}} \,Q_{20}^{({\rm intr})} \;. 
  \label{Q20spec_bh}
\end{equation}
The calculated charge rms radius $R_{\mathrm{rms}}$, the nuclear radius $R_{\rm N} = \sqrt{\frac{5}{3}R_{\mathrm{rms}}^2}$ \cite{Palffy2010},
and the charge quadrupole moment $Q_{20}$ in the body-fixed frame
$Q_{20}^{({\rm intr})}$ and in the laboratory frame
$Q_{20}^{({\rm spec})}$ together with experimental
data are given in \cref{tab_R_Q} for the studied Dy
isotopes $^{161,163}$Dy as well as stable isotopes of odd-proton
neighbour elements, namely \ce{^{159}_{65}Tb} and \ce{^{165}_{67}Ho}. The
numerical parameters used for the latter two nuclei are the same as
those reported in \cref{subsec_odd_nuc} for the $^{161,163}$Dy
isotopes. Despite a slight systematic overestimation for $R_{\rm c}$ and
$Q_{20}^{(\rm spec)}$, a good agreement is found for these four
prolate-deformed nuclei.
\newlength\origarrayrulewidth
\newlength\origtabcolsep
\let\origarraystretch\arraystretch

\setlength\origarrayrulewidth{\arrayrulewidth}
\setlength\origtabcolsep{\tabcolsep}

% …now make your global tweaks …
% \setlength{\arrayrulewidth}{0.25mm}
% \setlength{\tabcolsep}{4pt}
% \renewcommand{\arraystretch}{1.5}

\begingroup
  \setlength{\arrayrulewidth}{\origarrayrulewidth}
  \setlength{\tabcolsep}{\origtabcolsep}
  \renewcommand{\arraystretch}{\origarraystretch}
\begin{table*}[ht!]
    \centering
    \begin{tabular}{ccccccccccc}
        \hline\hline
        \multirow{2}{*}{Nucleus} & $K^{\pi}$ & \multicolumn{2}{c}{$R_{\rm N}$ (fm)} &&
        \multicolumn{2}{c}{ Rms charge radius $R_{\rm c}$ (fm)} & &
        \multicolumn{3}{c}{Charge quadrupole moment ($e$b)} \\
        \cline{3-4} \cline{6-7} \cline{9-11}
        & configuration & Emp. & HFBCS && HFBCS & Exp. &&
        $Q_{20}^{(\rm intr)}$ & $Q_{20}^{(\rm spec)}$ & Exp.~\cite{Stone16} \\
        \hline
        \ce{^{159}_{65}Tb} & $3/2^+[411]_p$ &
        6.583 & 6.704 && 5.258 & 5.06(15) \cite{Reich12_NDS113_Tb159} &&
        7.345 & 1.469 & +1.432(8) \\
        \hline
        \ce{^{161}_{66}Dy} &
        $5/2^+[411]_n$ & \multirow{3}{*}{6.607} & \multirow{3}{*}{6.730} && 5.277 & 5.197(6)
        \cite{Reich11_NDS112_Dy161} &&
        7.339 & 2.621 & +2.468(29) \\
        & $5/2^-[523]_n$ & & && 5.278 & ---
        && 7.348 & 2.624 & 2.506 \\
        & $3/2^-[521]_n$ & & && 5.279 & ---
        && 7.400 & 1.480 & 1.45 \\
        \hline
        %% \ce{^{162}_{66}Dy} & $0^+$ & & 5.281 & 5.2074(172) && & & \\
        \ce{^{163}_{66}Dy} & $5/2^-[523]_n$ & 6.631 & 6.754 && 5.296 & 5.2091(25)
        \cite{Reich10_NDS111_Dy163} &&
        7.644 & 2.730 & +2.648(21) \\
        \hline
        \ce{^{165}_{67}Ho} & $7/2^-[523]_p$ & 6.655 & 6.781 && 5.318 & 5.202(31)
        \cite{Singh24_NDS194_Ho165} && 7.766 & 3.624 & +3.58(2) \\
        \hline\hline
    \end{tabular}
    \caption{Root mean square charge radius $R_{\rm c}$ and charge quadrupole
      moment in the intrinsic frame $Q_{20}^{(\rm intr)}$ and
      laboratory frame $Q_{20}^{(\rm spec)}$ calculated in the
      Hartree--Fock--BCS nuclear-structure model compared with
      experimental data when available for the lowest-three bandhead
      states of rotational bands of $^{161}$Dy 
      isotopes, as well as ground-state values for $^{163}$Dy and the
      stable isotopes of odd-proton neighbour elements \ce{^{159}_{65}Tb} and
      \ce{^{165}_{67}Ho}. In addition, the empirical nuclear radius
      $R_{\rm N}$ (see \cref{subsec_res_nuc}) is given in the third column.} 
    \label{tab_R_Q}
\end{table*}
\endgroup

\subsection{Calculation of the magnetic dipole moment}
\label{subsec_mu}

Finally, in this last subsection, we investigate to what extent the
presented mean-field approach with self-consistent blocking is able to
reproduce the experimental data on the magnetic dipole moments.

Following Schwartz~\cite{Sc55}, we assume that the
response of the nucleus to an external electromagnetic field can be
described in a non-relativistic way by means of electric multipole
one-body operators of even rank and magnetic multipole one-body
operators of odd rank. As shown by Chemtob in Ref.~\cite{Chemtob69},
this means that one can to a good approximation neglect the
two-body interaction currents arising from meson exchange between
nucleons. Because of the above discussed axial symmetry of the
nuclear shape around the $z$ axis of the intrinsic frame, we need
only the component along this axis of the magnetic dipole moment
one-body operator $\mu_z$ as derived by Schwartz~\cite{Sc55}, namely 
\begin{equation}
  \mu_z = g^{(\ell)} \, \ell_z + g^{(s)} \, s_z \;,
  \label{muz_op}
\end{equation}
where $\ell_z$ and $s_z = \dfrac{\sigma_z}{2}$ are,
respectively, the components of the 
orbital and spin-$1/2$ angular {momentum operators} of a nucleon along the $z$
axis, and $g^{(\ell)}$, $g^{(s)}$ are, respectively, the orbital and
spin gyromagnetic ratios (or Landé $g$ factors). The values of these
constants depend on the isospin of the nucleon and are given in
\cref{tab_g}. 
\begin{table*}[ht!]
    \centering
    \begin{tabular}{ccc} 
    \hline \hline
     Nucleon & $\;\;g^{(\ell)}\;\;$ & $\;\;g^{(s)}\;\;$  \\
    \hline 
     $p$ & $1$ & $+5.585695$\\%5.5856946893(16)$  \\
     $n$ & $0$ & $-3.826085$\\%3.82608552(90)$  \\
    \hline \hline
    \end{tabular}
    \caption{Values with up to six decimal places of the orbital
      $g^{(\ell)}$ and spin $g^{(s)}$ gyromagnetic ratios (Land\'e
      factors) for protons and neutrons entering
        the magnetic dipole moment through
      \cref{muz_op} and implemented in the
      Skyrme--Hartree--Fock--BCS code used to produce the results of
      \cref{tab_mu}.} 
    \label{tab_g}
\end{table*}
Let $|\Phi\rangle$ denote the normalised many-body state of the odd-mass
nucleus in the configuration of interest, characterised by the quantum
numbers $K^\pi$ of the blocked nucleon. The expectation value of the
magnetic dipole-moment operator $\mu_z$ in the intrinsic (body-fixed)
frame, $
\mu^{(\mathrm{intr})} = \langle \Phi | \mu_z | \Phi \rangle$,
defines the intrinsic magnetic dipole moment, denoted $\mu_0^K$ in
Appendix~\ref{App_Intrinsic vs. spectroscopic multipole  moments}. When
$|\Phi\rangle$ is obtained as a Skyrme--Hartree--Fock--BCS solution for
the specified blocked configuration, $\mu^{(\mathrm{intr})}$
represents the magnetic dipole moment of the nucleus within the
Skyrme--Hartree--Fock--BCS framework.

To compare the calculated magnetic dipole moment with experiment, one
has, in the same way as for the quadrupole moment discussed in the previous
subsection, to perform a change of frame as explained in
detail in Appendix~\ref{App_Intrinsic vs. spectroscopic multipole
  moments}. More precisely, the magnetic dipole moment in the laboratory
frame (called spectroscopic) $\mu^{(\rm spec)}$ is proportional to the
magnetic dipole moment in the body-fixed frame (called intrinsic)
$\mu^{(\rm intr)}$ according to the relation
\begin{equation}
\mu^{(\rm spec)} = \dfrac{K}{K+1} \, \mu^{(\rm intr)} \equiv {\cal I}^{M_1}_{{\rm intr/spec}} \,\mu^{({\rm intr})}\,.
\end{equation}
To obtain this relation, we have applied \cref{eq::198} to
the bandhead state of a rotational band for which the total angular
momentum $I$ of the nucleus is equal to its projection $K$ on the
symmetry axis, where we have assumed that $K \ne 1/2$.

However, it is known from the unified-model picture of Bohr and
Mottelson~\cite{BM_V2} that a collective contribution $g_{\rm R}$ (called
collective gyromagnetic ratio) has to be added to the 
contribution $\mu^{(\rm intr)}$ from the intrinsic degrees of
freedom. It can be calculated in a microscopic 
approach relying on a perturbative treatment of the Hamiltonian in a
rotating frame according to~\cite{Sprung79,PBN68}
\begin{equation}
g_{\rm R} = \frac{\sum\limits_{k,\ell} \langle \varphi_\ell |\mu_{-}|
  \varphi_k \rangle \langle \varphi_k |j_{+}| \varphi_\ell \rangle
  (u_k \, v_\ell - u_\ell \, v_k)^2 / (E_k + E_\ell)}
{\sum\limits_{k,\ell} \langle \varphi_\ell |j_{-}| \varphi_k
  \rangle \langle \varphi_k |j_{+}| \varphi_\ell \rangle
  (u_k \, v_\ell - u_\ell \, v_k)^2 / (E_k + E_\ell)}\,, 
\label{gR}
\end{equation}
where the sums run over all states of the HFBCS single-particle
spectrum (for each nucleon type) except for blocked state and its pair
partner, where $u_k$ and $v_k$ are the usual BCS occupation amplitudes, and
$E_k$ are the quasiparticle energies (see Ref.~\cite{BMD15} for
details). The $j_{\pm} = j_{x} \pm i j_{y}$ are the standard angular
momentum ladder operators and the $\mu_{\pm} = g^{(\ell)} \, \ell_{\pm} +
g^{(s)} \, s_{\pm}$ operators are defined in an analogous way. Setting
\begin{align}
  \mu_{\rm intr}^{(\rm spec)} & = \frac{K}{K+1} \, \langle
  \Phi|\mu_z|\Phi\rangle
  \label{mu_intr} \\
  \mu_{\rm coll}^{(\rm spec)} & = \frac{K}{K+1} \, g_{\rm R} \;,
\label{mu_coll}
\end{align}
we therefore have the total magnetic moment in the laboratory frame
\begin{equation}
\mu_{\rm tot}^{(\rm spec)} = \mu_{\rm intr}^{(\rm spec)} + \mu_{\rm
  coll}^{(\rm spec)} \;.
\label{Eq31}
\end{equation}
\Cref{tab_mu} presents the calculated values of $\mu_{\rm tot}^{(\rm
  spec)}$ and its contributions $\mu_{\rm intr}^{(\rm spec)}$ and
$\mu_{\rm coll}^{(\rm spec)}$ compared with experimental data when
available for the lowest three bandhead states of $^{161,163}$Dy, as
well as the ground states of $^{159}$Tb and $^{165}$Ho.

\begin{table}[h]
\begin{tabular}{cccccc}
\hline\hline
\multirow{2}{*}{Nucleus} & \multirow{2}{*}{$K^{\pi}$ configuration} &
\multicolumn{4}{c}{Magnetic dipole moment $(\mu_{\rm N})$} \\
\cline{3-6}
& & $\mu_{\rm intr}^{(\rm spec)}$ & $\mu_{\rm coll}^{(\rm spec)}$ &
$\mu_{\rm tot}^{(\rm spec)}$ & Exp.~\cite{St05} \\
\hline
$^{159}$Tb & ${3/2}^+[411]_p$ & 1.886 & 0.283 & $+2.169$ & $+2.014(4)$ \\
\hline
$^{161}$Dy & ${5/2}^+[411]_n$ & $-0.656$ & 0.383 & $-0.274$ & $-0.480(3)$ \\
& ${5/2}^-[523]_n$ & 0.647 & 0.214 & $+0.861$ & $+0.594(3)$ \\
& ${3/2}^-[521]_n$ & $-0.639$ & 0.210 & $-0.428$ & $-0.403(4)$ \\
\hline
$^{163}$Dy & ${5/}2^-[523]_n$ & 0.663 & 0.128 & $+0.791$ & $+0.673(4)$ \\
& ${5/2}^+[411]_p$ & $-0.638$ & 0.334 & $-0.304$ & --- \\
& ${3/2}^-[521]_p$ & $-0.644$ & 0.169 & $-0.475$ & --- \\
\hline
$^{165}$Ho & ${7/2}^-[523]_p$ & 3.946 & 0.126 & $+4.072$ & $+4.17(3)$ \\
\hline \hline
\end{tabular}
\caption{\label{tab_mu}Magnetic dipole moment of the
lowest three bandhead states of $^{161,163}$Dy, as well as of the ground
states of $^{159}$Tb and $^{165}$Ho, calculated within the
Skyrme--Hartree--Fock--BCS approach and compared with experimental data
when available~\cite{St05}.}
\end{table}

In almost all cases for which experimental data are available, the
calculated intrinsic contribution $\mu_{\rm intr}^{(\rm spec)}$ (transformed into
the laboratory frame) underestimates the measured value and the
addition of the collective contribution improves the agreement with
experiment. This was already observed in Ref.~\cite{BMD15} over
a larger sample of nuclei. Keeping in mind the approximations made in
these calculations and that the parameters of the Skyrme EDF used here
were not fitted to reproduce this observable, we can say that a rather
good agreement is reached.

Now we are in a position to establish a relationship between the
different contributions (neutron and proton, orbital and spin) to the
magnetic moment of a nucleus and the different contributions to the 
the so-called ``convection current density''
$\vec{j}_q^{(\ell)}$ and the ``spin current density'' $\vec{j}_q^{(s)}$
According to Ref.~\cite{Sc55}, these currents can be expressed
respectively in terms of the local densities $\vec j_q(\vec R)$, \cref{Eq15}, and $\vec s_q(\vec R)$, \cref{Eq14}, as
\begin{align}
  \vec{j}_q^{(\ell)}(\vec{R}) & = g_q^{(\ell)} \, \dfrac{\mu_{\rm N}}i
  \sum_{k,\sigma} v_k^{\,2}\Big(\varphi_{k}^{*}(\vec{R},\sigma,q) \,
  \vec{\nabla} \varphi_k(\vec{R},\sigma,q)
  - \varphi_{k}(\vec{R},\sigma,q) \,
  \vec{\nabla} \varphi_k^*(\vec{R},\sigma,q)\Big)  \\
  & = 2\,g_q^{(\ell)} \mu_{\rm N}\,\vec j_q(\vec R)
\label{Eq45} \\
\vec{j}_q^{(s)}(\vec{R}) & = g_q^{(s)} \, \mu_{\rm N}
\vec{\nabla} \!\times\! \bigg(\sum_{k} v_k^{\,2} \, 
    [\varphi_{k}]^{\dagger}(\vec{R},q) \, \dfrac{\vec{\sigma}}2 \,
    [\varphi_{k}](\vec{R},q)\bigg)  \\
    & = \dfrac 1 2\, g_q^{(s)} \, \mu_{\rm N} \,
    \vec{\nabla}\!\times\!\vec s_q(\vec R) \;,
\label{Eq46}
\end{align}
where $\mu_{\rm N} = \dfrac{e\hbar}{2m}$ is the nuclear magneton. Using the expressions of \cref{Eq46,Eq45}, let
us calculate half the integrals $\displaystyle 
\int \vec R \times \vec j_q^{(s)}(\vec R) \, d^3R$ and
$\displaystyle \int \vec R \times \vec j_q^{(\ell)}(\vec R) \,
d^3R$. For the former we have
\begin{align}
  \dfrac 1 2\int \vec R \times \vec j_q^{(s)}(\vec R) \, d^3R & =
  \frac 1 2\, g_q^{(s)}\mu_{\rm N}
  \int \vec R \times \Big(\vec{\nabla}\times\vec s_q(\vec R)\Big) \,
  d^3R  \\
  & = \frac 1 4\, g_q^{(s)}\mu_{\rm N}
  \int \bigg[\vec{\nabla} (\vec R \cdot \vec s_q) -
    \Big(R\, \dfrac{\partial \vec s_q}{\partial R} + \vec s_q(\vec
    R)\Big)\bigg] \, d^3R  \\
  & = -\frac 1 4\, g_q^{(s)}\mu_{\rm N} \bigg[
  \int\bigg(\int_0^{\infty}\dfrac{\partial \vec s_q}{\partial R}
    \, R^3\,dR\bigg)d\Omega
  + \int \vec s_q(\vec R) \, d^3R \bigg] 
\end{align}
where $\displaystyle \int \cdots \, d\Omega$ denotes the double integral
over the two angles of the spherical coordinate system. 
Because the single-particle
states are partially-occupied bound states in the nucleus, their wave functions
exponentially decrease to 0 when $R \to \infty$, and consequently so do all the local densities of the Skyrme energy functional treatment, \cref{Eq14,Eq15,Eq16}, in particular the spin density $\vec
s_q(\vec R)$. Therefore one can perform an integration by parts over
$R$ 
\begin{equation*}
  \int_0^{\infty}\dfrac{\partial \vec s_q}{\partial R} \, R^3\,dR =
  -3\int_0^{\infty}\vec s_q(\vec R) \, R^2\,dR \;,
\end{equation*}
so that
\begin{align}
  \dfrac 1 2\int \vec R \times \vec j_q^{(s)}(\vec R) \, d^3R & =
  \dfrac 1 2 \, g_q^{(s)}\mu_{\rm N} \int \vec s_q(\vec R) \, d^3R 
  \\
  & = \dfrac 1 2 \, g_q^{(s)}\mu_{\rm N} \sum_k v_k^2 \int
    [\varphi_k]^{\dagger}(\vec R,q) \, \vec \sigma \, [\varphi_k](\vec
    R,q) \, d^3R  \\
    & = g_q^{(s)}\mu_{\rm N} \, {\sum_k}^{(q)} v_k^2 \,
    \langle\varphi_k|\frac{\vec{\sigma}}2|\varphi_k\rangle  \;,
\end{align}
where $\sum^{(q)} \cdots$ means that one sums only over single-particle
states having the charge state $q$. Finally we obtain, for the $z$
component:
\begin{equation}
  \label{int_js}
  \bigg[\dfrac 1 2\int \vec R \times \vec j_q^{(s)}(\vec R) \, d^3R\bigg]_z
  = g_q^{(s)}\mu_{\rm N} \, \langle\Phi|\frac{\sigma_z}2|\Phi\rangle_q \;,
\end{equation}
where the notation $\langle\Phi|\cdots|\Phi\rangle_q$ refers to the
contribution of nucleons of charge state $q$ to the expectation value
in the Slater determinant or BCS state $|\Phi\rangle$. 
Regarding the convection current, setting $\vec p =
-i\hbar\vec{\nabla}$ one has
\begin{align}
  \dfrac 1 2\int \vec R \times \vec j_q^{(\ell)}(\vec R) \, d^3R & =
  g_q^{(\ell)}\mu_{\rm N}
  \int \vec R \times \vec j_q(\vec R) \, d^3R  \\
  & = g_q^{(\ell)}\dfrac{e}{2m}
  \int \vec R \times \sum_k v_k^2 \, \mbox{Re}
    \Big([\varphi_k]^{\dagger}(\vec R,q) \,
    \vec p \, [\varphi_k](\vec R,q)\Big) \, d^3R  \\
  & = g_q^{(\ell)}\dfrac{e}{2m}\mbox{Re}\bigg[ \sum_k v_k^2 \, 
    \int \Big([\varphi_k]^{\dagger}(\vec R,q)\,
    (\vec R \times \vec p) \, [\varphi_k](\vec R,q)\Big) \, d^3R\bigg]
     \;.
\end{align}
Therefore, owing to the Hermitian character of the orbital angular
momentum operator $\vec \ell = \vec R \times \vec p$, one obtains for the $z$
component:
\begin{equation}
  \label{int_jl}
   \bigg[\dfrac 1 2\int \vec R \times \vec j_q^{(\ell)}(\vec R) \,
     d^3R\bigg]_z
   = g_q^{(\ell)}\mu_{\rm N}\,\langle\Phi|\ell_z|\Phi\rangle_q \,.
\end{equation}
The relations \cref{int_js} and \cref{int_jl} have been tested
numerically by computing each member separately: the integrals have
been performed by the Gauss-like quadratures discussed above with the
same number of mesh points, while the expectation values in the right
members have been calculated through matrix elements in the
axially-symmetric harmonic-oscillator basis. An agreement up to
$10^{-6}$ has been obtained.

Finally, let us apply the above definitions of the convection and spin
currents to the case of axially symmetric Skyrme--HF-BCS
solutions. The single-particle wave functions $\varphi_k(\vec
R,\sigma,q)$, with good quantum number $\Omega_k$, can then be put in
the following form, using cylindrical coordinates
$(\varrho,\varphi,z)$ where $\varrho$ is the radial coordinate and
$\varphi$ the polar angle: 
\begin{equation}
  \varphi_k(\vec R,\sigma,q) = \phi_k^{(\sigma)}(\varrho,z) \,
  \dfrac{e^{i\Lambda_{\sigma}\varphi}}{\sqrt{2\pi}}
\end{equation}
with $\phi_k^{(\sigma)}(\varrho,z)\in\mathbb R$, $\Lambda_{\sigma} =
\Omega_k - \sigma$ ($\sigma = \pm 1/2$) and the normalisation 
\begin{equation}
  \int \sum_{\sigma = \pm 1/2} \big|\varphi_k(\vec R,\sigma,q)\big|^2\,d^3R =
  \int_{-\infty}^{\infty}\bigg\{\int_0^{\infty}
  \bigg[\Big(\phi_k^{(1/2)}(\varrho,z)\Big)^2 +
    \Big(\phi_k^{(-1/2)}(\varrho,z)\Big)^2\bigg]\,\varrho\,d\varrho\bigg\}dz
  = 1 \,.
\end{equation}
The local density $\vec j_q(\vec R)$ defined by \cref{Eq15} can
be shown to be orthoradial and independent of the polar angle
\begin{equation}
\vec j_q(\vec R) = j_{q,\varphi}(\varrho,z) \, \vec e_{\!\varphi}
\end{equation}
where $\vec e_{\!\varphi} = \frac{\partial \vec R}{\partial
\varphi}/\big|\!\big|\frac{\partial \vec R}{\partial
\varphi}\big|\!\big|$ is the unit vector associated with the polar angle, and the
orthoradial component $j_{q,\varphi}(\varrho,z)$ can be expressed as
\begin{equation}
  j_{q,\varphi}(\varrho,z) = \dfrac{1}{2\pi\varrho} \sum_{k,\sigma =
    \pm 1/2} v_k^2 \, \Lambda_{\sigma}
  \Big(\phi_k^{(\sigma)}(\varrho,z)\Big)^2 \,.
  \label{j_phi_orthorad}
\end{equation}
It follows that the convection current density $\vec j_q^{(\ell)}$,
defined by \cref{Eq45}, becomes
\begin{equation}
  \vec j_q^{(\ell)}(\vec R) = j_{q,\varphi}^{(\ell)}(\varrho,z) \,
  \vec e_{\!\varphi} \,,
\end{equation}
where
\begin{equation}
  \label{jl_q_phi}
  j_{q,\varphi}^{(\ell)}(\varrho,z) =
  \dfrac{g_q^{(\ell)}\,\mu_{\rm N}}{\pi\varrho} \sum_{k} v_k^2 \, \bigg[
    \Big(\Omega_k-\frac{1}{2}\Big)\,\big(\phi_k^{(1/2)}(\varrho,z)\big)^2
    + \Big(\Omega_k+\frac{1}{2}\Big)\,\big(\phi_k^{(-1/2)}(\varrho,z)\big)^2
    \bigg] \,.
\end{equation}
It vanishes for neutrons since $g_n^{(\ell)} = 0$.

In contrast to the local density $\vec j_q(\vec R)$, the spin density
$\vec s_q(\vec R)$ with axial symmetry has a vanishing orthoradial
component 
\begin{equation}
  \vec s_q(\vec R) = s_{q,\varrho}(\varrho,z) \, \vec e_{\!\varrho} +
  s_{q,z}(\varrho,z)\, \vec e_{\!z} \,,
\end{equation}
where $\vec e_{\!\varrho} = \dfrac{\partial \vec R}{\partial \varrho}$
is the unit vector associated with the radial coordinate $\varrho$.
The radial component $s_{q,\varrho}$ and the $z$-component
$s_{s,z}$ are independent of the polar angle $\varphi$ and are shown
to have the following expressions: 
\begin{align}
  s_{q,\varrho}(\varrho,z) & = \frac{1}{\pi}\sum_k v_k^{\,2} \,
  \phi_k^{(1/2)}(\varrho,z) \, \phi_k^{(-1/2)}(\varrho,z) \;, \\
  s_{q,z}(\varrho,z) & = \frac{1}{2\pi}\sum_k v_k^{\,2} \,
  \Big[\big(\phi_k^{(1/2)}(\varrho,z)\big)^2 -
    \big(\phi_k^{(-1/2)}(\varrho,z)\big)^2\Big] \,.
\end{align}
It is worth noticing that the radial component of the spin local density
results from the mixing of spin projection in the single-particle states
$|\varphi_k\rangle$. From the above radial and $z$ components of $\vec s_q$ 
one deduces that the curl of $\vec s_q$ is orthoradial,
\begin{equation}
  \vec\nabla\times \vec s_q(\vec R) = \Big(\dfrac{\partial
    s_{q,\varrho}}{\partial z} - \dfrac{\partial
    s_{q,z}}{\partial \varrho}\Big) \, \vec e_{\!\varphi} \;,
\end{equation}
and so is the spin current density,
\begin{equation}
  \vec j_q^{(s)}(\vec R) = j_{q,\varphi}^{(s)}(\varrho,z) \, \vec e_{\!\varphi}\;,
\end{equation}
where
$j_{q,\varphi}^{(s)}(\varrho,z)$ is given by 
\begin{equation}
  \label{js_q_phi}
  j_{q,\varphi}^{(s)}(\varrho,z) = \frac{g_q^{(s)}\mu_{\rm N}}{2\pi}\sum_k v_k^{\,2} \bigg[
    \phi_k^{(1/2)}\Big(\dfrac{\partial \phi_k^{(-1/2)}}{\partial z}
    - \dfrac{\partial \phi_k^{(1/2)}}{\partial \varrho}\Big)
    + \phi_k^{(-1/2)}\Big(\dfrac{\partial \phi_k^{(-1/2)}}{\partial
      \varrho} + \dfrac{\partial \phi_k^{(1/2)}}{\partial z}\Big)
    \bigg] \,.
\end{equation}
 
\section{Atomic-structure framework and nuclear-size effects in spectroscopy \label{atomic-structure}}

\subsection{Atomic description} \label{Atomic description}
The $(\vec{j},\rho)$ source terms constitute the link between the nucleus and the electrons in the atom. As mentioned in the previous section, we have developed a realistic model that allows us to calculate these quantities on a microscopic scale. As we will see later, from $(\vec{j},\rho)$ and the Maxwell equations one can obtain the magnetic and electric potentials $(\vec{A},\phi)$ that will act on the electrons in the atom.

In this work, if not explicitly mentioned, we will only consider hydrogen-like ions (ions with a single electron) or muonic ions which are atomic hydrogen-like bound states formed by a negatively charged muon of mass $m_{\mu} \approx 207 m_e$ and a nucleus. As the net charge of the nucleus is large ($Z = 66$ in the case of a dysprosium atom), a relativistic description is required since the bound leptons start moving at a significant fraction of the speed of light with non-relativistic velocity $v \sim Z\alpha c$, with $\alpha$ the fine structure constant. We therefore need to solve the stationary Dirac equation for a single electron in the presence of external electric $\phi(\vec{r})$ and magnetic $\vec{A}(\vec{r})$ potentials due to the nucleus, which reads:
\begin{equation}\begin{aligned}
{}^{\rm at}H \psi &= W \psi \\
&:= \left( c \vec{\alpha} \cdot (\vec{p}+e\vec{A}(\vec{r})) -e\phi(\vec{r}) \mathds{1} + mc^2 \beta \right) \psi\\
&:=(T + V + mc^2 \beta) \psi \;,
\label{eq::Dirac}
\end{aligned}\end{equation}
where $\psi(\vec{r})=(u(\vec{r}),v(\vec{r}))$ is a four-component bispinor with $u$ and $v$ two-component spinors, $V \equiv -e\phi \mathds{1}$ with $e>0$ and $T \equiv  c \vec{\alpha} \cdot (\vec{p}+e\vec{A})$ is the kinetic energy operator. $\vec{\alpha}$, $\beta$ are the Dirac matrices \cite{Strange2005} and $\mathds{1}$ is the $4\times 4$ identity matrix. In the following ${}^{\rm at}H$ refers to the electronic or atomic Hamiltonian.

For the moment, let's ignore the external magnetic potential $(\vec{A}=\vec{0})$ and assume that the external electric potential is spherically symmetrical, i.e. that $\phi(\vec{r})=\phi(r)$. The Hamiltonian \cref{eq::Dirac} commutes with $\vec{\ell \,}^2$, $\vec{s \,}^2$, $\vec{j \,}^2$, and $j_z$ where $\vec{j}=\vec{\ell}+\vec{s}$ is the total angular momentum operator of the electron or muon. In other words, $\ell$, $s$, $j$, and $m_j$ are good quantum numbers (in addition to the principal quantum number $n$). As $s=\frac{1}{2}$ is fixed and does not change, it can be left unspecified. Therefore, energy and angular momentum eigenstates can be written in the form
\begin{equation}\begin{aligned}
\left\{\begin{array}{c}
  u_{n\kappa}(\vec{r}) = g_{n\kappa}(r) \chi_{\kappa}^{m_j}(\Omega) \\
  v_{n\kappa}(\vec{r}) = if_{n\kappa}(r) \chi_{-\kappa}^{m_j}(\Omega) \;,
\end{array}\right.
\end{aligned}\end{equation}
where $\chi$ are the spin spherical harmonics defined by
\begin{equation}\begin{aligned}
\chi_{\kappa}^{m_j}(\Omega)=
\left(
  \begin{array}{c}
    \braket{\ell,\frac{1}{2},m_j-\frac{1}{2},+\frac{1}{2}}{j,m_j} Y_{l,m=m_j-\frac{1}{2}}(\Omega) \\
    \braket{\ell,\frac{1}{2},m_j+\frac{1}{2},-\frac{1}{2}}{j,m_j} Y_{l,m=m_j+\frac{1}{2}}(\Omega) \\
  \end{array}
\right)
\;,
\end{aligned}\end{equation}
which satisfy the orthonormalisation conditions
\begin{equation}\begin{aligned}
\int \chi_{\kappa}^{m_j *}(\Omega) \chi_{\kappa'}^{m_{j'}}(\Omega) d\Omega = \delta_{\kappa \kappa'} \delta_{{m_{j} m_{j'}}} \;.
\end{aligned}\end{equation}
The radial functions $f_{n\kappa}(r)$ and $g_{n\kappa}(r)$ are solutions of the differential equations
\begin{equation}\begin{aligned}
\left\{
  \begin{array}{c}
    \frac{d g_{n\kappa}}{dr}=-\frac{\kappa+1}{r}g_{n\kappa} + \frac{1}{c\hbar}\left( W-V+mc^2\right)f_{n\kappa} \\
    \frac{d f_{n\kappa}}{dr}=+\frac{\kappa-1}{r}f_{n\kappa} - \frac{1}{c\hbar}\left( W-V-mc^2\right)g_{n\kappa} \;,
  \end{array}
\right.
\end{aligned}\end{equation}
with $m=m_e$ for an electron and $m=m_{\mu}$ for a muon. By definition we have $P_{n\kappa} \equiv r g_{n\kappa}$ (large component) and $Q_{n\kappa} \equiv r f_{n\kappa}$ (small component) that satisfy the normalisation condition $\int_0^{\infty} \left( P_{n\kappa}^2(r) + Q_{n\kappa}^2(r)\right) dr=1$.
The solutions of the radial Dirac equation are labeled by the quantum numbers $n\in\mathbb{N}^*$ (principal quantum number) and $\kappa \in \mathbb{Z}^*$. Alternatively, they may be labeled by the triplet $n\ell_j$, where $\ell \in \mathbb{N}$ is the orbital angular momentum and $j=\ell \pm \frac{1}{2}$ is the total angular momentum. $\kappa$ is determined by $\kappa=-\ell-1=-(j+\frac{1}{2})$ for $j=\ell+\frac{1}{2}$ and $\kappa=\ell=+(j+\frac{1}{2})$ for $j=\ell-\frac{1}{2}$.

\subsection{Multipole expansions of the electric and magnetic potentials}\label{Multipole}
In the following we will denote $\vec{R} = (R, \Theta, \Phi)$ and $\vec{r} = (r, \theta, \phi)$ as the spherical coordinates of the nucleus and the electron respectively and $C_q^{(k)}(\theta, \phi)$ designates the normalised spherical harmonics defined by $C_q^{(k)}(\theta, \phi)\equiv \sqrt{\frac{4\pi}{(2k+1)}}Y_{k,q}(\theta, \phi)$ and $\Omega\equiv(\theta, \phi)$. In the following, for convenience, in relation to the previous section, we will use different symbols to designate electric and magnetic multipole moments: $Q_{20}^{\mathrm{(intr)}} \equiv 2{\cal Q}^{(2)}_{0}$ and $\mu_{\mathrm{tot}}^{\mathrm{(intr)}} \equiv M_z^{(1)}$.
\subsubsection{Electric Potential}
The electric potential generated by the nucleus through its charge density ($\rho(\vec{R})=e\rho_p(\vec{R})$ with $\rho_p$ the proton density given by \cref{eq_rho_q}). Note that, in the following discussion, the symbol $\rho$ denotes the nuclear charge density and should not be confused with the total nucleon density used in the previous section. The potential in which the electron moves then has the form
\begin{equation}
\phi(\vec{r}) = \frac{1}{4\pi \epsilon_0} \int \frac{\rho (\vec{R})}{|\vec{r}-\vec{R}|} d^{3}R\;,
\label{eq::phi_from_rho}
\end{equation}
and $\phi$ is solution of the Poisson equation $\Delta \phi = -\frac{\rho}{\epsilon_0}$. By using the usual multipolar expansion
\begin{equation}
\frac{1}{\left|{\vec r}-{\vec R}\right|}=\sum_{k=0}^{\infty} \frac{r_<^{k}}{r_>^{k+1}} \sum_{q=-k}^{k} (-1)^q C^{(k)}_{-q}(\Theta, \Phi) C^{(k)}_{q}(\theta, \phi) \;,
\end{equation}
the electric potential takes the form
\begin{equation}\begin{aligned}
\phi(\vec{r}) &=\frac{1}{4\pi \epsilon_0} \sum_{k=0}^{\infty} \frac{1}{r^{k+1}} \sum_{q=-k}^{k} (-1)^q C_q^{(k)}(\theta, \phi) ^{\rm in}{{\cal Q}}^{(k)}_{q} (r) \\            &+ \frac{1}{4\pi \epsilon_0} \sum_{k=0}^{\infty} r^{k} \sum_{q=-k}^{k} (-1)^q C_q^{(k)}(\theta, \phi) ^{\rm ex}{{\cal Q}}^{(k)}_{q} (r) \\
&=\sum_{k=0}^{\infty} \sum_{q=-k}^{k} \phi_{k,q}(r) C_q^{(k)}(\theta, \phi) \\
&=\sum_{k=0}^{\infty} \phi_{k}({\vec r})  \;,
\label{eq::apdxmultipole}
\end{aligned}\end{equation}
where we have defined
\begin{equation}
^{\rm in}{{\cal Q}}^{(k)}_{q} (r) \equiv \int_{R=0}^{r} \rho(\vec{R}) R^k C_{q}^{(k)}(\Theta, \Phi) d^{3}R
\label{eq::correcQ}
\end{equation}
and
\begin{equation}
^{\rm ex}{{\cal Q}}^{(k)}_{q} (r) \equiv \int_{R=r}^{\infty} \frac{\rho(\vec{R})}{R^{k+1}} C_{q}^{(k)}(\Theta, \Phi) d^{3}R \;.
\end{equation}
Let us recall the usual definition of the multipole electric moment
\begin{equation}
{{\cal Q}}^{(k)}_{q} \equiv \int_{R=0}^{\infty} \rho(\vec{R}) R^k C_{q}^{(k)}(\Theta, \Phi) d^{3}R \;.
\label{eq::elecmultipole}
\end{equation}
It is worth noting that if the electronic wave function does not penetrate the nucleus (i.e. $R\ll r$), one has $^{\rm in}{{\cal Q}}^{(k)}_{q} (r)={{\cal Q}}^{(k)}_{q}$ $\forall\; r$ since ${{\cal Q}}^{(k)}_{q}= \lim_{r \rightarrow +\infty} {}^{\rm in}{{\cal Q}}^{(k)}_{q}(r)$.
\paragraph{Electric monopole}\label{Electric monopole}
The electric monopole term is spherically symmetric, and takes the form
\begin{equation}\begin{aligned}
\phi_{0} (\vec{r}) &= \phi_{0} (r) = \phi_{0,0}(r) = \frac{1}{4\pi \epsilon_0} {C}^{(0)}_0(\theta, \phi) \left( \frac{{}^{\rm in}{{\cal Q}}^{(0)}_0(r)}{r}   + {}^{\rm ex}{{\cal Q}}^{(0)}_0(r) \right) \\
&= \frac{1}{4\pi \epsilon_0} \left( \frac{{}^{\rm in}{{\cal Q}}^{(0)}_0(r)}{r}   + {}^{\rm ex}{{\cal Q}}^{(0)}_0(r) \right) \;.
\label{eq::correctionmonopole}
\end{aligned}\end{equation}
In the non-penetrating case we would have had
\begin{equation}
\phi_{0} (\vec{r}) = \phi_{0} (r) = \phi_{0,0}(r) = \frac{1}{4\pi \epsilon_0} \frac{{{\cal Q}}^{(0)}_0}{r} \;,
\label{eq::monopole_np}
\end{equation}
where ${{\cal Q}}^{(0)}_0$ is simply the total charge $Ze$ of the nucleus.

%The empirical Fermi distribution parameters are computed as follows:
%
%\[
%t = 2.30 \; \mathrm{fm}
%\]
%
%\[
%a = \frac{t}{4 \ln(3)}
%\]
%
%\[
%c = \sqrt{\frac{5}{3} R_{\text{rms}}^2 - \frac{7}{3} (\pi a)^2}
%\]
%
%\[
%\rho_{0} = \frac{3}{4 \pi \tilde{c}^3 \left(1 + \left(\frac{\pi a}{\tilde{c}}\right)^2\right)}
%\]
%
%\[
%\rho(r) = \frac{Ze  \rho_{0}}{1 + \exp\left(\frac{r - \tilde{c}}{a}\right)}
%\]
%
%Physical dimension $[\rho]=\mathrm{C}\times\mathrm{fm}^{-3}$.

To assess the impact of a microscopic calculation of the nuclear charge density on the hyperfine interactions, we will compare the results obtained with those derived from a semi-empirical description of the density, which is the well-known Fermi distribution. The latter is computed as follows
\begin{equation}
\rho(R) = \frac{\rho_{0}}{1 + \exp\left(\frac{R - \tilde{c}}{\tilde{a}}\right)} \;,
\label{Fermi_distribution}
\end{equation}
with
\begin{equation}
\rho_{0} = \frac{3}{4 \pi \tilde{c \,}^3 \left(1 + \left(\frac{\pi \tilde{a}}{\tilde{c}}\right)^2\right)} \;.
\end{equation}
The parameters $\tilde{a}$, $\tilde{c}$ have been taken from \cite{Palffy2010}.
\paragraph{Electric quadrupole}
Due to the symmetry of the nucleus, the only non-zero component of the quadrupole tensors is $q=0$ (see \cref{Nuclear description}, and we recall $Q_{20}^{\mathrm{(intr)}} \equiv 2{\cal Q}^{(2)}_{0}$). The electric quadrupole term of the electric potential is
\begin{equation}
\phi_{2} (\vec{r}) = \frac{1}{4\pi \epsilon_0} {C}^{(2)}_0(\theta, \phi) \left( \frac{{}^{\rm in}{{\cal Q}}^{(2)}_0(r)}{r^3}   + r^2 \; {}^{\rm ex}{{\cal Q}}^{(2)}_0(r) \right) \;.
\label{eq::correctionmonopole}
\end{equation}
Again, without penetration, we have
\begin{equation}\begin{aligned}\label{eq:phi20}
\phi_{2} (\vec{r}) &=\frac{{C}^{(2)}_0(\theta, \phi) }{4\pi \epsilon_0} \frac{{{\cal Q}}^{(2)}_0}{r^3} \\
&=\phi_{2,0}(r) {C}^{(2)}_0(\theta, \phi) \;.
\end{aligned}\end{equation}
It is important to note that this potential is no longer spherically symmetric.
\subsubsection{Magnetic potential}
The magnetic field produced by the nucleus and which acts on the electrons of the atom through the potential vector $\vec{A}(\vec{r})$ in the Dirac equation (\ref{eq::Dirac}) is fully characterised by the nuclear current density $\vec{j}(\vec{R})$ and is given by \cref{Eq45,Eq46}.

In what follows, we give the most important expressions that we will need in the following and refer the reader to the appendix \cref{App_Multipole expansion of the magnetic vector potential} for the demonstrations that are more complex and technical than for the electric potential.

From Biot and Savart's law, which gives the expression for the magnetic field $\vec{B}$ as a function of the current density $\vec{j}$, and using the definition of the vector potential $\vec{B}=\vec{\nabla} \wedge \vec{A}$ and the Coulomb gauge $(\vec{\nabla}\cdot \vec{A}=0)$, we obtain the magnetic equivalent of the scalar potential, \cref{eq::phi_from_rho},
\begin{equation}
\vec{A}(\vec{r}) = \frac{\mu_0}{4\pi} \int \frac{\vec{j} (\vec{R})}{|\vec{r}-\vec{R}|} d^{3}R\;.
\label{eq::A_from_j}
\end{equation}
As with the scalar potential, we can perform a multipole expansion of $\vec{A}$ which is expressed as
\begin{equation}\begin{aligned}
\vec{A}(\vec{r}) &=\frac{\mu_{0}}{4 \pi}\sum_{k=0}^{\infty} \sum_{q=-k}^{k} (-1)^q \frac{(-i)}{\hbar k}\left[\vec{L}{C}^{(k)}_q(\theta, \phi) \right] \frac{{}^{\rm in}{M}_{-q}^{(k)}(r)}{r^{k+1}} \\
&+ \frac{\mu_{0}}{4 \pi} \sum_{k=0}^{\infty} \sum_{q=-k}^{k} (-1)^q \frac{(-i)}{\hbar k}\left[\vec{L}{C}^{(k)}_q(\theta, \phi)\right]   {}^{\rm ex}{M}_{-q}^{(k)}(r) \; r^k \;,
\end{aligned}\end{equation}
with
\begin{equation}
{}^{\rm in}M^{(k)}_{q} (r) \equiv -\frac{i}{\hbar (k+1)} \int_{R=0}^{r} \left[\vec{L}{C}^{(k)}_q(\Theta, \Phi)\right]  \cdot \vec{j}(\vec{R}) R^k d^{3}R \;,
\end{equation}
and
\begin{equation}
{}^{\rm ex}M^{(k)}_{q} (r) \equiv -\frac{i}{\hbar (k+1)} \int_{R=r}^{\infty} \left[\vec{L}{C}^{(k)}_q(\Theta, \Phi)\right] \cdot \vec{j}(\vec{R}) \frac{1}{R^{k+1}} d^{3}R \;.
\end{equation}
In the above expressions, we have defined $\vec{L} \equiv -i\hbar \vec{R} \wedge \vec{\nabla}$. The usual definition of magnetic multipole moments is as follows
\begin{equation}
M^{(k)}_{q} \equiv  -\frac{i}{\hbar (k+1)} \int_{R=0}^{\infty} \left[\vec{L}{C}^{(k)}_q(\Theta, \Phi)\right]  \cdot \vec{j}(\vec{R}) R^k d^{3}R =  {}^{\rm in}M^{(k)}_{q} (r\rightarrow \infty) \;,
\end{equation}
and, again, corresponds to the case without penetration of the electron wave function into the nucleus (i.e. $R \ll r$). For the dipole term, which is the only magnetic multipole considered in this work, we obtain the following expressions (see Appendix~\ref{App_Magnetic dipole moment})
\begin{equation}
\vec{A}(\vec{r}) = \frac{\mu_{0}}{4 \pi} \frac{{}^{\rm in}\vec{M}^{(1)} (r) \wedge \vec{r}}{r^3} + \frac{\mu_{0}}{4 \pi} \left({}^{\rm ex}\vec{M}^{(1)}(r) \wedge \vec{r}\right) \;,
\end{equation}
with
\begin{equation}
{}^{\rm in}\vec{M}^{(1)} (r) = \frac{1}{2}\int_{R=0}^{r} \vec{R}  \wedge   \vec{j}(\vec{R}) d^{3}R
\end{equation}
and
\begin{equation}
{}^{\rm ex}\vec{M}^{(1)} (r) = \frac{1}{2}\int_{R=r}^{\infty} \frac{\vec{R} \wedge \vec{j}(\vec{R})}{R^3} d^{3}R \;.
\end{equation}
In the situation where the penetration of the electron wave function inside the nucleus is neglected, we obtain the well-known formula
\begin{equation}
\vec{A}(\vec{r}) = \frac{\mu_{0}}{4 \pi} \frac{{}^{\rm in}\vec{M}^{(1)} \wedge \vec{r}}{r^3} \;,
\end{equation}
with
\begin{equation}
\vec{M}^{(1)}  = \frac{1}{2}\int_{R=0}^{\infty} \vec{R}  \wedge   \vec{j}(\vec{R}) d^{3}R \;.
\label{Dipole_Magnetic_Moment}
\end{equation}

\subsection{Hyperfine interactions}\label{sec Hyperfine interactions}
The unperturbed electron Hamiltonian takes into account the monopole term of the scalar potential \cref{eq::correctionmonopole} that has spherical symmetry. The following terms of the multipole expansions of $\phi$ and $\vec{A}$, i.e. the quadrupole term of the scalar potential and the dipole term of the vector potential, contribute very little and can therefore be treated using first-order stationary perturbation theory. The perturbation Hamiltonian reads as
\begin{equation}\begin{aligned}
{}^{\rm at}H_{\rm hf}&= -e\phi_2 (\vec{r}) + e c \vec{\alpha} \cdot \vec{A}(\vec{r}) \\
&\equiv {}^{\rm at}H_{\rm hf,quad} + {}^{\rm at}H_{\rm hf, dip}\;.
\end{aligned}\end{equation}
Recall that the total angular momentum of the nucleus is generally represented by the symbol $I$ and is called the ''nuclear spin". For electrons in atoms, we clearly distinguish between the electron spin, $\vec{S}$, and the electron orbital moment, $\vec{L}$, and combine them to obtain the total angular momentum, $\vec{J}$. However, nuclei often act as a single entity with their own angular momentum $\vec{I}$. Each nuclear spin has a nuclear magnetic moment. This produces magnetic interactions with its surroundings, leading to the magnetic hyperfine interaction. As detailed in Appendix~\ref{App_Hyperfine interactions:_dipole}, for an atom or ion with nuclear spin $I$ and electronic total angular momentum $J$, the value of the splitting of the electronic energy level, which is characterised by quantum numbers $n \kappa$, is, to first order in ${}^{\rm at}H_{\rm hf}$ given by 
\begin{equation}\begin{aligned}
W_F &=\bra{n \kappa F M_F} {}^{\rm at}H_{\rm hf} \ket{n \kappa F M_F} \\
&=\bra{n \kappa F M_F} {}^{\rm at}H_{\rm hf, dip} \ket{n \kappa F M_F} + \bra{n \kappa F M_F} {}^{\rm at}H_{\rm hf,quad} \ket{n \kappa F M_F} \\
&=\frac{1}{2} a \tilde{K} + b \frac{\frac{3}{2}\tilde{K}(\tilde{K}+1)-2I(I+1)J(J+1)}{2I(2I-1)2J(2J-1)}\;,
\label{eq:hyperfine_def}
\end{aligned}\end{equation}
where $\tilde{K}=F(F+1)-I(I+1)-J(J+1)$, $a$ and $b$ are the magnetic dipole and electric quadrupole structure constants. $F$ is the quantum number of the total angular momentum $\vec{F} = \vec{I} + \vec{J}$.

To illustrate this, let us mention the work \cite{Crespo1996} that measured the transition between the $F=4$ and $F=3$ levels of the $1s_{1/2}$ ground state of the hydrogen-like ion ${}^{165}\mathrm{Ho}^{66+}$.

\subsubsection{Magnetic dipole hyperfine structure constant}
The first term in \cref{eq:hyperfine_def} is the magnetic dipole hyperfine interaction, which is responsible for the splitting of one fine structure atomic energy level into a hyperfine multiplet labeled by $F$. Consequently, if no other interaction influences the hyperfine levels, and considering the magnetic dipole interaction to be weak compared to the other terms of the electronic Hamiltonian, the energy interval between adjacent $F$ sub-levels is $E_{F}-E_{F-1}= a F$. This is the so-called Landé interval rule. As a result, the hyperfine levels are equidistant in energy. We will see later that if we include the quadrupole term this property is no longer true. To illustrate the importance of the non-equidistance of hyperfine energy levels—which makes it possible to address individual hyperfine transitions—one can cite, in the field of quantum computing, the implementation of quantum algorithms in molecular complexes \cite{Godfrin2017}.

Following the multipole expansion of the magnetic vector potential of the nucleus and its influence on the electrons, we obtain the magnetic dipole hyperfine constant $a$ of electrons, with quantum numbers $n$ and $\kappa$, to be equal to (see Appendix~\ref{App_Hyperfine interactions:_dipole})
\begin{equation}
a(\vec{j},\rho) = \mathcal{U}(\kappa,j) \int_0^{\infty} \left( \frac{1}{r^2} \int_{R=0}^{r} \left[\vec{R} \wedge \vec{j}(\vec{R})\right]_z d^{3}R + r\int_{R=r}^{\infty} \frac{\left[\vec{R} \wedge \vec{j}(\vec{R})\right]_z}{R^3} d^{3}R \right)  P_{n\kappa}(r)Q_{n\kappa}(r)dr \;,
\label{eq:definition_a}
\end{equation}
where $\vec{R}$ are the coordinates of nuclear current and charge distribution, while $r$ is the electronic radial coordinate. $P_{n\kappa}(r)$ and $Q_{n\kappa}(r)$ are, respectively, the large and small radial components of the Dirac wavefunction of the electrons. $\vec{j}(\vec{R})$ is the electric current distribution of the nucleus, as defined in \cref{eq:jcurrent}. The subscript $z$ on the vector product terms indicates that only the $z$ component needs to be considered, as it corresponds to the quantisation axis common to nucleons and electrons. Finally, $\mathcal{U}(\kappa,j)$ is a multiplicative factor depending only on the angular part of the electronic many-body wavefunction. For configurations with two or more electrons it requires the use of group theoretical tools to compute. For ions with one single electron or muon (as studied in this work) we have $\mathcal{U}(\kappa,j)\propto \frac{\kappa}{j(j+1)}$ (see Appendix~\ref{App_Hyperfine interactions:_dipole}). The dependence of $a$ on the charge distribution $\rho$ is indirect, via the calculation of $P_{n\kappa}$ and $Q_{n\kappa}$. Indeed, $P_{n\kappa}$ and $Q_{n\kappa}$ are solutions of the Dirac equation containing the electric monopole potential produced by $\rho$ (see \cref{Electric monopole}). Evidently, the accuracy of the value of $a$ depends on the quality of the nuclear model used to evaluate $\vec{j}$ and $\rho$.

The calculation of $(\vec{j},\rho)$ based on a microscopic description of the nucleus has been described in detail in \cref{Nuclear description}. As a reminder we have
\begin{equation}
\vec{j}(\vec{R}) = \vec{j}^{(l)}_{p}(\vec{R}) + \vec{j}^{(l)}_{n}(\vec{R}) + \vec{j}_{p}^{(s)}(\vec{R}) + \vec{j}_{n}^{(s)}(\vec{R}) \;, \label{eq:jcurrent}
\end{equation}
where $\vec{j}^{(l,s)}_{p,n}$ are the convection (or orbital) and spin currents of protons and neutrons respectively. Note that only the proton current as an electric charge current will produce an orbital magnetic moment.

If we assume a point magnetic dipole (i.e. no penetration inside the nucleus can occur and hereafter referred to as ``pd''), we have the following simplifications $(\forall \; r \gg R)$
\begin{equation}
a_{\rm pd} = \mathcal{U}(\kappa,j) M^{(1),(\mathrm{spec})}_z \left[\int_0^{\infty} \frac{P_{n\kappa}(r)Q_{n\kappa}(r)}{r^2} dr\right] \;.
\end{equation}
If we add up all the constants, we obtain for a one-electron system
\begin{equation}
a_{\rm pd} = -\frac{e}{4\pi \varepsilon_0} \frac{1}{c} \frac{\mu_{\rm N}}{a_0^2}  \frac{\kappa}{j (j+1)} \frac{1}{I} M^{(1),(\mathrm{spec})}_z  \left[\int_0^{\infty} \frac{P_{n\kappa}(r)Q_{n\kappa}(r)}{r^2} dr\right] \;.
\end{equation}
In the above expressions $M^{(1),(\mathrm{spec)}}_z$ is expressed in units of $\mu_{\rm N}$ and $\left[\int_0^{\infty} \frac{P_{n\kappa}(r)Q_{n\kappa}(r)}{r^2} dr\right]$ in units of $a_0^{-2}$. Moreover, $\mathcal{I}^{M_1}_{\mathrm{intr/spec}}$ is the conversion factor from intrinsic nuclear dipole moment to nuclear dipole moment measured in the laboratory (see \cref{Nuclear description} and Appendix~\ref{App_Intrinsic vs. spectroscopic multipole moments} \footnote{Note that the notation is different in the Appendix~\ref{App_Intrinsic vs. spectroscopic multipole moments}: we have $M^{(1),(\mathrm{spec})}_z \equiv M^{(\mathrm{spec})}_{10}$}.) leading to the definition of the spectroscopic nuclear dipole moment $M^{(1),(\mathrm{spec})}_z \equiv \mathcal{I}^{M_1}_{\mathrm{intr/spec}}  M^{(1)}_z$.

In the extreme case of a hydrogen atom we have $M^{(1)}_z=g_p^{(s)} I \mu_{\rm N}$ and $\mathcal{I}^{M_1}_{\mathrm{intr/spec}}=1$ with $I=1/2$, leading to \cite{WR2007}
\begin{equation}
a_{\rm pd} = -g_p^{(s)} \frac{\kappa}{j (j+1)} \left[\int_0^{\infty} \frac{P_{n\kappa}(r)Q_{n\kappa}(r)}{r^2} dr\right] \times 13074.7 \; \mathrm{MHz} \;.
\label{dipol_constant}
\end{equation}

Furthermore, any change in the radial (spherically-averaged) charge distribution of the nucleus, will change the radial components $P_{n\kappa}(r)$ and $Q_{n\kappa}(r)$ of the electrons. Therefore $P_{n\kappa}(r)$ and $Q_{n\kappa}(r)$ are not the same for a point (charge) nucleus and a finite size one. We call the radial components of the wavefunctions under the consideration of a point nucleus $P^{0}_{n\kappa}$ and $Q^{0}_{n\kappa}$. It allows us to define the magnetic dipole hyperfine constant for a point-charge nucleus,
\begin{equation}
a_{\rm point} = \mathcal{U}(\kappa,j) M^{(1),(\mathrm{spec})}_z \int_0^{\infty} \frac{P^{0}_{n\kappa}(r)Q^{0}_{n\kappa}(r)}{r^2} dr \;.\label{eq:apoint1st}
\end{equation}
\subsubsection{Electric quadrupole hyperfine structure constant}
The second term in the expression \cref{eq:hyperfine_def} is obtained by evaluating the matrix element $\bra{n \kappa F M_F}^{\rm at}H_{\rm hf,quad}\ket{n \kappa F M_F}.$ After fairly lengthy calculations, which are detailed in the appendix \cref{App_Hyperfine interactions:_quadrupole}, we obtain for the electric quadrupole structure constant
\begin{align}
b &= 2 \frac{e}{4\pi\epsilon_0}\left(\frac{2j-1}{2j+2}\right) \mathcal{I}^{E_2}_{\mathrm{intr/spec}}\times 
\label{eq:definition_bwp}\\
&\int_0^{\infty} \left( \frac{1}{r^3} \int_{R=0}^{r} \rho(\vec{R}) R^2 C_{0}^{(2)}(\Theta, \Phi) d^{3}R + r^2 \int_{R=r}^{\infty} \frac{\rho(\vec{R})}{R^{3}} C_{0}^{(2)}(\Theta, \Phi) d^{3}R \right) \left(P_{n \kappa}^2(r) + Q_{n \kappa}^2(r) \right)dr \;,\notag
\end{align}
where $\mathcal{I}^{E_2}_{\mathrm{intr/spec}}$ is the conversion factor from intrinsic nuclear quadrupole moment to nuclear quadrupole moment measured in the laboratory (see \cref{Nuclear description} and Appendix~\ref{App_Intrinsic vs. spectroscopic multipole moments} \footnote{Note that the notation is different in the Appendix~\ref{App_Intrinsic vs. spectroscopic multipole moments}: we have ${\cal Q}^{(2),(\mathrm{spec})}_{0} \equiv {\cal Q}^{(\mathrm{spec})}_{20}$}.).

If we assume a point electric quadrupole (i.e. no penetration inside the nucleus can occur and hereafter referred to as ``pq''), we have the following simplifications
\begin{align}
b_{\rm pq} &=2  \frac{e}{4\pi\epsilon_0} \left(\frac{2j-1}{2j+2}\right) \mathcal{I}^{E_2}_{\mathrm{intr/spec}}
\int_{R=0}^{\infty} \rho(\vec{R}) R^2 C_{0}^{(2)}(\Theta, \Phi) d^{3}R
\int_0^{\infty} \frac{\left(P_{n \kappa}^2(r) + Q_{n \kappa}^2(r) \right)}{r^3}
dr \label{eq:definition_bnp} \\
&=2{\cal Q}^{(2),(\mathrm{spec})}_{0} \left(\frac{2j-1}{2j+2}\right) \left[ \int_0^{\infty} \frac{\left(P_{n \kappa}^2(r) + Q_{n \kappa}^2(r) \right)}{r^3}
dr \right] \times 234.965 \; \mathrm{MHz}
\label{quadru_constant}
\;,
\end{align}
where the intrinsic quadrupole moment ${\cal Q}^{(2)}_{0} \equiv \int_{R=0}^{\infty} \rho(\vec{R}) R^2 C_{0}^{(2)}(\Theta, \Phi) d^{3}R$ is expressed in terms of $|e| \times$ barn and $\left[ \int_0^{\infty} \frac{\left(P_{n \kappa}^2(r) + Q_{n \kappa}^2(r) \right)}{r^3} dr \right]$ in terms of $a_0^{-3}$. Let us stress that this term cancels out for orbitals with $j=\frac{1}{2}$ i.e. $ns_{\frac{1}{2}}, n'p_{\frac{1}{2}}$ with $n>0$ and $n'>1$. The constants in the formulae of \cref{dipol_constant,quadru_constant} alone show that the quadrupole correction is much smaller than the dipole correction.

\subsection{Isotope Shift}\label{sec Isotope Shift}
Variations in the neutron numbers affects the distributions of both the nuclear charge and magnetisation within the nucleus. This is manifested when one considers a finite-size nuclear model, but is not present in a point-like model. Especially when one considers energy levels of $1s_{1/2}$ and $2p_{1/2}$ electrons, whose wavefunctions penetrate the nucleus and allow for probing of its internal structure. This leads to two effects, for a finite model, that can be represented as correction factors to the point model: (1) The Breit-Rosenthal (BR) effect \cite{Rosenthal1932} which results from the extension of the nucleus' charge distribution and (2) Bohr-Weisskopf (BW) effect \cite{Bohr1950} which is due to the extension of magnetisation in the nucleus.
\subsubsection{Breit-Rosenthal correction}Breit and Rosenthal first considered the effect of the finite distribution of charge within the nucleus on the electronic wavefunctions. By comparing a finite nuclear model with the point charge model, they found that this effect on the wavefunctions would lead to a shift in the hyperfine interaction energy. The correction factor due to the BR effect can be expressed analytically as
\begin{equation}
a_{\rm pd} = a_{\rm point}(1 + \varepsilon_{\mathrm{BR}}) \;,
\label{eq:BRdefinition}
\end{equation}
where $a_{\rm pd}$ is the hyperfine constant for a nuclear model of finite charge distribution but point-dipole nuclear magnetisation, $a_{\rm point}$ for a point-charge nucleus, and $\varepsilon_{\mathrm{BR}}$ the BR correction factor.
\subsubsection{Bohr-Weisskopf correction}The distribution of nuclear magnetisation over the nuclear volume leads to the so-called BW effect \cite{Buttgenbach1984}, after Bohr and Weisskopf, who were the first to study the shift of the magnetic dipole constant from a point-dipole model. The correction factor is written as
\begin{equation}
a = a_{\rm pd} (1 + \varepsilon_{\mathrm{BW}}) \; , \label{eq:BW}
\end{equation}
where $a$ is the hyperfine constant for a finite nucleus, $a_{\rm pd}$ is the hyperfine constant for a point-dipole nucleus, and $\varepsilon_{\mathrm{BW}}$ is the BW correction factor.
\subsubsection{Web of Hyperfine Anomalies}
Recall that $\varepsilon_{\mathrm{BR}}$ and $\varepsilon_{\mathrm{BW}}$ quantify the shift in the hyperfine constant from point-like to non-point-like models for the distributions of nuclear charge and magnetisation, respectively. Moreover, to denote the dependency structure linking nuclear charge/current models to the resulting BR/BW corrections and isotope shifts, we use the term web of hyperfine anomalies.

We can extend this notion to consider the difference in shifts between two isotopes $(1)$ and $(2)$, the isotope shifts, which we denote by
\begin{equation}\begin{aligned}
^1\Delta^2_{\mathrm{BR}} &=\varepsilon^{(1)}_{\mathrm{BR}} - \varepsilon^{(2)}_{\mathrm{BR}} \\
^1\Delta^2_{\mathrm{BW}} &=\varepsilon^{(1)}_{\mathrm{BW}} - \varepsilon^{(2)}_{\mathrm{BW}} \;.
\end{aligned}\end{equation}
Let us now turn our attention to the distributions used to model the nuclei. The point-like model of \cref{eq:apoint1st} is unique (w.r.t. the effect of the nucleus on the electronic wavefunctions) in its definition of $\mu_{\rm pd}$, $P^0_{n\kappa}(r)$ and $Q^0_{n\kappa}(r)$, while non-point-like distributions do not allow for a unique definition of these quantities. Using the general form of the electric current distribution given in \cref{eq:jcurrent} allows us to consider a range of different models, each of which carries its own set of correction factors and isotope shifts. Given the virtually infinite set of possible models, we instead consider a finite set of subsets; i.e. we consider grouping them into categories. For simplicity, consider those models where the terms of \cref{eq:jcurrent} are determined solely by either (i) mathematical functions, or (ii) numeric (realistic) data. Specifically, we consider the following models, $m$, where $m \in \lbrace{\mathrm{pt}, \mathrm{sp}, \mathrm{fm}, \mathrm{hf}\rbrace}$, corresponding to the following: a point charge \cref{eq::monopole_np} ($m=\mathrm{pt}$), a uniformly charged sphere ($m=\mathrm{sp}$), a Fermi function distribution ($m=\mathrm{fm}$), and our mean-field -- HFBCS -- approach ($m=\mathrm{hf}$).
We summarise these considerations in \cref{tab:hyperfineanomalies}.

\begingroup
  \setlength{\arrayrulewidth}{\origarrayrulewidth}
  \setlength{\tabcolsep}{\origtabcolsep}
  \renewcommand{\arraystretch}{\origarraystretch}
\begin{table}[htbp!]
\centering
\begin{tabular}{|c|c|c|}
    \hline
        \diagbox[width=15.5em]{$\vec{j}(\vec{R})$}{$\rho(\vec{R})$} & \textbf{Point charge} & $\underset{{m}\in\lbrace{\text{\rm sp, fm, hf}\rbrace}}{\textbf{Charge distribution model}}$ \\
        \hline
         \textbf{Point dipole} & $\mathtextover[c]{1}{\text{NN/AA}}$ \tikzmark{a} & \tikzmark{b}$\mathtextover[c]{(1-\varepsilon^{m}_{\text{BR}})}{(1-\varepsilon^{m}_{\text{BR}}) (1-\varepsilon_{\text{BW}})A}$\tikzmark{c} \\
        \hline
     \textbf{Current distribution model} & N/A & $\mathtextover[c]{(1-\varepsilon^{m}_{\text{BR}}) (1-\varepsilon_{\text{BW}})}{A(1-\varepsilon^{\text{model}}_{\text{BR}}) (1-\varepsilon_{\text{BW}})A}$\tikzmark{d} \\
    \hline
\end{tabular}
\begin{tikzpicture}[overlay, remember picture, shorten >=.5pt, shorten <=.5pt, transform canvas={yshift=0.2\baselineskip}]
    \draw [->] ([yshift=0pt]{pic cs:a}) -- ({pic cs:b});
    \draw [->] ({pic cs:c}) [bend left] to ({pic cs:d});
  \end{tikzpicture}
  \caption{Table summarising the composition of hyperfine anomalies due to increasingly realistic models for nuclear charge and current distributions. Note that: (i) $\varepsilon^{m}_{\mathrm{BR}}$ depends only on the chosen model for charge distribution, namely $P^{m}$, $Q^{m}$ and $\rho(\vec{R})$, and (ii) $\varepsilon^{m}_{\mathrm{BW}}$ depends both on the charge and current distribution models, namely $P^{m}$, $Q^{m}$ and $\vec{j}(\vec{R})$. Therefore a logical order of composing the effects is indicated by the arrows. Note that it is not physically meaningful to consider a nucleus having a point charge but finite magnetisation.
\label{tab:hyperfineanomalies}}
\end{table}
\endgroup

\subsubsection{Hyperfine Anomaly}
\paragraph{Magnetic Dipole}
Having defined the shift in magnetic hyperfine interaction energies due to different models of nuclear charge and magnetisation distributions, it is possible to compose these two effects by substituting \cref{eq:BRdefinition} into \cref{eq:BW}, resulting in
\begin{equation}
a = a_{\rm point} (1 + \varepsilon_{\mathrm{BR}})(1 + \varepsilon_{\mathrm{BW}}) \;. \label{eq:BRBW}
\end{equation}
Now, suppose that $a$ is to be obtained from a realistic (experimental or numerically simulated) nuclear model. In general, this is not feasible for complex atoms due to the difficulty in calculating the electronic radial wavefunctions with sufficient accuracy \cite{Buttgenbach1984}. In practice one may obtain $a$ experimentally (or computationally), but the uncertainties in calculating $a_{\rm point}$ are too large to reasonably produce the correction factors. Therefore, it is necessary to instead consider the differences of these effects between two isotopes of the same atom, referred to as the \textit{hyperfine anomaly}. In doing so, it is possible to effectively cancel out the uncertainties contained in the $a_{\rm point}$ terms. Explicitly, consider \cref{eq:BRBW} for isotopes $a^{(1)}$ and $a^{(2)}$
\begin{equation}\begin{aligned}
a^{(1)} &=a^{(1)}_{\rm point} (1 + \varepsilon^{(1)}_{\mathrm{BR}})(1 + \varepsilon^{(1)}_{\mathrm{BW}}) \\
a^{(2)} &=a^{(2)}_{\rm point} (1 + \varepsilon^{(2)}_{\mathrm{BR}})(1 + \varepsilon^{(2)}_{\mathrm{BW}}) \;.
\end{aligned}\end{equation}
Taking the ratio of these two quantities, we obtain
\begin{equation}
\frac{a^{(1)}}{a^{(2)}} = \frac{a^{(1)}_{\rm point} (1 + \varepsilon^{(1)}_{\mathrm{BR}})(1 + \varepsilon^{(1)}_{\mathrm{BW}})}{a^{(2)}_{\rm point} (1 + \varepsilon^{(2)}_{\mathrm{BR}})(1 + \varepsilon^{(2)}_{\mathrm{BW}})} \;. \label{eq:a1a2}
\end{equation}
Here, recall that the $a^{(i)}_{\rm point}$ is calculated from (to simplify the notation, we have used $\mu_{\rm N}^{(i)}$ instead of $M^{(1)}_z$ to specify the magnetic dipolar moment of the isotope $(i)$)
\begin{equation}\begin{aligned}
a^{(i)}_{\rm point} &=\frac{\mu_{\rm N}^{(i)}}{I^{(i)}}\mathcal{U}(\kappa,j) \int_0^{\infty} \frac{P^{0}_{n\kappa}(r)Q^{0}_{n\kappa}(r)}{r^2} dr \\
&=g_{I}^{(i)} \mu_{\rm N} \mathcal{U}(\kappa,j) \int_0^{\infty} \frac{P^{0}_{n\kappa}(r)Q^{0}_{n\kappa}(r)}{r^2} dr \;, \label{eq:apoint}
\end{aligned}\end{equation}
with $\mu_{\rm N}^{(i)} = g^{(i)} I^{(i)}_{\phantom{b}} \mu_{\rm N}$, $\mu_{\rm N}$ being the nuclear magneton and $g^{(i)}$ the gyromagnetic factor of the nucleus $(i)$ of spin $I^{(i)}$. Note that since $P^{0}_{n\kappa}(r)$ and $Q^{0}_{n\kappa}(r)$ are the wavefunctions calculated with a point nucleus model, they are the same for both isotopes. Likewise, since $\mathcal{U} (\kappa, j)$ only depends on the electrons, it is also identical for either isotope. Thus, substituting \cref{eq:apoint} into \cref{eq:a1a2} we obtain:
\begin{equation}\begin{aligned}
\frac{a^{(1)}}{a^{(2)}} &= \frac{g^{(1)}(1 + \varepsilon^{(1)}_{\mathrm{BR}})(1 + \varepsilon^{(1)}_{\mathrm{BW}})}{g^{(2)}(1 + \varepsilon^{(2)}_{\mathrm{BR}})(1 + \varepsilon^{(2)}_{\mathrm{BW}})} \;.
\end{aligned}\end{equation}
We define $W_F^{(i)} \equiv \bra{I^{(i)}I^{(i)}JJ}{}^{\rm at}H_{\rm hf, dip}\ket{I^{(i)}I^{(i)}JJ} = W_{F=I^{(i)}+J} = a_{\rm point}^{(i)} I^{(i)} J$, the hyperfine splitting of the state of maximal projection. Under the point nucleus approximation, the ratio in the hyperfine splitting $W_F^{(i)}$ between two isotopes should then be:
\begin{equation}\begin{aligned}
\frac{W_F^{(1)}}{W_F^{(2)}} &=\frac{a_{\rm point}^{(1)} I^{(1)}J}{a_{\rm point}^{(2)} I^{(2)}J} \\
&=\frac{g^{(1)}I^{(1)}_{\phantom{b}}}{g^{(2)}I^{(2)}_{\phantom{b}}}\\
&=\frac{\mu_{\rm N}^{(1)}}{\mu_{\rm N}^{(2)}} \;.
\end{aligned}\end{equation}
Therefore we define the hyperfine anomaly:
\begin{equation}
^1\Delta^2 = \frac{W_F^{(1)}}{W_F^{(2)}}\frac{\mu_{\rm N}^{(2)}}{\mu_{\rm N}^{(1)}} - 1,
\end{equation}
which is therefore always $0$ in a point nucleus model. However, considering a non-point nucleus model, we have
\begin{equation}\begin{aligned}
\frac{W_F^{(1)}}{W_F^{(2)}} &=\frac{a^{(1)} I^{(1)}J}{a^{(2)} I^{(2)}J} \\
&=\frac{a_{\rm point}^{(1)} I^{(1)}_{\phantom{b}}J}{a_{\rm point}^{(2)} I^{(2)}_{\phantom{b}}J}\times\frac{(1 + \varepsilon^{(1)}_{\mathrm{BR}})(1 + \varepsilon^{(1)}_{\mathrm{BW}})}{(1 + \varepsilon^{(2)}_{\mathrm{BR}})(1 + \varepsilon^{(2)}_{\mathrm{BW}})}\\
&=\frac{\mu_{\rm N}^{(1)}}{\mu_{\rm N}^{(2)}}\times\frac{(1 + \varepsilon^{(1)}_{\mathrm{BR}})(1 + \varepsilon^{(1)}_{\mathrm{BW}})}{(1 + \varepsilon^{(2)}_{\mathrm{BR}})(1 + \varepsilon^{(2)}_{\mathrm{BW}})}\;.
\end{aligned}\end{equation}
Therefore the full hyperfine anomaly can be written as
\begin{equation}
^1\Delta^2 = \frac{W_F^{(1)}}{W_F^{(2)}}\frac{\mu_{\rm N}^{(2)}}{\mu_{\rm N}^{(1)}} - 1 = \frac{(1 + \varepsilon^{(1)}_{\mathrm{BR}})(1 + \varepsilon^{(1)}_{\mathrm{BW}})}{(1 + \varepsilon^{(2)}_{\mathrm{BR}})(1 + \varepsilon^{(2)}_{\mathrm{BW}})} - 1
\end{equation}
To first order approximation, this becomes:
\begin{equation}
^1\Delta^2 = (1 + \varepsilon^{(1)}_{\mathrm{BR}} - \varepsilon^{(2)}_{\mathrm{BR}})(1 + \varepsilon^{(1)}_{\mathrm{BW}}  - \varepsilon^{(2)}_{\mathrm{BW}}) - 1 = (1+{^1\Delta^2}_{\mathrm{BR}})(1+{^1\Delta^2}_{\mathrm{BW}}) - 1
\end{equation}
where $^1\Delta_{E}^2 = \varepsilon^{(1)}_E - \varepsilon^{(2)}_E$ for $E\in \lbrace \mathrm{BR}, \mathrm{BW} \rbrace$.
\paragraph{Additional Contributions to the Isotope Shift}
With particular reference to the work of B{\"u}ttgenbach \cite{Buttgenbach1984}, let us now consider the hyperfine anomaly from the perspective of experiment. We have thus far followed a constructive approach: We obtained $^1\Delta^2$ starting from a theoretical description of the electronic wavefunctions, $P_{n\kappa}(r)$ and $Q_{n\kappa}(r)$ and the nuclear charge and current distributions, $\vec{j}(\vec{R})$. However, these quantities are not accessible by experiment. In practice, a goal is to investigate the nuclear current distributions through (spectroscopic) measurements of the hyperfine constant. These may then be compared with existing models in order to infer details about the nuclear structure. Therefore, through the hyperfine anomaly it becomes possible to investigate isotopic effects on nuclear structure, and thereby test the validity of theoretical models of nuclear structure. Additionally, recall that only electrons in the $s$ and $p_{1/2}$ orbitals have wavefunctions with non-zero probability amplitude within the nucleus. These are the only contact terms contributing to the BW Effect, and thus the hyperfine anomaly may also be used to identify contact and non-contact contributions to the hyperfine interaction.

In practice, however, the BR and BW Effects are not the only ones that contribute to the experimentally measured hyperfine anomaly, $^1\Delta^2_{\rm exp}$.

Additional isotopic variations manifest themselves most prominently in mass corrections, due to the finite mass of the nucleus (keep in mind that the relativistic corrections are already included in the relativistic electronic wavefunctions, $P_{n\kappa}(r)$ and $Q_{n\kappa}(r)$). Accounting for the motion of the nucleus-electron system about the common center of mass will result in reduced mass corrections to (i) the electronic wavefunctions, $(1 + m_e/m_{\rm N})^{-3}$ with $m_{\rm N} \approx A m$ and (ii) the electronic orbital $g$-factor. However, it is important to note that the magnitude of these effects scales inversely with the mass of the nucleus. This makes it possible to neglect them when considering heavy elements, such as the lanthanides.

Finally, another set of contributions appears due to higher-order corrections to the magnetic hyperfine interaction energy. Indeed, the energy correction, as given in \cref{eq:hyperfine_def}, is only a first-order approximation (assuming $J$ to be a good quantum number). Higher-order terms (off-diagonal terms in the hyperfine interaction Hamiltonian) exist due to effective hyperfine interactions between different fine structure levels, via the nucleus. These interactions make $J$ no longer a good quantum number. Thus second-order perturbation theory is needed to account for this.
\paragraph{Electric Quadrupole}\label{Electric Quadrupole}
As in the magnetic case, we investigate how the finite size of the nucleus affects the quadrupole correction to the hyperfine interaction. For this purpose, we introduce the quantity
\begin{equation}
{\Delta B}^{m} \equiv \frac{b_{\mathrm{np}}^{\mathrm{pt}} - b_{\mathrm{wp}}^{m}}{b_{\mathrm{np}}^{\mathrm{pt}}} ;,\label{eq:deltaBm}
\end{equation}
where $m \in \lbrace{\mathrm{pt}, \mathrm{sp}, \mathrm{fm}, \mathrm{hf}\rbrace}$. This dimensionless ratio quantifies the penetration corrections for different nuclear-structure models, ranging from the simplest point-charge model (pt) to the most sophisticated Hartree–Fock–BCS model (hf). The superscript $m$ indicates that the quadrupole constant is evaluated using the radial wave functions $P_{n\kappa}$ and $Q_{n\kappa}$ obtained from the spherically symmetric monopole potential \cref{eq::correctionmonopole}, generated by the nuclear charge distribution of the corresponding model (see \cref{Fig06}). Specifically, we consider a point charge \cref{eq::monopole_np} ($m=\mathrm{pt}$), a uniformly charged sphere ($m=\mathrm{sp}$), a Fermi distribution ($m=\mathrm{fm}$), and an HFBCS distribution ($m=\mathrm{hf}$). The subscript $\mathrm{n}$ specifies whether electron penetration into the nucleus is taken into account. When penetration is included ($n=\mathrm{wp}$), the quadrupole constant $b$ is computed using \cref{eq:definition_bwp}. When penetration is neglected ($n=\mathrm{np}$), one has $b = b_{\rm pq}$ and \cref{eq:definition_bnp} is used instead.

To quantify the effect of the finite nucleus size on the quadrupole hyperfine correction, we also define the quantity
\begin{equation}
{\Delta B}_{\mathrm{np}}^{\mathrm{wp},m} \equiv \frac{b_{\mathrm{np}}^{m}-b_{\mathrm{wp}}^{m}}{b_{\mathrm{np}}^{m}} \;.
\end{equation}
The latter quantity gives us information about the influence of the model used to describe the charge distribution of the nucleus (through the radial wave functions) on the correction of the quadrupole moment due to penetration into the nucleus. Note that ${\Delta B}^{\mathrm{pt}}={\Delta B}_{\mathrm{np}}^{\mathrm{wp},\mathrm{pt}}$.

In order to quantify the impact on the quadrupole constant of a completely microscopic description (HFBCS) compared with the Fermi model of the nucleus which is the most accurate of the non-microscopic models, we define the quantity
\begin{equation}
\Delta B \equiv \frac{b_{\mathrm{wp}}^{\mathrm{fm}}-b_{\mathrm{wp}}^{\mathrm{hf}}}{b_{\mathrm{wp}}^{\mathrm{fm}}} \;.
\end{equation}

Finally, the isotopic effect on $b$ for dysprosium is quantified by evaluating the quantity
\begin{equation}
{\Delta B}_{161}^{163} \equiv \frac{b_{\mathrm{wp}}^{\mathrm{hf}}(163)-b_{\mathrm{wp}}^{\mathrm{hf}}(161)}{b_{\mathrm{wp}}^{\mathrm{fm}}(161)} \;.
\end{equation}

\section{Results and discussions}\label{Results and discussions}
\subsection{Nucleon density distributions and nuclear currents}\label{subsec_res_nuc}

\begin{figure*}[ht!]
\begin{center}
\includegraphics[width=.9\textwidth,angle=0]{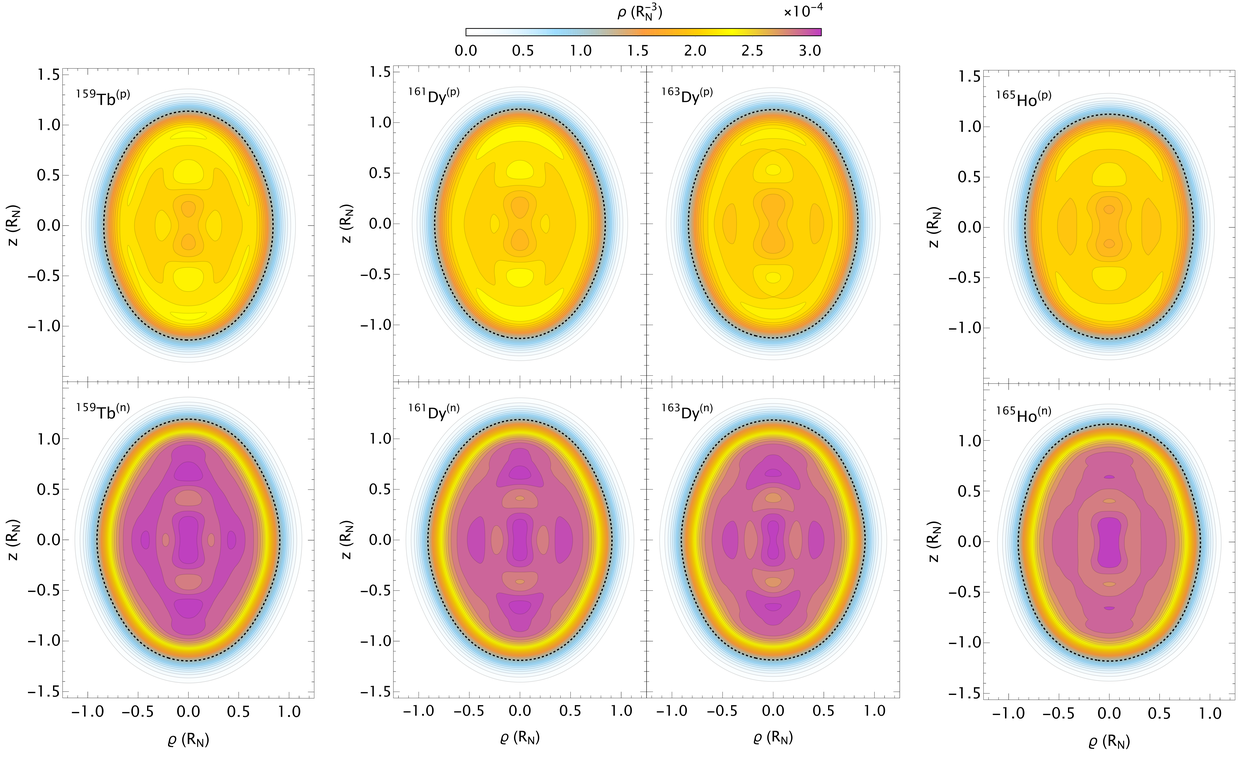}
\caption{Nucleon density distributions, $\rho(\vec{R})$, ($(p)$ above: protons, $(n)$ below: neutrons) of $^{159}_{65}$Tb (left), $^{161}_{66}$Dy and $^{163}_{66}$Dy (center), and $^{165}_{67}$Ho (right) on logarithmically spaced contour curves. The figures are plotted in signed cylindrical coordinates $(z,\varrho)$, on a planar slice through the $z-$axis. The spatial coordinates are in units of $R_{\rm N}$, the empirical nuclear radius (see text) for each isotope, leading to a density distribution in units of $R_{\rm N}^{-3}$. The contours represent curves of \sam{equivalent}{equal} density, with numerical values as indicated by the legend above. The dashed curve indicates the value of the spherically-averaged monopole distribution at a distance of $1 \times R_{\rm N}$, representing the outer shell of the nucleus.}\label{Fig03}
\end{center}

\end{figure*}
In this section, lengths are normalised by considering the nucleus as a homogeneously charged sphere whose radius is given by \begin{equation}R_{\rm N}=R_{\rm N}^{\text{(emp)}}=\sqrt{\frac{5}{3}R^2_{\mathrm{rms}}},\end{equation} where $R_{\mathrm{rms}}$ is the root mean square (rms)
charge radius of the nucleus given by the empirical
formula~\cite{Palffy2010}
\begin{equation}
  R_{\mathrm{rms}} = (0.836 A^{1/3} + 0.57)\mbox{ fm.}
  \label{R_rms}
\end{equation}
The resulting values of $R_{\rm N}$ (of the order of 6.6~fm
for the considered nuclei) are displayed in \cref{tab_R_Q}.

Here we discuss the results of nuclear-structure calculations obtained
using the microscopic model described in \cref{Nuclear
  description}. In particular, we focus on the nucleon densities
$\rho_q(\vec{R})$ as well as the convection (orbital)
$\vec{j}_{q}^{(\ell)}(\vec{R})$ and spin $\vec{j}_{q}^{(s)}(\vec{R})$
current densities for neutrons ($q=n$) and protons ($q=p$), as
functions of the nuclear coordinates $\vec{R}$. Our analysis
emphasises the two dysprosium isotopes \ce{^{161}_{66}Dy} and
\ce{^{163}_{66}Dy}, while also including complementary results for
\ce{^{159}_{65}Tb} and \ce{^{165}_{67}Ho} for the sake of
\sam{completeness}{comparison with odd-$Z$ nuclei}.

% Since at the end of the self-consistent mean-field calculation all the single-particle wave functions of the neutron and proton are known, it is possible to calculate any local density, such as neutron and proton densities and currents, and any global quantity, such as electric and magnetic multipole moments.

In \cref{Fig03} we show the proton (row of upper panels) and neutron (row of lower panels) densities $\rho_p(\vec R)$ and $\rho_n(\vec R)$ expressed in signed cylindrical coordinates $(z,\varrho,\varphi),\ \text{with }\varrho\in\mathbb{R},\ \phi\in[0,\pi),\ z\in\mathbb{R}$, the slice along $\varphi=0$ is shown,  and represented with logarithmically spaced contour curves for $^{159}_{65}$Tb (left), $^{161}_{66}$Dy and $^{163}_{66}$Dy (center), and $^{165}_{67}$Ho (right).

Contrary to a simplified model, in which such densities would be constant within a certain radius and fall off exponentially beyond a certain constant radius, one notices that nucleon densities show some structure in the nuclear interior and quickly decrease to 0 as the distance to the center-of-mass (at $z=\varrho = 0$) increases. Moreover one notices a deformed pattern of the isodentity contours, revealing a prolate shape for the four considered nuclei. Finally, it is worth noting that, particularly in the inner region of the nuclei, there are clear differences in the proton and neutron densities.

% In the same way, one can investigate the behaviour of the proton and neutron current densities shown respectively on Figs. \cref{Fig04} and \cref{Fig05}.

\begin{figure*}[h]
  \includegraphics[width=\textwidth,angle=0]{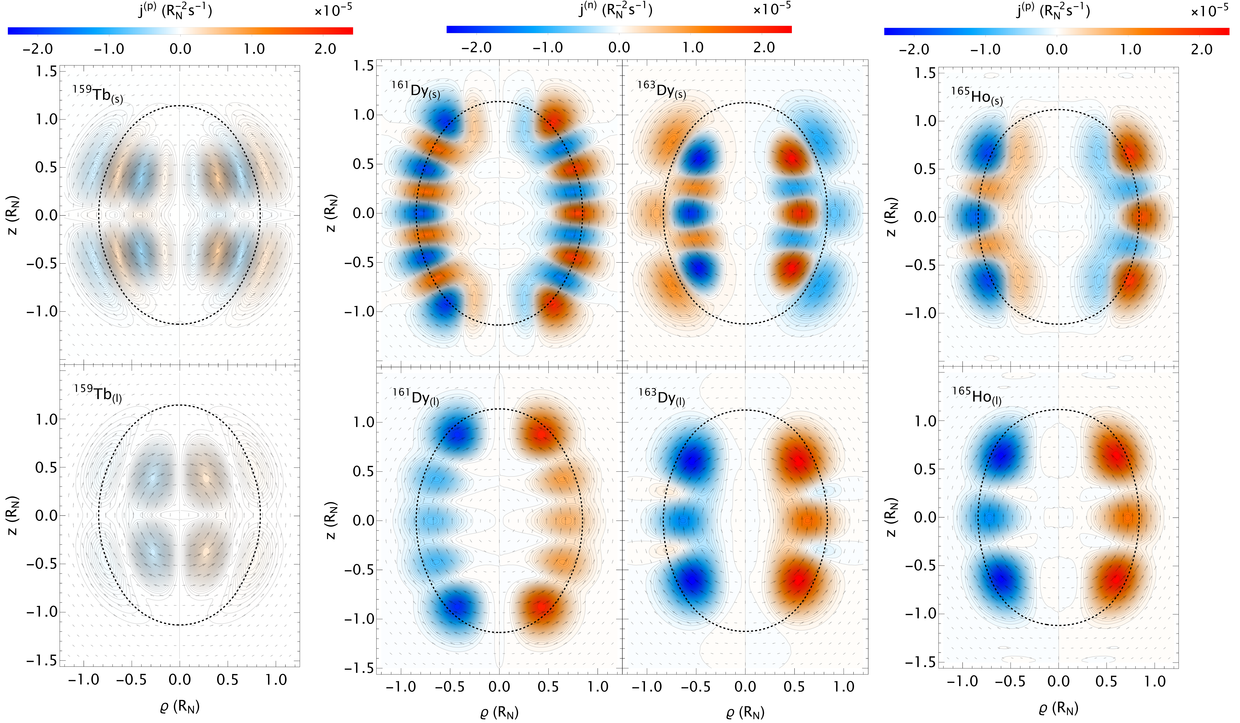}
  \caption{(color online) Two-dimensional color maps of the orthoradial component of
    the spin current density $j_{q,\varphi}^{(s)}(\varrho,z)$ (top row)
    and orbital current density $j_{q,\varphi}^{(\ell)}(\varrho,z)/g_{\bar q}^{(\ell)}$
    (bottom row) for the charge state $q$ identical to the one of the
    unpaired nucleon in the ground states of the two odd-proton nuclei
    $^{159}$Tb and $^{165}$Ho and the two odd-neutron nuclei
    $^{161}$Dy and $^{163}$Dy. \label{fig_currents_same_isospin}}
\end{figure*}

\begin{figure*}[h]
  \includegraphics[width=\textwidth,angle=0]{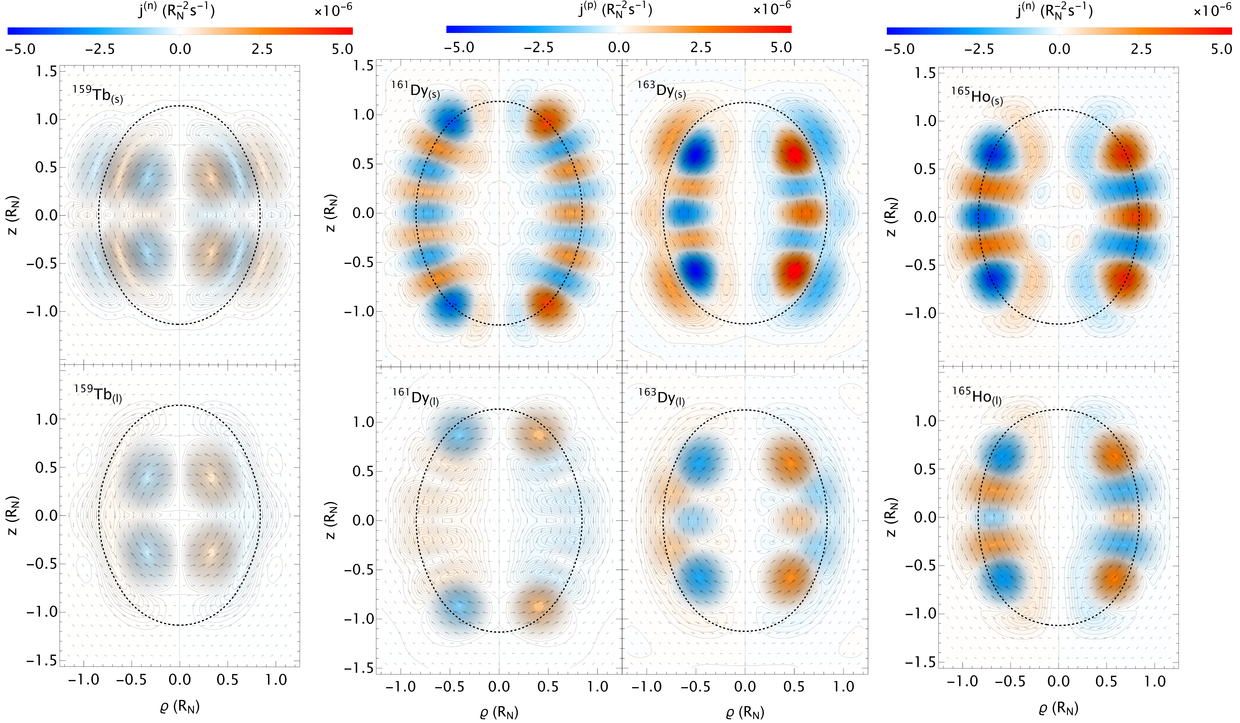}
  \caption{(color online) Two-dimensional color maps of the orthoradial component of
    the spin current density $j_{\bar q,\varphi}^{(s)}(\varrho,z)$ (top row)
    and orbital current density $j_{\bar q,\varphi}^{(\ell)}(\varrho,z)/g_{\bar q}^{(\ell)}$
    (bottom row) for the charge state $\bar q$ different from the one of the
    unpaired nucleon in the ground states of the two odd-proton nuclei
    $^{159}$Tb and $^{165}$Ho and the two odd-neutron nuclei
    $^{161}$Dy and $^{163}$Dy.
     \label{fig_currents_opposite_isospin}}
\end{figure*}

Next we investigate the structure of the proton and neutron current densities depending on the charge state $q$ of the unpaired nucleon. In \cref{fig_currents_same_isospin}, the orthoradial components of the spin current (defined by \cref{js_q_phi}) and the orbital current (defined by \cref{jl_q_phi} setting $g_n^{(\ell)}$ to $1$, for illustrative and comparative purposes, as otherwise the neutron orbital current is vanishing for the neutrons) are represented as two-dimensional color maps of the currents for the nucleons having charge state $q$ the same as that of the unpaired nucleon, whereas \cref{fig_currents_opposite_isospin} shows the currents for the nucleons having charge state $\bar q$ different from that of the unpaired nucleon. For a nucleus with an odd number of neutrons and an even number of protons, such as $^{161}$Dy or $^{163}$Dy, we have $q = n$ and $\bar q = p$, whereas for $^{159}$Tb and $^{165}$Ho we have $q = p$ and $\bar q = n$.
This choice has been made because the time-odd local densities $\vec s_q$ and $\vec j_q$ involved, respectively, in the spin current and the convection current, are dominated by the contribution of the unpaired nucleon. Taking the example of the spin current density, we can write the orthoradial component for the charge states $q$ as
\begin{align}
  j_{q,\varphi}^{(s)} & = j_{\rm odd,\varphi}^{(s)} +
  j_{q-\rm core,\varphi}^{(s)}
\end{align}
where the subscript ``odd'' refers to the contribution of the unpaired nucleon while the subscript ``$q$-core'' corresponds to the contribution of the remaining nucleons (called ``core'') of the same charge state as the unpaired nucleon. By definition the current for the charge state $\bar q$ receives only contributions from the core (nucleons with isospin opposite to the one of the unpaired nucleon). If the blocking procedure to describe odd nuclei were not self-consistent, the core contribution to the time-odd local densities would vanish and so would $j_{q-\rm core,\varphi}^{(s)}$ and $j_{q-\rm core,\varphi}^{(\ell)}$.

\begin{figure*}[h]
  \includegraphics[width=\textwidth,angle=0]{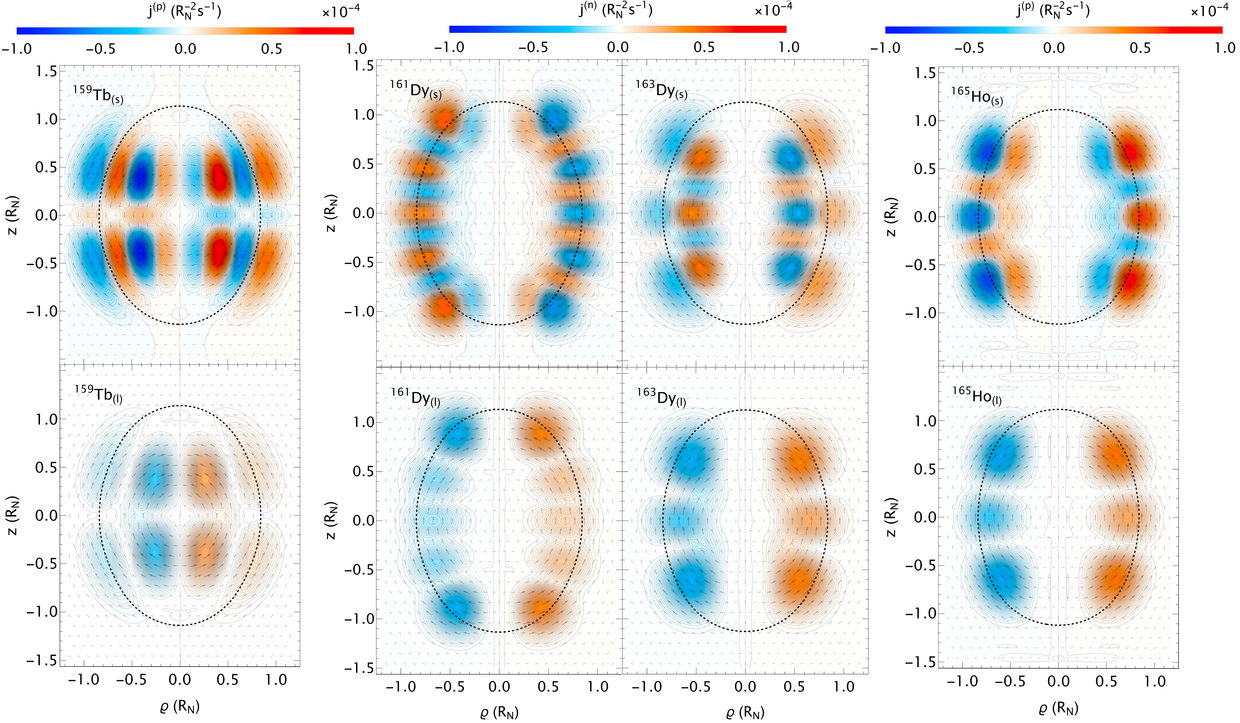}
  \caption{(color online) Two-dimensional color maps of the orthoradial component of
    the spin current density $j_{\rm odd,\varphi}^{(s)}(\varrho,z)$ (top row)
    and orbital current density $j_{\rm odd,\varphi}^{(\ell)}(\varrho,z)/g_q^{(\ell)}$
    (bottom row) of the unpaired nucleon in the ground states of the two odd-proton nuclei
    $^{159}$Tb and $^{165}$Ho and the two odd-neutron nuclei
    $^{161}$Dy and $^{163}$Dy. \label{fig_currents_blocked}}
\end{figure*}

\begin{figure*}[h]
  \includegraphics[width=\textwidth,angle=0]{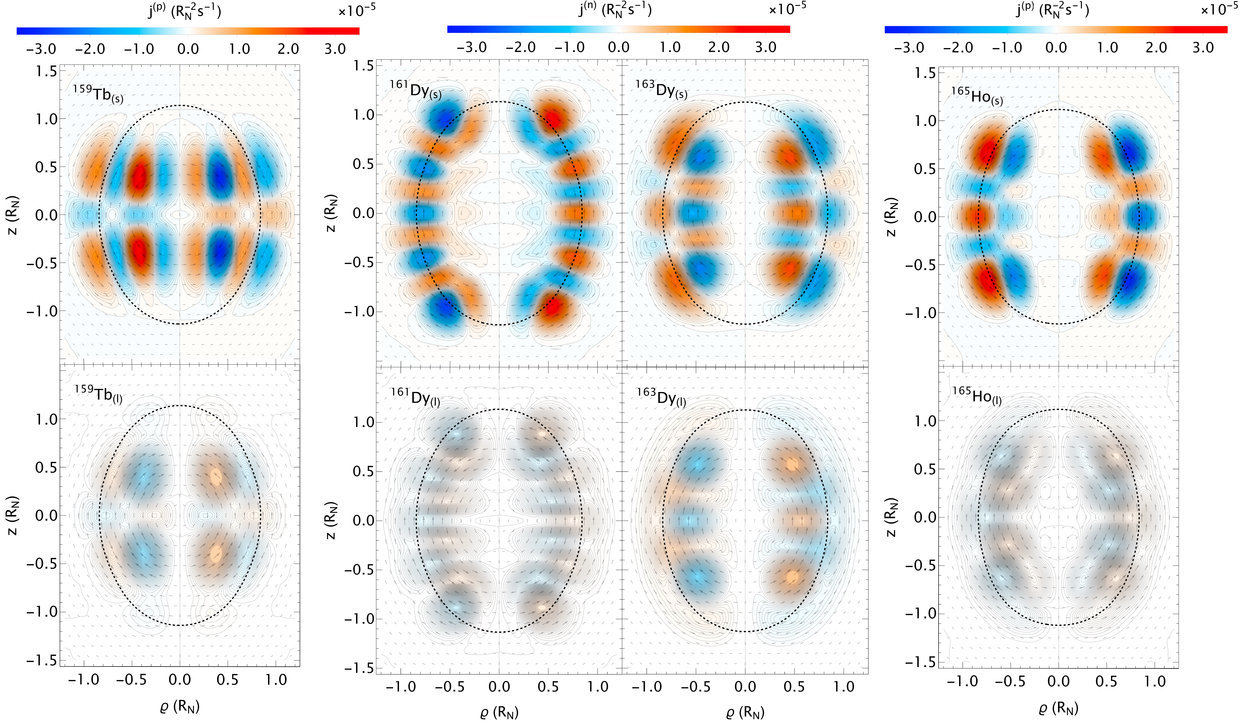}
  \caption{(color online) Two-dimensional color maps of the orthoradial component of
    the spin current density $j_{q-\rm core,\varphi}^{(s)}(\varrho,z)$ (top row)
    and convection current density $j_{q-\rm core,\varphi}^{(\ell)}(\varrho,z)$
    without the $g_q^{(\ell)}$ factor
    (bottom row) of the
    core with same isospin as the unpaired nucleon in the ground states of the two odd-proton nuclei
    $^{159}$Tb and $^{165}$Ho and the two odd-neutron nuclei
    $^{161}$Dy and $^{163}$Dy.
     \label{fig_currents_core_same_isospin}}
\end{figure*}

As can be seen in \cref{fig_currents_blocked,fig_currents_core_same_isospin}, displaying the contributions to
both types of currents from, respectively, the unpaired nucleon and the core nucleons with the same isospin as the unpaired one, one always find the latter weaker than the former by a factor of 3 for the spin current and by a factor of 10 for the orbital current. Moreover the core contribution to the spin current is of the opposite sign to that of the unpaired nucleon.
This means that the gross features of the currents result from the structure of the wave function of the unpaired nucleon. Since this wave function is dominated by a single deformed harmonic-oscillator component (in cylindrical coordinates), one can attempt to explain the number of local extrema of $j_{q,\varphi}^{(\ell)}$ and $j_{q,\varphi}^{(s)}$ in terms of the Nilsson quantum numbers $\Omega^{\pi}[N n_z \Lambda]$ of the unpaired nucleon. Within this approximation, the projection $\sigma = \Omega - \Lambda$ of spin angular momentum on the symmetry axis becomes a good quantum number and so do $N$, $n_z$ and $\Lambda$. In \cref{jl_q_phi,js_q_phi} the sum over $k$ is reduced to the term corresponding to the unpaired nucleon, for which $\phi_k^{(-\sigma)}(\varrho,z) = 0$ and $\phi_k^{(\sigma)}(\varrho,z) \approx \psi_{Nn_z\Lambda}(\varrho,z)$, a cylindrical harmonic-oscillator wave function defined by
\begin{equation}
  \psi_{Nn_z\Lambda}(\varrho,z) = \mathcal N H_{n_z}(\xi)
  \eta^{\Lambda/2}L_{n_r}^{(\Lambda)}(\eta)e^{-(\xi^2+\eta)/2} \,,
  \label{HO_wf}
\end{equation}
where $\mathcal N$ is a normalisation coefficient, $H_{n_z}$ is the Hermite polynomial of degree $n_z$ and $L_{n_r}^{(\Lambda)}$ is the generalised Laguerre polynomial of degree $n_r = (N-n_z-\Lambda)/2$, $\xi = z \,\sqrt{m\omega_z/\hbar}$ and $\eta = \varrho^2\,m\omega_{\bot}/\hbar$. Among the four studied nuclei, the $^{159}$Tb nucleus in its ground-state has a dominant Nilsson configuration of $3/2^+\,[411]$, such that $n_r = 1$, while the other three nuclei have $n_r = 0$ in their respective dominant Nilsson configurations. The study of local extrema of the resulting expressions of the orthoradial components of $\vec \nabla \times \vec s_q$ and $\vec j_q$ shows that there are $n_z + 1$ local maxima with respect to $z$ in both functions, whose $z$ values correspond to the roots of $H_{n_z+1}(\xi)$; whereas in the $\varrho$ direction, the spin current has $n_r+1$ maxima and the convection current has 2 minima and 2 maxima for the cases at play. \Cref{tab_extrema_HO_approx} summarises these properties applied to each of the four studied {nuclei}.
\begin{table}[h]  
  \begin{tabular}{cccccc}
    \hline \hline
    Nucleus & $\Omega^{\pi} [Nnz\Lambda]_q$ & $n_r$ &
    Maxima in the $z$ direction &
    \multicolumn{2}{c}{Maxima, minima in the $\rho$ direction} \\
    \cline{5-6}
    & & & & $j_{\rm odd,\varphi}^{(s)}/g_q^{(s)}$ & $j_{\rm
      odd,\varphi}^{(\ell)}/g_q^{(\ell)}$ \\
    \hline
    $^{159}$Tb & $3/2^+[411]_p$ & 1 & 2 & 2, 2 & 2 \\
    $^{165}$Ho & $7/2^-[523]_p$ & 0 & 3 & 1, 1 & 1 \\
    $^{161}$Dy & $5/2^+[642]_n$ & 0 & 5 & 1, 1 & 1 \\
    $^{163}$Dy & $5/2^-[523]_n$ & 0 & 3 & 1, 1 & 1 \\
    \hline\hline
    \end{tabular}
    \caption{Number of extrema along the $z$ and $\varrho$ directions
    of the spin and convection currents without Landé factors in the
    approximation of the unpaired nucleon in its leading Nilsson
    configuration $\Omega^{\pi}
    [Nnz\Lambda]_q$. \label{tab_extrema_HO_approx}}
\end{table}
This matches the number of extrema in the exact currents presented in \cref{fig_currents_same_isospin}.

% On the figures, the quantity $\vec{R} \wedge \vec{j}^{(l,s)}_{p,n}(\vec{R})$ represents the magnetic dipole moment density \cref{Dipole_Magnetic_Moment}.
%It is used to visualise and quantify nuclear magnetism on a microscopic scale. 
In addition, depending on the type of current (orbital or spin), we can identify the origin of the magnetism at the scale of the nucleus. As mentioned above, the neutron orbital current does not contribute to nuclear magnetism.

Although this single-particle model of the unpaired nucleon in a harmonic potential falls short
in fully capturing the complex interactions between nucleons in the
nucleus and the resulting currents, it offers a useful foundation for
understanding the structure of current distributions. For instance, in
the case of ${}^{163}$Dy, which has 97 neutrons, simple predictions
from the nuclear shell model {without residual interaction}
suggest that the unpaired neutron occupies the $1h_{9/2}$
\sam{state}{subshell} \cite{degroote2020}, which is the dominant
contribution in the expansion of the {actual} single-particle state in the spherical harmonic-oscillator basis. On the contrary, experimental data shows that the ground state of ${}^{163}$Dy corresponds to a $K^\pi = 5/2^-$ band \cite{St05}. 

\begin{figure}[htb!]
    \centering
    \includegraphics[width=.6\textwidth]{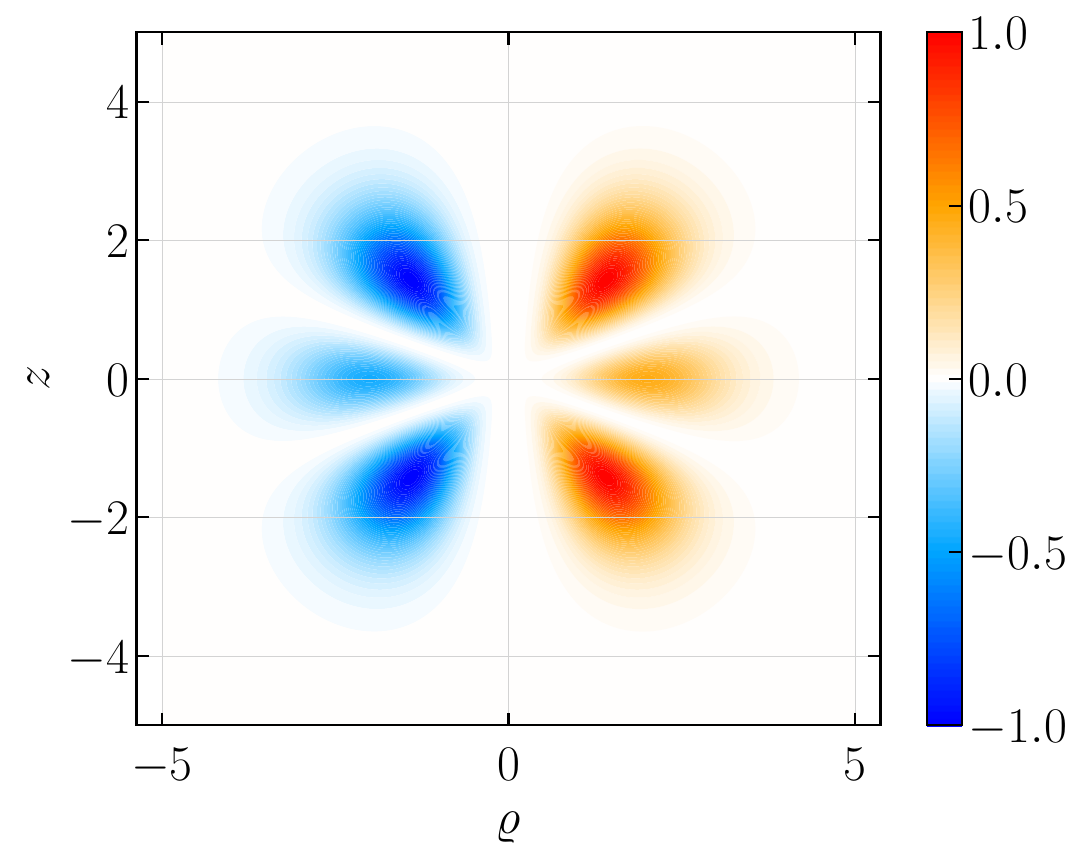}
    \caption{(color online) Normalised current distribution for the $3D$ spherical harmonic oscillator eigenstate corresponding to the quantum state $\ket{1h_{9/2}\; m_j=-5/2}$, which is a component of the single particle ground state of ${}^{163}$Dy according to the shell model. The figure is plotted in signed cylindrical coordinates $(z,\varrho)$, on a planar slice through the $z-$axis. The contours represent curves of \sam{equivalent}{equal} density, with numerical values as indicated by the legend.
    \label{fig::spher_harm_3DHO_163Dy}}
\end{figure}

\Cref{fig::spher_harm_3DHO_163Dy} shows the normalised current
distributions \footnote{$i\left(\psi_{n,\ell}^*\nabla\psi_{n,\ell} -
  \psi_{n,\ell}\nabla\psi_{n,\ell}^*\right)$} calculated for the $3D$
spherical harmonic oscillator eigenstate, which corresponds to the
quantum state $\ket{n{\ell j}\; m_j=-5/2}$. This state can be expressed as a combination of orbital ($m_{\ell}$) and spin ($m_s$) components
\begin{equation}
\ket{n{\ell j}\; m_j=-5/2} = \frac{1}{\sqrt{11}}\left(\sqrt{3}\ket{n\ell\;m_{\ell}=-2\;m_s=-1/2} - 2\sqrt{2}\ket{n\ell\;m_{\ell}=-3\;m_s=+1/2}\right).
\end{equation}
In this expression, $n=6$ and $\ell=5$, which correspond to the $1h$ orbital state. The Clebsch-Gordan coefficients in the equation represent the relative contributions of two specific combinations of the orbital angular momentum ($m_{\ell}$) and spin angular momentum ($m_s$) to form the total angular momentum state $m_j=-5/2$.

The current distribution closely resembles the one obtained for ${}^{163}$Dy. Similar qualitative considerations can be applied to the other isotopes and elements investigated in this study, even though, in some cases, one should also account for configuration mixing and other non-trivial effects. For example, the single particle spectra for both the neutrons and the protons in ${}^{162}$Dy, obtained from the HFBCS calculations, are shown in \cref{fig::sp_spectra_Dy162}. The Nilsson quantum numbers of the corresponding harmonic oscillator states are also indicated.%, along with the occupation probabilities of the states, and we can observe the configuration mixing in the single-particle states.

\begin{figure*}[ht!]
\begin{center}
\includegraphics[width=.6\textwidth,trim={0 3.1cm 0 3.1cm},clip]{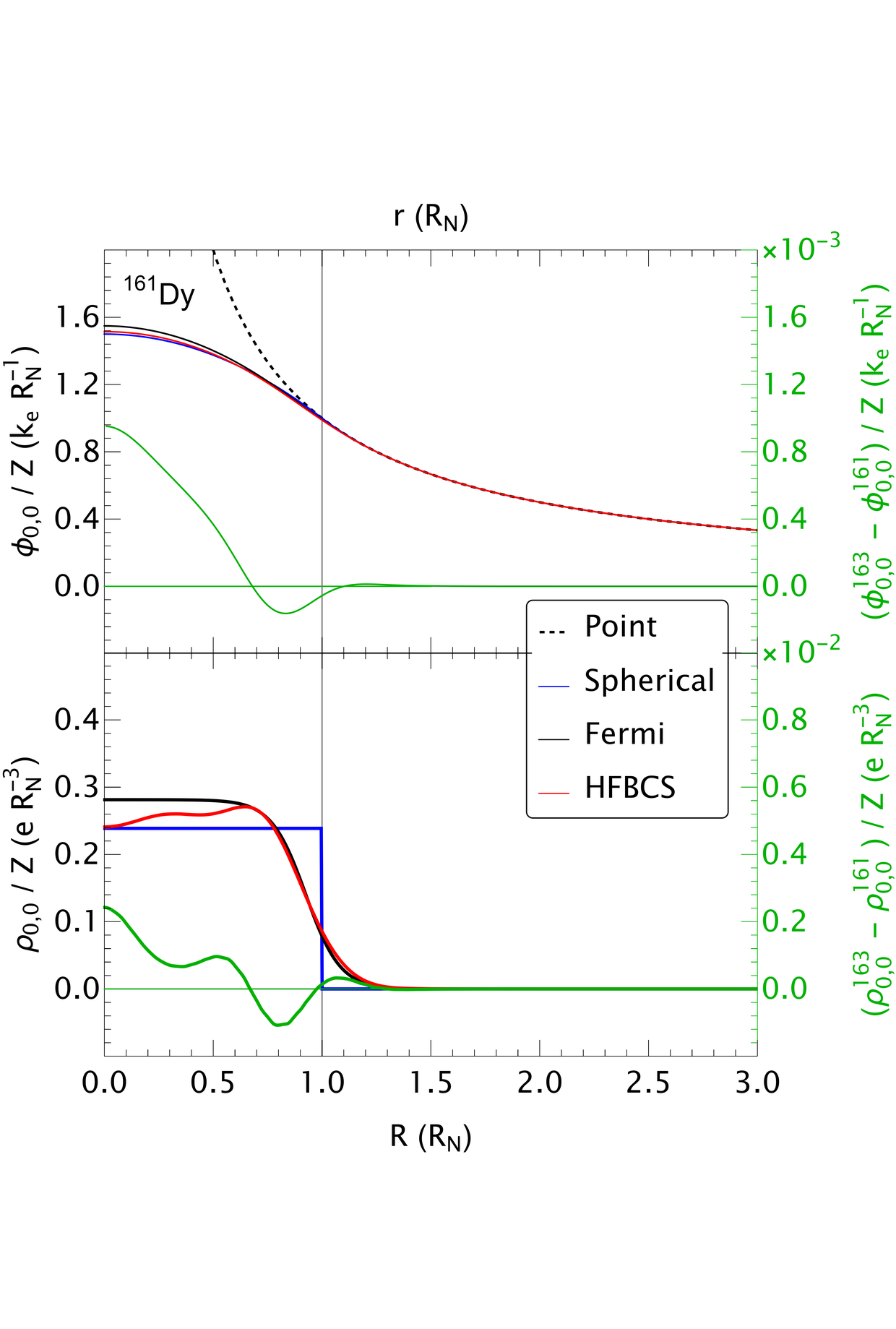}
\caption{(color online) Bottom part: proton monopole density (as a function of nuclear radial coordinate $R$) obtained for the $^{161}$Dy nucleus from a uniform sharp-edge (blue), a Fermi-function type (black) and the distribution resulting from the here presented HFBCS type calculation with self-consistent blocking. Upper part: monopole potential (as a function of the electronic radial coordinate $r$) experienced by the electron resulting from the different approaches for the density. The Coulomb potential generated by a nuclear point-charge density is shown by the dotted line and $k_e=e/4\pi\varepsilon_0$. The green lines show the difference in density (bottom) and potential (upper) between the two dysprosium isotopes obtained with the HFBCS model.}
\label{Fig06}
\end{center}
\end{figure*}

On the upper part of \cref{Fig06} we have shown the monopole electric potentials, $\phi_{0,0}$, relating to the effect of the spatial distribution of protons within the nuclei of $^{161}$Dy. Due to the spherical symmetry of the monopole, it is given in terms of the electronic radial component $r$, by averaging over the angular components. Here, and throughout the text unless specified otherwise, $r$ is expressed in units of the nuclear radius, $R_{\rm N}$, thus giving the potential in units of $R_{\rm N}^{-1} e/4\pi \varepsilon_0$. Each curve describes a different model for the monopole density distribution. Note how outside of the nuclear radius at $R/R_{\rm N} = 1$ the potentials are all effectively equivalent, and the differences in the HFBCS calculations for each isotope is also minor (shown by the green curve and the corresponding $y-$axis on the righthand side). The lower part of the figure shows distributions of the monopolar nuclear charge densities, $\rho_{0,0}(R) = \frac{1}{4\pi}\int \rho(\vec{R})d\Omega$, relating to the radial distribution of protons within the nuclei of isotope $^{161}$Dy. Due to the spherical symmetry of the monopole, it is given in terms of the nuclear radial component $R$, by averaging over the angular components.  Here, and throughout the text unless specified otherwise, $R$ is used in units of the nuclear radius, $R_{\rm N}$, thus giving the monopole in units of $R_{\rm N}^{-3}$. Note how, due to Coulomb repulsion of the protons, the HFBCS model suggests a lower density of protons at the center of the nucleus that increases unevenly towards a maximum at the outer edge. The results are almost equivalent for the two isotopes, with only a small difference in the HFBCS model near $R = 0$ (see green curve and the corresponding $y-$axis on the righthand side). The dashed vertical line represents the surface of the nucleus at $R = 1 \times R_{\rm N}$, to which the other curves are normalised.

\begin{figure}[t!]
    \centering
    \begin{subfigure}[b]{0.45\textwidth}
        \centering
        \includegraphics[trim=0cm 2.75cm 0cm 3.25cm, clip, width=\textwidth]{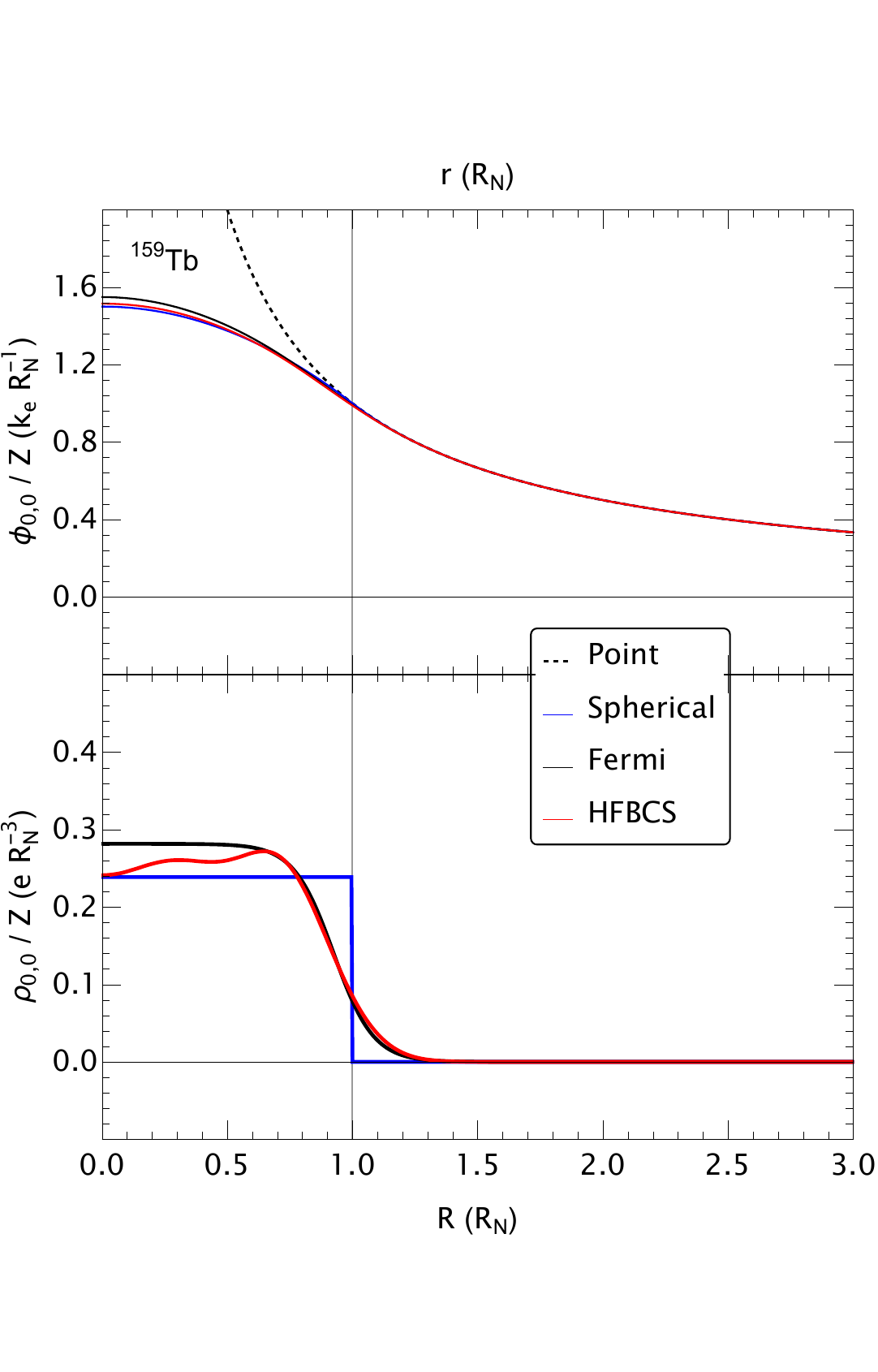}
        \caption{$^{159}$Tb with $R_{\rm{rms}}^{\rm{(emp)}}$.}
        \label{fig::Tb_monopoles_emp}
    \end{subfigure}
    \hfill
    \begin{subfigure}[b]{0.45\textwidth}
        \centering
        \includegraphics[trim=0cm 2.75cm 0cm 3.25cm, clip, width=\textwidth]{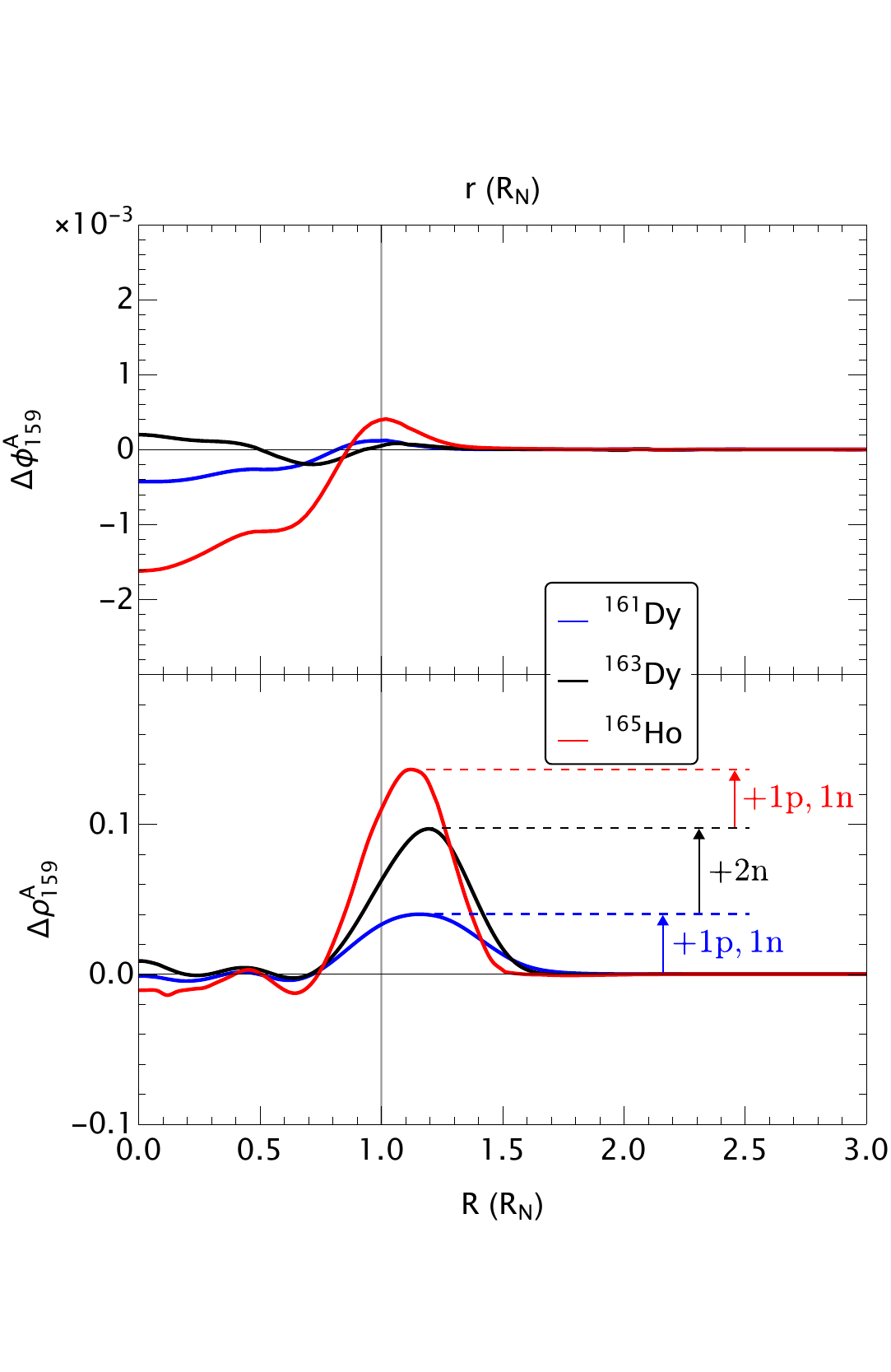}
        \caption{Differences relative to $^{159}$Tb.}
        \label{fig::radial_densities_compar}
    \end{subfigure}
\caption{(color online) $Z$-normalised  (radial, spherically averaged) proton monopole charge densities, $\rho_{0,0}(R) =\frac{1}{4\pi} \int \rho(\vec{R})d\Omega$ (lower panels), and the corresponding monopole electron potentials (upper panels) for $^{159}$Tb. The radial coordinate is shown as $r/R_{\rm N}$ for the potentials seen by atomic electrons (upper panels) and as $R/R_{\rm N}$ for the nuclear charge densities (lower panels), normalised with respect to the empirical nuclear radius $R_{\rm N}$. The densities are in units of $e/R_{\rm N}^3$; potentials are in units of $k_e/R_{\rm N}$, with $k_e \equiv \tfrac{e}{4\pi\epsilon_0}$. HFBCS-type (red) calculations are compared with a uniform sphere (blue), a Fermi distribution (black), and a point-charge Coulomb potential (dotted). In (b), differences relative to $^{159}$Tb are shown for $^{161}$Dy (blue), $^{163}$Dy (black) and $^{165}$Ho (red), using each isotope’s own $R_{\rm N}$ for normalisation. 
}
\label{fig:ch2:Tb_relative_monopoles}
\end{figure}

In \cref{fig::Tb_monopoles_emp}, we show the monopole potentials and charge densities for $^{159}$Tb after normalising by $Z$, the total charge of the nucleus. The radial coordinates are plotted in units of $R_{\rm N}$. One finds, as expected, that the monopole potentials and charge densities for $^{159}$Tb \sam{bare}{bear} a strong similarity to those of $^{161}$Dy at a first glance. Using
\begin{align}\label{eq:ch2:relative_monopole_error}
    \Delta \rho_{159}^A &= \frac{\rho_{0,0}^{(A)} - \rho_{0,0}^{(159)}}{\rho_{0,0}^{(159)}},  \\
    \rm{and} \; \Delta \phi_{159}^A &= \frac{\phi_{0,0}^{(A)} - \phi_{0,0}^{(159)}}{\phi_{0,0}^{(159)}},
\end{align} one can take a closer look and observe the compared radial spherically averaged charge densities of the isotopes ${}^{161}$Dy, ${}^{163}$Dy and ${}^{165}$Ho, relative to ${}^{159}$Tb, in \cref{fig::radial_densities_compar}. They are shown as $Z$-normalised and rescaled to be expressed in units of the respective nuclear radii $eR_{\rm N}^{-3}$. Note that the addition of 2 nucleons has the effect of seemingly concentrating the monopolar charge densities near the nuclear surface (nuclear radius, $r=R_{\rm N}$). Since $R_{\rm N}$ increases with $A$, the charge distributions are also more spread out for the heavier isotopes, as expected. One may consider that for these heavy nuclei with densely-packed, practically incompressible, cores having filled inner shells and charge densities not varying significantly between isotopes, the additional nucleons appear instead to be bound near the surface in the unfilled outer shells.

% \begin{figure}[htb!]
% \centering
% \includegraphics[width=.6\textwidth, trim={0cm 3cm 0cm 3cm},clip]{compar.pdf}
% \caption{(lower panel) Radial spherically averaged charge densities of ${}^{161}$Dy, ${}^{163}$Dy and ${}^{165}$Ho relative to ${}^{159}$Tb. The densities are $Z$-normalised and rescaled to be expressed in units depending on the respective nuclear radii $R_{\rm N}$ of the isotope. The radial coordinate is in units of $R_{\rm N}$ and the charge distribution is in units of $Z/R_{\rm N}^3$. (top panel) The same relative difference in potentials stemming from the charge distributions, in units of $k_e/R_{\rm N}$. $k_e \equiv \tfrac{e}{4\pi\epsilon_0}$ is the Coulomb constant.
% \label{fig::radial_densities_compar}}
% \end{figure}
%%%%%%%%%%%%%%%%%%%%%%%%%%%%%%%%%%%%%%%%%%%%%%%%%

We now turn our attention to the magnetic dipole moments of the isotopes $^{161}$Dy, $^{163}$Dy, $^{159}$Tb, and $^{165}$Ho. The magnetic dipole moments were obtained by integrating the magnetic dipole densities over the volume of the nucleus, \begin{align}
\rho_{1,0}(R) = \frac{1}{2} \sqrt{\frac{3}{\pi}} \int \cos(\Theta) \rho(\vec{R})d\Omega.
\end{align}
The magnetic dipole density contributions are given by the sum of the proton and neutron spin densities and the proton orbital current density (the neutron orbital current density makes no contribution since its Land\'e $g$ factor is zero).

In \cref{Fig07}, we show the magnetic dipole densities for each
isotope. The radial part of the dipole density is plotted in units of
$R_{\rm N}^{-3}$, with the distributions of the magnetic
\sam{dipolar}{dipole} radial density contributions normalised by the
total magnetic moment of the respective isotope $\mu_{\rm N}^{(A)}$, as
obtained from the HFBCS model. Across these rare-earth isotopes, the
magnetic dipole densities are predominantly governed by the spin currents of the unpaired nucleons. For dysprosium (even \(Z\), odd \(N\)), the neutron (spin) currents dominate—most clearly when considering \(^{163}\mathrm{Dy}\), which has two additional neutrons compared to \(^{161}\mathrm{Dy}\). In contrast, for terbium and {holmium (odd \(Z\), even \(N\)), the dominant contributions arise from the proton spin and orbital currents. In all cases, consistent with this picture of the unpaired nucleons, the dipole-density curves exhibit peaks close to the (spherically averaged) nuclear radius $R_{\rm N}$. This is consistent with the current distributions observed in \crefrange{fig_currents_core_same_isospin}{Fig06}, as the nuclear magnetism originates mostly from the unpaired nucleon that is naturally more free to move in the lower density outer shell of the nucleus as opposed to the densely packed inner core. Moreover, comparing the Dy isotopes shows that adding only two neutrons can drive a complete sign reversal of the dipole densities, keeping in mind that $\mu_{\rm N}^{(161)}$ and $\mu_{\rm N}^{(163)}$ carry different signs.

\begin{figure}[ht!]
    \centering
    \begin{subfigure}[b]{0.45\textwidth}
        \centering
        \includegraphics[trim=0cm 3cm 0cm 4cm, clip, width=\textwidth]{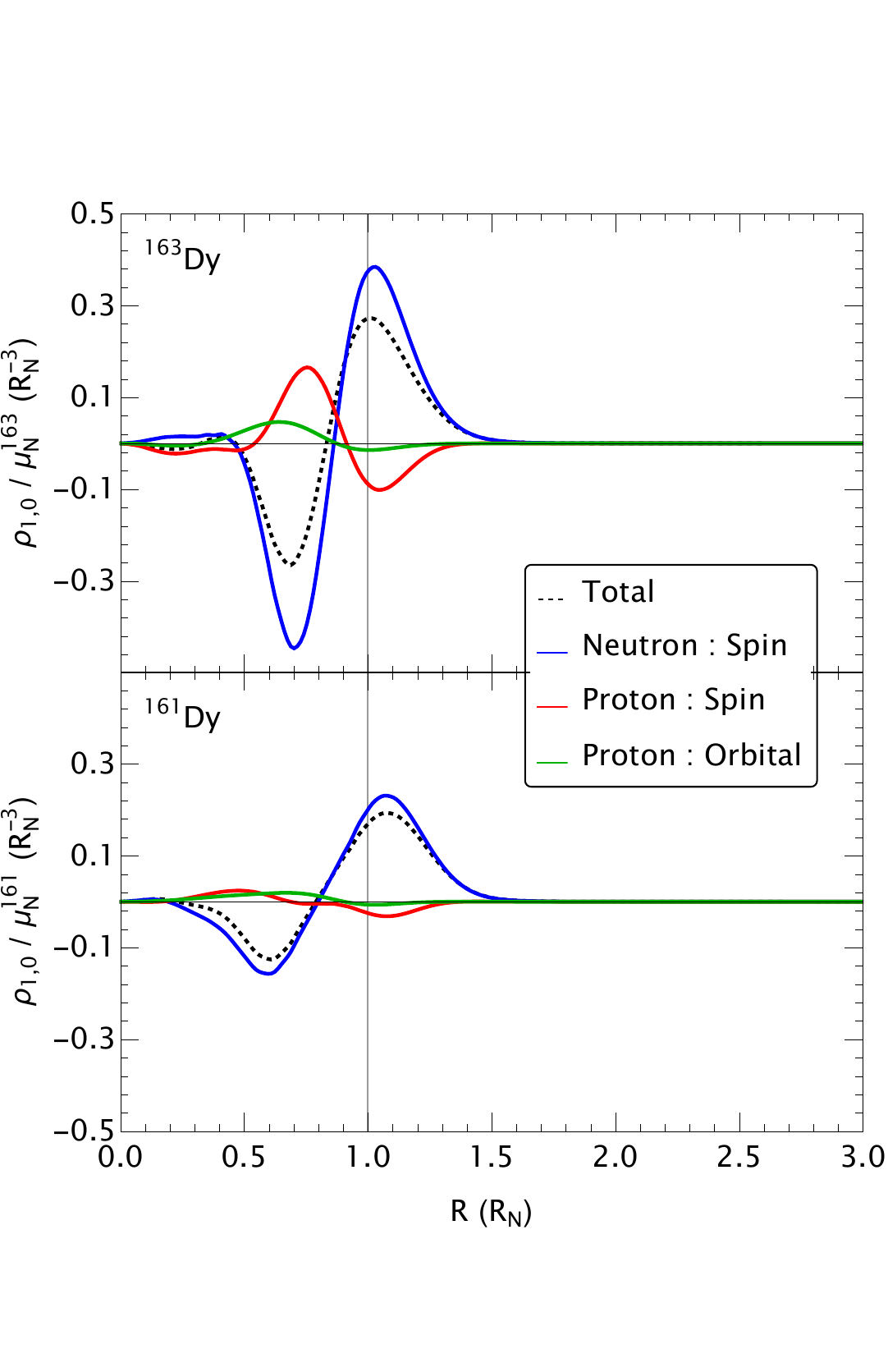}
        \caption{$^{161}$Dy and $^{163}$Dy.}
        \label{fig:ch2:Dy_dipoles}
    \end{subfigure}
    \hfill
    \begin{subfigure}[b]{0.45\textwidth}
        \centering
        \includegraphics[trim=0cm 3cm 0cm 4cm, clip, width=\textwidth]{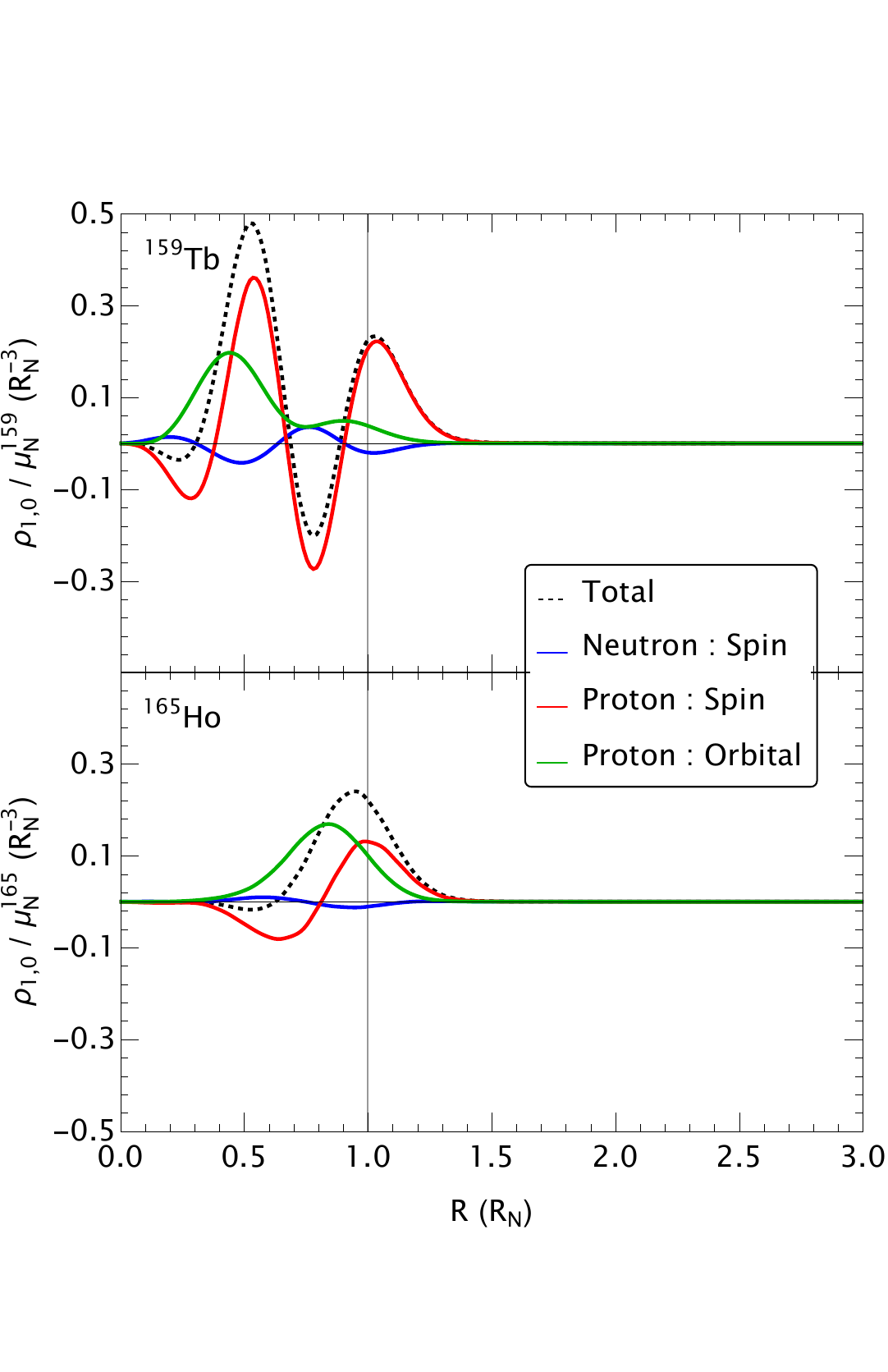}
        \caption{$^{159}$Tb and $^{165}$Ho.}
        \label{fig:ch2:TbHo_dipoles}
    \end{subfigure}
    \caption{(color online) Radial distributions of the magnetic dipolar density contributions, $\rho_{1,0}(R)$, relating to the spatial distribution of protons and neutrons within the nuclei of isotopes: (a) $^{163}$Dy (upper panel) and $^{161}$Dy (lower panel), and (b) $^{159}$Tb (upper panel) and $^{165}$Ho (lower panel). By spherically averaging over the angular components, it is given only in terms of the radial component $R$. $R$ is used in units of the nuclear radius, $R_{\rm N}$, thus giving the dipole in units of $\mu_{\rm N}\,R_{\rm N}^{-3}$. The dashed vertical line represents the surface of the nucleus at $R = 1 \times  R_{\rm N}$, to which the other curves are normalised. Moreover, the dipole densities are normalised by $\mu_{\rm N}^{(A)}$ for each isotope such that they lie on the same scale. Each curve describes a different contribution to the dipole density distribution: (Dashed) The total dipole density, (Blue) the neutron spins, (Red) the proton spins, and (Green) the proton orbital currents. }

    \label{Fig07}
\end{figure}

% \begin{figure}[ht!]
% \begin{center}
% \includegraphics[width=.6\textwidth,trim={0 3cm 0 3cm},clip]{N5_dipole_distribution_600dpi}
% \caption{Radial distributions of the magnetic dipolar density contributions, $\rho_{1,0}(R) = \int \frac{1}{2} \sqrt{\frac{3}{\pi}} \cos(\Theta) \rho(\vec{R})d\Omega$, relating to the spatial distribution of protons and neutrons within the nuclei of isotopes $^{163}$Dy (up) and $^{161}$Dy (down). By spherically averaging over the angular components, it is given only in terms of the radial component $R$. $R$ is used in units of the nuclear radius, $R_{\rm N}$, thus giving the dipole in units of $\mu_{\rm N}$.$R_{\rm N}^{-3}$. The dashed vertical line represents the surface of the nucleus at $R = 1 \times  R_{\rm N}$, to which the other curves are normalised. Each curve describes a different contribution to the dipole density distribution: (Dashed) The total dipole density, (Blue) the neutron spins, (Red) the proton spins, and (Green) the proton orbital currents. Note that the dipole densities are dominated by the contributions from the neutron spins, particularly for isotope $^{163}$Dy with an additional $2$ neutrons. Furthermore, it is interesting to see that there are peaks in the curves lying close to the outer surface of the (spherically averaged) nuclear radius. Comparison of the curves for each isotope show that the difference in only $2$ neutrons in the nucleus leads to a complete change of sign of the dipole densities
% contributing towards the dipole moment.}
% \label{Fig07}
% \end{center}
% \end{figure}

Quadrupole radial potentials, $\phi_{2,0}(r)$ produced by the spatial distributions of protons within the spherically deformed nuclei of isotopes $^{159}$Tb (left), $^{161}$Dy (center, black) and $^{163}$Dy (center, red) and $^{165}$Ho (right) are shown in \cref{Fig08}. By spherically averaging over the angular components, it is given only in terms of the radial component $r$. $r$ is used in units of the nuclear radius, $R_{\rm N}$, thus giving the potential in units of $e R_{\rm N}^{-3}/4\pi \varepsilon_0$. The dashed vertical line represents the surface of the nucleus at $R = R_{\rm N}$, to which the other curves are normalised. Note the strong similarities in the curves for each dysprosium isotope, since the neutrons do not contribute towards the quadrupole moment, and they have the same number of protons, yet a slightly higher maximum for $^{163}$Dy. We may remark that the relative differences in the quadrupole potentials at the nuclear radius ($r=1$) for the different elements (Tb, Dy, Ho) are consistent with that of their experimentally measured quadrupole moments (cf. \cref{tab_R_Q}). Moreover, in comparison with the monopole potentials (cf. upper panels of \cref{Fig06,fig:ch2:Tb_relative_monopoles}) where Gauss's Law ensures that the potentials due to all nuclear models converge to the same value outside of the nucleus ($r \geqslant1$), the quadrupole potentials are not constrained in the same way and exhibit observable differences at, and just outside, the nuclear radius.

%%%%%%%%%%%%%%%%%%%%%%%%%%%%%%%%%%%%%%%%%%%%%%%%%%
\begin{figure}[t!]
    \centering   
    \begin{subfigure}[b]{0.32\textwidth}
        \centering
        \includegraphics[width=\textwidth]{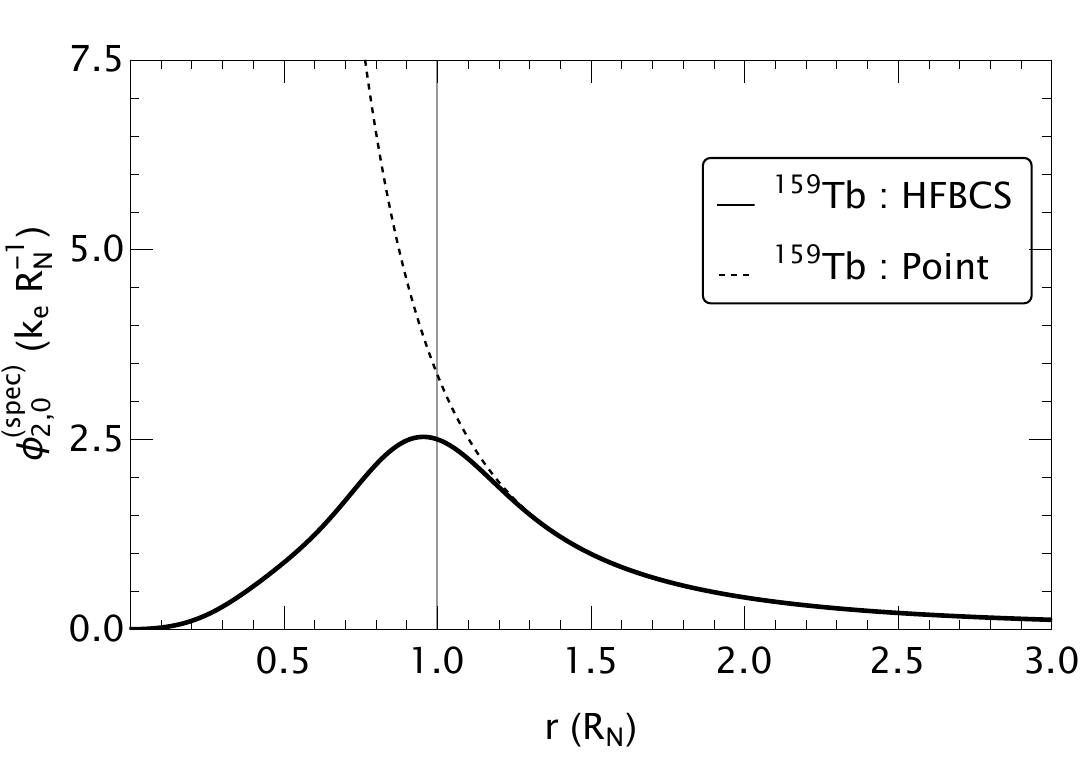}
        \caption{$^{159}$Tb.}
        \label{fig::Tb_quadrupoles}
    \end{subfigure}
    \hfill
    \begin{subfigure}[b]{0.32\textwidth}
        \centering
            \includegraphics[width=\textwidth]{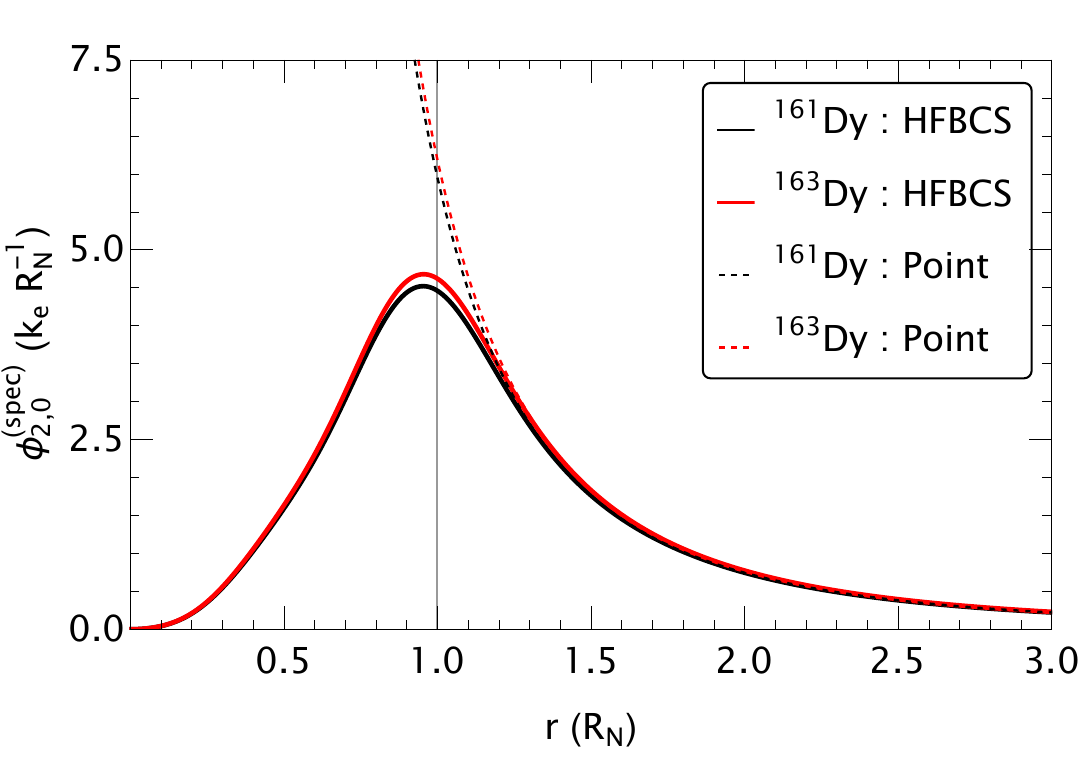}
            \caption{$^{161}$Dy and $^{163}$Dy.}
            \label{fig::Dy_quadrupoles}
        \end{subfigure}
    \hfill
    \begin{subfigure}[b]{0.32\textwidth}
        \centering
        \includegraphics[width=\textwidth]{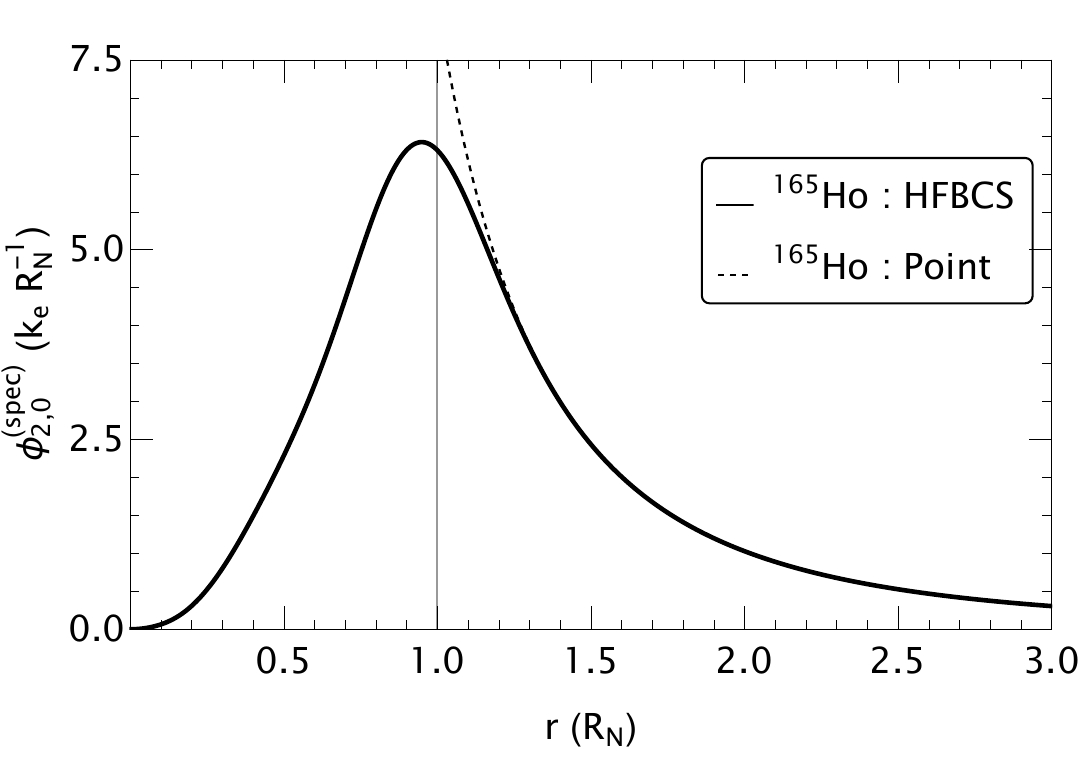}
        \caption{$^{165}$Ho.}
        \label{fig::Ho_quadrupoles}
    \end{subfigure}
    \caption{(color online) The radial part of the spectroscopic quadrupole contribution to the nuclear electric potential, $\phi_{2,0}^{(\rm spec)}$ as defined in \cref{eq:phi20}, in units of $k_{e}/ R_{\rm N}$, for (a) $^{159}$Tb (black), (b) $^{161}$Dy (black) and $^{163}$Dy (red) and (c) $^{165}$Ho (black), as obtained from our HFBCS calculations, compared with the corresponding potentials obtained for the quadrupole potential of a point-like model $(\frac{{\cal Q}_{0}^{(2)}}{r^3})$ (dashed). Here, $k_e \equiv \tfrac{e}{4\pi\epsilon_0}$ is the Coulomb constant.}
    \label{Fig08}
\end{figure}
% \begin{figure*}[ht!]
% \begin{center}
% \includegraphics[width=0.65\textwidth,angle=0]{N6_quadrupole_distribution_600dpi}
% \caption{Radial part of the quadrupole contribution to the nuclear electric potential for $^{161}$Dy (black) and $^{163}$Dy (red), as obtained from our HFBCS calculation, compared with the corresponding potentials obtained for a point-charge nucleus. $k_e=e/4\pi\varepsilon_0$. {\color{red}Add Ho and Tb, Also use the figure with Qspec.}}
% \label{Fig08}
% \end{center}
% \end{figure*}
%%%%%%%%%%%%%%%%%%%%%%%%%%%%%%%%%%%%%%%%%%%%%%%%%%
%%

\subsection{Results combining nuclear and atomic descriptions}
In the following, unless otherwise stated, atomic units are used: $m_e=e=4\pi\epsilon_0=\hbar=1$.

\subsubsection{Charge corrections}
\Cref{fig:e1} shows the radial parts (large $P_{n\kappa}$ and small $Q_{n\kappa}$ components) of the Dirac wave functions of the first two electronic orbitals $1s_{1/2}$ and $2p_{1/2}$ of $^{163}$Dy$^{65+}$ $(Z=66)$. They have been obtained for different models of the nucleus (monopole charge distributions): point charge localised in the center of the nucleus (Point), spherical distribution (Spherical), Fermi distribution (Fermi) and finally the microscopic distribution calculated in the HFBCS model. In all situations, the three models Spherical, Fermi and HFBCS give very similar results to each other. On the other hand, we note that the results obtained with the point charge model deviate significantly from those of the other models for distances smaller than the radius of the nucleus $R_{\rm N}$ (represented by a vertical line on the graphs). In a way, these electronic wavefunctions converging to the same behaviour outside of the nuclear radius is an illustration of Gauss's theorem in electrostatics. To show the effect of a microscopic description (HFBCS) compared with the empirical Fermi model usually used in atomic physics, the difference between the HFBCS and Fermi models, multiplied by a factor of 20, is presented in green. As expected, this difference is only visible inside the nucleus and is very small.

\begin{figure}[ht!]
\centering
\includegraphics[width=.95\textwidth, trim={0 5.5cm 0 5.5cm},clip]{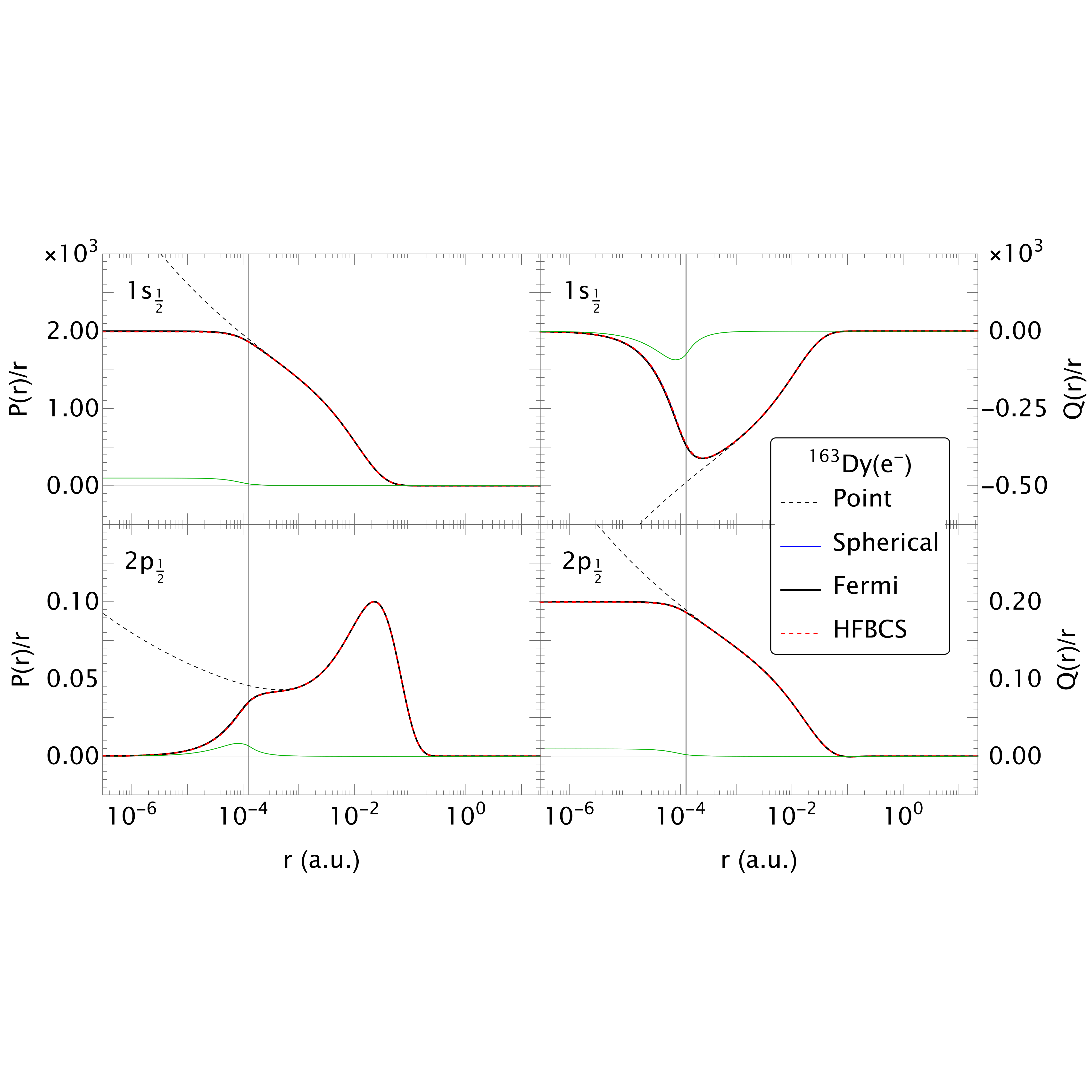}
\caption{(color online) Large ($P_{n\kappa}$) and small ($Q_{n\kappa}$) components of the radial wave functions for the 1$s_{1/2}$ and 2$p_{1/2}$ orbitals of $^{163}$Dy$^{65+}$. The dashed black and red curves represent the Point and HFBCS models respectively, while the solid black and blue curves represent the Fermi and Spherical models respectively. The solid green line represents the difference between the Fermi and HFBCS models multiplied by 20. Vertical lines indicate the position of the nuclear radius $R_{\rm N}$. As there are very tiny differences, the results for $^{161}$Dy$^{65+}$ are given in the appendix {(cf. \cref{fig:a1})}. Atomic units are used.} \label{fig:e1}
\end{figure}

\Cref{fig:e2} is similar to \cref{fig:e1}, but for a muon. Compared with the electron case, it is clear that the density of presence inside the nucleus is much greater. As in the case of the electron, the deviation from the point-charge model also occurs in the vicinity of the nuclear radius. Finally, the difference between the Fermi and HFBCS models is much larger than in the electronic case. This makes muonic atoms interesting physical systems for probing microscopic properties at the nuclear scale. Muonic atoms can also be used to probe the distribution of the magnetic field inside the nucleus, as we will see later.

\begin{figure}[ht!]
\centering
\includegraphics[width=.95\textwidth, trim={0 5.5cm 0 5.5cm},clip]{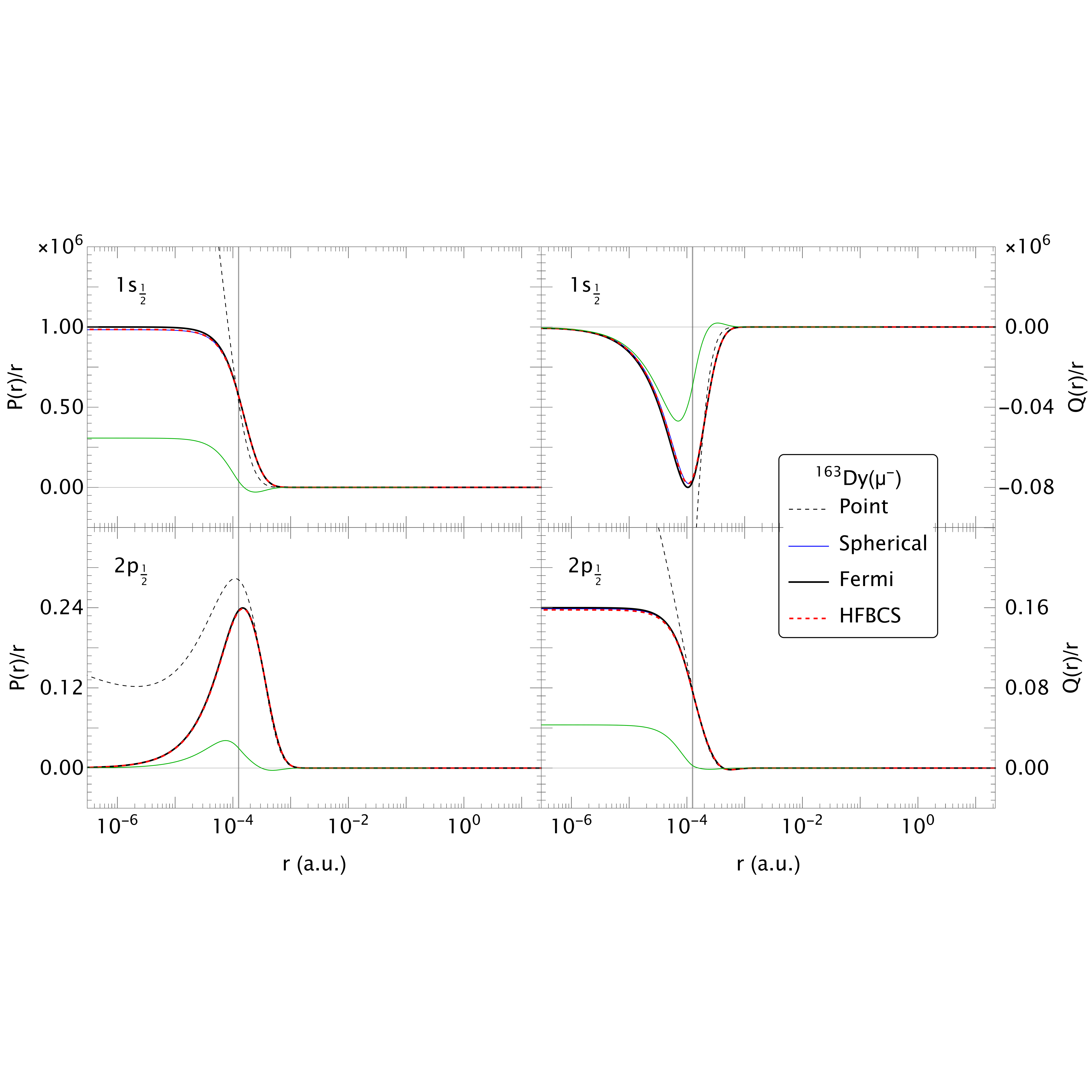}
\caption{(color online) Same as \cref{fig:e1} but for a muonic hydrogen-like ion of $^{163}$Dy$^{65+}$ . As there are very tiny differences, the results for $^{161}$Dy$^{65+}$ are given in the appendix{ (cf. \cref{fig:a2})}. Atomic units are used.} \label{fig:e2}
\end{figure}

For the point-charge model, there is an exact scaling law for the wave function. According to Dirac's equation \cite{WR2007}, we have $P_{n\kappa}(r)=a_0^{-3/2}P_{n\kappa}(a_0^{-1}r)$ and $Q_{n\kappa}(r)=a_0^{-3/2}Q_{n\kappa}(a_0^{-1}r)$ with the Bohr radius given by $a_0=\frac{4\pi\varepsilon_0 \hbar^2}{m e^2}$. If $m=m_{\mu}=207 m_e$ then $a_0=1/207$ in atomic units. To check our numerical procedure for solving the Dirac equation, we have verified in \cref{fig:e3} that this scaling law is satisfied by the electron and muon wavefunctions obtained in the point-charge model. For the Point model, we check that $a_0^{-3/2}P^{(e)}_{\mathrm{pt}}(a_0^{-1}r)$ and $P^{(\mu)}_{\mathrm{pt}}(r)$ are identical. The same is true for $a_0^{-3/2}Q^{(e)}_{\mathrm{pt}}(a_0^{-1}r)$ and $Q^{(\mu)}_{\mathrm{pt}}(r)$. We also note that the latter is no longer fulfilled at all if we use a realistic nucleus model such as HFBCS.

\begin{figure}[ht!]
\centering
\includegraphics[width=.95\textwidth, trim={0 5.5cm 0 5.5cm},clip]{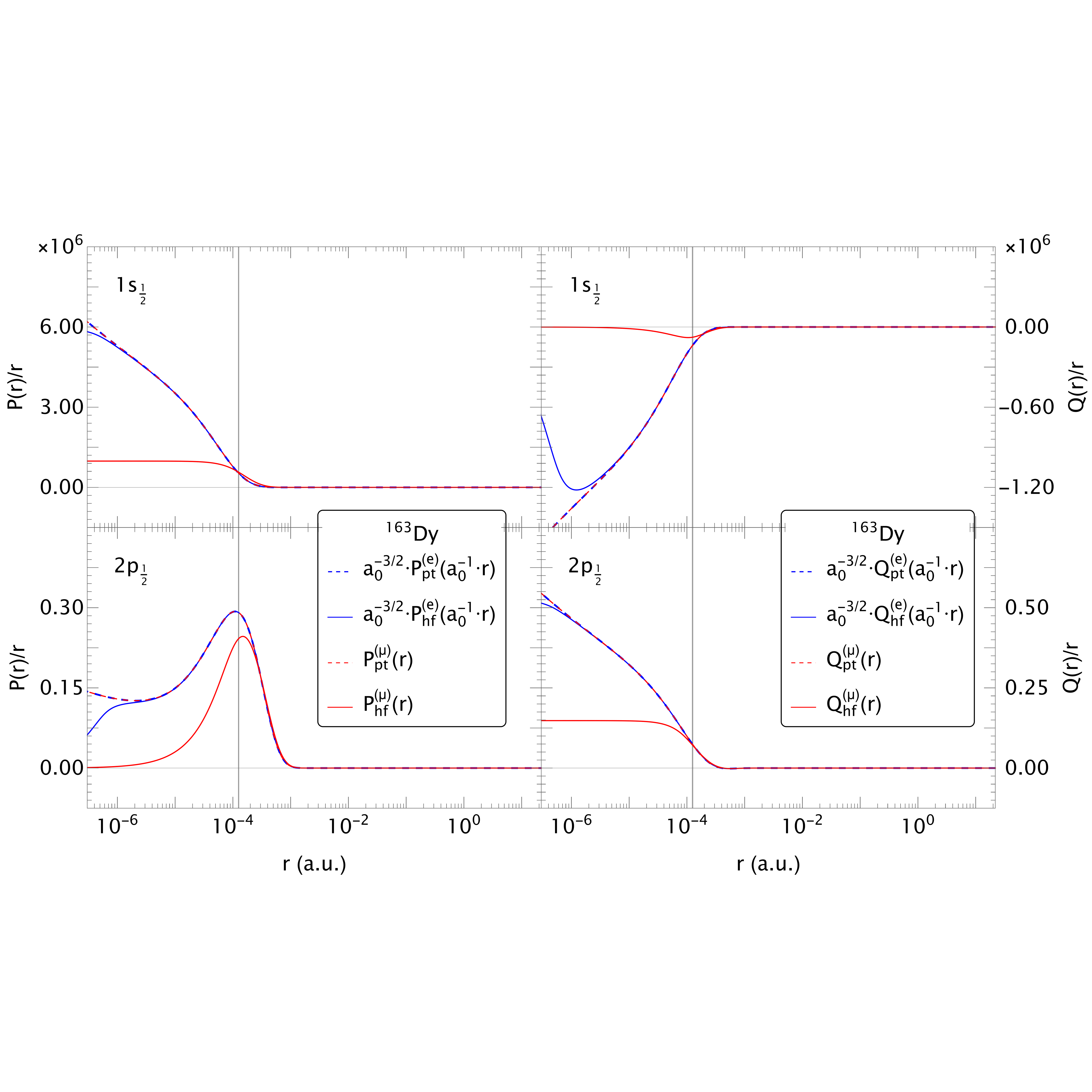}
\caption{(color online) Large ($P_{n\kappa}$) and small ($Q_{n\kappa}$) components of the radial wave functions for the 1$s_{1/2}$ and 2$p_{1/2}$ orbitals of $^{163}$Dy$^{65+}$. Comparison between electron (blue) and muon (red) for Point (dashed) and HFBCS (solid) models. The results for the electron are rescaled using the Bohr radius for muon, $a_0=1/207$ au. Vertical lines indicate the position of the nuclear radius $R_{\rm N}$. As there are very tiny differences, the results for $^{161}$Dy$^{65+}$ are given in the appendix {(cf. \cref{fig:a3})}. Atomic units are used.} \label{fig:e3}
\end{figure}

The aim of \cref{fig:e4} is to show the influence {on
  orbital energies} of the nucleus model used\sam{ on orbital
  energies}. To do this, monopolar shifts in the energy levels of
electronic orbitals of $^{163}$Dy$^{65+}$ (from $1s_{1/2}$ to
$4f_{7/2})$ are shown for different nuclear models: Point (pt),
Spherical (sp), Fermi (fm) and HFBCS (hf). We have defined the
following quantities: $\Delta E_{m_1,i}^{m_2} \equiv
\frac{E_i^{m_2}-E_i^{m_1}}{E_i^{m_1}}$ with
$m_2\in \lbrace{\mathrm{sp,fm,hf}\rbrace}$ and $m_1\in\lbrace{\mathrm{pt,fm}\rbrace}$ and $\Delta
E_{161,i}^{163} \equiv \frac{E_i^{163}-E_i^{161}}{E_i^{161}}$ with $i$
running from $1s_{1/2}$ to $4f_{7/2}$. In the upper panel of the
figure, the relative difference with the point charge is shown. This relative difference is always very small, {of} the order
of $0.01\%$, and is comparable for the three more realistic nuclear
models. Moreover, for the same model, \sam{almost only the $s$ states
  are affected}{the affected states are almost exclusively the $s$
  states}, and the effect decreases with the quantum number $n$. The
$p_{1/2}$ states are minimally affected, which is due to the greater
penetration of $s$ orbitals inside the nucleus, where there are strong
deviations from the point-charge model. Moreover, as $n$ increases,
the \sam{overlap}{amplitude} of the wave function \sam{inside}{in} the nuclear {region} decreases.

In order to estimate the effect of a microscopic modeling of the
nucleus (HFBCS model) \sam{versus a more conventional}{as compared to a} macroscopic description (Fermi model), the lower panel shows the relative energy difference between the two models defined by $\Delta E_{\mathrm{hf},i}^{\mathrm{fm}}$. This quantity is very small, always less than $0.001\%$. The same trends can be observed as those discussed for the upper panel.

Finally, the influence of the isotope on the energy shift of the orbitals is also studied by evaluating $\Delta E_{161,i}^{163} \equiv \frac{E_i^{163}-E_i^{161}}{E_i^{161}}$. This quantity is shown in the lower panel, and its value is always less than $0.0001\%$. As before, the same trends can be observed as those discussed for the upper panel.

\begin{figure}[hb!]
    {\centering
    \begin{subfigure}[t]{0.49\textwidth}
        \centering
        \includegraphics[width=\textwidth, trim={7cm 3.5cm 7cm 3.5cm}, clip]{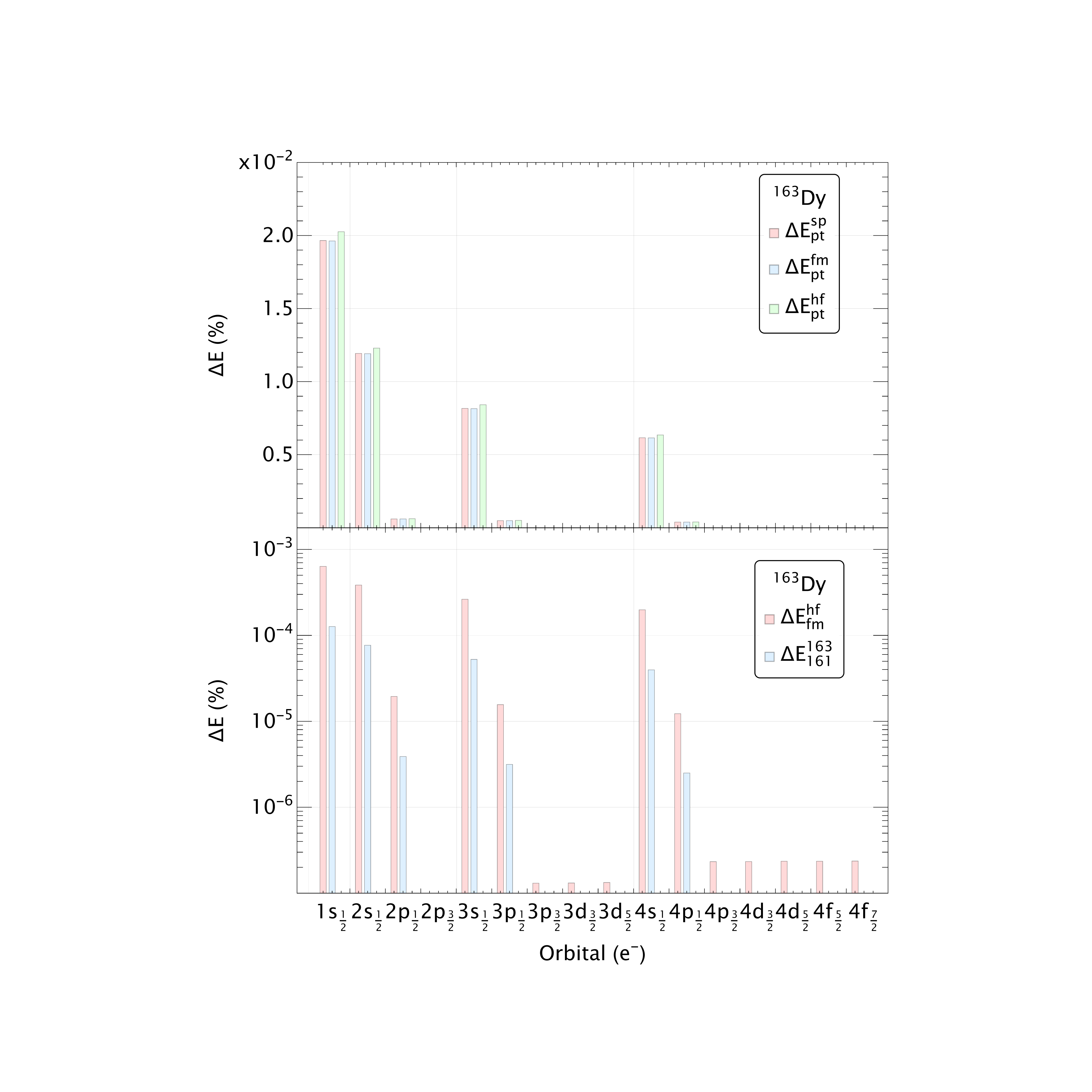}
        \caption{Electronic orbitals of ${}^{163}$Dy$^{65+}$.
        }
        \label{fig:e4}
    \end{subfigure}
    \hfill
    \begin{subfigure}[t]{0.49\textwidth}
        \centering
        \includegraphics[width=\textwidth, trim={7cm 3.5cm 7cm 3.5cm}, clip]{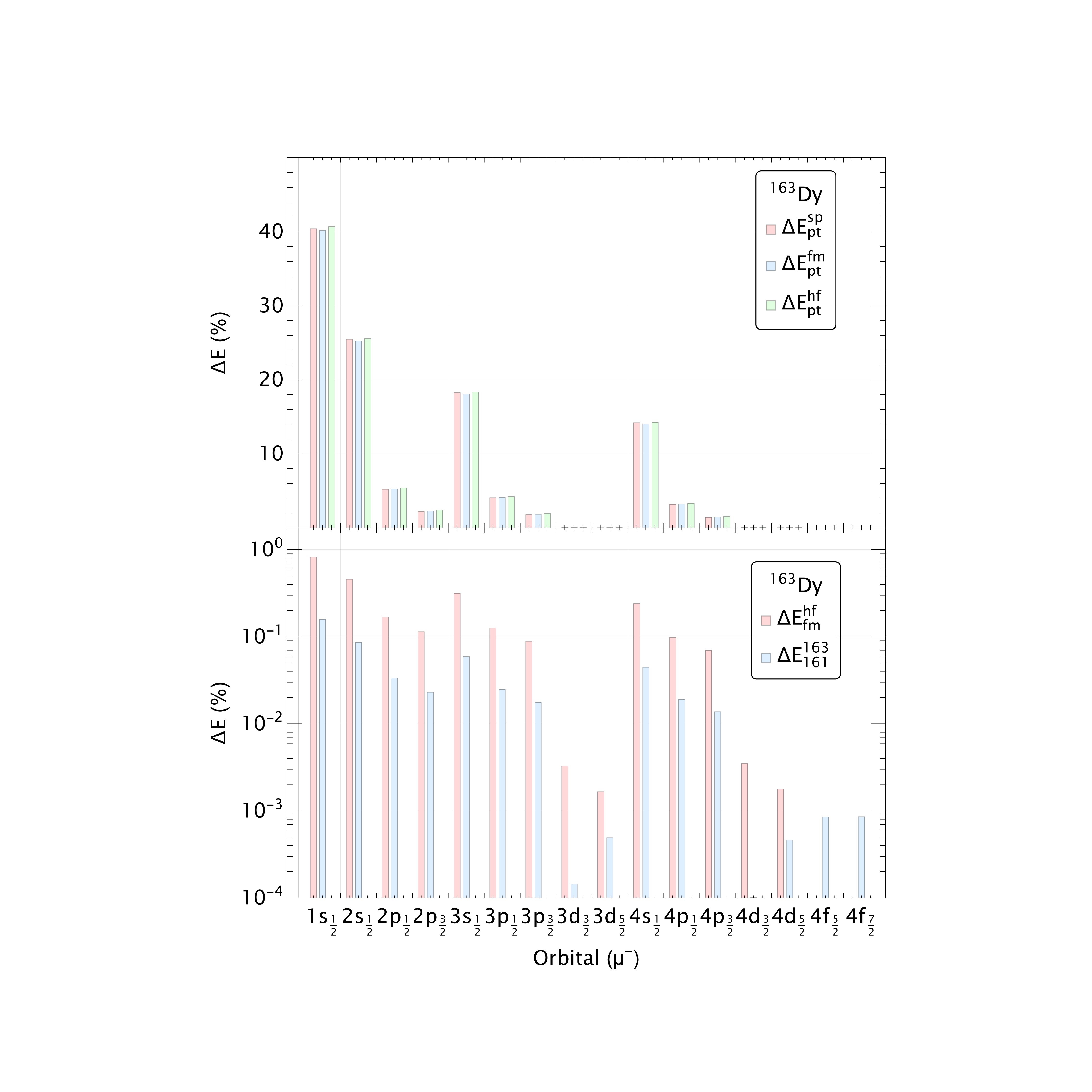}
        \caption{Muonic orbitals of ${}^{163}$Dy$^{65+}(\mu^-)$.}
        \label{fig::monopole_energy_levels_muon}
        \label{fig:e5}
    \end{subfigure}}
    \caption{(color online) Monopolar shifts in the energy levels of electronic orbitals of $^{163}$Dy$^{65+}$ (from $1s_{1/2}$ to $4f_{7/2})$ in different nuclear models, Point (pt), Spherical (sp), Fermi (fm) and HFBCS (hf). We have defined, $\Delta E_{m_1,i}^{m_2} \equiv \frac{E_i^{m_2}-E_i^{m_1}}{E_i^{m_1}}$ with $m_2\mathrm{=sp,fm,hf}$ and $m_1\mathrm{=pt,fm}$ and $\Delta E_{161,i}^{163} \equiv \frac{E_i^{163}-E_i^{161}}{E_i^{161}}$ with $i$ running from $1s_{1/2}$ to $4f_{7/2}$. This latter quantity was evaluated using the energies of the HFBCS model. As there are very small differences, the results for $^{161}$Dy$^{65+}$ are given in the appendix {(cf. \cref{fig:a4})}.}
\end{figure}

% \begin{figure}[ht!]
% \centering
%  \includegraphics[width=\textwidth, trim={7cm 3.5cm 7cm 3.5cm}, clip]{04_163Dy_e_monopoles_600dpi}
% \caption{(color online) Monopolar shifts in the energy levels of electronic orbitals of 163^{163}Dy65+^{65+} (from 1s1/21s_{1/2} to 4f7/2)4f_{7/2}) in different nuclear models, point (pt), spherical (sp), Fermi (fm) and HFBCS (hf). We have defined, ΔEm2m1,i≡Em2i−Em1iEm1i\Delta E_{m_1,i}^{m_2} \equiv \frac{E_i^{m_2}-E_i^{m_1}}{E_i^{m_1}} with m2=sp,fm,hfm_2\mathrm{=sp,fm,hf} and m1=pt,fmm_1\mathrm{=pt,fm} and ΔE163161,i≡E163i−E161iE161i\Delta E_{161,i}^{163} \equiv \frac{E_i^{163}-E_i^{161}}{E_i^{161}} with ii running from 1s1/21s_{1/2} to 4f7/24f_{7/2}. This latter quantity was evaluated using the energies of the HFBCS model. As there are very tiny differences, the results for 161^{161}Dy65+^{65+} are given in the appendix (cf. Fig A4).}  \label{fig:e4}
% \end{figure}

\Cref{fig:e5} presents the \sam{same resulting calculations}{results
of the same type of calculations} as {in~}\cref{fig:e4}, but for a muonic atom. As expected, the major difference is the very significant increase in the value of the quantities studied. $\Delta E_{\mathrm{pt},i}^{m_2}$ with $m_2\in\lbrace{\mathrm{sp,fm,hf}\rbrace}$ is of the order of ten percent, reaching over 40\% for the $1s_{1/2}$ orbital. The same observation can be made for the muonic $p$ orbitals, but with the addition of a noticeable deviation for the $p_{3/2}$ states. We can therefore test the description of the nucleus on orbital energies. Moreover, the relative difference between HFBCS and Fermi models is no longer negligible, and can reach 1\% for the $1s_{1/2}$ orbital. Finally, a less negligible isotopic effect is observed with $\Delta E_{161,i}^{163} $of the order of 0.01\%. 

% \begin{figure}[ht!]
% \centering
% \includegraphics[scale=0.20,angle=0]{05_163Dy_m_monopoles_600dpi}
% \caption{(color online) Same as \cref{fig:e4}) but for a muonic hydrogen-like ion. As there are very tiny differences, the results for 161^{161}Dy65+^{65+} are given in the appendix (cf. Fig A5).}  \label{fig:e5}
% \end{figure}

\Cref{fig:e6} shows the Breit-Rosenthal correction obtained for different nuclear models and orbitals. This quantity is denoted by
$\varepsilon_{\mathrm{BR}}$, as defined in \cref{eq:BRdefinition}, and is shown in the upper panel. The different nuclear models all exhibit approximately the same behaviour and results. As expected, $\varepsilon_{\mathrm{BR}}$ is much larger for $ns_{1/2}$ orbitals, with a value of around 5\%. It is much smaller for $np_{1/2}$ orbitals and practically \sam{non-existent}{zero} for $np_{3/2}$ orbitals and the others. When we compare the value of $\varepsilon_{\mathrm{BR}}$ as a function of $n$ for $s_{1/2}$ and and $p_{1/2}$ orbitals, we see that it is almost constant.

On the lower panel of the figure, the quantity $\Delta \varepsilon_{\mathrm{fm}}^{\mathrm{hf}} \equiv \frac{\varepsilon_{\mathrm{BR}}^{\mathrm{hf}}-\varepsilon_{\mathrm{BR}}^{\mathrm{fm}}}{\varepsilon_{\mathrm{BR}}^{\mathrm{fm}}}$ is represented. It quantifies the extent of the corrections made by a microscopic model (i.e. HFBCS) compared with a more macroscopic one, namely the Fermi model. The correction is of the order of one percent for all orbitals except for the $np_{3/2}$ orbitals, where it reaches more than 5\%. The lower panel of the figure also shows the quantity $ \Delta \varepsilon_{161}^{163} \equiv \frac{\varepsilon_{\mathrm{BR}}^{\mathrm{hf}}(163)-\varepsilon_{\mathrm{BR}}^{\mathrm{hf}}(161)}{\varepsilon_{\mathrm{BR}}^{\mathrm{hf}}(161)}$. It provides information on the influence of the isotope on the Breit-Rosenthal correction.
The isotopic effect is of the order of $0.2\%$ for all orbitals except for the $np_{3/2}$ orbitals, where it reaches about $1\%$.

\begin{figure}[ht!]
\centering
\includegraphics[width=.66\textwidth, trim={5.8cm 5cm 5.8cm 5cm},clip]{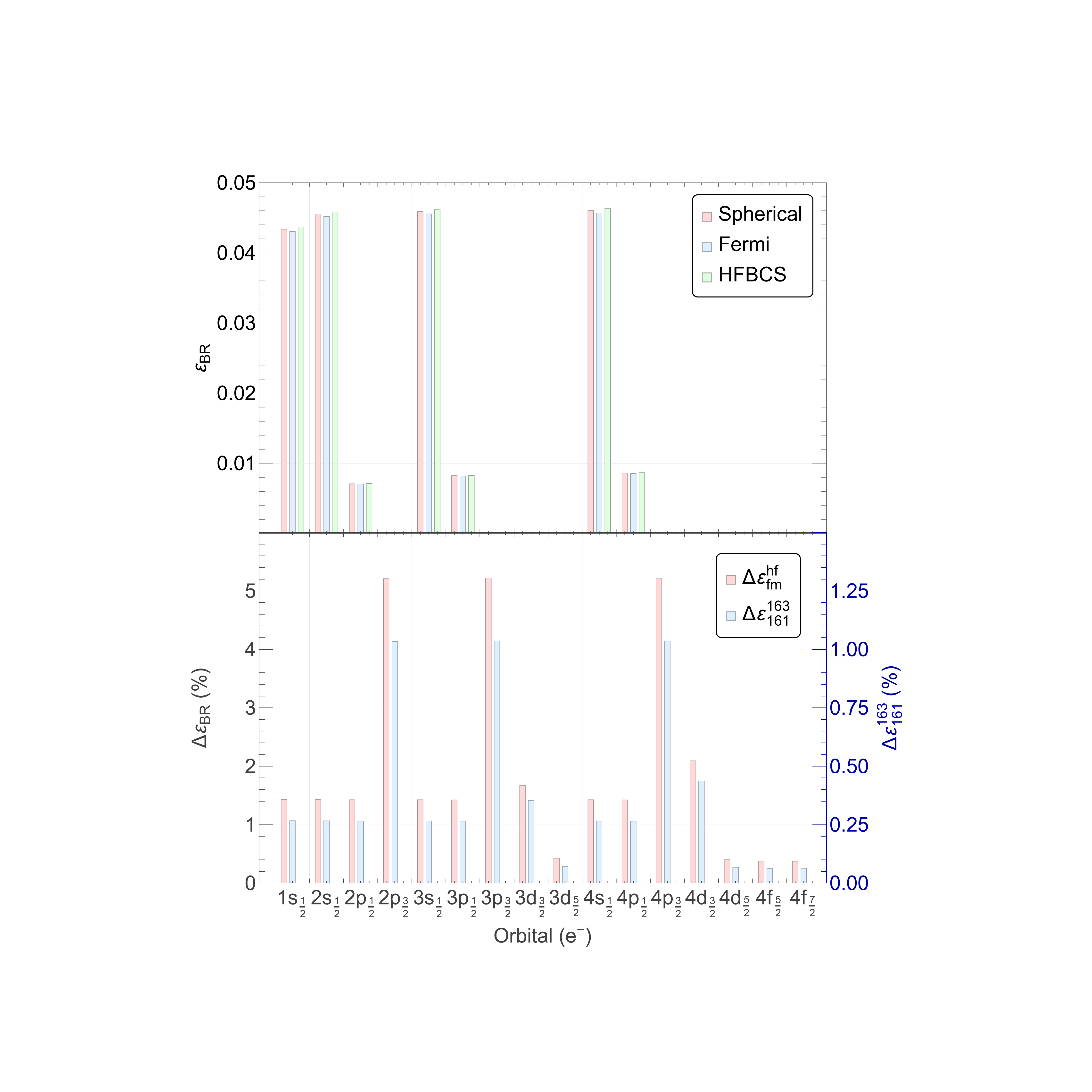}
\caption{(color online) Breit-Rosenthal correction $\varepsilon_{\mathrm{BR}}$ for $^{163}$Dy$^{65+}$ in different nuclear models: Spherical, Fermi and HFBCS. We have defined, $\Delta \varepsilon_{\mathrm{fm}}^{\mathrm{hf}} \equiv \frac{\varepsilon_{\mathrm{BR}}^{\mathrm{hf}}-\varepsilon_{\mathrm{BR}}^{\mathrm{fm}}}{\varepsilon_{\mathrm{BR}}^{\mathrm{fm}}}$ (reading on the left axis) and $\Delta \varepsilon_{161}^{163} \equiv \frac{\varepsilon_{\mathrm{BR}}^{\mathrm{hf}}(163)-\varepsilon_{\mathrm{BR}}^{\mathrm{hf}}(161)}{\varepsilon_{\mathrm{BR}}^{\mathrm{hf}}(161)}$ (reading on the right axis). The results for $^{161}$Dy$^{65+}$ are given in the appendix {(cf. \cref{fig:a6})}.}  \label{fig:e6}
\end{figure}

\Cref{fig:e7} is identical to \cref{fig:e6}, but pertains to a muonic atom. As in the electronic case, the different models yield similar results. What is very different is the value of the correction, which is close to $100\%$ for $ns_{1/2}$ orbitals. This value is more than 20 times higher than the one obtained in the electronic case. This increase is a direct consequence of the greater penetration of wave functions into the \sam{core}{nucleus}. The difference between $ns_{1/2}$ and $np_{1/2}$ is smaller in the muonic case, and the corrections to the $np_{3/2}$ orbitals are no longer negligible.

In the case of $\Delta \varepsilon_{\mathrm{fm}}^{\mathrm{hf}}$ and
$\Delta \varepsilon_{161}^{163}$, as can be seen in the lower panel,
the situation is very different between the electronic and muonic
cases. {In hydrogen-like Dy, the $\ell$-dependence of the BR effect is
  controlled by the orbital’s spatial extent relative to the nucleus:
  for electrons, increasing $\ell$ drives the probability density
  rapidly outward via centrifugal effects, so that near $R_{\rm N}$ the
  wavefunctions from different models and isotopes converge toward one
  another much faster than their respective BR effects decay. Hence
  the quantities $\Delta \varepsilon_{\mathrm{fm}}^{\mathrm{hf}}$ and
  $\Delta \varepsilon_{161}^{163}$ remain roughly constant as
  functions of $\ell$ since the numerators decay to zero faster than
  the denominators. For muons, the much smaller Bohr radius keeps the
  wavefunctions concentrated closer to the nuclear surface, so model-
  and isotope-dependent differences in wavefunctions remain large and
  decay more slowly than the BR effect as functions of $\ell$. Hence
  the denominators in $\Delta \varepsilon_{\mathrm{fm}}^{\mathrm{hf}}$
  and $\Delta \varepsilon_{161}^{163}$ decay to zero faster than the
  numerator terms, resulting in monotonically increasing fractional
  effects with $\ell$. Focusing on the electronic case, the largest
  relative BR signals arise for $j=\tfrac{3}{2}$
  ($p_{3/2},d_{3/2}$). Their absolute interior amplitude is modest,
  while $(\Delta\Psi/\Psi)$ is strongly model {and} isotope dependent, so the observable BR effect is comparatively large.}

\begin{figure}[ht!]
\centering
\includegraphics[width=.75\textwidth, trim={5.8cm 5cm 5.8cm 5cm},clip]{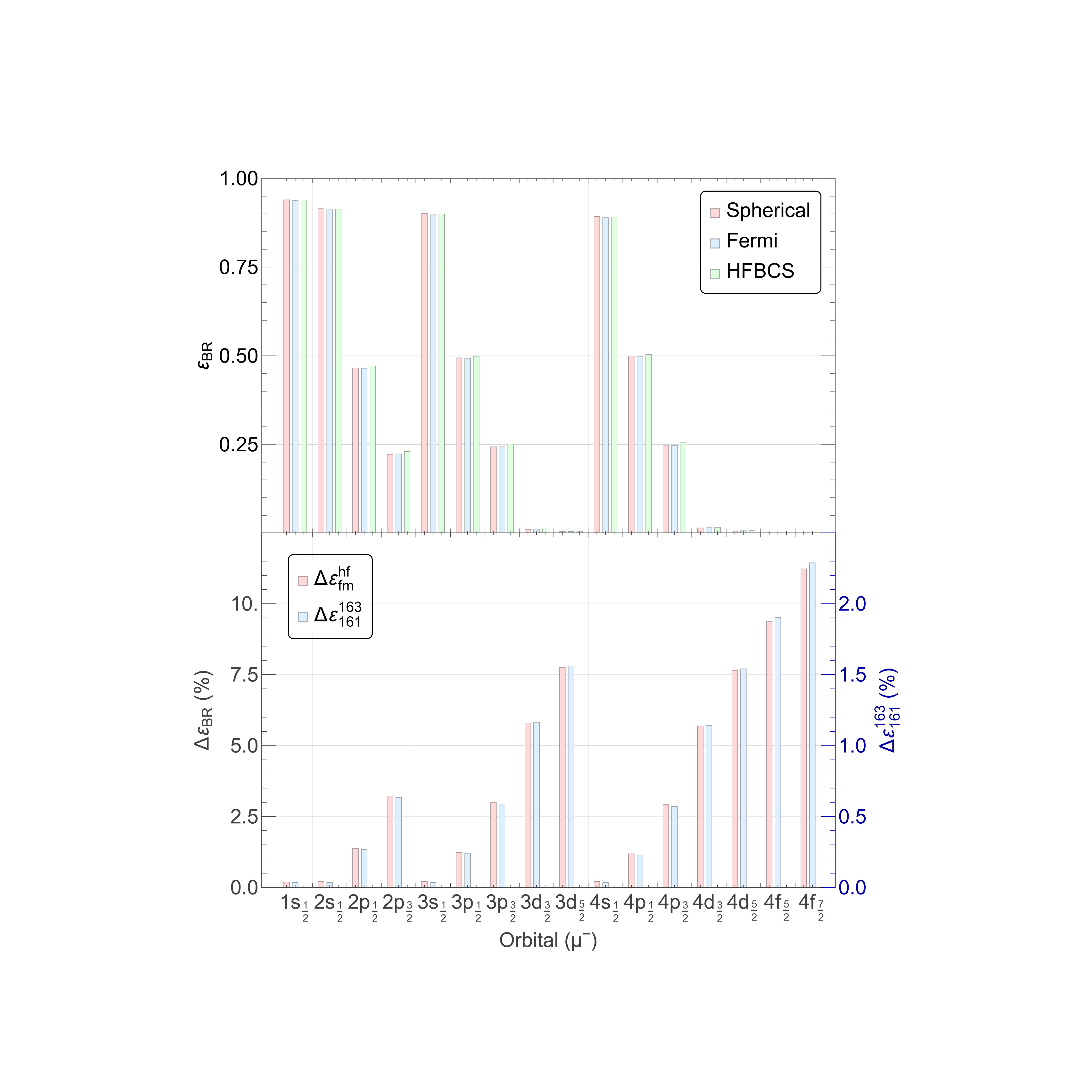}
\caption{(color online) Same as \cref{fig:e6} but for a muonic hydrogen-like ion. The results for $^{161}$Dy$^{65+}$ are given in the appendix {(cf. \cref{fig:a7})}.}  \label{fig:e7}
\end{figure}

{ It is interesting to note however, that model {discrepancies} for the $s$ states are of the same order of magnitude as QED and nuclear polarisation effects \cite{Vandeleur2025} ($\approx 10-100$ eV), this hints at $p_{1/2}$ states being more favorable for studying the effect of the internal nuclear structure as the effect is relatively one order of magnitude larger, all the while $\varepsilon_\text{BR}$ is only half the absolute value, relegating QED and nuclear \sam{polriation}{polarisation} effects down in importance.}

\subsubsection{Magnetic corrections}

Let us now analyse the results for the Bohr-Weisskopf correction defined in \cref{eq:BW} and denoted by the quantity $\varepsilon_{\rm BW}$.
\Cref{fig:e8} shows the contributions of the various orbital and spin currents ($j_p^{(\ell)}$, $j_n^{(\ell)}$, $j_p^{(s)}$ and $j_n^{(s)}$) to the $\varepsilon_{\mathrm{BW}}$ term for the orbitals $1s_{1/2}$ to $4f_{7/2}$ for the two hydrogen-like isotopes $^{161}$Dy$^{65+}$ and $^{163}$Dy$^{65+}$, in the upper and lower panels respectively.

We can clearly see that for all orbitals of both isotopes, the dominant contribution is provided by the neutron spin current. In accordance with preceding discussions, this observation can be attributed to both isotopes having an even number of paired protons and a single unpaired neutron. Moreover, all of the corrections provided by this current are positive, despite the two isotopes having magnetic moments of opposite signs. The correction terms increase positively since the absolute magnitudes in the denominators all decrease in size. We also observe that for both isotopes the two proton contributions, induced by the aforementioned neutron currents, are negative and that the contribution associated with the orbital motion is almost non-existent. This leads us to conclude that it is the nuclear magnetism due to the spins of the nucleons that acts on the atomic electrons through the hyperfine interaction. The strongest corrections affect the $s$ orbitals, which penetrate deeper into the nucleus. The total correction is around $2\%$, which is far from negligible. Finally, another noteworthy comparison between the two isotopes is that the magnitudes of the spin contributions of protons and neutrons are larger for $^{163}$Dy than for $^{161}$Dy. However, since the proton contributions are positive while the neutron contributions are negative, we can see that the total correction remains approximately unchanged.

\begin{figure}[ht!]
\centering
\includegraphics[width=.9\textwidth, trim={3cm 0cm 3cm 0cm},clip]{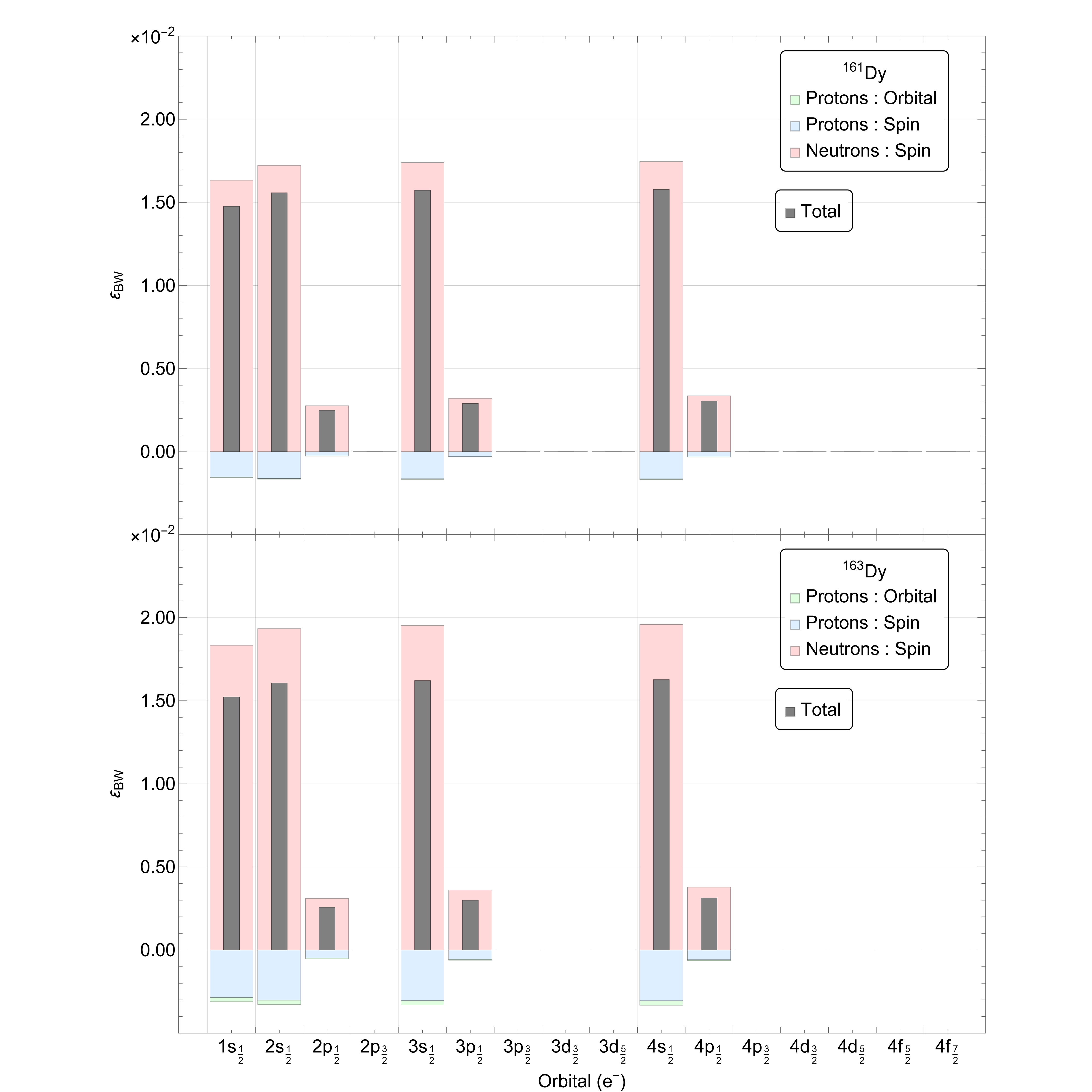}
\caption{(color online) Bohr-Weisskopf correction $\varepsilon_{\mathrm{BW}}$ for $^{161}$Dy$^{65+}$ (upper panel) and $^{163}$Dy$^{65+}$ (lower panel). Only the HFBCS model is used. The various magnetic contributions (proton and neutron spins and proton orbital motion) are shown, together with the total correction.}  \label{fig:e8}
\end{figure}

\Cref{fig:e9} is similar to \cref{fig:e8}, but for a muonic ion. As
expected, the main difference is the very significant increase in the
value of the quantities studied. Indeed, the correction is increased
by a factor of $50$ compared with the electronic case and is of the
same order as that obtained from the calculation without correction
(i.e. close to $100\%$){, \sam{coherent}{consistent} with the order
  of magnitude calculated for other elements
  \cite{Vandeleur2025}. Moreover, these results fall within the same
  order of magnitude as recent experimental results for neighbouring
  lanthanides, providing corroborating evidence for our calculations
  \cite{Persson2025}.} Overall, the conclusions are the same as for
the electronic case, although there is one significant difference: the
$np_{1/2}$ orbitals are just as strongly affected as the $ns_{1/2}$
orbitals. As in the case of the electron, we can see that it is the
nuclear magnetism \sam{due to}{induced by} the spins of the nucleons that acts on the atomic muons through the hyperfine interaction. 

\begin{figure}[ht!]
\centering
\includegraphics[width=.9\textwidth, trim={3cm 0cm 3cm 0cm},clip]{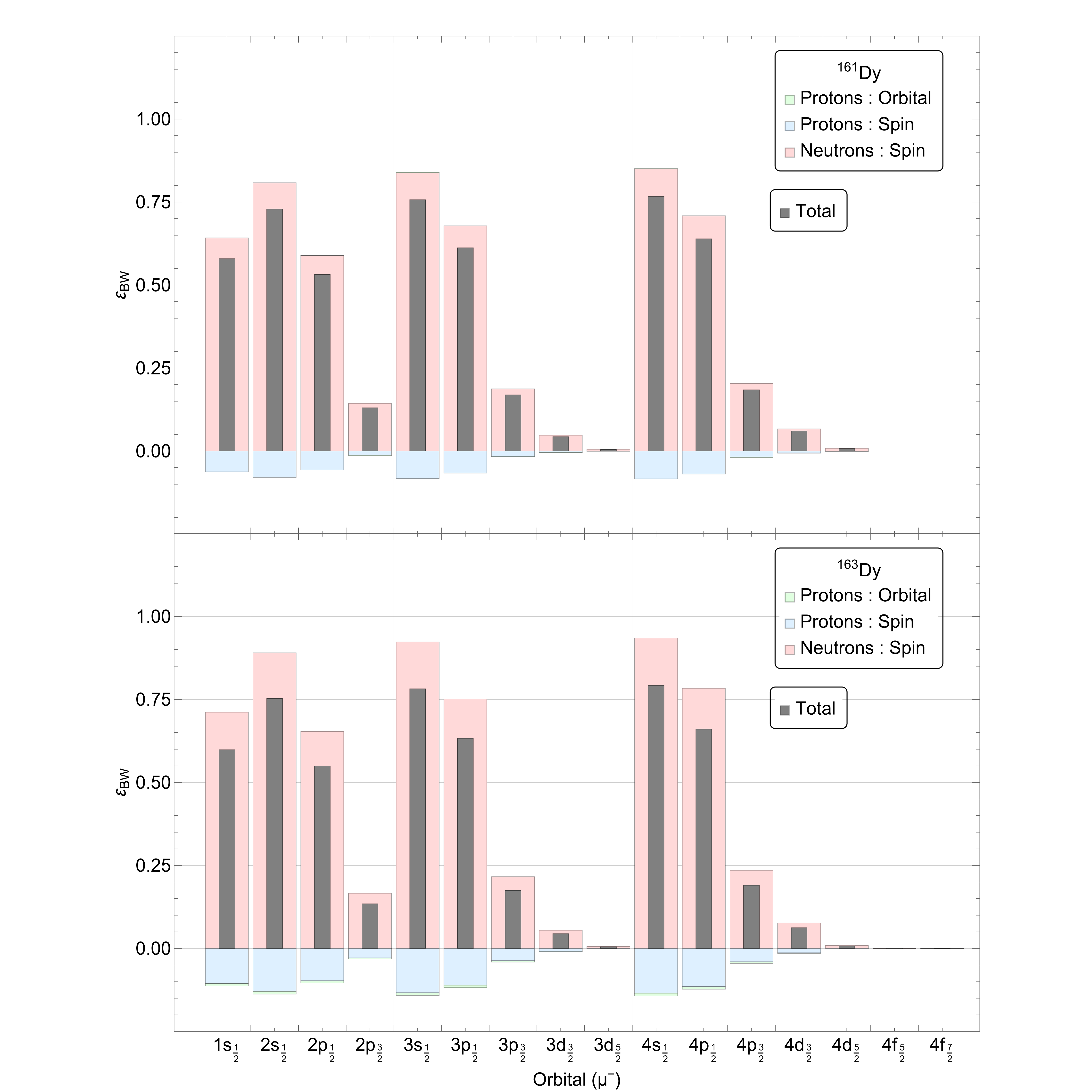}
\caption{(color online) Same as \cref{fig:e8} but for muonic hydrogen-like ions.}  \label{fig:e9}
\end{figure}

\Cref{fig:e10} shows the BR and BW corrections on the same graph for the two isotopes as a function of the principal quantum number $n$ for the $s_{1/2}$, $p_{1/2}$ and $p_{3/2}$ orbitals. Isotopic effects are very small. They can only be observed for the BW correction and the $s_{1/2}$ and $p_{1/2}$ orbitals. The behaviour of $\varepsilon_{\rm BR}$ and $\varepsilon_{\rm BW}$ as a function of $n$ is almost constant from $n=2$ onward.

\begin{figure}[ht!]
\centering
\includegraphics[width=.66\textwidth, trim={0cm 5cm 0cm 5cm},clip]{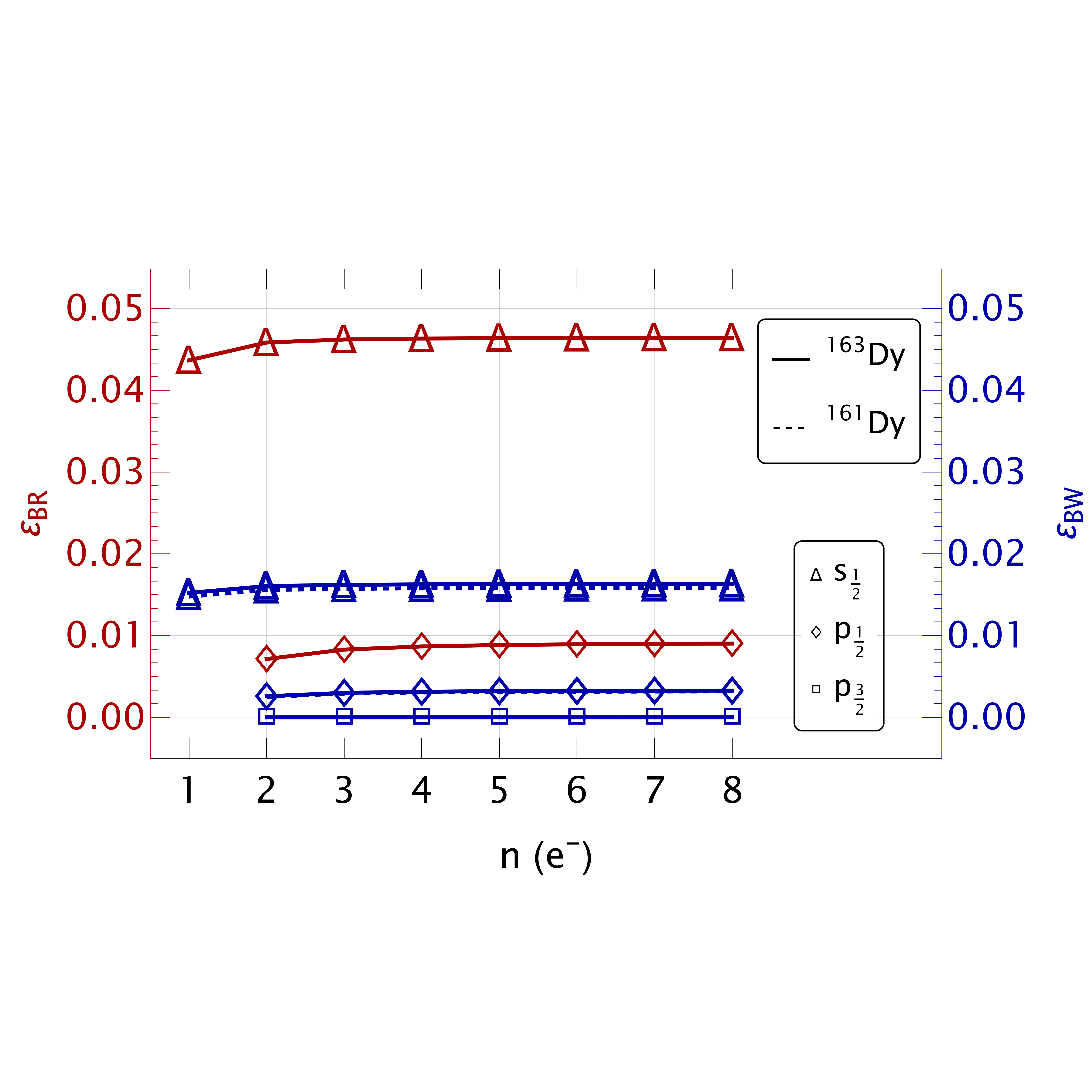}
\caption{(color online) Breit-Rosenthal (in red and reading on the left axis) and Bohr-Weisskopf (in blue and reading on the right axis) corrections for $^{163}$Dy (solid lines) and $^{161}$Dy (dashed lines). The results are plotted as a function of quantum number $n$ for the orbitals $ns_{1/2}$ (triangles), $np_{1/2}$ (diamonds) and $np_{3/2}$ (squares). It is important to note that, with the exception of the Bohr-Weisskopf correction for the $s_{1/2}$ and $p_{1/2}$ orbitals, where there is a slight difference for $^{161}$Dy$^{65+}$ and $^{163}$Dy$^{65+}$, the other curves are superimposed for both isotopes. For $p_{3/2}$, as the corrections are very small, the straight line with square symbols actually corresponds to four superimposed curves including the two corrections and the two isotopes.}  \label{fig:e10}
\end{figure}

\Cref{fig:e11} is similar to \cref{fig:e10}, but for a muonic ion. As expected, the isotopic effects are greater than in the electronic case, particularly for the $p_{1/2}$ orbitals. In contrast to the electronic case, the values of the BR and BW corrections are almost similar for the three types of orbitals. This was not the situation for the electrons, where the BR corrections were much larger than those for the BW case. Finally, the behaviour as a function of $n$ is less constant than in the electronic case, with significant variation for small values of $n$.

\begin{figure}[ht!]
\centering
\includegraphics[width=.66\textwidth, trim={0cm 4.5cm 0cm 5.5cm},clip]{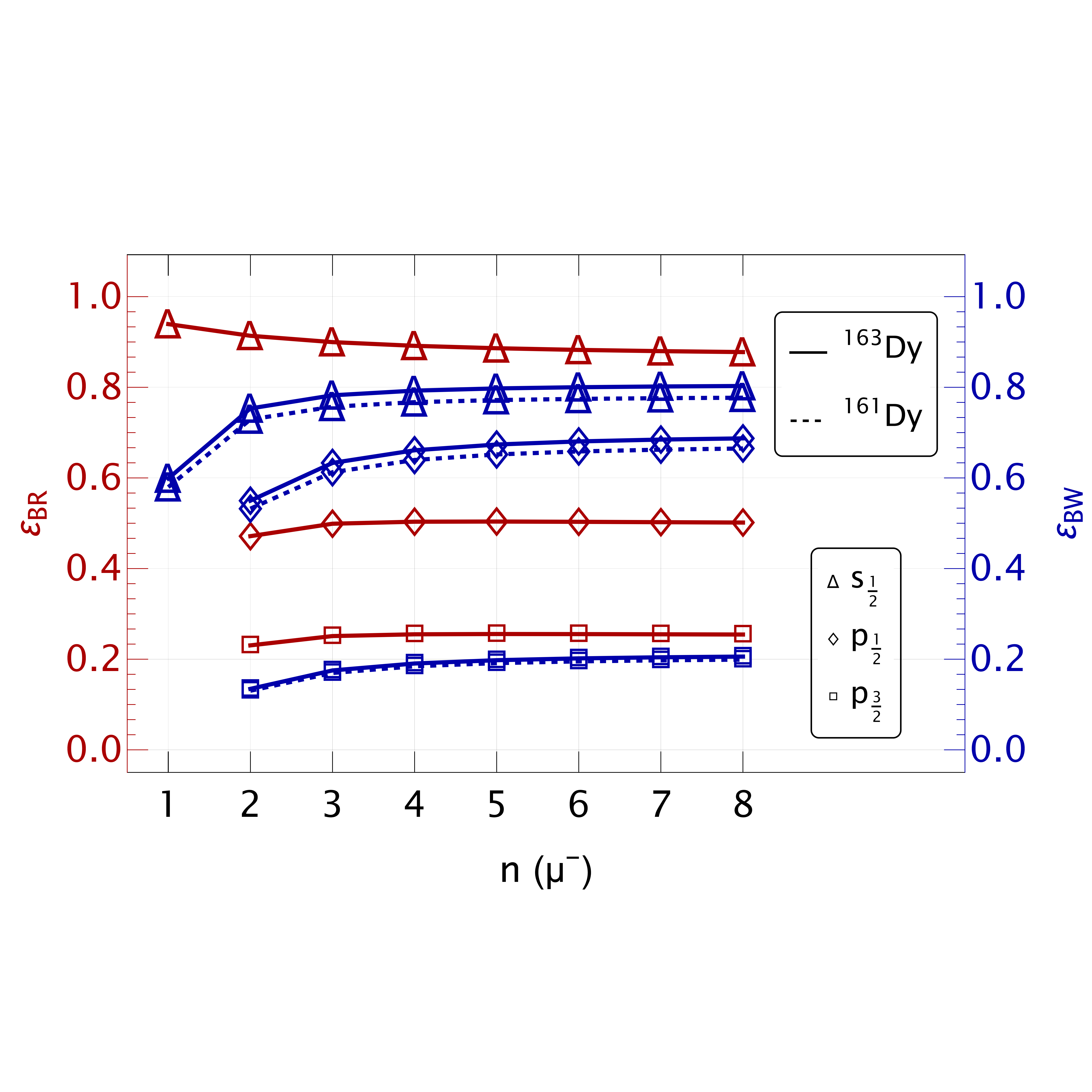}
\caption{(color online) Same as \cref{fig:e10} but for muonic hydrogen-like ions.}  \label{fig:e11}
\end{figure}

\subsubsection{Hyperfine anomaly}

The hyperfine anomaly for both dysprosium isotopes is shown in \cref{fig:e12}, where both electronic and muonic cases are shown on the same graph. As expected, the hyperfine anomaly is most prominent in the $j=1/2$ states, and especially the $s$ states, showcasing the importance of the Fermi contact term \cite{Persson2023hyperfine}. For example, for the $s$ orbitals of muonic ions, we can reach values of $\Delta_{163}^{161}$ larger than $10\%$. Moreover, it is still of the order of $1\%$ for $j=3/2$ states in the muonic case, whereas it is practically nonexistent for the electronic case in this situation, demonstrating the increased sensitivity of the muonic ions to the internal nuclear structure beyond the Fermi contact term.

\begin{figure}[ht!]
\centering
\includegraphics[width=.75\textwidth, trim={0cm 5cm 0cm 5cm},clip]{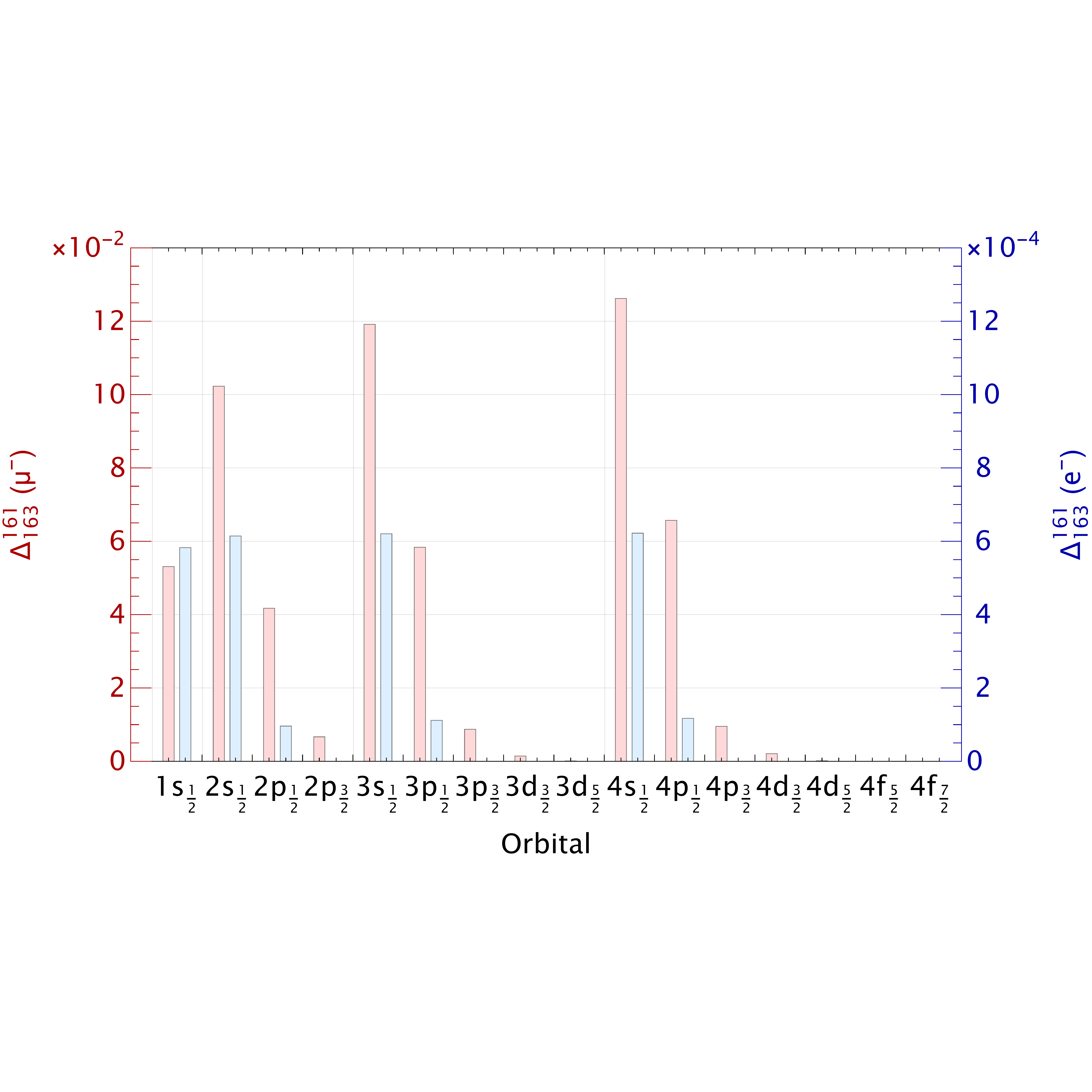}
\caption{(color online)  The hyperfine anomaly, $\Delta_{163}^{161} \equiv \frac{\left[(1+\varepsilon_{\mathrm{BR}})\times(1+\varepsilon_{\mathrm{BW}})\right]^{161}}
{\left[(1+\varepsilon_{\mathrm{BR}})\times(1+\varepsilon_{\mathrm{BW}})\right]^{163}}-1$, is depicted for muonic (in red and reading on the left axis) and electronic (in blue and reading on the right axis) hydrogen-like ions for the orbitals from $1s_{1/2}$ to $4f_{7/2}$.}  \label{fig:e12}
\end{figure}

\Cref{fig:e13} shows the hyperfine anomaly as a function of the
principal quantum number $n$ again for the electronic as well as the muonic case. It can be seen that for muonic ions the
value of $\Delta_{163}^{161}$ increases with $n$ and reaches a plateau
for $n>6$ whereas the behaviour is effectively constant for electronic
ions. \sam{The latter}{This result} gives credence to the customary
assumption that the Fermi contact term and hence the hyperfine anomaly
is $n$-independent. However, for muonic ions, it becomes apparent
{that} the hyperfine anomaly is heavily dependent on $n$, with
e.g. the anomaly for the $1s_{1/2}$ state being of the order of $5\%$,
less than half of that for $n\geqslant2$. Given that a lot of muonic
studies focus near the ground $1s_{1/2}$ state
\cite{Persson2023hyperfine, Freeman1984}, this observation indicates
that simply importing the electronic-atom intuition to the muonic
sector can lead to sizeable systematic biases when extracting nuclear
properties from hyperfine data. More precisely, any analysis that constrains nuclear magnetisation radii or higher-order nuclear moments using only (or predominantly) $1s_{1/2}$ muonic transitions may underestimate the impact of nuclear-structure effects on levels with larger $n$. This is particularly relevant when extracting the hyperfine anomaly from $s$--$s$ transitions, as in the electronic case.
\cite{PrezGalvn2007}\sam{, this}{. This} $n$-dependence needs to be taken into account.

Furthermore, these results demonstrate the relevance of our model for the first-principles calculation of the Fermi contact term for specific shells, which is commonly treated as a free parameter extracted from experimental data.
\begin{figure}[ht!]
\centering
\includegraphics[width=.66\textwidth, trim={0cm 5cm 0cm 5cm},clip]{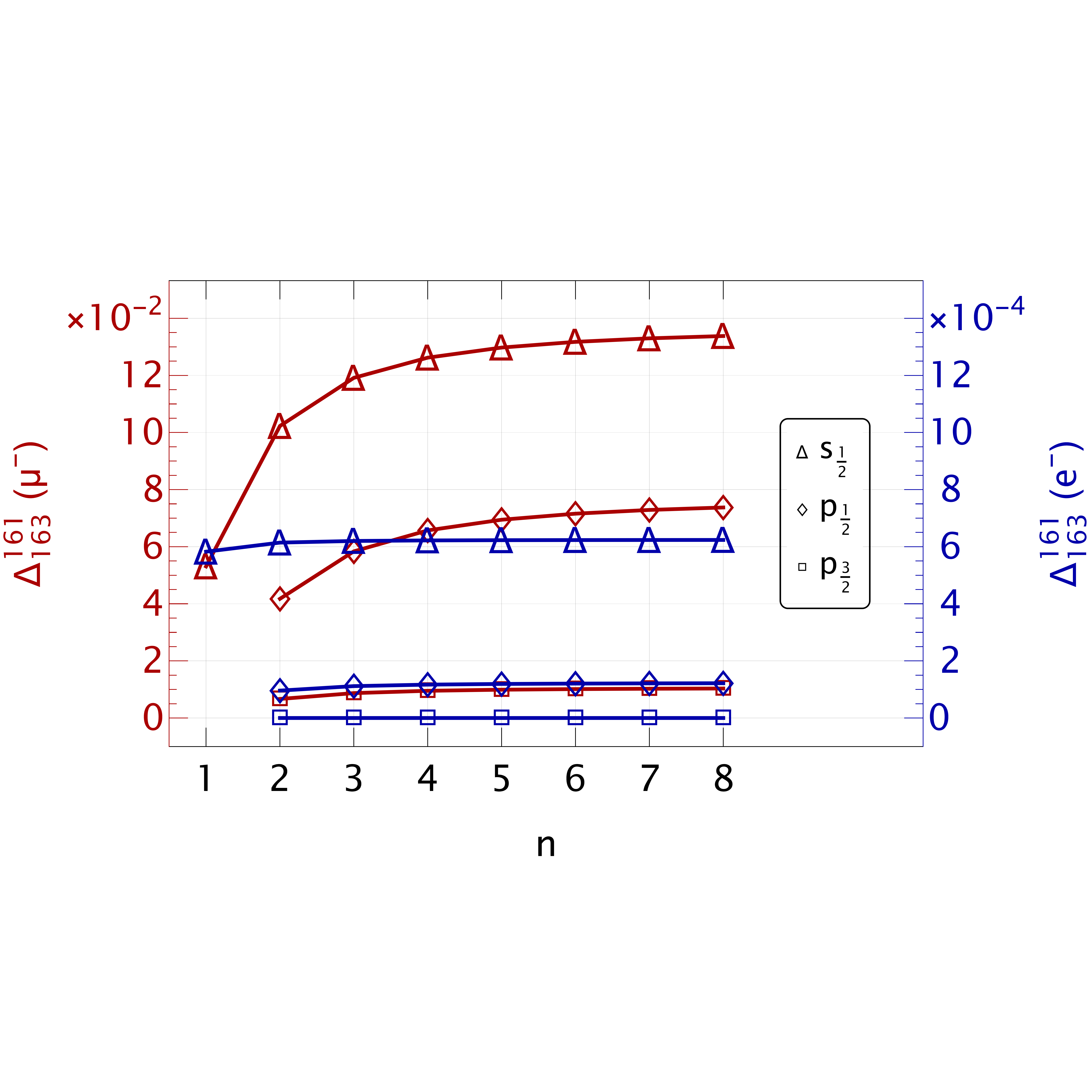}
\caption{(color online)  The quantity $\Delta_{163}^{161} \equiv \frac{\left[(1+\varepsilon_{\mathrm{BR}})\times(1+\varepsilon_{\mathrm{BW}})\right]^{161}}
{\left[(1+\varepsilon_{\mathrm{BR}})\times(1+\varepsilon_{\mathrm{BW}})\right]^{163}}-1$ is depicted for muonic (in red and reading on the left axis) and electronic (in blue and reading on the right axis) hydrogen-like ions as a function of quantum number $n$ for the orbitals $ns_{1/2}$ (triangles), $np_{1/2}$ (diamonds) and $np_{3/2}$ (squares).}  \label{fig:e13}
\end{figure}

{ An important point for these hyperfine magnetic corrections in the
  muonic case is that they become {significantly}
  larger than QED and nuclear polarisation corrections to the
  hyperfine constant, which were calculated to be only around $0.1 \%$ of
  that of the Bohr-Weisskopf correction \cite{Vandeleur2025} for
  example. \sam{Meaning}{This means that} experimentally obtained hyperfine anomalies mainly only originate from the finite size effects.}

\subsubsection{Correction of the quadrupole term}
In the following we will discuss the results obtained for the hyperfine quadrupole term given by the formulas \cref{eq:definition_bwp} and \cref{eq:definition_bnp}. \Cref{fig:e14} shows the three quantities introduced in \cref{Electric Quadrupole}.

For ${\Delta B}^{m}$, defined in \cref{eq:deltaBm} and shown in the upper panel of the figure, we see that: (i) the four models give almost identical results; (ii) The $np_{3/2}$ orbitals make by far the largest contribution and; (iii) for $m=\mathrm{pt}$, ${\Delta B}^{\mathrm{pt}}$ quantifies the quadrupole moment correction due to penetration into the nucleus. This is of the order of $0.006\%$ for the $np_{3/2}$ orbitals. Let us recall that the $s_{1/2}$ and $p_{1/2}$ contributions to the quadrupole term are identically zero due to spherical symmetry of the orbital wavefunctions.

In the lower panel of the figure we have plotted ${\Delta B}$, which measures the relative deviation between the results of the microscopic model and those from the nuclear model with a Fermi distribution, which is the most accurate of the non-microscopic models. We can clearly see that the relative difference is extremely small, of the order of $10^{-6}\%$. So the use of a simple model like that of Fermi is well-suited to this situation. Let us point out that the choice of nuclear model only affects the electronic wavefunctions that appear in the integral used to define the quadrupole constant ($b$) of \cref{eq:definition_bwp}. In that same integral, the nuclear charge distribution must always be taken from the HFBCS calculation, because the other empirical nuclear models considered here are by definition spherically symmetric. As a consequence, $\Delta B$ does not have a sound physical definition; it is introduced solely as a numerical benchmark to test whether one really needs to solve the Dirac equation with the full HFBCS nuclear structure, or whether it is sufficient to use the HFBCS charge distribution only in the post-processing stage, together with atomic wavefunctions obtained from a simpler nuclear model.

The lower panel also shows ${\Delta B}_{161}^{163}$ which measures the isotopic effect on the hyperfine quadrupolar correction. This quantity is calculated using the most sophisticated nuclear model (HFBCS) and includes the quadrupole moment correction due to the penetration into the nucleus (i.e. $b=b_{\mathrm{wp}}^{\mathrm{hf}}$). It is interesting to see that this quantity is virtually constant (at $~4\%$) across all the electronic orbitals presented.

\begin{figure}[ht!]
\centering
\includegraphics[width=.75\textwidth, trim={5.8cm 4.8cm 5.8cm 5cm},clip]{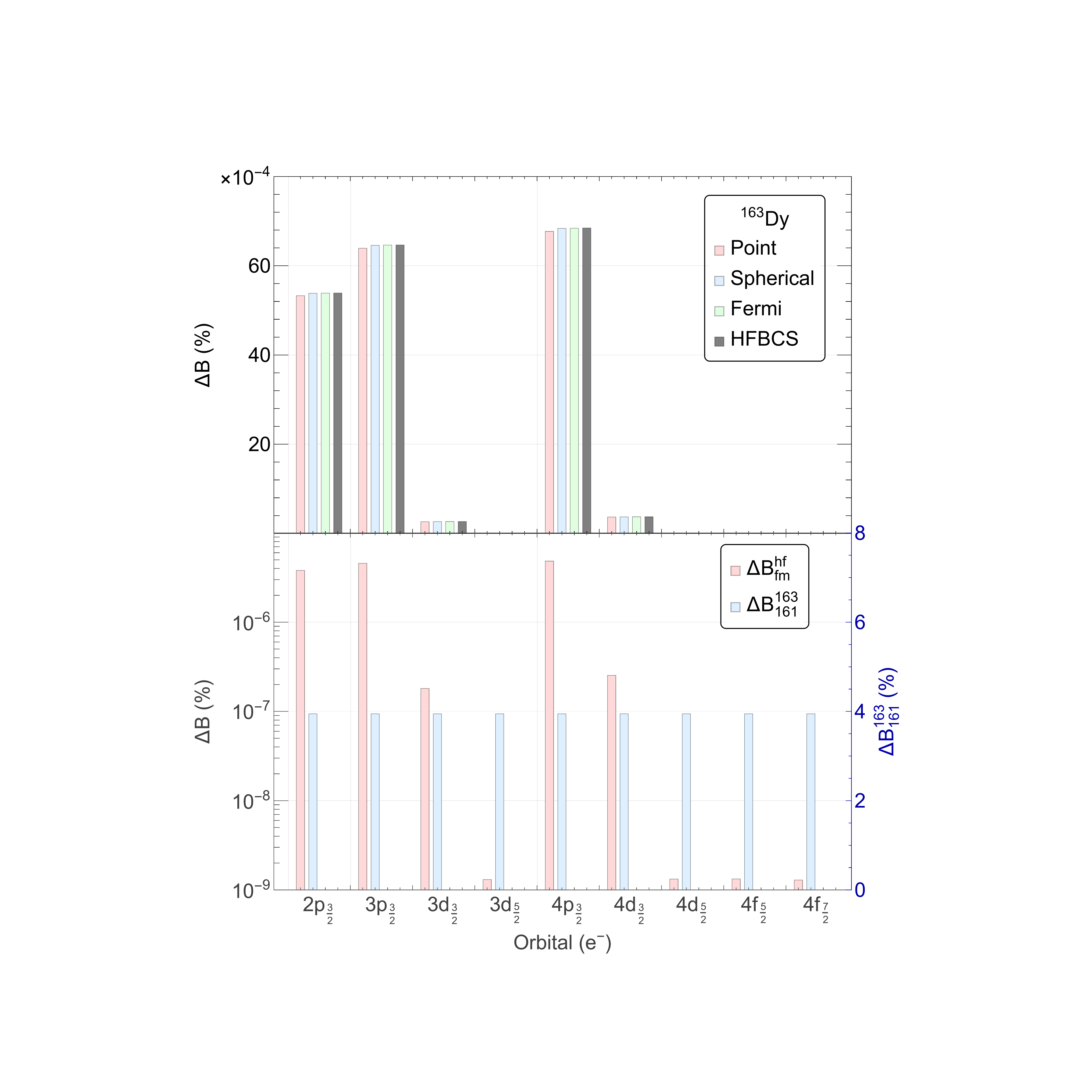}
\caption{(color online)  Quadrupole shifts for electronic $^{163}$Dy$^{65+}$. (Upper panel) ${\Delta B}^{m} \equiv \frac{b_{\mathrm{np}}^{\mathrm{pt}}-b_{\mathrm{wp}}^{m}}{b_{\mathrm{np}}^{\mathrm{pt}}}$ with $m\in\left\{\mathrm{pt}, \mathrm{sp}, \mathrm{fm}, \mathrm{hf}\right\}$. The subscript indices $\mathrm{np}$ and $\mathrm{wp}$ represent calculations with no penetration and with penetration, respectively (see text for explanation). (Lower panel) Two quantities are depicted: ${\Delta B}\equiv \frac{b_{\mathrm{wp}}^{\mathrm{fm}}-b_{\mathrm{wp}}^{\mathrm{hf}}}{B_{\mathrm{wp}}^{\mathrm{fm}}}$ for $^{163}$Dy$^{65+}$ (in red and reading on the left axis) and ${\Delta B}_{161}^{163} \equiv \frac{b_{\mathrm{wp}}^{\mathrm{hf}}(163)-b_{\mathrm{wp}}^{\mathrm{hf}}(161)}{b_{\mathrm{wp}}^{\mathrm{hf}}(161)}$ (in blue and reading on the right axis). Note that the $s_{1/2}$ and $p_{1/2}$ contributions are zero. The results for electronic $^{161}$Dy$^{65+}$ are given in the appendix {(cf. \cref{fig:a14})}.}  \label{fig:e14}
\end{figure}

{ For electronic ions, the quadrupole shifts are approximately $10^{-3}\%$ for $p_{3/2}$ states, which is considerably larger than the dipole corrections (BR and BW) for the same states. Moreover, the quadrupole contribution to the hyperfine splitting of the $2p_{3/2}$ state (transition $F=4 \rightarrow F=3$) in $^{163}$Dy$^{65+}$ is around $24 \, \mathrm{meV}$, nearly eight times larger than the dipole contribution ($\approx 3 \, \mathrm{meV}$). Nonetheless, this quadrupole shift of $10^{-3}\%$ of the total transition energy is still difficult to measure for electronic ions, as it would amount to a difference of the order of $\approx 10 \, \mathrm{meV}$.}

\begin{figure}[ht!]
\centering
\includegraphics[width=.75\textwidth, trim={5.8cm 4.5cm 5.8cm 5cm},clip]{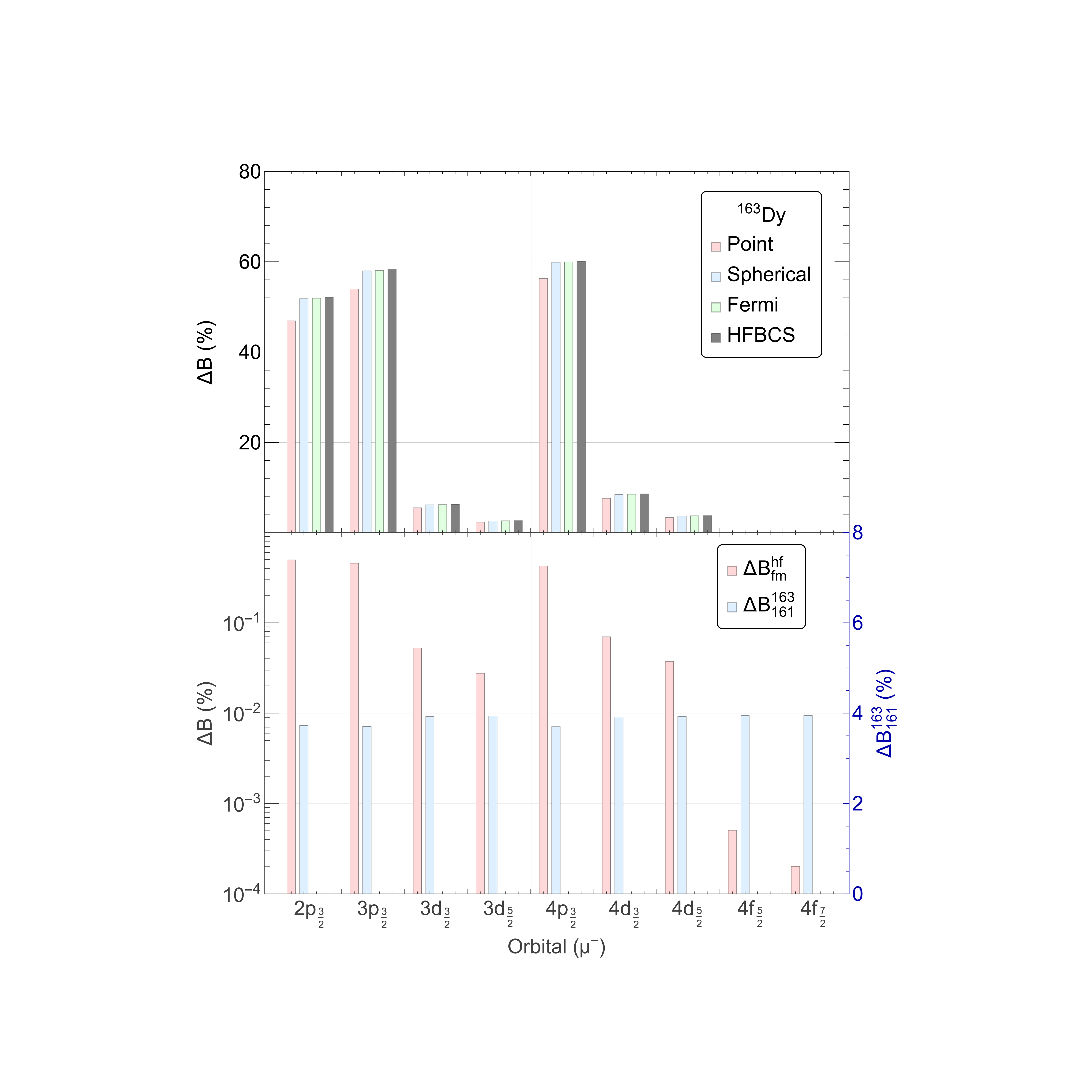}
\caption{(color online) Same as \cref{fig:e14} but for muonic $^{163}$Dy$^{65+}$. The results for muonic $^{161}$Dy$^{65+}$ are given in the appendix {(cf. \cref{fig:a15})}.}  \label{fig:e15}
\end{figure}

{ In contrast, the quadrupole shifts for the muonic orbitals of
  $^{163}$Dy$^{65+}(\mu^-)$, as seen in \cref{fig:e15}, are much
  larger: around $50\%$ for $p_{3/2}$ states, \sam{$5-10\%$}{$5$ to
    $10\%$} for $d_{3/2}$ states, and $5\%$ for $d_{5/2}$
  states. These values can be compared to the magnetic hyperfine
  corrections for the same states, which are \sam{around
    $1-2\%$}{between $1\%$ and $2\%$}. Here, the quadrupole
  contribution to the hyperfine splitting of the $2p_{3/2}$ state in
  the muonic ion is \sam{about $100 \, \mathrm{keV}$}{of the order of
    hundreds of {keV}}, consistent with experimental data \cite{Povel1973,Michel2017,Persson2025}, and significantly higher than the dipole contribution ($\approx 100s \, \mathrm{eV}$). Therefore, we may note that the quadrupole interaction is the dominant factor in the hyperfine splitting of the muonic $2p_{3/2}$ state, making the quadrupole shift measurable in this case, with an energy shift {of} the order of tens of keV. { Recent studies have also concluded that the QED and second-order corrections to the quadrupolar constant in muonic atoms are of the order of $0.1-1 \, \mathrm{keV}$ \cite{Michel2019}, placing both the finite size effects and the latter in the same ballpark for their effect on muonic quadrupolar splittings.} It is also interesting to compare the shifts between the Fermi and HFBCS models ($\Delta B ^{\rm hf}_{\rm fm}$) for the muonic ions in the lower panel of \cref{fig:e15}, which show an increase of roughly 5 orders of magnitude from the electronic case in \cref{fig:e14}. This renders it experimentally relevant, as the corresponding energy shift falls within the experimental precision of the order of eVs \cite{Persson2025}. As for the isotopic effect, ${\Delta B}_{161}^{163}$, we can see the increased sensitivity of the muonic orbitals to the orbital quantum number that was not visible in the electronic case.}

\begin{figure}[ht!]
\centering
\includegraphics[width=.75\textwidth, trim={5.8cm 4.8cm 5.8cm 5cm},clip]{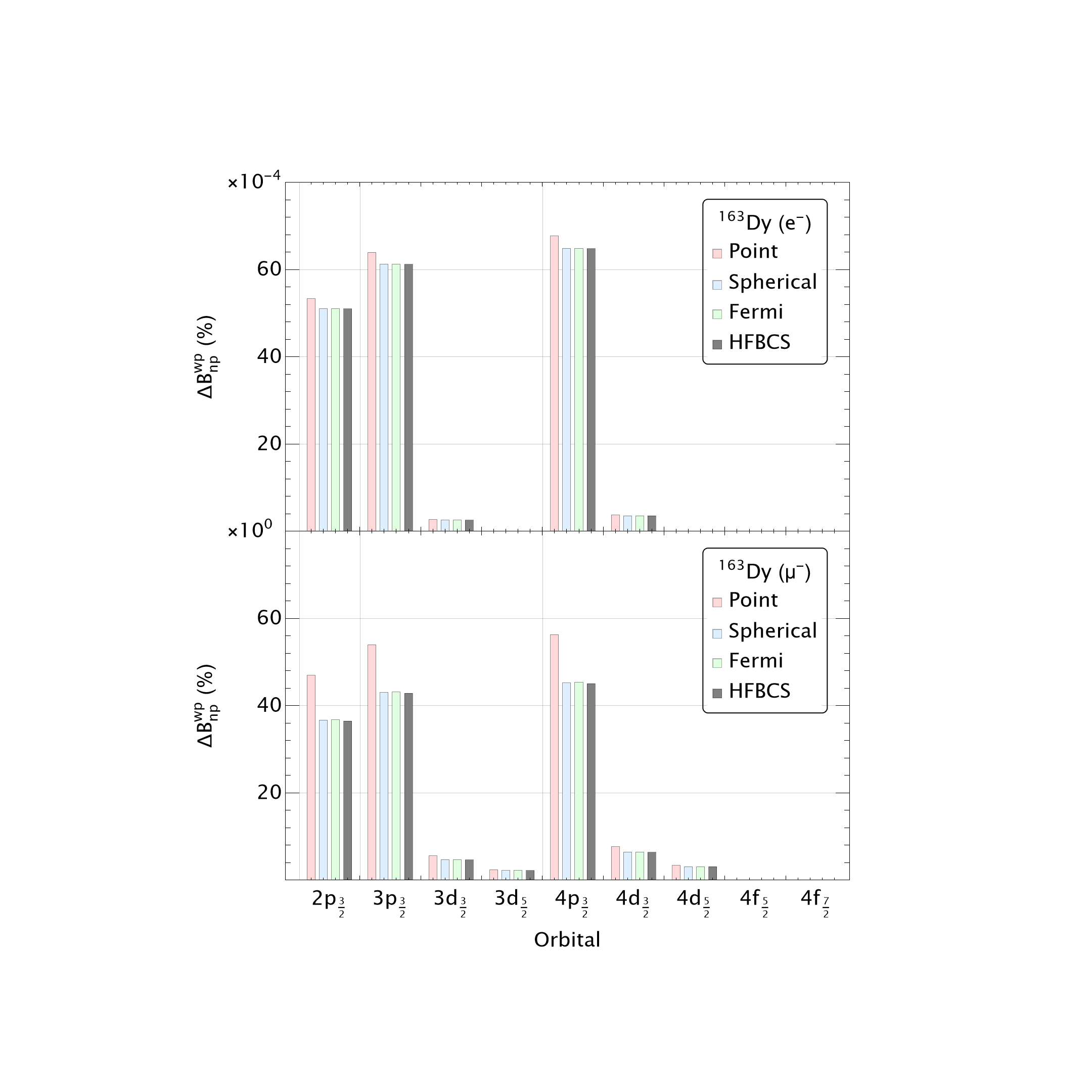}
\caption{(color online) Relative change of the quadrupole constant $\Delta B_{\mathrm{np}}^{\mathrm{wp}} \equiv \frac{b_{\mathrm{wp}}-b_{\mathrm{np}}}{b_{\mathrm{np}}}$ for each model ($\mathrm{pt}, \mathrm{sp}, \mathrm{fm}, \mathrm{hf}$) calculated with penetration ($\rm wp$) and with no penetration ($\rm np$) for electronic (Upper panel) and muonic (Lower panel) $^{163}$Dy$^{65+}$, respectively. Note that the $s_{1/2}$ and $p_{1/2}$ contributions are zero.}  \label{fig:e18}
\end{figure}

{ As a complement, \cref{fig:e18} illustrates how the penetration of the electronic and muonic wavefunctions into the nucleus of $^{163}$Dy affects the quadrupole constant, by comparing the results with those obtained for a point-like quadrupole. The magnitude of this relative difference reflects the significance of finite-size effects on the quadrupole interaction, as opposed to a simplistic point-charge approximation. For the electronic case, the deviation reaches at most about $10^{-4}$, rendering it negligible from an experimental standpoint. In contrast, for the muonic configuration, the effect is roughly $10^4$ times larger, while the absolute quadrupole interaction itself is also orders of magnitude greater, making it a quantitatively relevant contribution.}

{ As observed in \cref{fig:e14,fig:e15}, the relative difference of
  the quadrupole constant between the two dysprosium isotopes is
  effectively constant around \sam{$\approx 4\%$}{$4\%$}, with
  deviations from that value becoming more apparent \sam{for}{in} the muonic
  case. This constant corresponds naturally to the relative difference
  of the calculated quadrupole constants for the two isotopes.} 

{ In order to resolve the finer changes due to the internal nuclear structure, a quantity similar to the magnetic hyperfine anomaly can be defined, namely the quadrupole hyperfine anomaly. The latter is further calculated for both electronic and muonic ions, as defined by
\begin{equation}
    {_Q\Delta^{A}_B} \equiv \frac{b(A)}{b(B)} \frac{{\cal Q}_{0}^{(2),\,\text{(spec)}}(B)}{{\cal Q}_{0}^{(2),\,\text{(spec)}}(A)} - 1,
\end{equation}    
where $b(A)$ and $b(B)$ are the quadrupole hyperfine constants for isotopes $A$ and $B$, ${\cal Q}_{0}^{(2),\,\text{(spec)}}(A)$ and ${\cal Q}_{0}^{(2),\,\text{(spec)}}(B)$ are their respective nuclear quadrupole moments.} 

\begin{figure}[ht!]
\centering
\includegraphics[width=.66\textwidth, trim={0cm 5cm 0cm 5cm},clip]{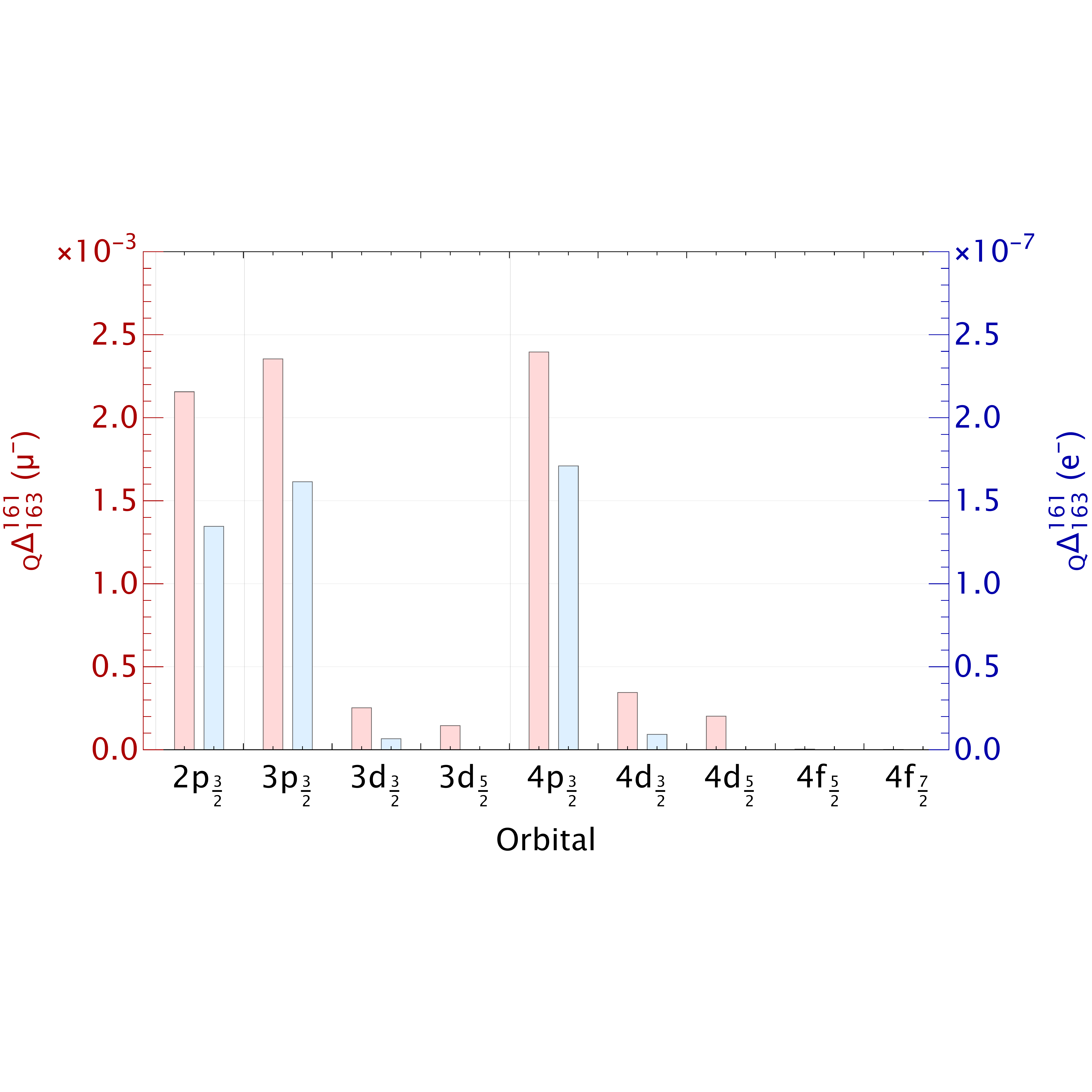}
\caption{(color online) The ratio by isotope of the shift in quadrupole moment from the simplest $(\mathrm{np}, \mathrm{pt})$ to most complex $(\mathrm{wp}, \mathrm{hf})$ model is defined by: ${}_{Q}\Delta_{163}^{161}=\frac{\Delta B(161)}{\Delta B(163)}-1$ where $\Delta B(A)\equiv \frac{b_{\mathrm{wp}}^{\mathrm{hf}}(A)}{b_{\mathrm{np}}^{\mathrm{pt}}(A)}$. Results for the electronic and muonic ions are shown in blue and red, with readings on the right and left axes respectively. Note that the $s_{1/2}$ and $p_{1/2}$ contributions are zero.} \label{fig:e16}
\end{figure}

{ Unlike the magnetic hyperfine anomaly, the quadrupole anomaly does not need to be decomposed into two separate contributions. The magnetic anomaly arises from both (i) the finite nuclear charge distribution and (ii) the finite nuclear current (magnetisation) distribution, whereas the quadrupole anomaly is sourced solely by the finite charge distribution. As shown in \cref{fig:e16,fig:e17}, our results suggest that the anomaly is of the order of $10^{-7}$ for the electronic ion and $10^{-3}$ for the muonic ion. Coupled with the much larger quadrupole hyperfine interaction in the muonic case, this makes the quadrupole hyperfine anomaly observable for the muonic ion but not for the electronic ion.}

\begin{figure}[ht!]
\centering
\includegraphics[width=.66\textwidth, trim={0cm 5cm 0cm 5cm},clip]{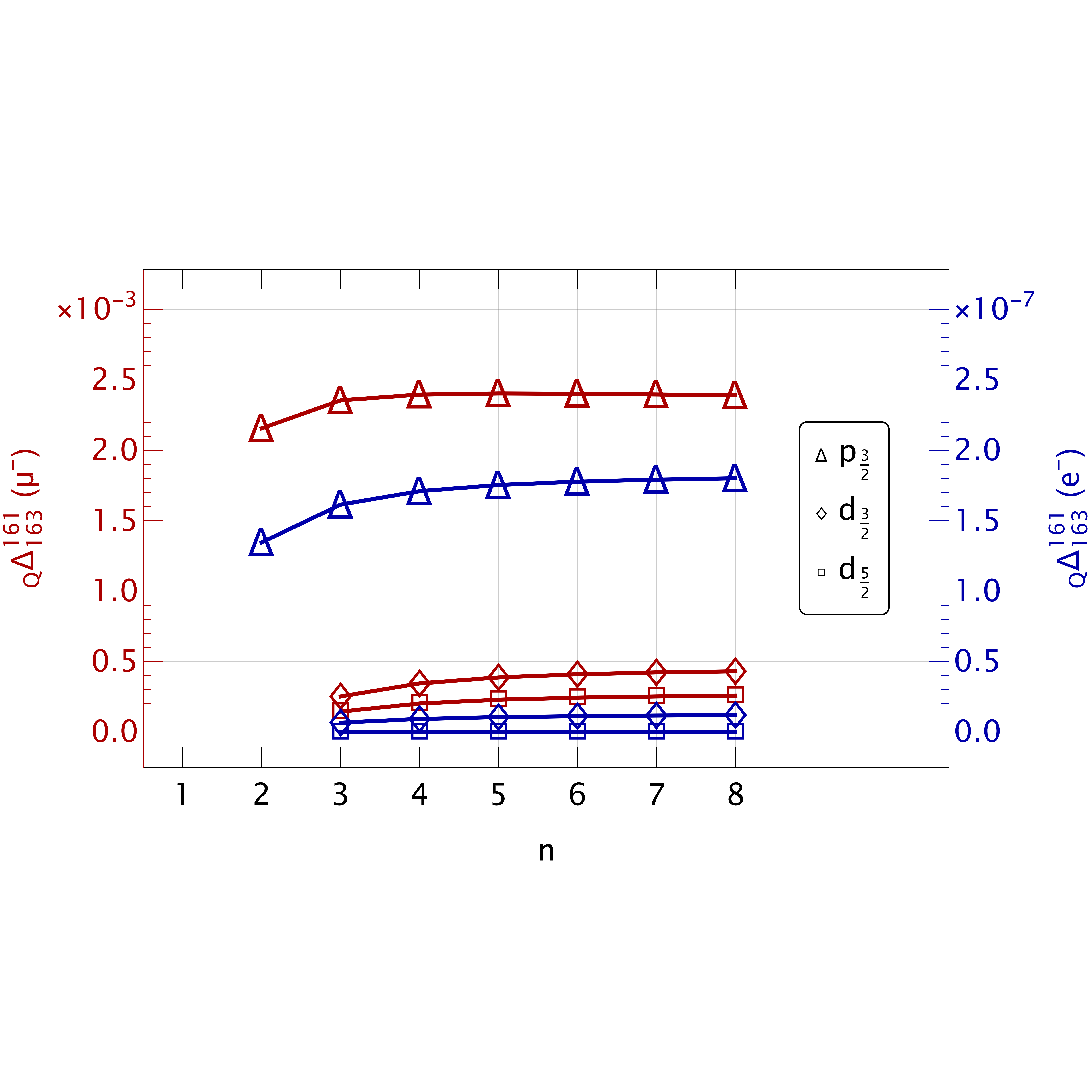}
\caption{(color online) Same as \cref{fig:e16} but ${}_{Q}\Delta_{163}^{161}$ is plotted as a function of quantum number $n$ for the orbitals $p_{3/2}$ (triangles), $d_{3/2}$ (diamonds) and $d_{5/2}$ (squares).}  \label{fig:e17}
\end{figure}

The dependence of the quantity ${}_Q\Delta^{161}_{163}$ on the principal quantum number $n$ for $p_{3/2}$ and $d_{5/2}$ states is illustrated in \cref{fig:e17}. Similar to the magnetic hyperfine anomaly, the quadrupole anomaly is fairly constant with $n$, with a slight dip for the $2p_{3/2}$ state. The behaviour is consistent across both electronic and muonic ions, aside from the difference of four orders of magnitude.

\subsubsection{ Remark regarding the choice of the empirical nuclear radius}

{ As a last remark, we would like to point out the impact that the choice of $R_\text{rms}$ (and consequently $R_{\rm N}=\sqrt{\frac{5}{3}R_\text{rms}^2}$) has on the computed results. For example, instead of using the empirical formula $R_{\mathrm{rms}} = (0.836 A^{1/3} + 0.57)$ fm \cite{Palffy2010}, the root-mean-square radius ($R_{\text{rms}} = \sqrt{\ev{R^2}}$) of both the uniform spherical distribution and the Fermi distribution of \cref{Fermi_distribution} can be redefined to match the $R_{\text{rms}}$ of the HFBCS model. The results are shown in Appendix~\ref{subsec:RRMS}. Once all charge distributions are renormalised to the HFBCS $R_{\text{rms}}$, the electronic binding energies from the Fermi and HFBCS models agree at the level of $10^{-6}\%$, so that $R_{\text{rms}}$ essentially fixes the spectrum and a renormalised Fermi model is sufficient \cite{Palffy2010}. For the electronic ion, this means that the HFBCS value of $R_{\text{rms}}$ obtained via our method effectively encapsulates the nuclear model and can be used as the sole nuclear-structure input in the atomic calculations, providing a computationally simplified implementation of our approach. In contrast, for muonic ions residual differences of order $10^{-3}\%$ remain, showing that their energy levels are sensitive to nuclear-structure details beyond $R_{\text{rms}}$ and thus still provide a valuable probe for discriminating between different nuclear models.}

\section{Conclusions}\label{Conclusions}
In this work we have presented a comprehensive microscopic treatment of hyperfine interactions in heavy hydrogen-like ions, explicitly bridging nuclear and atomic length scales within a single, internally consistent framework. Starting from self-consistent nuclear structure calculations within the Skyrme--Hartree--Fock framework supplemented with BCS pairing and self-consistent blocking, we obtained, via this  microscopic approach involving a quantum-mechanical treatment of the structure of the nucleus, proton and neutron densities and currents for the prolate deformed rare-earth nuclei \(^{159}\mathrm{Tb}\), \(^{161}\mathrm{Dy}\), \(^{163}\mathrm{Dy}\), and \(^{165}\mathrm{Ho}\). %\cite{Bender2003,Ring-Schuck,Sk59,VB72,DFT84,CBH97,CBH98,LBB07,BBD09,PC22}. 
These microscopic densities provide the source terms for a multipole expansion of the nuclear electric and magnetic fields entering the Dirac equation for a bound electron or muon. By solving the Dirac equation with finite-size nuclear potentials, we analysed finite-size effects for both charge (Breit--Rosenthal) and current/magnetisation (Bohr--Weisskopf) corrections to hyperfine structure constants. %\cite{RosenthalBreit1932,BohrWeisskopf1950,Buttgenbach1984,Stone16}. 
At the same time, the same nuclear quantities that reproduce measured spectroscopic quadrupole and magnetic dipole moments were employed directly to compute hyperfine splittings and hyperfine anomalies between the two dysprosium isotopes, providing a unified approach in which atomic and nuclear ingredients are treated on equal footing.

The Skyrme--HF+BCS \sam{(HFBCS)}{} calculations predict substantially prolate deformed shapes for all nuclei studied here. Proton and neutron densities exhibit significant differences in the nuclear interior, reflecting underlying shell structure. The corresponding current densities show non-trivial loop structures that depend on the occupancy of valence orbitals. In \(^{161}\mathrm{Dy}\) and \(^{163}\mathrm{Dy}\), the unpaired neutron dominates the intrinsic magnetisation, whereas in \(^{159}\mathrm{Tb}\) and \(^{165}\mathrm{Ho}\) it is the unpaired proton. Induced currents in the even--even core, caused by the breaking of time-reversal symmetry generated by the unpaired nucleon,  are an order of magnitude smaller, but contribute with opposite sign, highlighting the importance of core polarisation and demonstrating that a realistic Bohr--Weisskopf correction must resolve the underlying nucleonic current components.

When these microscopic densities are used as inputs to the Dirac
equation for hydrogen-like electronic ions, finite-size effects on the
electronic wave functions are moderate. Breit--Rosenthal charge
corrections to the magnetic hyperfine constant \(a\) are at the level
of a few percent for \(ns_{1/2}\) states and are much smaller for
\(np_{1/2}\) states, consistent with the relativistic contact-density
systematics. Bohr--Weisskopf corrections are dominated by spin
currents: the neutron spin current provides the leading contribution
to \(a\) in Dy isotopes, while induced proton spin and orbital
currents give smaller, often opposite, contributions. For the
electronic ions considered here the net BW correction is at the
$2\,\%$ level, while the resulting hyperfine anomalies
(vanishing for a point-like nucleus) remain below \(10^{-4}\) for
electronic states. These findings reinforce the broader point that, with the ever-improving precision of modern experimental platforms, nuclear magnetisation and charge
distributions must be included consistently alongside relativistic
atomic structure to avoid biased extraction of nuclear moments from
spectroscopic data.% \cite{WR2007,Jentschura2011,Roberts2021}.
%{\color{red}Sounds like bootstrapping: using a nuclear-structure model to extract nuclear moment from experimental data?}

The situation changes dramatically for muonic ions, where the Bohr radius is much smaller and the lepton wave function has substantial overlap with the nuclear interior. Finite-size corrections become decisive: the BR correction exceeds \(40\,\%\) for the \(1s_{1/2}\) muon orbital and the BW correction approaches \(100\,\%\) for \(s\)-states. Hyperfine anomalies can exceed \(10\,\%\) for muonic \(s\)-states and reach the \(10^{-4}\) level in quadrupole interactions. These large anomalies trace directly to the radial overlap of the muon wave function with the nuclear currents, turning muonic spectroscopy into a highly sensitive probe of the spatial distribution of nuclear magnetisation and charge. This enhanced sensitivity is precisely what makes muonic ions an exceptional tool to benchmark microscopic nuclear models. Because the muon samples the nuclear interior, the dominant contributions to the muonic hyperfine splittings depend on the \textit{radial profiles} of the charge density and magnetisation density rather than only on their integrated moments. In practice, different nuclear structure descriptions that reproduce the same spectroscopic dipole moment and quadrupole moment can still predict measurably different Bohr--Weisskopf and Breit--Rosenthal corrections once folded with the sharply peaked muonic wave functions. The resulting hyperfine anomalies therefore act as differential observables that are directly sensitive to core polarisation, the balance of spin and orbital currents, and the spatial distribution of the valence-nucleon magnetisation. This creates a stringent consistency test: a nuclear model must simultaneously account for electromagnetic moments \textit{and} the muon-weighted overlap integrals that govern the hyperfine shifts. In this sense, systematic measurements of muonic ions on sets of isotopes (and across different muonic orbitals) provide a controlled way to disentangle deformations, single-particle configurations, and induced core currents, offering a powerful discriminator among effective interactions and beyond-mean-field refinements. Moreover, we find that the electric quadrupole interaction---often negligible in electronic hydrogen-like ions---becomes the dominant hyperfine contribution for muonic \(p_{3/2}\) and \(d\)-states: for \(^{163}\mathrm{Dy}^{65+}(\mu^-)\) the quadrupole shift of the \(2p_{3/2}\) state is of order \(100~\mathrm{keV}\), dwarfing the dipole contribution. In all cases, the calculated spectroscopic magnetic dipole moments \(\mu_{\mathrm{tot}}^{(\mathrm{spec})}\) and electric quadrupole moments \(Q_{20}^{(\mathrm{spec})}\) agree with available experimental data from atomic-beam magnetic resonance measurements, supporting the consistency of the nuclear inputs.% \cite{St05,Stone16,degroote2020}.

Beyond the specific isotopes studied, the present results highlight a general conclusion: precision hyperfine spectroscopy of heavy ions increasingly demands a genuinely combined atomic--nuclear treatment. Finite magnetisation (Bohr--Weisskopf) and finite-size (Breit--Rosenthal) effects are not small add-ons but leading systematics once one targets high accuracy, particularly in highly charged systems where relativistic enhancement increases sensitivity to the nuclear interior. %\cite{Volotka2012,Volotka2013,Roberts2021}. 
In this sense, the framework developed here is well aligned with the broader trajectory from early atomic time standards \cite{ESSEN1955} to modern precision platforms, including directions where nuclear structure is central rather than incidental, such as the \textsuperscript{229}Th low-energy nuclear transition and the nuclear-clock program. %\cite{PeikOkhapkin2015,Campbell2011,Seiferle2019,Morgan2025}.

\section{Perspectives}\label{Perspectives}
Our fully microscopic approach demonstrates the feasibility and utility of treating the nucleus and the bound electron or muon within a unified quantum-mechanical framework. By disentangling contributions of different nucleonic currents, quantifying finite-size effects, and predicting hyperfine anomalies for both electronic and muonic systems, we have shown that, in particular, muonic spectroscopy of heavy hydrogen-like ions can provide stringent tests of nuclear models. In particular, the large anomalies and sizeable quadrupole shifts predicted for muonic ions define clear experimental targets capable of discriminating between competing microscopic descriptions of nuclear magnetisation and deformation, especially when combined with isotope-chain systematics and hyperfine-anomaly measurements that help separate electronic-structure and nuclear-distribution effects \cite{RosenthalBreit1932,BohrWeisskopf1950,Roberts2021}.
Several extensions follow naturally. On the atomic side, incorporating radiative QED corrections and other higher-order effects will enable direct, quantitative confrontation with forthcoming high-precision data in the heaviest systems \cite{Volotka2012,Volotka2013}. On the nuclear side, exploring other effective interactions (e.g.\ Gogny forces) and systematically comparing EDF parameterisations will refine uncertainty estimates for magnetisation distributions and their induced hyperfine signatures \cite{DG80,GCP10,PC22}. The formalism can also be generalised to multi-electron ions: specific differences of hyperfine splittings in H-like and Li-like (or other few-electron) ions can cancel significant electronic correlations and isolate nuclear-structure contributions, providing an experimentally accessible pathway to nuclear magnetisation profiles with reduced atomic-theory systematics \cite{Volotka2012,Volotka2013}.
Finally, the same microscopic understanding of hyperfine couplings is increasingly relevant beyond metrology, for example in nuclear-spin and molecular-spin platforms for quantum technologies, where controlling hyperfine interactions and their material dependence is essential \cite{Thiele2014,Godfrin2017,Moreno2018,Yu2025}. Overall, combining high-precision hyperfine spectroscopy---especially in muonic systems---with the kind of microscopic calculations presented here will deepen our understanding of the interplay between atomic and nuclear physics and open new avenues for probing nuclear structure with unprecedented sensitivity.

\section{Acknowledgements}
This work was funded by the French National Research Agency (ANR) through the Programme d’Investissement d’Avenir under contract ANR-11-LABX-0058 NIE and ANR-17-EURE-0024 within the Investissement d’Avenir program ANR-10-IDEX-0002-02. J-G.H. also acknowledges QUSTEC funding from the European Union’s
Horizon 2020 research and innovation program under the Marie Sklodowska-Curie Grant Agreement No. 847471. The authors would like to acknowledge the High Performance Computing Center of the University of Strasbourg for supporting this work by providing scientific support and access to computing resources. Part of the computing resources were funded by the Equipex Equip@Meso project (Programme Investissements d’Avenir) and the CPER Alsacalcul/Big Data. Finally, we are grateful for the helpful discussions with J. Meyer.

\section{Appendices}

\subsection{Intrinsic vs. spectroscopic multipole moments}
\label{App_Intrinsic vs. spectroscopic multipole moments}
\subsubsection{General expressions}

The components of the spectroscopic multipole moment operators ${{\cal M}}^\textrm{(spec)}_{\lambda}$ of rank $\lambda$
(expressed in the laboratory reference frame), are related to the components of the intrinsic multipole moment operators ${{\cal M}}^{\textrm{(\rm intr)}(\theta)}_{\lambda}$
(expressed in the reference frame of the nucleus) 
via the well known formula for spherical tensor operators \cite{VMK88}
\begin{equation}\label{labvsintformultipoleoperators}
{{\cal M}}^\textrm{(spec)}_{\lambda\mu}
=
\sum_{\nu=-\lambda}^{\lambda} 
{\cal D}^{\lambda*}_{\mu\nu} (\theta) \ {{\cal M}}^{\textrm{(\rm intr)}(\theta)}_{\lambda\nu}  ,
\end{equation}
where ${\cal D}^{\lambda}_{\mu\nu} (\theta)$
represents the Wigner functions at Euler angles $\theta=(\alpha, \beta,\gamma)$ \cite{VMK88}, specifying the orientation of the intrinsic frame with respect to
the laboratory frame.

In the framework of the nuclear collective model, the total Hamiltonian $H$ of the nuclear system is composed of a collective $H_\textrm{Coll}$, and an intrinsic part $H_\textrm{intr}$
\begin{equation}
H = H_\textrm{Coll} + H_\textrm{intr} .
\end{equation}
The collective part of the Hamiltonian reads
\begin{equation}
H_\textrm{Coll} = \sum_{k=1}^3 \frac{\hbar^2}{2\mathcal{J}_k} {\mathfrak R}^2_k ,
\end{equation}
where $k=1,2,3$ label the intrinsic axes, and $\vec{{\mathfrak R}}$ represents the angular momentum of the core, considered as a quantum rotor and $\mathcal{J}_k$ are the components of the inertia tensor.

The total spin $\vec{I}$ of the nucleus is obtained by coupling the core angular momentum $\vec{{\mathfrak R}}$ with the intrinsic angular momentum $\vec{J}$, i.e.
\begin{equation}
\vec{I} = \vec{{\mathfrak R}} + \vec{J} .
\end{equation}
Reverting this expression to $\vec{{\mathfrak R}} = \vec{I} - \vec{J}$, one easily obtains
\begin{equation}
H_\textrm{Coll} = \sum_{k=1}^3 \frac{\hbar^2}{2\mathcal{J}_k} I^2_k 
                           - \sum_{k=1}^3 \frac{\hbar^2}{\mathcal{J}_k} I_k J_k
                           +  \sum_{k=1}^3 \frac{\hbar^2}{2\mathcal{J}_k} J^2_k .
\end{equation}
In this expression, the first term in the right member is usually called the rotation Hamiltonian $H_\textrm{Rot}$, 
the second is the (Coriolis) coupling term $H_\textrm{Coupl}$, and the last term is called the recoil term $H_\textrm{Recoil}$.
The recoil term acts on the intrinsic degrees of freedom, and is therefore usually incorporated in the intrinsic part of the Hamiltonian.

Assuming adiabaticity (the intrinsic structure follows intimately the rotational motion of the core) will allow for expressing the total nuclear wave functions essentially as a product of an intrinsic and a rotational wave function. This means that the coupling term $H_\textrm{Coupl}$ can be ignored.
Stretching effects due to centrifugal forces will also be neglected, so that it will be sufficient to only consider $H_\textrm{Rot}$.

Furthermore we will assume axial symmetry of the intrinsic Hamiltonian. This assumption leads to the fact that $K$, the projection of the total angular momentum $\vec{I}$ on the symmetry axis of the nucleus (chosen as the ``3'' axis),
is a good quantum number. 
Note that since the system (the nucleus) is isolated in space, the total angular momentum is always conserved because of isotropy in space
(the total Hamiltonian is rotationally invariant), thus $I$ is a good quantum number.
Also, its projection onto the ``$z$'' laboratory axis, the quantum number $M$, is a good quantum number as well.
For the rotational part, $H_\textrm{Rot}$, axial symmetry of the nuclear shape leads to the symmetric-rotor case 
$\mathcal{J}_1 = \mathcal{J}_2 \equiv \mathcal{J}$. We will not consider the third component $\frac{\hbar^2}{2\mathcal{J}_3} I^2_3$ in $H_\textrm{Rot}$
since states that would be rotated around the 3-axis cannot be distinguished in a collective rotation
(it is meaningless to speak about a rotation of a quantum core about its symmetry axis, just like it would be impossible, for a macroscopic object with a carefully polished surface, to distinguish optically a rotation about such an axis). Also, the angular momentum vector of the core, $\vec{{\mathfrak R}}$, is perpendicular to the symmetry axis. We consider therefore
\begin{equation}
H_\textrm{Rot} = \sum_{k=1}^2 \frac{\hbar^2}{2\mathcal{J}} I^2_k
                          = \frac{\hbar^2}{2\mathcal{J}} (I^2 - I^2_3) .
\end{equation}
 
We will also assume that the intrinsic Hamiltonian is invariant with respect to a rotation of 180$^\circ$ around an axis perpendicular to its symmetry axis. 
Conventionally, we use this axis to be the intrinsic ``2'' axis (all axes perpendicular to the symmetry axis being, in this respect, equivalent).
This assumption leads to the fact that states with $K \neq 0$ are doubly degenerate.
In the following, and if not otherwise stated, we will use only positive values of $K$ ($K \geqslant 0$).

\bigskip 

We shall use in the following the normalised and symmetrised wave functions, as given, e.g., in Ref. \cite{Rowe70,BMII75} ($\xi$ denotes the set of intrinsic variables):

\begin{equation}\label{normsymwavef}
\begin{array}{ll}
\langle \alpha\beta\gamma ; \xi \vert \widetilde{IMK} \rangle
\equiv
\psi^I_{MK}
= & \displaystyle{\frac{1}{\sqrt{2(1+\delta_{K0})}}}
    \displaystyle{\sqrt{\frac{2I+1}{8\pi^2}}}
    \Big\lbrace
    {\cal D}^{I*}_{MK} (\alpha\beta\gamma) \Phi_K(\xi)  \\[+6mm]
  & + (-)^{I+K} {\cal D}^{I*}_{M-K} (\alpha\beta\gamma) \Phi_{\bar{K}}(\xi)
\Big\rbrace ; \qquad K \geqslant 0.
\end{array}   
\end{equation}
In this expression we have introduced the intrinsic wave functions $\Phi_K$
and $\Phi_{\bar{K}}$ according to their development onto a basis of good
intrinsic angular momentum \cite{BMII75}:
\begin{equation}
\Phi_K = \sum_J C_J \Phi_{JK}
\ee
and 
\begin{equation}
\Phi_{\bar{K}} 
\equiv {{\cal R}}^{-1}_i \Phi_K \equiv e^{i \pi {J}_2} \Phi_K
= \sum_J (-)^{J+K} C_J \Phi_{J,-K}.
\end{equation}
We use the same notation as Bohr and Mottelson \cite{BMII75}
for the operator describing a rotation by the angle $\pi$ around the ``$2$" axis,
namely ${{\cal R}}_i \equiv e^{-i \pi {J}_2}$.

In what follows, we will also make use of the ket notation
\begin{equation}\label{kbar}
\vert \bar{K} \rangle 
= {{\cal R}}^{-1}_i \vert K \rangle
= {{\cal R}}^{\dagger}_i \vert K \rangle .
\end{equation}

Moreover when using symmetrised bras and kets as in \cref{normsymwavef}, these
will be specified with a tilda symbol. If they are not symmetrised, the tilda
symbol is omitted.

Further notations are used, for normalised wave functions (depending on whether they are symmetrised or
not, and whether one adds intrinsic wave functions or not) :
\begin{align}
\langle \alpha\beta\gamma \vert IMK \rangle
&\equiv
\sqrt{\frac{2I+1}{8\pi^2}} {\cal D}^{I*}_{MK} (\alpha\beta\gamma) ,\\
\langle \alpha\beta\gamma ; \xi \vert IMK \rangle
&\equiv
\sqrt{\frac{2I+1}{8\pi^2}} {\cal D}^{I*}_{MK} (\alpha\beta\gamma) \Phi_K(\xi)  ,\\
\langle \alpha\beta\gamma \vert \widetilde{IMK} \rangle
&= 
\frac{1}{\sqrt{2(1+\delta_{K0})}}
\sqrt{\frac{2I+1}{8\pi^2}}
\Big\lbrace
{\cal D}^{I*}_{MK} (\alpha\beta\gamma)
+
(-)^{I+K} {\cal D}^{I*}_{M-K} (\alpha\beta\gamma) 
\Big\rbrace ,
\end{align}
and for the kets we use 
\begin{equation}\label{symket}
\vert \widetilde{IMK} \rangle
\equiv
\frac{1}{\sqrt{2(1+\delta_{K0})}}
\Big\lbrace
\vert IMK \rangle
+
(-)^{I+K} \vert IM-K \rangle 
\Big\rbrace  ; \qquad K \geqslant 0  .
\end{equation}

Using \cref{labvsintformultipoleoperators} and integrating over all possible orientations of the intrinsic reference frame
(i.e. integrating over Euler angles), one can evaluate the matrix elements of the multipole moment operator as
\begin{equation}\begin{aligned}
\langle \widetilde{I_2 M_2 K_2} \vert  {{\cal M}}^{\textrm{(spec)}}_{\lambda\mu} \vert \widetilde{I_1 M_1 K_1} \rangle
& = \sum_{\nu=-\lambda}^{\lambda} \int d\theta \langle \widetilde{I_2 M_2 K_2} \vert  {\cal D}^{\lambda*}_{\mu\nu} (\theta) 
{{\cal M}}^{\textrm{(intr)}(\theta)}_{\lambda\nu}  \vert \widetilde{I_1 M_1 K_1} \rangle \\  & = \frac{1}{\sqrt{2(1+\delta_{K_1 0})}  
              \sqrt{2(1+\delta_{K_2 0})}} \sum_{\nu=-\lambda}^{\lambda}
          \Big[  (a)_\nu + (-)^{I_2+K_2} \ (b)_\nu \\ 
          & + (-)^{I_1+K_1}  \ (c)_\nu  +  (-)^{I_1+K_1}(-)^{I_2+K_2} \ (d)_\nu \Big] .
\end{aligned}\end{equation}
In this expression one has
\begin{equation}\begin{aligned}
(a)_\nu &=
          \langle K_2 \vert 
          {{\cal M}}^{\textrm{(\rm intr)}(\theta)}_{\lambda\nu}     
	  \vert K_1  \rangle 
          \int d\theta \ \langle I_2 M_2 K_2 \vert \theta \rangle \
          {\cal D}^{\lambda*}_{\mu\nu} (\theta) \
          \langle \theta \vert I_1 M_1 K_1  \rangle   \\
    &=
          \langle K_2 \vert 
          {{\cal M}}^{\textrm{(\rm intr)}(\theta)}_{\lambda\nu}     
	  \vert K_1  \rangle 
	  \sqrt{\frac{2I_1+1}{8\pi^2}}
	  \sqrt{\frac{2I_2+1}{8\pi^2}}
          \int d\theta \ {\cal D}^{I_2}_{M_2 K_2} (\theta)   \
          {\cal D}^{\lambda*}_{\mu\nu} (\theta)  \
          {\cal D}^{I_1*}_{M_1 K_1} (\theta) ,
\end{aligned}\end{equation}
which can be evaluated using the  formula  \cite{VMK88}
\begin{equation}
\int d\theta \ {\cal D}^{I_3}_{M_3K_3} (\theta)  \  {\cal D}^{I_2*}_{M_2K_2} (\theta) \  {\cal D}^{I_1*}_{M_1K_1}  (\theta)
 =
 \frac{8\pi^2}{2I_3+1} (I_1 M_1 I_2 M_2 \vert I_3 M_3)  (I_1 K_1 I_2 K_2 \vert I_3 K_3).
\end{equation}
As a result, the term $(a)_\nu$ can be written as
\begin{equation}
(a)_\nu = \langle K_2 \vert 
      {{\cal M}}^{\textrm{(\rm intr)}(\theta)}_{\lambda\nu}     
      \vert K_1  \rangle 
      \sqrt{2I_1+1}
      \frac{(I_1 M_1 \lambda \mu \vert I_2 M_2)}{\sqrt{2I_2+1}}
      (I_1 K_1 \lambda \nu \vert I_2 K_2) .      
\end{equation}
In a very similar way, we get for the other terms :
\begin{equation}
(b)_\nu = \langle \bar{K}_2 \vert 
      {{\cal M}}^{\textrm{(\rm intr)}(\theta)}_{\lambda\nu}     
      \vert K_1 \rangle 
      \sqrt{2I_1+1}
      \frac{(I_1 M_1 \lambda \mu \vert I_2 M_2)}{\sqrt{2I_2+1}}
      (I_1 K_1 \lambda \nu \vert I_2 \ -K_2)  ,   
\end{equation}
and
\begin{equation}
(c)_\nu = \langle K_2 \vert 
      {{\cal M}}^{\textrm{(\rm intr)}(\theta)}_{\lambda\nu}     
      \vert \bar{K}_1  \rangle 
      \sqrt{2I_1+1}
      \frac{(I_1 M_1 \lambda \mu \vert I_2 M_2)}{\sqrt{2I_2+1}}
      (I_1 \ -K_1 \lambda \nu \vert I_2 K_2) ,      
\end{equation}
and finally
\begin{equation}
(d)_\nu = \langle \bar{K}_2 \vert 
      {{\cal M}}^{\textrm{(\rm intr)}(\theta)}_{\lambda\nu}     
      \vert \bar{K}_1  \rangle 
      \sqrt{2I_1+1}
      \frac{(I_1 M_1 \lambda \mu \vert I_2 M_2)}{\sqrt{2I_2+1}}
      (I_1 \ -K_1 \lambda \nu \vert I_2 \ -K_2) .
\end{equation}

\bigskip

Grouping these results together, one obtains
\begin{equation}\begin{aligned}\label{firstx}
\langle \widetilde{I_2 M_2 K_2} \vert  {{\cal M}}^{\textrm{(spec)}}_{\lambda\mu} \vert \widetilde{I_1 M_1 K_1} \rangle
&= \displaystyle{\frac{1}{2}}
            \displaystyle{\sqrt{\frac{2I_1+1}{(1+\delta_{K_1 0})
				                (1+\delta_{K_2 0})}}}
            \frac{(I_1 M_1 \lambda \mu \vert I_2 M_2)}{\sqrt{2I_2+1}}  \\
         &\times \displaystyle{\sum_{\nu=-\lambda}^{\lambda}}  \Big[  
            (I_1 K_1 \lambda \nu \vert I_2 K_2)      
	    \langle K_2 \vert 
            {{\cal M}}^{\textrm{(\rm intr)}(\theta)}_{\lambda\nu}     
            \vert K_1  \rangle  \\[+3mm]
	 & +  (-)^{I_2+K_2} \ 
            (I_1 K_1 \lambda \nu \vert I_2 \ -K_2)      
	    \langle \bar{K}_2 \vert 
            {{\cal M}}^{\textrm{(\rm intr)}(\theta)}_{\lambda\nu}
            \vert K_1  \rangle  \\[+3mm]
	 & +  (-)^{I_1+K_1} \   
            (I_1 \ -K_1 \lambda \nu \vert I_2 K_2)      
	    \langle K_2 \vert 
            {{\cal M}}^{\textrm{(\rm intr)}(\theta)}_{\lambda\nu}     
            \vert \bar{K}_1  \rangle  \\[+3mm]
	 & +  (-)^{I_1+K_1}(-)^{I_2+K_2} \ 
            (I_1 \ -K_1 \lambda \nu \vert I_2 \ -K_2)      
	    \langle \bar{K}_2 \vert 
            {{\cal M}}^{\textrm{(\rm intr)}(\theta)}_{\lambda\nu}
            \vert \bar{K}_1  \rangle 
	    \Big] .    
\end{aligned}\end{equation}

Now, on the other hand, one has directly
\begin{equation}\begin{aligned}
\langle \widetilde{I_2 M_2 K_2} \vert  {{\cal M}}^{\textrm{(spec)}}_{\lambda\mu} \vert \widetilde{I_1 M_1 K_1} \rangle
&= \displaystyle{\frac{1}{2}}
            \displaystyle{\frac{1}{\sqrt{(1+\delta_{K_1 0})
				         (1+\delta_{K_2 0})}}} \\[+3mm]
         &\times \Big[ 
            \langle I_2 M_2 K_2 \vert 
	    {{\cal M}}^{\textrm{(spec)}}_{\lambda\mu} 
	    \vert I_1 M_1 K_1 \rangle \\[+3mm]
	 & +  (-)^{I_2+K_2} \ 
            \langle I_2 M_2 \ -K_2 \vert 
	    {{\cal M}}^{\textrm{(spec)}}_{\lambda\mu} 
	    \vert I_1 M_1 K_1 \rangle \\[+3mm]
	 & +  (-)^{I_1+K_1} \   
            \langle I_2 M_2 K_2  \vert 
	    {{\cal M}}^{\textrm{(spec)}}_{\lambda\mu} 
	    \vert I_1 M_1 \ -K_1 \rangle \\[+3mm]
	 & +  (-)^{I_1+K_1}(-)^{I_2+K_2} \ 
            \langle I_2 M_2 \ -K_2 \vert 
	    {{\cal M}}^{\textrm{(spec)}}_{\lambda\mu} 
	    \vert I_1 M_1 \ -K_1 \rangle            
	    \Big] ,
\end{aligned}\end{equation}
which can be transformed using the Wigner-Eckart theorem
%
%\footnotesize
\begin{equation}
\langle I_2 M_2 K_2 \vert 
{{\cal M}}^{\textrm{(spec)}}_{\lambda\mu} 
\vert I_1 M_1 K_1 \rangle
=
\frac{(-1)^{2\lambda}}{\sqrt{2I_2+1}} 
(I_1 M_1 \lambda \mu \vert I_2 M_2)
\langle I_2 K_2 \Vert 
{{\cal M}}^{\textrm{(spec)}}_{\lambda} 
\Vert I_1 K_1 \rangle 
\end{equation}
\normalsize
into (note that since $\lambda$ is integer here, we have $(-1)^{2\lambda}=1$)
\begin{equation}\begin{aligned}
\langle \widetilde{I_2 M_2 K_2} \vert  {{\cal M}}^{\textrm{(spec)}}_{\lambda\mu} \vert \widetilde{I_1 M_1 K_1} \rangle
&= \displaystyle{\frac{1}{2}}
            \displaystyle{\frac{1}{\sqrt{(1+\delta_{K_1 0})
				         (1+\delta_{K_2 0})}}}
	    \displaystyle{\frac{(I_1 M_1 \lambda \mu \vert I_2 M_2)}
	                       {\sqrt{2I_2+1}}} \\[+3mm]
         &\times \Big[ 
            \langle I_2 K_2 \Vert 
	    {{\cal M}}^{\textrm{(spec)}}_{\lambda} 
	    \Vert I_1 K_1 \rangle \\[+3mm]
	 & +  (-)^{I_2+K_2} \ 
            \langle I_2 \ -K_2 \Vert 
	    {{\cal M}}^{\textrm{(spec)}}_{\lambda} 
	    \Vert I_1 K_1 \rangle \\[+3mm]
	 & +  (-)^{I_1+K_1} \   
            \langle I_2 K_2 \Vert 
	    {{\cal M}}^{\textrm{(spec)}}_{\lambda} 
	    \Vert I_1 \ -K_1 \rangle \\[+3mm]
	 & +  (-)^{I_1+K_1}(-)^{I_2+K_2} \ 
            \langle I_2 \ -K_2 \Vert 
	    {{\cal M}}^{\textrm{(spec)}}_{\lambda} 
	    \Vert I_1 \ -K_1 \rangle            
	    \Big] .
\end{aligned}\end{equation}
Furthermore, one has by definition
\begin{equation}\begin{aligned}\label{xintermed}
\langle \widetilde{I_2 K_2} \Vert 
{{\cal M}}^{\textrm{(spec)}}_{\lambda}
\Vert \widetilde{I_1 K_1} \rangle
        &\equiv
            \displaystyle{\frac{1}{2}}
            \displaystyle{\frac{1}{\sqrt{(1+\delta_{K_1 0})
				         (1+\delta_{K_2 0})}}}\\[+3mm] 
	&\times \Big[ 
            \langle I_2 K_2 \Vert 
	    {{\cal M}}^{\textrm{(spec)}}_{\lambda} 
	    \Vert I_1 K_1 \rangle \\[+3mm]
	 & +  (-)^{I_2+K_2} \ 
            \langle I_2 \ -K_2 \Vert 
	    {{\cal M}}^{\textrm{(spec)}}_{\lambda} 
	    \Vert I_1 K_1 \rangle \\[+3mm]
	 & +  (-)^{I_1+K_1} \   
            \langle I_2 K_2 \Vert 
	    {{\cal M}}^{\textrm{(spec)}}_{\lambda} 
	    \Vert I_1 \ -K_1 \rangle \\[+3mm]
	 & +  (-)^{I_1+K_1}(-)^{I_2+K_2} \ 
            \langle I_2 \ -K_2 \Vert 
	    {{\cal M}}^{\textrm{(spec)}}_{\lambda} 
	    \Vert I_1 \ -K_1 \rangle            
	    \Big] ,  
\end{aligned}\end{equation}
allowing to write 
\begin{equation}\label{xwigneck}
\langle \widetilde{I_2 M_2 K_2} \vert  {{\cal M}}^{\textrm{(spec)}}_{\lambda\mu}  \vert \widetilde{I_1 M_1 K_1} \rangle  
         = \displaystyle{\frac{(I_1 M_1 \lambda \mu \vert I_2 M_2)}
	                        {\sqrt{2I_2+1}}}
             \langle \widetilde{I_2 K_2} \Vert 
             {{\cal M}}^{\textrm{(spec)}}_{\lambda\mu}
             \Vert \widetilde{I_1 K_1} \rangle ,
\end{equation}
which can actually be viewed as a generalisation of the Wigner-Eckart theorem for symmetrised states.

We now identify 
expression \cref{xwigneck} with \cref{firstx} to
get the reduced laboratory matrix elements expressed with the help of
intrinsic matrix elements as :
\begin{equation}\begin{aligned}
\langle \widetilde{I_2 K_2} \Vert 
{{\cal M}}^{\textrm{(spec)}}_{\lambda}
\Vert \widetilde{I_1 K_1} \rangle
        &= \displaystyle{\frac{1}{2}}
            \displaystyle{\sqrt{\frac{2I_1+1}{(1+\delta_{K_1 0})
				                (1+\delta_{K_2 0})}}}  \\[+3mm]
         &\times \displaystyle{\sum_{\nu=-\lambda}^{\lambda}}  \Big[  
            (I_1 K_1 \lambda \nu \vert I_2 K_2)      
	    \langle K_2 \vert 
            {{\cal M}}^{\textrm{(\rm intr)}(\theta)}_{\lambda\nu}     
            \vert K_1  \rangle  \\[+3mm]
	 & +  (-)^{I_2+K_2} \ 
            (I_1 K_1 \lambda \nu \vert I_2 \ -K_2)      
	    \langle \bar{K}_2 \vert 
            {{\cal M}}^{\textrm{(\rm intr)}(\theta)}_{\lambda\nu}
            \vert K_1  \rangle  \\[+3mm]
	 & +  (-)^{I_1+K_1} \   
            (I_1 \ -K_1 \lambda \nu \vert I_2 K_2)      
	    \langle K_2 \vert 
            {{\cal M}}^{\textrm{(\rm intr)}(\theta)}_{\lambda\nu}     
            \vert \bar{K}_1  \rangle  \\[+3mm]
	 & +  (-)^{I_1+K_1}(-)^{I_2+K_2} \ 
            (I_1 \ -K_1 \lambda \nu \vert I_2 \ -K_2)      
	    \langle \bar{K}_2 \vert 
            {{\cal M}}^{\textrm{(\rm intr)}(\theta)}_{\lambda\nu}
            \vert \bar{K}_1  \rangle 
	    \Big] . 
\end{aligned}\end{equation}
Using the selection rules for the Clebsh-Gordan coefficients, since $K_1$ and
$K_2$ are fixed, the sum over the index $\nu$ disappears. 

The first coefficient is non zero only for $K_1+\nu=K_2$, and therefore this
imposes $\nu=-K_-$, where $K_- \equiv K_1-K_2$. The second coefficient is non zero only for $K_1+\nu=-K_2$, 
or $\nu=-K_+$, where \mbox{$K_+ \equiv K_1+K_2$}. In the same way, the third coefficient demands $\nu=K_+$, and finally the last one requires $\nu=K_-$. This leads to the expression 
\begin{equation}\begin{aligned}\label{fourterms}
\langle \widetilde{I_2 K_2} \Vert 
{{\cal M}}^{\textrm{(spec)}}_{\lambda}
\Vert \widetilde{I_1 K_1} \rangle
        &=\displaystyle{\frac{1}{2}}
            \displaystyle{\sqrt{\frac{2I_1+1}{(1+\delta_{K_1 0})
				                (1+\delta_{K_2 0})}}}  \\[+3mm]
         &\times \Big[  
            (I_1 K_1 \lambda -K_- \vert I_2 K_2)      
	    \langle K_2 \vert 
            {{\cal M}}^{\textrm{(\rm intr)}(\theta)}_{\lambda, -K_-}     
            \vert K_1  \rangle  \\[+3mm]
	 & +  (-)^{I_2+K_2} \ 
            (I_1 K_1 \lambda -K_+ \vert I_2 -K_2)      
	    \langle \bar{K}_2 \vert 
            {{\cal M}}^{\textrm{(\rm intr)}(\theta)}_{\lambda, -K_+}
            \vert K_1  \rangle  \\[+3mm]
	 & +  (-)^{I_1+K_1} \   
            (I_1 -K_1 \lambda K_+ \vert I_2 K_2)      
	    \langle K_2 \vert 
            {{\cal M}}^{\textrm{(\rm intr)}(\theta)}_{\lambda K_+}     
            \vert \bar{K}_1  \rangle  \\[+3mm]
	 & +  (-)^{I_1+K_1} (-)^{I_2+K_2} \ 
            (I_1 -K_1 \lambda K_- \vert I_2 -K_2)      
	    \langle \bar{K}_2 \vert 
            {{\cal M}}^{\textrm{(\rm intr)}(\theta)}_{\lambda K_-}
            \vert \bar{K}_1  \rangle  \ \Big] . 
\end{aligned}\end{equation}

We will now demonstrate that the first and the last terms in this expression
are identical and, in the same way, that 
the second and third terms are equal.

For this purpose we shall make use of the following property of the 
Clebsh-Gordan coefficients \cite{VMK88}:
\begin{equation}
(A a B b \vert C c) = (-)^{A+B-C} (A -a \ B -b \vert C -c) .
\end{equation}

With the help of this expression and \cref{kbar}, 
the last term can be written as
\begin{equation}\begin{aligned}\label{interm1}
&\phantom{=} (-)^{I_1 + K_1} (-)^{I_2 + K_2} (-)^{I_1 + \lambda -I_2} \
    (I_1 K_1 \lambda -K_- \vert I_2 K_2)      
    \langle \bar{K}_2 \vert 
    {{\cal M}}^{\textrm{(\rm intr)}(\theta)}_{\lambda K_-}
    \vert \bar{K}_1  \rangle  \\[+3mm]
&= (-)^{I_1 + K_1} (-)^{I_2 + K_2} (-)^{I_1 + \lambda -I_2} \
    (I_1 K_1 \lambda -K_- \vert I_2 K_2)
    \langle K_2 \vert 
    {{\cal R}}_{i}
    {{\cal M}}^{\textrm{(\rm intr)}(\theta)}_{\lambda K_-}
    {{\cal R}}^{\dagger}_{i}
    \vert K_1  \rangle .
\end{aligned}\end{equation}

Now, for any 
irreducible spherical tensor operator
${{\cal T}}_{\lambda}$, we can write the transformation
rule under the action of the unitary transformation ${{\cal R}}_{i}$ as \cite{BMI69}
\begin{equation}\begin{aligned}
{{\cal R}}_{i}
{{\cal T}}_{\lambda \mu}
{{\cal R}}^{\dagger}_{i}
   &=\sum_\nu {\cal D}^{\lambda}_{\nu\mu} (0,\pi,0)
       {{\cal T}}_{\lambda \nu}  \\[+3mm]
   &=\sum_\nu d^{\lambda}_{\nu\mu} (\pi)
       {{\cal T}}_{\lambda \nu}  \\[+3mm]
   &=\sum_\nu (-)^{\lambda-\mu} \delta_{\mu, -\nu}
       {{\cal T}}_{\lambda \nu}  \\[+3mm]
   &=(-)^{\lambda-\mu} \
       {{\cal T}}_{\lambda -\mu} .
\end{aligned}\end{equation}

Applying this expression to the multipole operator 
${{\cal M}}^{\textrm{(\rm intr)}(\theta)}_{\lambda}$ allows to express \cref{interm1} as
\begin{equation}\begin{aligned}
&\phantom{=} (-)^{I_1 + K_1} (-)^{I_2 + K_2} (-)^{I_1 + \lambda -I_2} \
    (I_1 K_1 \lambda -K_- \vert I_2 K_2)  
    (-)^{\lambda-K_-}    
    \langle K_2 \vert 
    {{\cal M}}^{\textrm{(\rm intr)}(\theta)}_{\lambda, -K_-}
    \vert K_1  \rangle  \\[+3mm]
&= (-)^{I_1 + K_1} (-)^{I_2 + K_2} (-)^{I_1 + \lambda -I_2} \
    (-)^{\lambda - K_1 + K_2}    
    (I_1 K_1 \lambda -K_- \vert I_2 K_2)  
    \langle K_2 \vert 
    {{\cal M}}^{\textrm{(\rm intr)}(\theta)}_{\lambda, -K_-}
    \vert K_1  \rangle  \\[+3mm]
&= (-)^{I_1 + K_1 + I_2 + K_2 + I_1 + \lambda -I_2 + \lambda - K_1 + K_2}    
    (I_1 K_1 \lambda -K_- \vert I_2 K_2)  
    \langle K_2 \vert 
    {{\cal M}}^{\textrm{(\rm intr)}(\theta)}_{\lambda, -K_-}
    \vert K_1  \rangle  \\[+3mm]
&= (-)^{2(I_1 + K_2)}    
    (I_1 K_1 \lambda -K_- \vert I_2 K_2)  
    \langle K_2 \vert 
    {{\cal M}}^{\textrm{(\rm intr)}(\theta)}_{\lambda, -K_-}
    \vert K_1  \rangle 
\end{aligned}\end{equation}
since $\lambda$ is integer.

Now, $K$ is the projection of $I$ onto the ``3" axis of the intrinsic reference
frame, and therefore the phase factor $(-)^{2(I_1 + K_2)}$ is always equal to
unity, thus terminating the proof that the first and the last term in 
\cref{fourterms} are equal.

In the same way, the second term in \cref{fourterms} can be written as
\begin{equation}\begin{aligned}
&\phantom{=} (-)^{I_2+K_2} \ 
    (I_1 K_1 \lambda -K_+ \vert I_2 -K_2)      
    \langle \bar{K}_2 \vert 
    {{\cal M}}^{\textrm{(\rm intr)}(\theta)}_{\lambda, -K_+}
    \vert K_1  \rangle  \\[+3mm]
&= (-)^{I_2+K_2} \ 
    (-)^{I_1 + \lambda - I_2}
    (I_1 -K_1 \lambda K_+ \vert I_2 K_2)      
    \langle K_2 \vert 
    {{\cal R}}_{i}
    {{\cal M}}^{\textrm{(\rm intr)}(\theta)}_{\lambda, -K_+}
    \vert K_1  \rangle  \\[+3mm]
&= (-)^{I_1+K_2+\lambda} \ 
    (I_1 -K_1 \lambda K_+ \vert I_2 K_2)      
    (-)^{\lambda + K_+}   
    \langle K_2 \vert 
    {{\cal M}}^{\textrm{(\rm intr)}(\theta)}_{\lambda K_+}
    {{\cal R}}_{i}
    \vert K_1  \rangle  \\[+3mm]
&= (-)^{I_1+K_2+\lambda} \ 
    (I_1 -K_1 \lambda K_+ \vert I_2 K_2)      
    (-)^{\lambda + K_+}   
    \langle K_2 \vert 
    {{\cal M}}^{\textrm{(\rm intr)}(\theta)}_{\lambda K_+}
    {{\cal R}}^{2}_{i}
    \vert \bar{K}_1  \rangle  \\[+3mm]
&= (-)^{I_1 + K_1 + 2 K_2 + 2 I_1} \ 
    (I_1 -K_1 \lambda K_+ \vert I_2 K_2)      
    \langle K_2 \vert 
    {{\cal M}}^{\textrm{(\rm intr)}(\theta)}_{\lambda K_+}
    \vert \bar{K}_1  \rangle  \\[+3mm]
&= (-)^{2(I_1 + K_2)} (-)^{I_1 + K_1} \ 
    (I_1 -K_1 \lambda K_+ \vert I_2 K_2)      
    \langle K_2 \vert 
    {{\cal M}}^{\textrm{(\rm intr)}(\theta)}_{\lambda K_+}
    \vert \bar{K}_1  \rangle  \\[+3mm]
&= (-)^{I_1 + K_1} \ 
    (I_1 -K_1 \lambda K_+ \vert I_2 K_2)      
    \langle K_2 \vert 
    {{\cal M}}^{\textrm{(\rm intr)}(\theta)}_{\lambda K_+}
    \vert \bar{K}_1  \rangle
\end{aligned}\end{equation}
since ${{\cal R}}^{2}_{i}=(-)^{2I}$ \cite{BMII75}, and since 
$(-)^{2(I_1 + K_2)}=1$ for the reason explained above. Thus we have
demonstrated that the second and the third terms in \cref{fourterms}
are equal.

Finally we can write \cref{fourterms} as
\begin{equation}\begin{aligned}\label{finalreduced}
\begin{array}{ll}
\!\!\!\!\!\! \langle \widetilde{I_2 K_2} \Vert 
{{\cal M}}^{\textrm{(spec)}}_{\lambda}
\Vert \widetilde{I_1 K_1} \rangle
          & =  \displaystyle{\sqrt{\frac{2I_1+1}{(1+\delta_{K_1 0})
				                (1+\delta_{K_2 0})}}}  \\[+5mm]
          & \times \Big[   
            (I_1 K_1 \lambda K_2-K_1 \vert I_2 K_2)  \    
	    \langle K_2 \vert 
            {{\cal M}}^{\textrm{(\rm intr)}(\theta)}_{\lambda, K_2-K_1}     
            \vert K_1  \rangle  \\[+5mm] 
	  &  +   (-)^{I_1+K_1}   
            (I_1 -K_1 \lambda K_2+K_1 \vert I_2 K_2)  \  
	    \langle K_2 \vert 
            {{\cal M}}^{\textrm{(\rm intr)}(\theta)}_{\lambda, K_2+K_1}     
            \vert \bar{K}_1  \rangle \ \Big] . \!\!\!\!\!\! 
\end{array}      
\end{aligned}\end{equation}

\subsubsection{Electric quadrupole moment}

Let us consider as one of the most important applications the electric quadrupole moment operator (or simply quadrupole moment operator).
In this case, one has $\lambda=2$, and we will denote the corresponding 
tensor by $Q$ instead of ${\cal M}$. 

We will define the intrinsic quadrupole moment, generally depending on
$K$, as \cite{Rowe70}
\begin{equation}
Q^K_0 
\equiv 
\frac{1}{e} \sqrt{\frac{16 \pi}{5}}
\langle K \vert {Q}^{(\rm intr)}_{2 0}  \vert K \rangle ,
\end{equation}
where $e$ stands for the elementary charge, and the spectroscopic quadrupole moment as \cite{Rowe70}
\begin{equation}
Q^K 
\equiv 
\frac{1}{e} \sqrt{\frac{16 \pi}{5}}
\langle \widetilde{I M=I K} \vert 
Q^{(\rm spec)}_{20}
\vert \widetilde{I M=I K} \rangle .
\end{equation}

Using the Wigner-Eckart theorem \cref{xwigneck} 
and the result \cref{finalreduced}, one gets immediately
\begin{equation}
\begin{array}{ll}
Q^K & = \frac{1}{(1+\delta_{K0})} \Big[  
               (II20 \vert II) (IK20 \vert IK) {Q}^K_0  \\[+5mm]
           & + (II20 \vert II) (I \ -K \ 2 \ 2K \vert IK) (-)^{I+K} \
               \frac{1}{e} \sqrt{\frac{16 \pi}{5}}
               \langle K \vert 
	       {Q}^{(\rm intr)}_{2,2K}  
	       \vert \bar{K} \rangle	       
	                                 \Big] .
\end{array}      
\label{eq188}
\end{equation}

Owing to this expression, one can discuss separately the following cases :

\begin{itemize}
  \item If $K=0$, it is easy to see that the first and the second term in this
        expression are identical. This is so, because one must have
	\begin{equation}
	   (-)^I \Phi_{\bar{0}} = \Phi_0
	\end{equation}
	to ensure that for $K=0$ the wave function \cref{normsymwavef}
	is not identically vanishing. 
	As a consequence, one has
	\begin{equation}
	   {Q}^{K=0} = (II20 \vert II) (I020 \vert I0) {Q}^{K=0}_0 .
	\end{equation}
  \item If $K > 1$, only the first term has to be considered because
        the second index $2K$ of the quadrupole tensor cannot be greater than 2
	in absolute value. Therefore one gets
	\begin{equation}
	   {Q}^{K > 1}  = (II20 \vert II) (I120 \vert I1) {Q}^{K > 1}_0 .
	\end{equation}
   \item The previous cases can be cast into the general formula
	      \begin{equation}
	        {Q}^{K=0 \, \textrm{or} \, K > 1}  = (II20 \vert II) (IK20 \vert IK) {Q}^{K=0 \, \textrm{\bf or} \, K > 1}_0 .
	     \end{equation}
	     The product of the two Clebsh-Gordan coefficients in this formula can be simplified further with the help of
	     the following expressions, that can be found in \cite{VMK88}:
	      \footnotesize
	      \begin{equation}\begin{aligned}
	        (A a B b \vert C C) &=  \delta_{a+b,C} \, (-)^{A-a} \\ & \times  \Big\lbrack \frac{(2C+1)! (A+B-C)! (A+a)! (B+b)! }{(A+B+C+1)! (A-B+C)! (-A+B+C)! (A-a)! (B-b)! } \Big\rbrack^{1/2}    
	                                        ,
	     \end{aligned}\end{equation}
	      \normalsize
	     for $C = c$,	     
	     and
	      \footnotesize
	      \begin{equation}\begin{aligned}
	        (A a B b \vert A + B - 2 ; a+b) &= \Big\lbrack \frac{2A (2A-1) 2B (2B-1)}{2 (2A+2B-2)(2A+2B-1) } \Big\rbrack^{1/2} 
             \\
&\times \Big\lbrack  {2A \choose A-a}  {2B \choose B-b} {2A+2B-4 \choose A+B-a-b-2}  \Big\rbrack^{-1/2}   \\
&\times \Big\lbrack  {2A-2  \choose A-a}     {2B-2  \choose B+b}- 2 {2A-2  \choose A-a-1}  {2B-2  \choose B+b-1}   \\
&+   {2A-2  \choose A-a-2}  {2B-2  \choose B+b-2} \Big\rbrack
	     \end{aligned}\end{equation}
	      \normalsize
	     for $C = A + B - 2$.	     
	     
	     One obtains then the well known result
	     \begin{equation}
	        {Q}^{K=0 \, \textrm{or} \, K > 1}  = \frac{3K^2 - I(I+1)}{(I+1) (2I+3)} {Q}^{K=0 \, \textrm{or} \, K > 1}_0 .
	     \label{eq:195}
         \end{equation}
	     One can notice the remarkable result that for nuclei with spin $I=0$ (thus having $K=0$), the spectroscopic quadrupole moment automatically vanishes,
	     although the intrinsic quadrupole moment could be well different from zero. This expresses the fact that the latter is actually not an observable, for example, for the ground state of even-even nuclei, for which $I=0$ is verified.
   \item If $K = 1/2$ or $K = 1$ the full expression in \cref{eq188} has
        in principle to be considered. Note however that for even-even nuclei the case $K = 1/2$
	cannot occur, because $K$ must be integer since $I$ itself is integer.
\end{itemize}

\subsubsection{Magnetic dipole moment}

Let us consider as another important application the magnetic dipole moment operator (or simply dipole moment operator).
In this case, one has $\lambda=1$, and we will denote the corresponding 
tensor by ${M}$ instead of ${\cal M}$. 

Following, e.g., Rowe \cite{Rowe70} we define the intrinsic dipole moment, in the state
$ \vert K \rangle$, as
\begin{equation}
\mu^K_0 
\equiv 
\sqrt{\frac{4 \pi}{3}}
\langle K \vert {{M}}^{(\rm intr)}_{1 0}  \vert K \rangle
\end{equation}
and the spectroscopic dipole moment in the state $\vert \widetilde{IM=IK}\rangle$ as
\begin{equation}
\mu^K 
\equiv 
\sqrt{\frac{4 \pi}{3}}
\langle \widetilde{I M=I K} \vert 
{M}^{\rm (spec)}_{10}
\vert \widetilde{I M=I K} \rangle .
\end{equation}

Using again the Wigner-Eckart theorem in \cref{xwigneck} 
and the result of \cref{finalreduced}, one gets in this case
\begin{equation}
\begin{array}{ll}
\mu^K & = \frac{1}{(1+\delta_{K0})} \Big[  
               (II10 \vert II) (IK10 \vert IK) \mu^K_0  \\[+5mm]
           & + (II10 \vert II) (I \ -K \ 1 \ 2K \vert IK) (-)^{I+K} \
              \sqrt{\frac{4 \pi}{3}}
               \langle K \vert 
	       {{M}}^{(\rm intr)}_{1,2K}  
	       \vert \bar{K} \rangle	       
	                                 \Big] .
\end{array}      
\end{equation}

The Clebsh-Gordan coefficients appearing in the first term of this expression are given by \cite{VMK88}
\begin{equation}
   (II10 \vert II)  = \Big\lbrack \frac{I}{I+1} \Big\rbrack^{1/2}
\end{equation}
and
\begin{equation}
   (IK10 \vert IK)  = K \Big\lbrack \frac{1}{I(I+1)} \Big\rbrack^{1/2} ,
\end{equation}
and a similar discussion as for the electric quadrupole moment finally leads to the conclusion that for
$K=0$ or $K > 1/2$, one gets
\begin{equation}
   \mu^{K=0 \, \textrm{or} \, K > 1/2} = \frac{K}{I+1} \ \mu^K_0 .
\label{eq::198}
\end{equation}
To close this appendix, let us recall that to obtain the spectroscopic magnetic dipole and electric quadrupole moments of the bandhead state of a rotational band with $K \notin \{1/2,1\}$ one has to set $I=K$ in \cref{eq:195,eq::198}.  
\subsection{Multipole expansion of the magnetic vector potential}\label{App_Multipole expansion of the magnetic vector potential}
\subsubsection{General case}
We start from the form of the vector potential in the Coulomb gauge, given by
\begin{equation}
\label{eq::A_j}
\vec{A}(\vec{r}) = \frac{\mu_0}{4\pi} \int \frac{\vec{j}(\vec{R})}{|\vec{r} - \vec{R}|} d^3R.
\end{equation}

Given a set of normalised spherical harmonics $\{C_{q}^{(k)}(\theta, \phi)\}_{k,q}$, one can define the following normalised vector spherical harmonics, for $k\geqslant1$,
\begin{align}
\vec{Y}_{q}^{(k)} &= \vec{e}_r C_{q}^{(k)}(\theta, \phi), \\
\vec{\Psi}_{q}^{(k)} &= \frac{1}{\sqrt{k(k+1)}} \left[r \vec{\nabla} C_{q}^{(k)}(\theta, \phi)\right], \\
\vec{\Phi}_{q}^{(k)} &=  \frac{i}{\hbar\sqrt{k(k+1)}} \left[\vec{L} C_{q}^{(k)}(\theta, \phi)\right].
\end{align}

It should be emphasised that while the definition of $\vec{Y}_q^{(k)}$ remains unchanged for $(k=0,q=0)$, both $\vec{\Psi}_0^{(0)}$ and $\vec{\Phi}_0^{(0)}$ are identically zero. Any vector field $\vec{V}(\vec{r})$ can be decomposed into a series of vector spherical harmonics as
\begin{equation}
\vec{V}(\vec{r}) = \sum_{k=0}^{\infty} \sum_{q=-k}^{k} \left({^{(r)}V}_{q}^{(k)}(r) \vec{Y}_{q}^{(k)} + {^{(1)}V}_{q}^{(k)}(r) \vec{\Psi}_{q}^{(k)} + {^{(2)}V}_{q}^{(k)}(r) \vec{\Phi}_{q}^{(k)}\right).
\end{equation}

However,
\begin{equation}
\vec{\nabla} \cdot \vec{V} = \left(\frac{d\left[{{^{(r)}V}}_{q}^{(k)}\right]}{dr}+\frac{2}{r}{{^{(r)}V}}_{q}^{(k)}-\frac{k(k+1)}{r}{^{(1)}V}_{q}^{(k)}\right)C_{q}^{(k)}(\theta, \phi),
\end{equation}
and $\vec{A}$ respects the Coulomb gauge condition $\vec{\nabla} \cdot \vec{A} = 0$, therefore, $\vec{A}$ can be expressed into a series of $\vec{\Phi}_{q}^{(k)}$ only, taking the form
\begin{equation}
\vec{A}(\vec{r}) = \frac{\mu_0}{4\pi} \sum_{k=1}^{\infty} \sum_{q=-k}^{k} \frac{i}{\hbar\sqrt{k(k+1)}} \left[\vec{L} C_{q}^{(k)}(\theta, \phi)\right] A_{q}^{(k)} (r) .
\end{equation}

Using the orthogonality relation
\begin{equation}
\frac{-1}{\hbar^2{k(k+1)}} \int_{\theta, \phi} \left[\vec{L} C_{q'}^{(k')}(\theta, \phi)\right] \cdot \left[\vec{L} C_{-q}^{(k)}(\theta, \phi)\right] d^2\Omega = (-1)^q \delta_{k,k'}\delta_{q,q'},
\end{equation}
one obtains
\begin{equation}
\label{eq::VSHA}
A_{q}^{(k)}(r) = \frac{4\pi}{\mu_0} \frac{i}{\hbar\sqrt{k(k+1)}} (-1)^q \int_{\theta, \phi} \left[\vec{L} C_{-q}^{(k)}(\theta, \phi)\right]\cdot \vec{A}(\vec{r})d^2\Omega.
\end{equation}

The multipolar expansion of $\frac{1}{|\vec{r} - \vec{R}|}$ in \cref{eq::A_j},
\begin{equation}
\begin{aligned}
{\vec A}(\vec{r}) &= \frac{\mu_0}{4\pi}   \sum_{k=0}^{\infty} \frac{1}{r^{k+1}} \sum_{q=-k}^{k} (-1)^q  C^{(k)}_q(\theta,\phi) \int_{R=0}^r  C^{(k)}_{-q}(\Theta,\Phi) \vec{j}(\vec{R}) R^{k} d^3R,\\
& + \frac{\mu_0}{4\pi}  \sum_{k=0}^{\infty} r^k \sum_{q=-k}^{k} (-1)^q  C^{(k)}_q(\theta,\phi) \int_{R=r}^\infty  C^{(k)}_{-q}(\Theta,\Phi) \vec{j}(\vec{R}) \frac{1}{R^{k+1}} d^3R,
\end{aligned}
\end{equation}
can be inserted in \cref{eq::VSHA}, and, computing the effect of each component of $\vec{L}$ on $C_{-q}^{(k)}$, one can prove that
\begin{equation}
\left(\int_{\theta, \phi} (-1)^{q'} C_{q'}^{(k')}(\theta, \phi)\cdot \left[\vec{L} C_{-q}^{(k)}(\theta, \phi)\right] d^2\Omega \right) C^{(k')}_{-q'}(\Theta,\Phi) = \vec{L} C^{(k)}_{-q}(\Theta,\Phi),
\end{equation}
therefore,
\begin{equation}
\begin{aligned}
A_{q}^{(k)}(r) &= \frac{i}{\hbar\sqrt{k(k+1)}} \frac{1}{r^{k+1}} (-1)^q \int_{R=0}^r  \left[\vec{L} C_{-q}^{(k)}(\Theta, \Phi)\right] \cdot \vec{j}(\vec{R})  R^{k} d^3R,\\
& +  \frac{i}{\hbar\sqrt{k(k+1)}} r^k (-1)^q  \int_{R=r}^\infty  \left[\vec{L} C_{-q}^{(k)}(\Theta, \Phi)\right] \cdot \vec{j}(\vec{R})\frac{1}{R^{k+1}} d^3R.
\end{aligned}
\end{equation}

To follow convention, we can write the multipolar expansion of the vector potential as
\begin{equation}
\begin{aligned}
\vec{A}(\vec{r}) = & \frac{\mu_0}{4\pi} \sum_{k=1}^{\infty} \sum_{q=-k}^{k} (-1)^q \frac{(-i)}{\hbar k}\left[\vec{L}  C_{q}^{(k)} (\theta, \phi) \right] {^{\rm in}M_{-q}^{(k)}} (r) \frac{1}{r^{k+1}} \\
& + \frac{\mu_0}{4\pi} \sum_{k=1}^{\infty} \sum_{q=-k}^{k} (-1)^q \frac{(-i)}{\hbar k}\left[\vec{L}  C_{q}^{(k)} (\theta, \phi) \right] {^{\rm ex}M_{-q}^{(k)}} (r) r^k,
\end{aligned}
\end{equation}
with
\begin{equation}
{^{\rm in}M_{q}^{(k)}} (r) \equiv - \frac{i}{\hbar (k+1)} \int_{R=0}^{r} \left[\vec{L}  C_{q}^{(k)} (\Theta, \Phi) \right] \cdot \vec{j} (\vec{R}) R^k d^3R,
\end{equation}
and
\begin{equation}
{^{\rm ex}M_{q}^{(k)}} (r) \equiv - \frac{i}{\hbar (k+1)} \int_{R=r}^{\infty} \left[\vec{L}  C_{q}^{(k)} (\Theta, \Phi) \right] \cdot \vec{j} (\vec{R}) \frac{1}{R^{k+1}} d^3R.
\end{equation}

\subsubsection{Magnetic dipole moment}\label{App_Magnetic dipole moment}

Truncating the previous expansion to the first non-vanishing term ($k=1$) yields
\begin{equation}
\begin{aligned}
\label{eq::A_dipole_wp_tensor}
\vec{A}(\vec{r}) &= \frac{\mu_0}{4\pi} \sum_{q=-1}^{1} (-1)^q \frac{(-i)}{\hbar} \left[ \vec{L}C_{q}^{(1)}(\theta, \phi) \right] \frac{^{\rm in}M_{-q}^{(1)}(r)}{r^{2}} \\&+ \frac{\mu_0}{4\pi} \sum_{q=-1}^{1} (-1)^q \frac{(-i)}{\hbar} \left[ \vec{L}C_{q}^{(1)}(\theta, \phi) \right] {^{\rm ex}M_{-q}^{(1)}(r)} r \;.
\end{aligned}
\end{equation}

For an arbitrary vector $\vec{v}^{(1)}$, one can show that
\begin{equation}\label{eq:LCcross}
[\vec{r} \wedge \vec{v}^{(1)}]_{\lambda}= \frac{-i}{\hbar} \left[r \vec{L}C^{(1)}_{\lambda}(\theta,\phi)\right] \cdot \vec{v}^{(1)}.
\end{equation}

We recall that for vectors $\vec{v}$, represented as first-rank tensors $\vec{v}^{(1)}$, the scalar product is defined as
\begin{equation}
\vec{v}^{(1)} \cdot \vec{u}^{(1)} = \sum_{\lambda} (-1)^\lambda v_{\lambda}^{(1)} u_{-\lambda}^{(1)}.
\end{equation}

Moreover, it is important to remember $\vec{L}$ commutes with any function of $r$, from there, one can look at the following scalar product:
\begin{equation}\label{eq:dipolcross}
\begin{aligned}
\vec{A}(\vec{r}) \cdot \vec{v}^{(1)} &= \frac{\mu_{0}}{4 \pi r^3}\sum_{q=-1}^{1} (-1)^q \frac{(-i)}{\hbar} \left[r \vec{L}C_{q}^{(1)}(\theta, \phi) \right] \frac{^{\rm in}M_{q}^{(1)}(r)}{r^{3}}\cdot\vec{v}^{(1)} \\&\qquad + \frac{\mu_{0}}{4 \pi} \sum_{q=-1}^{1} (-1)^q \frac{(-i)}{\hbar} \left[r \vec{L}C_{q}^{(1)}(\theta, \phi) \right] {^{\rm ex}M_{-q}^{(1)}(r)} \cdot\vec{v}^{(1)} \\
&= \frac{\mu_{0}}{4 \pi r^3}\sum_{q=-1}^{1} (-1)^q [\vec{r} \wedge \vec{v}^{(1)}]_q {^{\rm in}M^{(1)}_{-q}}(r)+ \frac{\mu_{0}}{4 \pi} \sum_{q=-1}^{1} (-1)^q  [\vec{r} \wedge \vec{v}^{(1)}]_q {^{\rm ex}M^{(1)}_{-q}}(r)\\
&= \frac{\mu_{0}}{4 \pi r^3} {^{\rm in}\vec{M}}^{(1)}(r)\cdot(\vec{r} \wedge \vec{v}^{(1)}) + \frac{\mu_{0}}{4 \pi} {^{\rm ex}\vec{M}}^{(1)}(r)\cdot(\vec{r} \wedge \vec{v}^{(1)}) \\
&= \left(\frac{\mu_{0}}{4 \pi r^3} ({^{\rm in}\vec{M}}^{(1)}(r) \wedge \vec{r}) + \frac{\mu_{0}}{4 \pi} ({^{\rm ex}\vec{M}}^{(1)}(r) \wedge \vec{r}) \right)\cdot\vec{v}^{(1)}.
\end{aligned}
\end{equation}

Therefore, one obtains
\begin{equation}
\label{eq::A_dipole_wp}
\vec{A}(\vec{r}) = \frac{\mu_0}{4\pi} \frac{^{\rm in}\vec{M}^{(1)}(r) \wedge \vec{r}}{r^3} + \frac{\mu_0}{4\pi} \left( ^{\rm ex}\vec{M}^{(1)}(r) \wedge \vec{r} \right),
\end{equation}
with, also using \cref{eq:LCcross},
\begin{equation}
^{\rm in}\vec{M}^{(1)}(r) = \frac{1}{2} \int_{R=0}^{r} \vec{R} \wedge \vec{j}(\vec{R}) d^3R
\end{equation}
and
\begin{equation}
^{\rm ex}\vec{M}^{(1)}(r) = \frac{1}{2} \int_{R=r}^{\infty} \frac{\vec{R} \wedge \vec{j}(\vec{R})}{R^3} d^3R.
\end{equation}
\subsection{Hyperfine interactions}\label{Hyperfine interactions}
\subsubsection{Dipole}\label{App_Hyperfine interactions:_dipole}

One can rewrite \cref{eq::A_dipole_wp_tensor,eq::A_dipole_wp} as
\begin{equation}
\vec{A}(\vec{r}) \equiv \frac{\mu_0}{4\pi} \sum_{q=-1}^{1} (-1)^q \frac{(-i)}{\hbar} \left[ \vec{L}C_{q}^{(1)}(\theta, \phi) \right] f_{M,-q}(r),
\end{equation}
\begin{equation}
\vec{f}_{M}(r) =  \frac{1}{2} \left(\frac{1}{r^{2}}\int_{R=0}^{r} \vec{R} \wedge \vec{j}(\vec{R}) d^3R + r \int_{R=r}^{\infty} \frac{\vec{R} \wedge \vec{j}(\vec{R})}{R^3}d^3R \right),
\end{equation}
and insert it in ${}^{\rm at}H_{\rm hf, dip} = -ec\vec{\alpha}\cdot\vec{A}(\vec{r})$,
\begin{equation}
\begin{aligned}
\mel{n_a\kappa_aFM_F}{{}^{\rm at}H_{\rm hf, dip}}{n_b\kappa_bFM_F} &= \\ -iec\frac{\mu_0}{4\pi\hbar} \sum_{q=-1}^{1} (-1)^q &\mel{n_a\kappa_am_aFM_F}{ \left[ \vec{\alpha} \cdot \vec{L}C_{q}^{(1)}(\theta, \phi) \right] f_{M,-q}(r)}{n_b\kappa_bm_bFM_F}
\end{aligned}
\end{equation}

The Wigner-Eckart theorem states that
\begin{equation}
\label{eq::wigner_eckart}
\bra{FM_F} T^{(k)}_q \ket{F'M_{F'}} =  (-1)^{F-M_F} \begin{pmatrix} F & k & F' \\ -M_F & q & M_{F'} \end{pmatrix} \langle F || T^{(k)} || F' \rangle,
\end{equation}
where $T^{(k)}_q$ is a tensor operator of rank $k$ and $q$ is the projection of the operator in the $z$-axis. $\langle F || T^{(k)} || F' \rangle$ is the reduced matrix element of the operator, independent of $M_F$ and $M_{F'}$.

Using \cref{eq::wigner_eckart}, one can obtain that
\begin{equation}
\begin{aligned}
\mel{I M_I}{f_{M,-q}(r)}{I M_I'} &= (-1)^{I-M_I} \begin{pmatrix} I & 1 & I \\ -M_I & -q & M_{I'} \end{pmatrix} \langle I || f_{M}(r) || I \rangle
\\ &= (-1)^{I-M_I} \begin{pmatrix} I & 1 & I \\ -M_I & -q & M_{I'} \end{pmatrix} \begin{pmatrix} I & 1 & I \\ -I & -q & I \end{pmatrix}^{-1} \mel{II}{{f_{M,-q}}(r)}{II}.
\end{aligned}
\end{equation}

Given two irreducible tensor operators ${T}^{(k)}$ and ${U}^{(k)}$ of rank $k$, acting, respectively on the electronic and nuclear spaces, one can compute the matrix element of their scalar product as
\begin{equation}
\begin{aligned}
\label{eq::psrdx}
\bra{(J_aI)FM_F}& {T}^{(k)} \cdot {U}^{(k)}   \ket{(J_bI)F'M_{F'}} \\&= (-1)^{J_b+I+F} \delta_{F,F'} \delta_{M_F,M_{F'}} \begin{Bmatrix} J_b & I & F \\ I & J_a & k \end{Bmatrix} \langle{J_a} || {T}^{(k)} ||{J_b} \rangle \langle {I}|| {U}^{(k)} || {I}\rangle.
\end{aligned}
\end{equation}

Therefore, since only $f_{M,0}(r)$ is non-zero, one can write
\begin{equation}
\label{eq::hf_dip_FMF}
\begin{aligned}
&\mel{n_a\kappa_aFM_F}{ \left[ \vec{\alpha} \cdot \vec{L}C_{q}^{(1)}(\theta, \phi) \right] f_{M,0}(r)}{n_b\kappa_bFM_F}  \\ &= (-1)^{J_b+I+F} \delta_{F,F'} \delta_{M_F,M_{F'}} \delta_{q,0}\begin{Bmatrix} J_b & I & F \\ I & J_a & 1 \end{Bmatrix} \\ &\times\langle{n_a\kappa_a} || \vec{\alpha} \cdot \vec{L}C_{q}^{(1)}(\theta,\phi)  \mel{II}{f_{M,0}(r)}{II} ||{n_b\kappa_b} \rangle \begin{pmatrix} I & 1 & I \\ -I & 0 & I \end{pmatrix}^{-1}.
\end{aligned}
\end{equation}

One notices that
\begin{equation}
\begin{aligned}
\label{eq::IJ}
\bra{(J_aI)FM_F}& \vec{I} \cdot \vec{J}   \ket{(J_bI)F'M_{F'}} = \\& (-1)^{J_b+I+F} \delta_{F,F'} \delta_{M_F,M_{F'}} \begin{Bmatrix} J_b & I & F \\ I & J_a & 1 \end{Bmatrix} \langle{J_a} || \vec{J} ||{J_b} \rangle \langle {I}|| \vec{I} || {I}\rangle.
\end{aligned}
\end{equation}

Since,
\begin{equation}
\label{eq::jjj}
\langle j || \vec{j} || j' \rangle = \sqrt{j(2 j+1)(j+1)} \delta_{j j^{\prime}},
\end{equation}
one sees \cref{eq::IJ} vanishes for states with different $F$, $I$ or $\kappa$ (equiv. $J$). Furthermore, comparing \cref{eq::hf_dip_FMF,eq::IJ}, one can see that for diagonal matrix elements, the magnetic dipole hyperfine interaction is proportional to $\vec{I} \cdot \vec{J}$,
\begin{equation}
\label{eq::aIJ}
\mel{n_a\kappa_aFM_F}{{}^{\rm at}H_{\rm hf, dip}}{n_a\kappa_aFM_F} = a \mel{n_a\kappa_aFM_F}{\vec{I} \cdot \vec{J}}{n_a\kappa_aFM_F},
\end{equation}
and
\begin{equation}
\label{eq::a}
a \equiv  -iec \frac{\mu_0}{4\pi\hbar} \frac{\langle{n_a\kappa_a} || \vec{\alpha} \cdot \vec{L}C_{0}^{(1)} \mel{II}{f_{\text{M},0}(r)}{II} ||{n_a\kappa_a} \rangle}{\begin{pmatrix} I & 1 & I \\ -I & 0 & I \end{pmatrix} \langle{J_a} || \vec{J} ||{J_a} \rangle \langle {I}|| \vec{I} || {I}\rangle}.
\end{equation}

This expression can be readily computed in the case of a single electron. One can calculate the matrix element $\mel{n_a\kappa_am_a}{ \left[ \vec{\alpha} \cdot \vec{L}C_{0}^{(1)}(\theta, \phi) \right] {^I f_{M}(r)}}{n_b \kappa_b m_b}$, where ${^I f_{M}(r)} \equiv \mel{II}{f_{M,0}(r)}{II}$, by noticing that $\vec{L}$ commutes with any function of $r$, and by extension so does $\vec{\alpha}\cdot\vec{L}$

\begin{equation}\left[ \vec{\alpha} \cdot \vec{L}C_{0}^{(1)}(\theta, \phi) \right] {^I f_{M}(r)} = {^I f_{M}(r)}\left[ \vec{\alpha} \cdot \vec{L}C_{0}^{(1)}(\theta, \phi) \right]  = \vec{\alpha} \cdot \vec{L} \left[C_{0}^{(1)} {^I f_{M}(r)} \right],\end{equation}
and
\begin{equation}
\vec{\alpha} \cdot \vec{L} \left[C_{0}^{(1)}(\theta, \phi)  \ket{n \kappa m} \right] = \left[ \vec{\alpha} \cdot \vec{L} C_{0}^{(1)}(\theta, \phi)  \right] \ket{n \kappa m} + C_{0}^{(1)}(\theta, \phi)  \left[\vec{\alpha} \cdot \vec{L} \ket{n \kappa m}\right],
\end{equation}
since $\vec{L}$ is a differential operator this is simply the Leibniz product rule.

Thus,
\begin{equation}
\begin{aligned}
\bra{n_a \kappa_a m_a} \vec{\alpha}\cdot\vec{L}\left[{C}^{(1)}_0(\theta, \phi) ^I f_{M}(r)\right] \ket{n_b \kappa_b m_b} &= \\ \left[\bra{n_a \kappa_a m_a} \vec{\alpha}\cdot\vec{L}\right]\left[{C}^{(1)}_0(\theta, \phi) ^I f_{M}(r) \ket{n_b \kappa_b m_b}\right] &- 
\left[\bra{n_a \kappa_a m_a} ^I f_{M}(r) {C}^{(1)}_0(\theta, \phi) \right]\left[\vec{\alpha}\cdot\vec{L} \ket{n_b \kappa_b m_b} \right]
\end{aligned}
\end{equation}
Moreover,
\begin{equation}
\vec{{\alpha}}\cdot\vec{L} \ket{n \kappa m} =
\begin{pmatrix}
0 & \vec{\sigma}\cdot\vec{L} \\
\vec{\sigma}\cdot\vec{L} & 0
\end{pmatrix}
\frac{1}{r}\left(\begin{array}{cc}
P_{n \kappa}(r) \chi_{\kappa m}(\theta, \phi) \\
i Q_{n \kappa}(r) \chi_{-\kappa m}(\theta, \phi)
\end{array}\right) =
\frac{1}{r}\left(\begin{array}{cc}
\hbar(\kappa - 1)i Q_{n \kappa}(r) \chi_{-\kappa m}(\theta, \phi) \\
\hbar(-\kappa - 1)P_{n \kappa}(r) \chi_{\kappa m}(\theta, \phi)
\end{array}\right),
\end{equation}
since $(\vec{\sigma}\cdot\vec{L}+\hbar) \chi_{\kappa m}(\theta, \phi) = -\hbar \kappa \chi_{\kappa m}(\theta, \phi)$.

Hence, one can write
\begin{equation}
\begin{aligned}
\mel{n_a\kappa_a}{ \left[ \vec{\alpha} \cdot \vec{L}C_{0}^{(1)}(\theta, \phi) \right] {^I f_{M}(r)}}{n_b \kappa_b} &= \\-i\hbar (\kappa_a+\kappa_b)\bra{\kappa_a m_a} {C}^{(1)}_0(\theta, \phi) \ket{-\kappa_b m_b}& \int_{r=0}^\infty {^I f_{M}(r)} \left( P_{n_a\kappa_a}(r)Q_{n_b\kappa_b}(r) + Q_{n_a\kappa_a}(r)P_{n_b\kappa_b}(r)\right) dr.
\end{aligned}
\end{equation}

Or, alternatively,
\begin{equation}
\begin{aligned}
\langle{n_a\kappa_am_a} || \left[ \vec{\alpha} \cdot \vec{L}C_{0}^{(1)}(\theta, \phi) \right] {^I f_{M}(r)} ||{n_b \kappa_b m_b}\rangle &= \\-i\hbar (\kappa_a+\kappa_b)\langle{\kappa_a} ||{C}^{(1)}(\theta, \phi)||{-\kappa_b} \rangle &\int_{r=0}^\infty {^I f_{M}(r)}\left( P_{n_a\kappa_a}(r)Q_{n_b\kappa_b}(r) + Q_{n_a\kappa_a}(r)P_{n_b\kappa_b}(r)\right) dr.
\end{aligned}
\end{equation}

One can now calculate the hyperfine splitting due to the magnetic dipole interaction, using \cref{eq::hf_dip_FMF},
\begin{equation}\label{eq::rdxC}
\left\langle\kappa_{a}\left\|C^{(k)}\right\| \kappa_{b}\right\rangle=(-1)^{J_{a}} i \sqrt{\left[J_{a}\right]\left[J_{b}\right]}\left(\begin{array}{ccc}
J_{a} & J_{b} & k \\
-1 / 2 & 1 / 2 & 0
\end{array}\right) \Pi\left(L_{a}+k+L_{b}\right)
\end{equation}
with
\begin{equation}
[x] \equiv 2x+1, \quad \Pi(l)\equiv\left\{\begin{array}{l}
1, \text { if } l \text { is even } \\
0, \text { if } l \text { is odd }
\end{array}\right. .
\end{equation}
This means that
\begin{equation}
\langle{\kappa_a} \| {C}^{(1)}(\theta, \phi) \| {-\kappa_a}\rangle = - \frac{1}{2}\sqrt{\frac{[J_a]}{J_a(J_a+1)}}.
\end{equation}

Finally,
\begin{equation}
\label{eq::II}
\begin{pmatrix}
I & 1 & I \\
-I & 0 & I
\end{pmatrix}
=\sqrt{\frac{I}{(I + 1)[I]}},
\end{equation}
thus, recalling \cref{eq::a,eq::jjj,eq::aIJ},
\begin{equation}
\begin{aligned}
a = ec &\frac{\mu_0}{2\pi}\frac{\kappa_a}{IJ_a(J_a+1)} \int_{r=0}^\infty
\ev{f_{M,0}}{II} P_{n_a\kappa_a}(r)Q_{n_a\kappa_a}(r)  dr.\\
= ec &\frac{\mu_0}{2\pi}\frac{\kappa_a}{IJ_a(J_a+1)} \mathcal{I}^{M_1}_{\mathrm{intr/spec}} \int_{r=0}^\infty
\frac{1}{2}\biggl( \frac{1}{r^2}\int_{R=0}^{r} \vec{R} \wedge \vec{j}(\vec{R}) d^3R \biggr.\\  &\biggl. + r \int_{R=r}^{\infty} \frac{\vec{R} \wedge \vec{j}(\vec{R})}{R^3}d^3R \biggr) P_{n_a\kappa_a}(r)Q_{n_a\kappa_a}(r) dr,
\end{aligned}
\end{equation}
where $\mathcal{I}^{\mathrm{M1}}_{\mathrm{spec/intr}}$ is the conversion factor from the intrinsic magnetic dipole to the laboratory one.

Ultimately, using \cref{eq::IJ,eq::jjj,eq::aIJ}, and
\begin{equation}
\begin{aligned}
\begin{Bmatrix} J & I & F \\ I & J & 1 \end{Bmatrix} &= (-1)^{F+I+J+1} \frac{2\left(I(I+1)+J(J+1)-F(F+1)\right)}{\sqrt{2I(2I+1)(2I+2)2J(2J+1)(2J+2)}}
\\&= (-1)^{F+I+J}\,\frac{F(F+1)-I(I+1)-J(J+1)}{2\langle{J}||\vec{J}||{J}\rangle\langle{I}||\vec{I}||{I}\rangle},
\end{aligned}
\end{equation}
one obtains that
\begin{equation}
\label{eq::hf_dip_aK}
\mel{n_a\kappa_aFM_F}{{}^{\rm at}H_{\rm hf, dip}}{n_a\kappa_aFM_F} = \frac{a}{2} {\tilde K}_a
\end{equation}
where ${\tilde K}_a\equiv F(F+1)-I(I+1)-J_a(J_a+1)$.

While \cref{eq::hf_dip_aK} is a general expression for the magnetic dipole hyperfine interaction, it is particularly useful for the case of a single electron, where $a$ can be computed explicitly with relative ease. In the case of multiple electrons, the expression of $a$ is more complex, but the general form remains the same, and referring to \cref{eq::wigner_eckart,eq::jjj,eq::a,eq::II}, one has
\begin{equation}
\begin{aligned}
a &= -iec \frac{\mu_0}{4\pi\hbar} \frac{\langle{n_a\kappa_a} || \vec{\alpha} \cdot \vec{L}C_{0}^{(1)}(\theta, \phi)  \mel{II}{f_{M,0}(r)}{II} ||{n_a\kappa_a} \rangle}{I\sqrt{J_a(2J_a+1)(J_a+1)}}\\&= -iec \frac{\mu_0}{4\pi\hbar} \frac{\langle{n_a\kappa_a(J_a)m_J=J_a} | \vec{\alpha} \cdot \vec{L}C_{0}^{(1)} (\theta, \phi) \mel{II}{f_{M,0}(r)}{II} |{n_a\kappa_a(J_a)m_J=J_a} \rangle}{IJ_a}\\&\equiv -iec \frac{\mu_0}{4\pi\hbar} \frac{\ev{T_{M,0}^{(1)}(\vec{r})}{n_aJ_aJ_aII}}{IJ_a}.
\end{aligned}
\end{equation}
It can be seen that, in the case one does not consider the electronic penetration inside the nucleus (i.e. $r\gg R$), $\vec{f}_{M}(r)$ reduces to only one coordinate
\begin{equation}
f_{M}(r) = \frac{1}{2 r^{2}}\int_{R=0}^{+\infty} {\left[\vec{R} \wedge \vec{j}(\vec{R})\right]}_z d^3R = \frac{M_z^{(1)}}{r^2},
\end{equation}
with $M_z^{(1)}$ the intrinsic nuclear magnetic moment, consequently,
\begin{equation}
\begin{aligned}
a &= -iec \frac{\mu_0}{4\pi\hbar} \frac{\mel{II}{M_z^{(1)}}{II}\langle{n_a\kappa_a(J_a)m_J=J_a} | \vec{\alpha} \cdot \vec{L}C_{0}^{(1)}(\theta, \phi) r^{-2} |{n_a\kappa_a(J_a)m_J=J_a} \rangle}{IJ_a}\\&= -iec \frac{\mu_0}{4\pi\hbar} M_z^{(1),(\mathrm{spec})}   \frac{\ev{\vec{\alpha} \cdot \vec{L}C_{0}^{(1)}(\theta, \phi) r^{-2} }{n_aJ_aJ_a}}{J_a},
\end{aligned}
\end{equation}
with $M_z^{(1),(\mathrm{spec})} \equiv \mathcal{I}_{\mathrm{intr/spec}}^{M_1} M_z^{(1)}$ the spectroscopic nuclear magnetic moment.

Or, in the case of the single electron again, one can write,
\begin{equation}
\begin{aligned}
a = |e| c  \frac{\mu_0}{2\pi} M_z^{(1),(\mathrm{spec})} \frac{\kappa_a}{IJ_a(J_a+1)}\int_{r=0}^\infty\frac{1}{r^2} P_{n_a\kappa_a}(r)Q_{n_a\kappa_a}(r) dr.
\end{aligned}
\end{equation}
\subsubsection{Quadrupole}\label{App_Hyperfine interactions:_quadrupole}
Starting from the multipole expansion of the quadrupolar nuclear potential
\begin{equation}
{}^{\rm at}H_{\rm hf,quad} = e \phi_{2}(\vec{r}),
\end{equation}
with
\begin{equation}
\phi_{2} (\vec{r}) = \frac{1}{4\pi \epsilon_0} f_{{\cal Q},0}(r)
\end{equation}
and
\begin{equation}
f_{{\cal Q},0}(r) = \frac{1}{2r^3}\int_{R=0}^{r} \rho(\vec{R})R^2(3\cos^2(\Theta) - 1) d^3R + r^2 \int_{R=r}^{\infty} \frac{\rho(\vec{R})(3\cos^2(\Theta) - 1)}{2R^3}d^3R,
\end{equation}
similarly to \cref{eq::hf_dip_FMF}, one can write
\begin{equation}
\begin{aligned}
&\mel{n_a \kappa_a F M_F}{{}^{\rm at}H_{\rm hf,quad}}{n_b \kappa_b F M_F} =\\
&(-1)^{J_b+I+F} \delta_{F,F'} \delta_{M_F,M_{F'}} \begin{Bmatrix} J_b & I & F \\ I & J_a & 2 \end{Bmatrix} \langle{n_a\kappa_a} || C_{0}^{(2)} (\theta,\phi) \mel{II}{f_{{\cal Q},0}(r)}{II} ||{n_b\kappa_b} \rangle \begin{pmatrix} I & 2 & I \\ -I & 0 & I \end{pmatrix}^{-1},
\end{aligned}
\end{equation}
with, in the case of a single electron,
\begin{equation}
\begin{aligned}
&\langle{n_a\kappa_a}||{ C_{0}^{(2)}(\theta, \phi) {\ev{f_{{\cal Q}}(r)}{II}}}||{n_b \kappa_b} \rangle= \\ &\langle{\kappa_a} ||{C}^{(2)}(\theta, \phi)||{\kappa_b} \rangle \int_{r=0}^\infty \ev{f_{{\cal Q},0}(r)}{II}\left( P_{n_a\kappa_a}(r)P_{n_b\kappa_b}(r) + Q_{n_a\kappa_a}(r)Q_{n_b\kappa_b}(r)\right) dr.
\end{aligned}
\end{equation}

The use of the following relations
\begin{align}
\Sixj{I}{J}{F}{J}{I}{2} &= (-1)^{I+J+F} \Threej{J}{2}{J}{-J}{0}{J}\Threej{I}{2}{I}{-I}{0}{I}\frac{6{\tilde K}({\tilde K} + 1) - 8J(J + 1)I(I + 1)}{2I(2I - 1)2J(2J - 1)}, \\
\Threej{I}{2}{I}{-I}{0}{I} &= \sqrt{\frac{2I(2I-1)}{[I](2I+2)(2I+3)}}, \\
\langle{\kappa_a} \| {C}^{(2)}(\theta, \phi) \| {\kappa_a}\rangle &\stackrel{\textrm{\cref{eq::rdxC}}}{=}-\frac{1}{4}\sqrt{\frac{(2J-1)[J](2J+3)}{J(J+1)}},
\end{align}
allows one to write
\begin{equation}
\ev{{}^{\rm at}H_{\rm hf,quad}}{n_a\kappa_aFM_F} = \frac{b}{2} \frac{3{\tilde K}_a({\tilde K}_a+1)-4J_a(J_a+1)I(I+1)}{2I(2I-1)2J_a(2J_a-1)},
\end{equation}
with
\begin{equation}
\begin{aligned}
b &= -\frac{e}{4\pi\epsilon_0}\left(\frac{2J_a-1}{J_a+1}\right)\int_{r=0}^\infty \ev{f_{{\cal Q}}(r)}{II}\left( P^2_{n_a\kappa_a}(r) + Q^2_{n_a\kappa_a}(r)\right) dr \\
&= -\frac{e}{4\pi\epsilon_0} \left(\frac{2J_a-1}{J_a+1}\right)\mathcal{I}^{E_2}_\text{intr/spec}  \\ &\times \int_{r=0}^\infty \biggl(\frac{1}{r^3}\int_{R=0}^{r} \rho(\vec{R})R^2(3\cos^2(\Theta) - 1) d^3R \biggr. \\ &+ \biggl. r^2 \int_{R=r}^{\infty} \frac{\rho(\vec{R})(3\cos^2(\Theta) - 1)}{R^3}d^3R\biggr) \left( P^2_{n_a\kappa_a}(r) + Q^2_{n_a\kappa_a}(r)\right) dr,
\end{aligned}
\end{equation}
where $\mathcal{I}^{E_2}_\text{intr/spec}$ is the conversion factor from the intrinsic quadrupole moment to the laboratory one.

Just as for the magnetic dipole hyperfine interaction, in the case of multiple electrons, the expression of $b$ is more complex to compute,
\begin{equation}
\begin{aligned}
b &= -\frac{e}{4\pi\epsilon_0}\sqrt{\frac{2J(2J-1)}{[J](2J+2)(2J+3)}} \langle{\kappa_a} ||{C}^{(2)}(\theta, \phi)||{\kappa_b} \rangle \\ & \times \int_{r=0}^\infty \ev{4f_{{\cal Q}}(r)}{II}\left( P^2_{n_a\kappa_a}(r) + Q^2_{n_a\kappa_a}(r)\right) dr.
\end{aligned}
\end{equation}

Finally, in the case where one does not consider the electronic penetration inside the nucleus (i.e. $r\gg R$), $f_{{\cal Q}}(r)$ reduces to
\begin{equation}
4f_{{\cal Q}}(r) = \frac{2}{r^3}\int_{R=0}^{+\infty} \rho(\vec{R})R^2(3\cos^2(\Theta) - 1) d^3R = \frac{2{\cal Q}^{(2),(\rm spec)}_0}{r^3},
\end{equation}
where $Q$ is the intrinsic nuclear quadrupole moment, and consequently
\begin{equation}
\begin{aligned}
b &= \frac{e}{4\pi\epsilon_0}  \sqrt{\frac{2J(2J-1)}{[J](2J+2)(2J+3)}} \langle{\kappa_a} ||{C}^{(2)}(\theta, \phi)||{\kappa_b} \rangle\ev{2{\cal Q}}{II} \int_{r=0}^\infty \frac{1}{r^3}\left( P^2_{n_a\kappa_a}(r) + Q^2_{n_a\kappa_a}(r)\right) dr\\
&= \frac{2e}{4\pi\epsilon_0} {\cal Q}^{(2),(\rm spec)}_0 \sqrt{\frac{2J(2J-1)}{[J](2J+2)(2J+3)}} \langle{\kappa_a} ||{C}^{(2)}(\theta, \phi)||{\kappa_b} \rangle \int_{r=0}^\infty \frac{1}{r^3}\left( P^2_{n_a\kappa_a}(r) + Q^2_{n_a\kappa_a}(r)\right) dr,
\end{aligned}
\end{equation}
with ${\cal Q}^{(2),(\rm spec)}_0 \equiv \ev{\cal Q}{II}=\mathcal{I}_{\mathrm{intr/spec}}^{E_2} {\cal Q}^{(2),(\rm intr)}_0$ the spectroscopic nuclear quadrupole moment.

Or, in the case of the single electron again, one can write,
\begin{equation}
\begin{aligned}
b = -\frac{e}{4\pi\epsilon_0} {\cal Q}^{(2),(\rm spec)}_0 \left(\frac{2J_a-1}{J_a+1}\right)\int_{r=0}^\infty \frac{1}{r^3} \left( P^2_{n_a\kappa_a}(r) + Q^2_{n_a\kappa_a}(r)\right) dr.
\end{aligned}
\end{equation}

\subsection{Note on the definition of the nuclear radius}

\label{subsec:RRMS}

As a last remark, we would like to point out the impact the choice of $R_\text{rms}$ (and consequently \mbox{$R_{\rm N}=\sqrt{\frac{5}{3}R_\text{rms}^2}$}) has on the computed results. For example, instead of using the empirical \mbox{$R_{\mathrm{rms}} = (0.836 A^{1/3} + 0.57)$ fm} \cite{Palffy2010}, the root-mean-square radius ($R_{\text{rms}} = \sqrt{\ev{R^2}}$) of both the uniform spherical distribution and the Fermi distribution \cref{Fermi_distribution} can be redefined to match the $R_{\text{rms}}$ of the HFBCS model.

In \cref{fig::Tb_monopoles_emp_2} the empirical nuclear radius was used, while in \cref{fig:ch2:Tb_monopoles_Rrms} the HFBCS-derived radius was used. Therefore, the $x$-abscissae of these two figures are not to scale, but instead show the effect of the calculated nuclear radius on the relative differences between the different nuclear models. The monopole charge densities are shown in the lower panels, with the monopole potentials in the upper panels. The HFBCS model results are compared with the uniform sphere, Fermi-function, and point-charge Coulomb potentials. We can observe that after normalising by the HFBCS-derived radius, the monopole potential, and the charge density at the nuclear surface are in better agreement with the Fermi-function model. 

The recalculated binding energies for both electronic and muonic ions are summarised in \cref{tab:combined_energies} and can be compared with the results in \cref{tab::163Dy_energy_levels} and \cref{tab::mu_163Dy_energy_levels} showing the binding energies where the empirical $R_{\mathrm{rms}}$ has been chosen.

By performing this renormalisation, one observes that the energy shifts between the Fermi distribution and the HFBCS model for the electronic ion are of the order of $10^{-6}\%$, corresponding to a difference of hundreds of $\rm \mu$eV. This indicates that, for electronic hydrogen-like ions, the $R_{\text{rms}}$ is the dominant factor in determining energy levels, as previously noted in \cite{Palffy2010}. Consequently, one can use the HFBCS model to determine the $R_{\text{rms}}$ of the nucleus and calculate the energy levels using the Fermi distribution, renormalised with the same $R_{\text{rms}}$. This approach yields accurate results while reducing computational costs.

%%%%%%%%%%%%%%%%%%%%%%%%%%%%%%%%%%%%%%%%%%
\begin{figure}[t!]
    \centering
    \begin{subfigure}[b]{0.45\textwidth}
        \centering
        \includegraphics[trim=0cm 2.75cm 0cm 3.25cm, clip, width=\textwidth]{159Tb_monopole_distribution_vec.pdf}
        \caption{$^{159}$Tb with $R_{\rm{rms}}^{\rm{(empirical)}}$.}
        \label{fig::Tb_monopoles_emp_2}
    \end{subfigure}
    \hfill
    \begin{subfigure}[b]{0.45\textwidth}
        \centering
        \includegraphics[trim=0cm 2.75cm 0cm 3.25cm, clip, width=\textwidth]{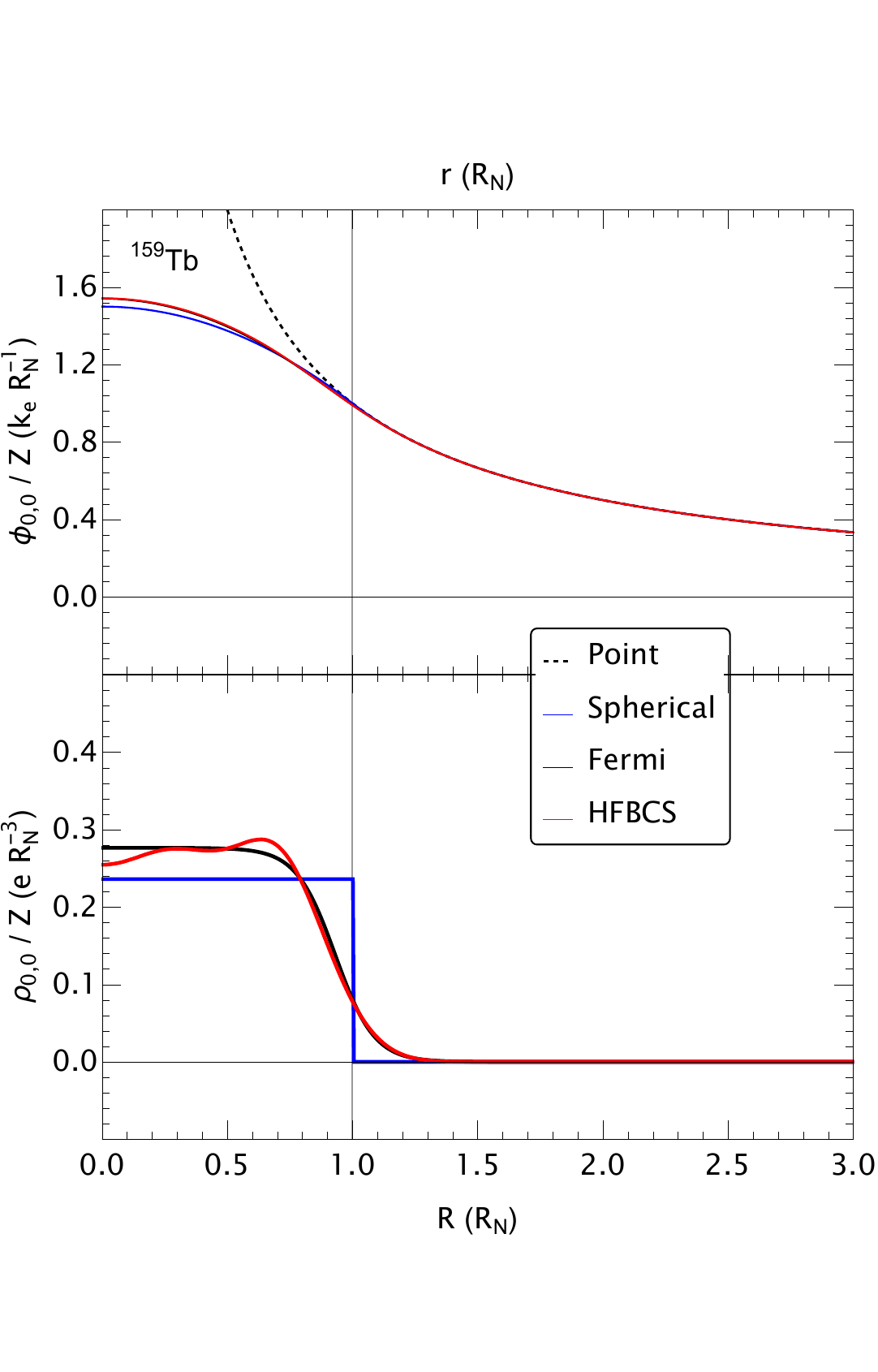}
        \caption{$^{159}$Tb with $R_{\rm{rms}}^{\rm{(HFBCS)}}$.}
        \label{fig:ch2:Tb_monopoles_Rrms}
    \end{subfigure}
\caption{(color online) $Z$-normalised  (radial, spherically averaged) proton monopole charge densities (lower panels) and the corresponding monopole electron potentials (upper panels) for $^{159}$Tb. The radial coordinate is shown as $r/R_{\rm N}$ for the potentials seen by atomic electrons (upper panels) and as $R/R_{\rm N}$ for the nuclear charge densities (lower panels), normalised with respect to the model-specific nuclear radius: in (a) $R_{\rm N}$ is the empirical radius and in (b) the HFBCS-derived radius for $^{159}$Tb. The densities are in units of $e/R_{\rm N}^3$; potentials are in units of $k_e/R_{\rm N}$, with $k_e \equiv \tfrac{e}{4\pi\epsilon_0}$. HFBCS-type (red) calculations are compared with a uniform sphere (blue), a Fermi distribution (black), and a point-charge Coulomb potential (dotted).
}
\label{fig::Tb_monopoles_radii}
\end{figure}
%%%%%%%%%%%%%%%%%%%%%%%%%%%%%%%%%%%%%%%%%%

The energy shifts between the Fermi distribution and the HFBCS model for the muonic ion are of the order of $10^{-3}\%$, corresponding to a difference of hundreds of eV, which is still significant. This means that the $R_{\text{rms}}$ is not the only factor necessary in determining the energy levels of the muonic ion, and their binding energies still represent a valuable tool for discriminating between different nuclear models.

\begin{table}[h]
\centering
\begin{tabular}{|l|cccc|}
\hline
$n\ell_j$ & $E^\text{pt}$ & $E^\text{sp}$ & $E^\text{fm}$ & $E^\text{hf}$ \\
\hline
$1s_{1/2}$ & -2321.4948 & -2321.0385 & -2321.0392 & -2321.0245 \\
$2s_{1/2}$ & -589.6305 & -589.5601 & -589.5603 & -589.5580 \\
$2p_{1/2}$ & -589.6305 & -589.6269 & -589.6269 & -589.6268 \\
$2p_{3/2}$ & -552.6315 & -552.6315 & -552.6315 & -552.6315 \\
$3s_{1/2}$ & -257.8394 & -257.8184 & -257.8184 & -257.8177 \\
$3p_{1/2}$ & -257.8394 & -257.8382 & -257.8382 & -257.8381 \\
$3p_{3/2}$ & -246.8293 & -246.8293 & -246.8293 & -246.8293 \\
$3d_{3/2}$ & -246.8293 & -246.8293 & -246.8293 & -246.8293 \\
$3d_{5/2}$ & -243.5797 & -243.5797 & -243.5797 & -243.5797 \\
$4s_{1/2}$ & -143.2890 & -143.2802 & -143.2802 & -143.2799 \\
$4p_{1/2}$ & -143.2890 & -143.2884 & -143.2884 & -143.2884 \\
$4p_{3/2}$ & -138.6699 & -138.6699 & -138.6699 & -138.6699 \\
$4d_{3/2}$ & -138.6699 & -138.6699 & -138.6699 & -138.6699 \\
$4d_{5/2}$ & -137.2922 & -137.2922 & -137.2922 & -137.2922 \\
$4f_{5/2}$ & -137.2922 & -137.2922 & -137.2922 & -137.2922 \\
$4f_{7/2}$ & -136.6220 & -136.6220 & -136.6220 & -136.6220 \\
\hline
\end{tabular}
\caption{Binding energies in Hartrees for $n\ell_{j}$ states of the ${}^{163}\text{Dy}^{65+}$ ion. $E^\text{m}$ denotes the relativistic energy values (excluding the rest mass) calculated by solving the Dirac equation for a (m=pt) point nucleus, (m=sp) sphere model, (m=fm) Fermi model, and (m=hf) spherically averaged HFBCS model.}
\label{tab::163Dy_energy_levels}
\end{table}

\begin{table}[ht]
\centering
\begin{tabular}{|l|cccc|}
\hline
$n\ell_j$ & $E^\text{pt}$ & $E^\text{sp}$ & $E^\text{fm}$ & $E^\text{hf}$ \\
\hline
$1s_{1/2}$ & -479680.4 & -285823.2 & -286916.6 & -284560.7 \\
$2s_{1/2}$ & -121832.8 & -90810.8 & -91087.4 & -90671.6 \\
$2p_{1/2}$ & -121832.8 & -115503.4 & -115443.9 & -115250.1 \\
$2p_{3/2}$ & -114187.8 & -111674.2 & -111596.0 & -111468.7 \\
$3s_{1/2}$ & -53276.2 & -43553.2 & -43650.9 & -43513.4 \\
$3p_{1/2}$ & -53276.2 & -51122.5 & -51106.6 & -51042.3 \\
$3p_{3/2}$ & -51001.3 & -50101.0 & -50076.4 & -50032.1 \\
$3d_{3/2}$ & -51001.3 & -50980.2 & -50978.1 & -50976.4 \\
$3d_{5/2}$ & -50329.8 & -50322.2 & -50320.8 & -50319.9 \\
$4s_{1/2}$ & -29607.2 & -25412.2 & -25456.7 & -25395.6 \\
$4p_{1/2}$ & -29607.2 & -28664.8 & -28658.5 & -28630.5 \\
$4p_{3/2}$ & -28652.8 & -28250.6 & -28240.1 & -28220.4 \\
$4d_{3/2}$ & -28652.8 & -28640.1 & -28638.8 & -28637.8 \\
$4d_{5/2}$ & -28368.1 & -28363.4 & -28362.6 & -28362.1 \\
$4f_{5/2}$ & -28368.1 & -28368.1 & -28368.1 & -28368.1 \\
$4f_{7/2}$ & -28229.6 & -28229.6 & -28229.6 & -28229.6 \\
\hline
\end{tabular}
\caption{Same as in \cref{tab::163Dy_energy_levels} but for the $n\ell_{j}$ states of the muonic ${}^{163}\text{Dy}^{65+}(\mu^-)$ ion.}
\label{tab::mu_163Dy_energy_levels}
\end{table}

\begin{table}[ht]
\centering

\begin{subtable}[t]{0.45\textwidth}
\centering
\begin{tabular}{|l|ccc|}
\hline
$n\ell_j$ & $E^\text{sp}$ & $E^\text{fm}$ & $E^\text{hf}$ \\
\hline
$1s_{1/2}$ & -2321.0242 & -2321.0249 & -2321.0245 \\
$2s_{1/2}$& -589.5579 & -589.5581 & -589.5580 \\
$2p_{1/2}$ & -589.6268 & -589.6268 & -589.6268 \\
$2p_{3/2}$ & -552.6315 & -552.6315 & -552.6315 \\
$3s_{1/2}$ & -257.8177 & -257.8178 & -257.8177 \\
$3p_{1/2}$  & -257.8381 & -257.8381 & -257.8381 \\
$3p_{3/2}$  & -246.8293 & -246.8293 & -246.8293 \\
$3d_{3/2}$  & -246.8293 & -246.8293 & -246.8293 \\
$3d_{5/2}$  & -243.5797 & -243.5797 & -243.5797 \\
$4s_{1/2}$  & -143.2799 & -143.2799 & -143.2799 \\
$4p_{1/2}$ & -143.2884 & -143.2884 & -143.2884 \\
$4p_{3/2}$  & -138.6699 & -138.6699 & -138.6699 \\
$4d_{3/2}$  & -138.6699 & -138.6699 & -138.6699 \\
$4d_{5/2}$ & -137.2922 & -137.2922 & -137.2922 \\
$4f_{5/2}$ & -137.2922 & -137.2922 & -137.2922 \\
$4f_{7/2}$  & -136.6220 & -136.6220 & -136.6220 \\
\hline
\end{tabular}
\caption{Binding energies for the electronic ${}^{163}\text{Dy}^{65+}$ ion.
\label{tab:energies_Rrms}}
\end{subtable}\hfill
\begin{subtable}[t]{0.45\textwidth}
\centering
\begin{tabular}{|l|ccc|}
\hline
$n\ell_j$ & $E^\text{sp}$ & $E^\text{fm}$ & $E^\text{hf}$ \\
\hline
$1s_{1/2}$ & -283536.8 & -284602.4 & -284560.7 \\
$2s_{1/2}$ & -90405.7 & -90677.2 & -90671.6 \\
$2p_{1/2}$ & -115315.7 & -115256.5 & -115250.1 \\
$2p_{3/2}$ & -111551.3 & -111473.7 & -111468.7 \\
$3s_{1/2}$ & -43419.0 & -43515.2 & -43513.4 \\
$3p_{1/2}$ & -51060.0 & -51044.4 & -51042.3 \\
$3p_{3/2}$ & -50058.0 & -50033.7 & -50032.1 \\
$3d_{3/2}$ & -50978.7 & -50976.5 & -50976.4 \\
$3d_{5/2}$ & -50321.5 & -50320.0 & -50319.9 \\
$4s_{1/2}$ & -25352.5 & -25396.3 & -25395.6 \\
$4p_{1/2}$ & -28637.6 & -28631.4 & -28630.5 \\
$4p_{3/2}$ & -28231.5 & -28221.1 & -28220.4 \\
$4d_{3/2}$ &-28639.2 & -28637.9 & -28637.8 \\
$4d_{5/2}$ & -28363.0 & -28362.1 & -28362.1 \\
$4f_{5/2}$ &-28368.1 & -28368.1 & -28368.1 \\
$4f_{7/2}$ & -28229.6 & -28229.6 & -28229.6 \\
\hline
\end{tabular}
\caption{Binding energies for the muonic ${}^{163}\text{Dy}^{65+}(\mu^-)$ ion.\label{tab:energies_muons_Rrms}}
\end{subtable}

\caption{Binding energies in Hartrees for different $n\ell_{j}$ states of the ${}^{163}\text{Dy}^{65+}$ ion, with spherical and Fermi distributions normalised for the same $R_\text{rms}$ values as the HFBCS model.\label{tab:combined_energies}}

\end{table}

\subsection{Results for $^{161}$Dy}

\begin{figure}[ht!]
\centering
\includegraphics[width=.95\textwidth, trim={0 5.5cm 0 5.5cm},clip]{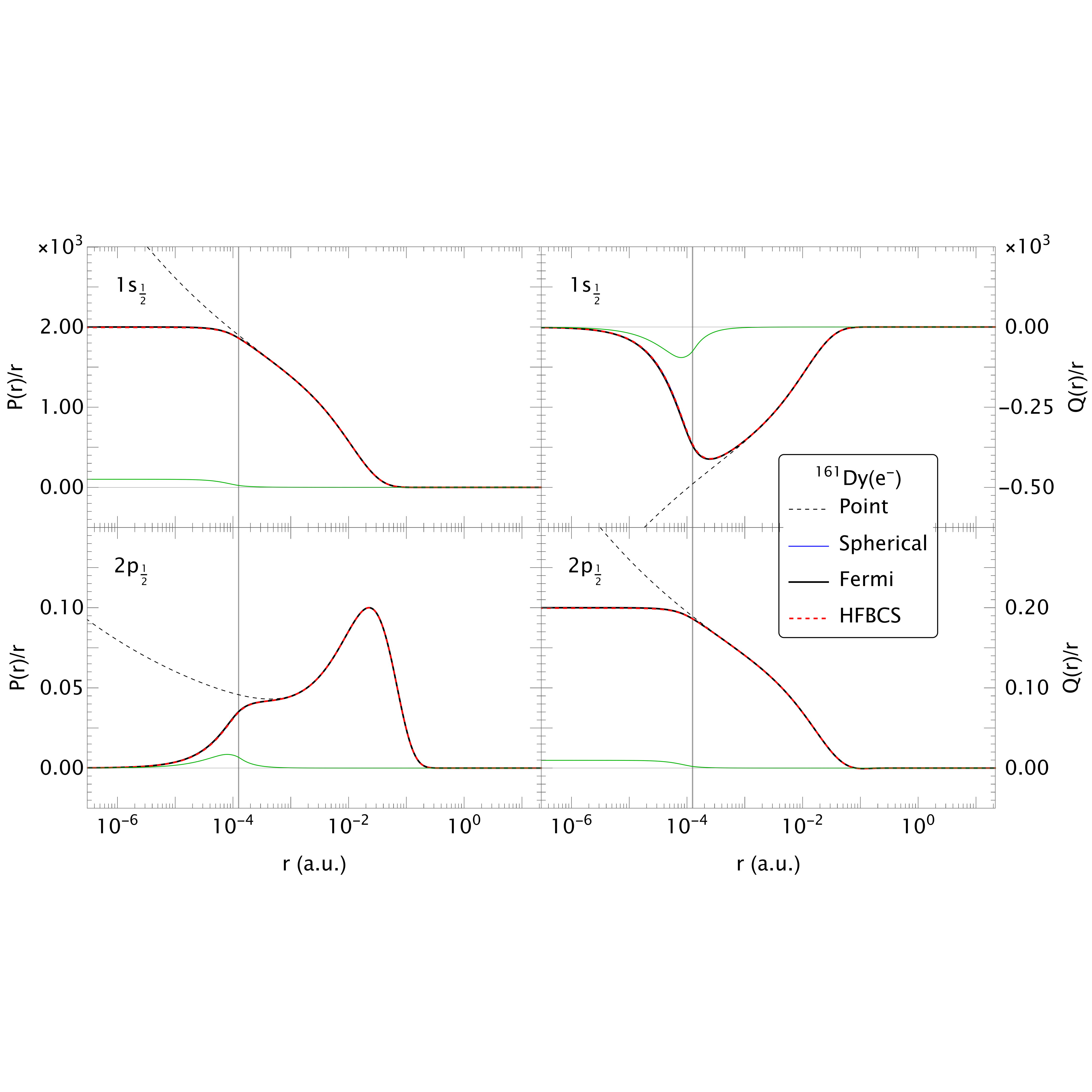}
\caption{(color online) Large ($P_{n\kappa}$) and small ($Q_{n\kappa}$) components of the radial wave functions for the 1$s_{1/2}$ and 2$p_{1/2}$ orbitals of $^{161}$Dy$^{65+}$. The dashed black and red curves represent the Point and HFBCS models respectively, while the solid black and blue curves represent the Fermi and Spherical models respectively. The solid green line represents the difference between the Fermi and HFBCS models multiplied by 20. Vertical lines indicate the position of the nuclear radius $R_{\rm N}$. Atomic units are used.} \label{fig:a1}
\end{figure}

\begin{figure}[ht!]
\centering
\includegraphics[width=.95\textwidth, trim={0 5.5cm 0 5.5cm},clip]{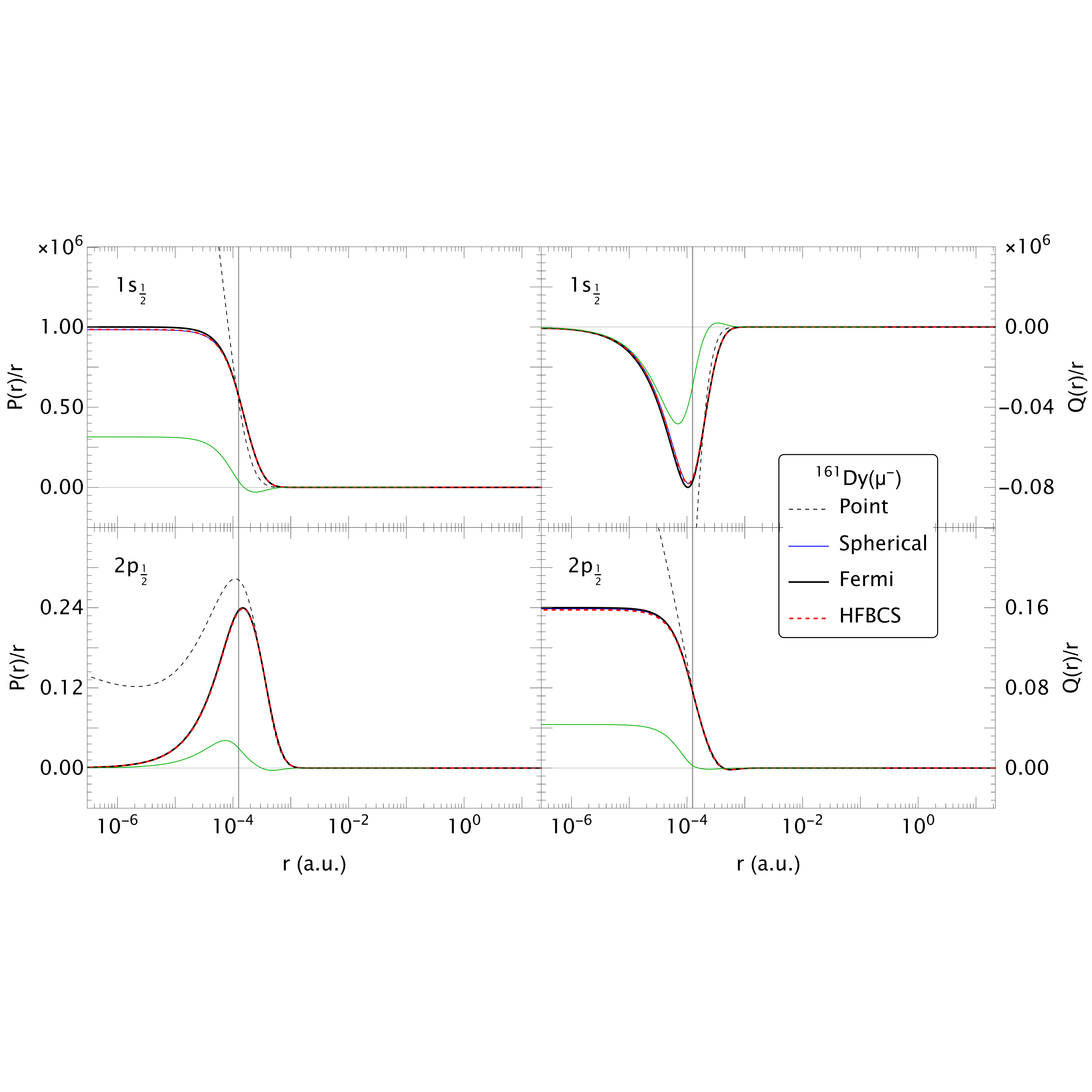}
\caption{(color online) Same as \cref{fig:e1} but for a muonic hydrogen-like ion of $^{161}$Dy$^{65+}$. Atomic units are used.} \label{fig:a2}
\end{figure}

\begin{figure}[ht!]
\centering
\includegraphics[width=.95\textwidth, trim={0 5.5cm 0 5.5cm},clip]{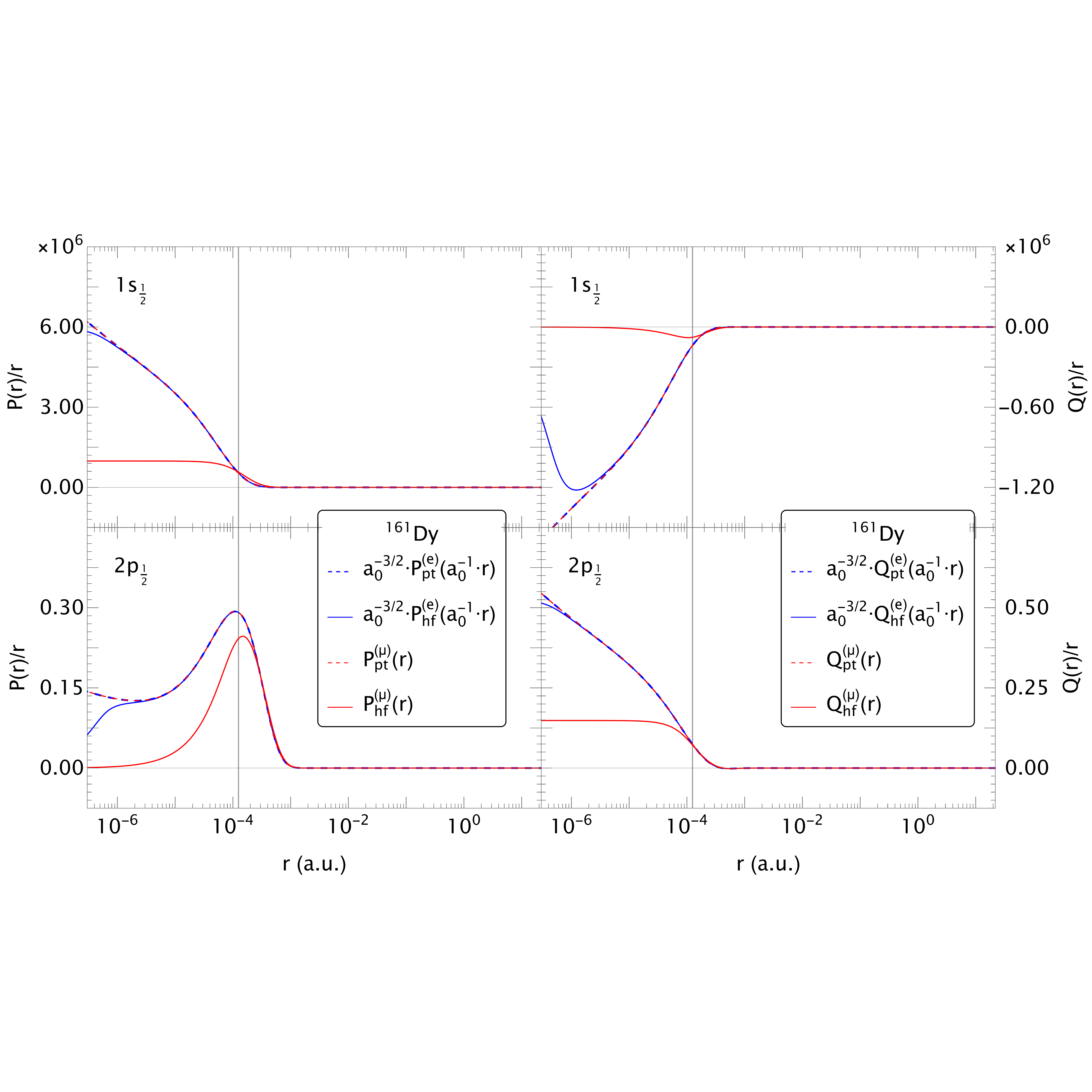}
\caption{(color online) Large ($P_{n\kappa}$) and small ($Q_{n\kappa}$) components of the radial wave functions for the 1$s_{1/2}$ and 2$p_{1/2}$ orbitals of $^{161}$Dy$^{65+}$. Comparison between electron (blue) and muon (red) for Point (dashed) and HFBCS (solid) models. The results for the electron are rescaled using the Bohr radius for muon, $a_0=1/207$ au. Vertical lines indicate the position of the nuclear radius $R_{\rm N}$. \label{fig:a3}}
\end{figure}

\begin{figure}[hb!]
    {\centering
    \begin{subfigure}[t]{0.49\textwidth}
        \centering
        \includegraphics[width=\textwidth, trim={7cm 3.5cm 7cm 3.5cm}, clip]{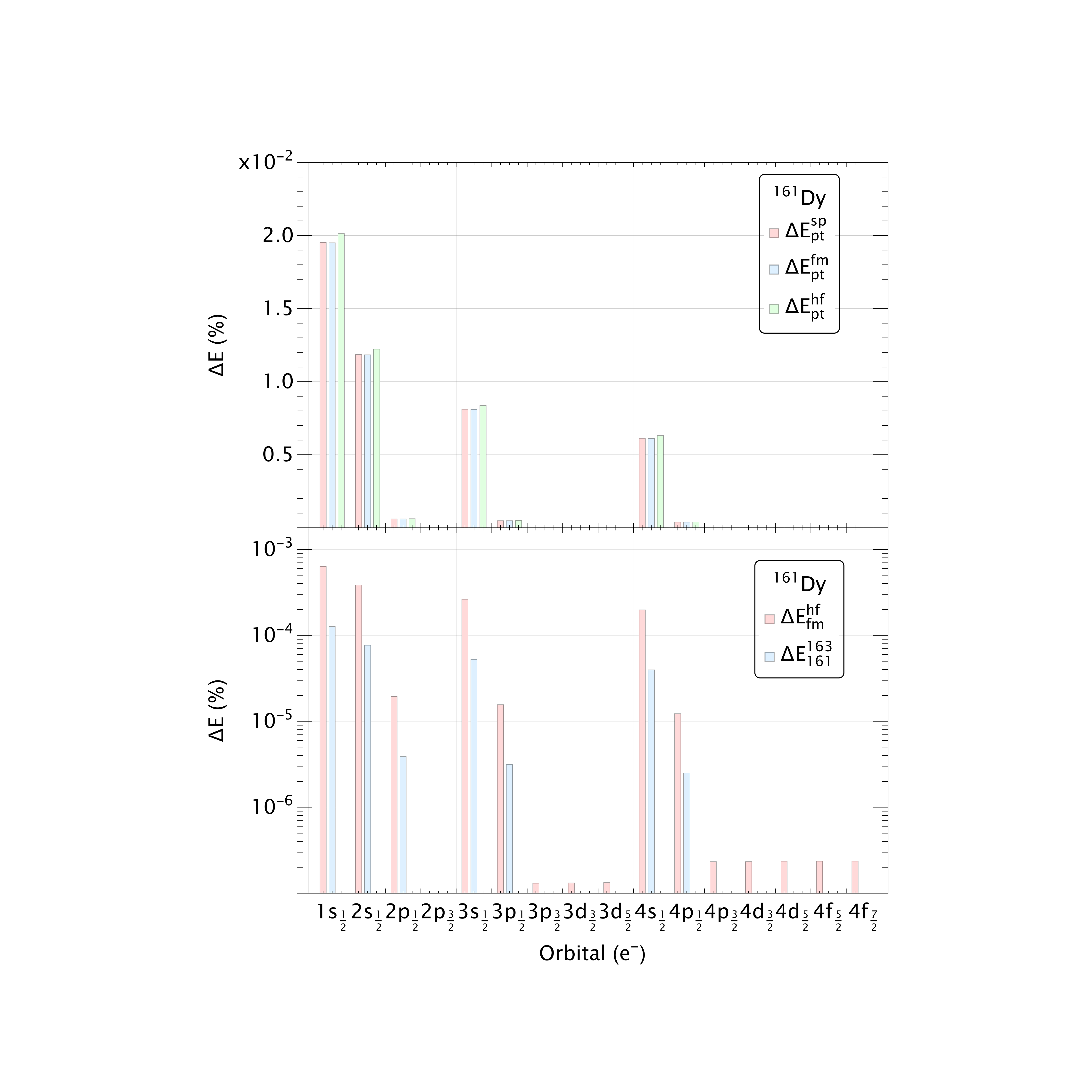}
        \caption{Electronic orbitals of ${}^{161}$Dy$^{65+}$.
        }
        \label{fig:a4}
    \end{subfigure}
    \hfill
    \begin{subfigure}[t]{0.49\textwidth}
        \centering
        \includegraphics[width=\textwidth, trim={7cm 3.5cm 7cm 3.5cm}, clip]{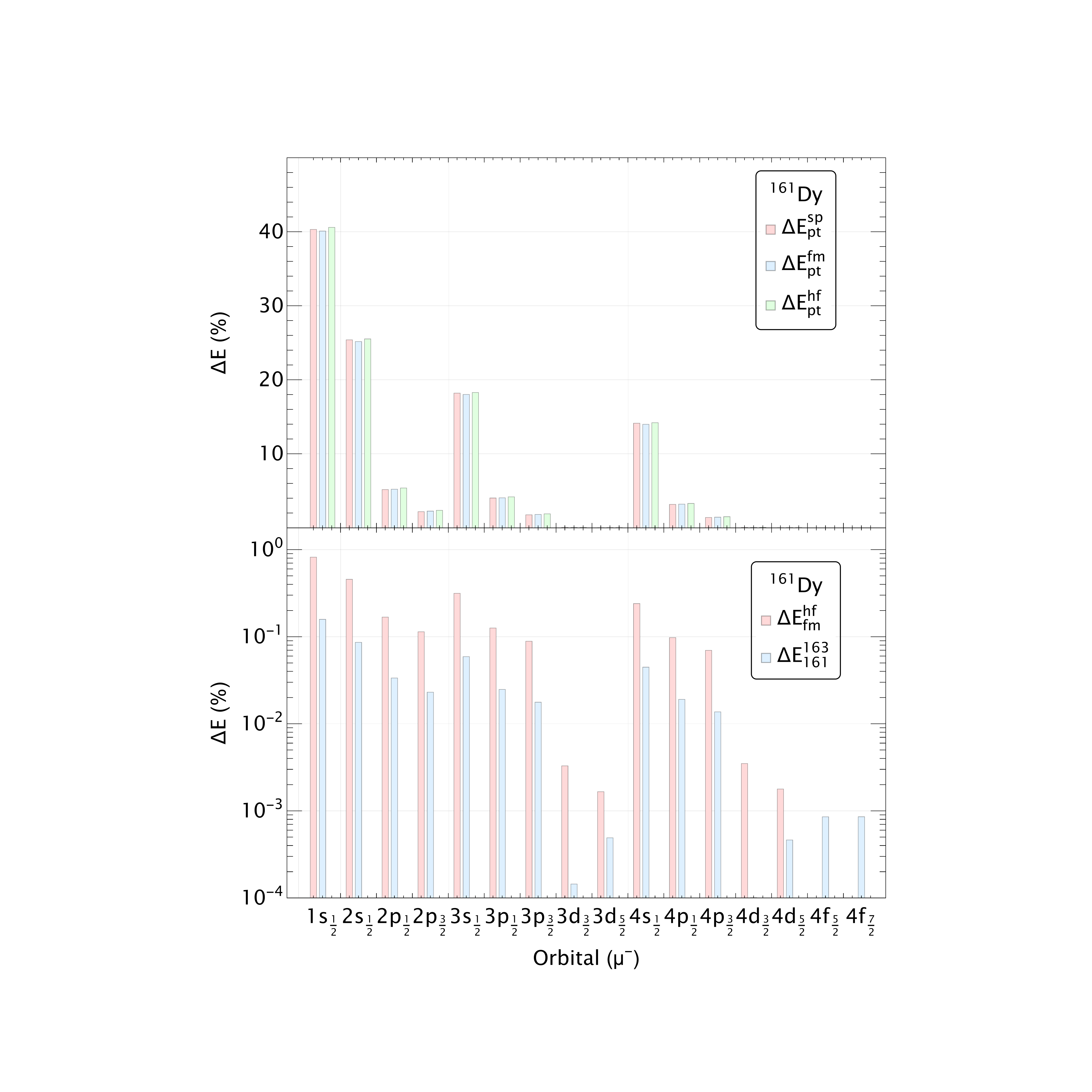}
        \caption{Muonic orbitals of ${}^{161}$Dy$^{65+}(\mu^-)$.}
        \label{fig:a5}
    \end{subfigure}}
    \caption{(color online) Monopolar shifts in the energy levels of electronic orbitals of $^{161}$Dy$^{65+}$ (from $1s_{1/2}$ to $4f_{7/2})$ in different nuclear models, Point (pt), Spherical (sp), Fermi (fm) and HFBCS (hf). We have defined, $\Delta E_{m_1,i}^{m_2} \equiv \frac{E_i^{m_2}-E_i^{m_1}}{E_i^{m_1}}$ with $m_2\in\lbrace{\mathrm{sp,fm,hf}\rbrace}$ and $m_1\in\lbrace{\mathrm{pt,fm}\rbrace}$ and $\Delta E_{163,i}^{161} \equiv \frac{E_i^{161}-E_i^{163}}{E_i^{163}}$ with $i$ running from $1s_{1/2}$ to $4f_{7/2}$. This latter quantity was evaluated using the energies of the HFBCS model.}
\end{figure}

\begin{figure}[ht!]
\centering
\includegraphics[width=.66\textwidth, trim={5.8cm 5cm 5.8cm 5cm},clip]{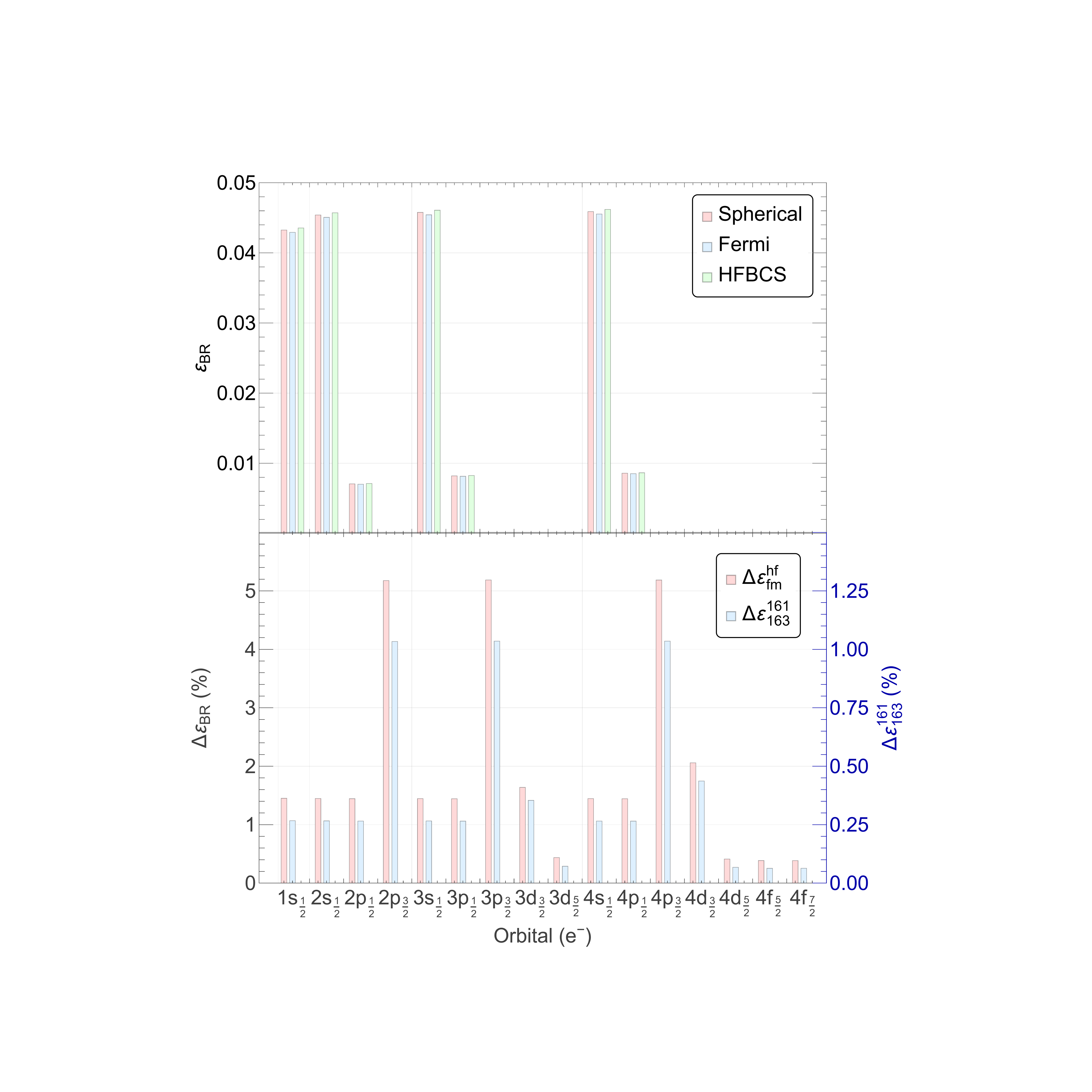}
\caption{(color online) (Upper panel) Breit-Rosenthal correction $\varepsilon_{\mathrm{BR}}$ for $^{161}$Dy$^{65+}$ in different nuclear models: Spherical, Fermi and HFBCS. (Lower panel) We have defined, $\Delta \varepsilon_{\mathrm{fm}}^{\mathrm{hf}} \equiv \frac{\varepsilon_{\mathrm{BR}}^{\mathrm{hf}}-\varepsilon_{\mathrm{BR}}^{\mathrm{fm}}}{\varepsilon_{\mathrm{BR}}^{\mathrm{fm}}}$ (reading on the left axis) and $\Delta \varepsilon_{163}^{161} \equiv \frac{\varepsilon_{\mathrm{BR}}^{\mathrm{hf}}(161)-\varepsilon_{\mathrm{BR}}^{\mathrm{hf}}(163)}{\varepsilon_{\mathrm{BR}}^{\mathrm{hf}}(163)}$ (reading on the right axis).}  \label{fig:a6}
\end{figure}

\begin{figure}[ht!]
\centering
\includegraphics[width=.75\textwidth, trim={5.8cm 5cm 5.8cm 5cm},clip]{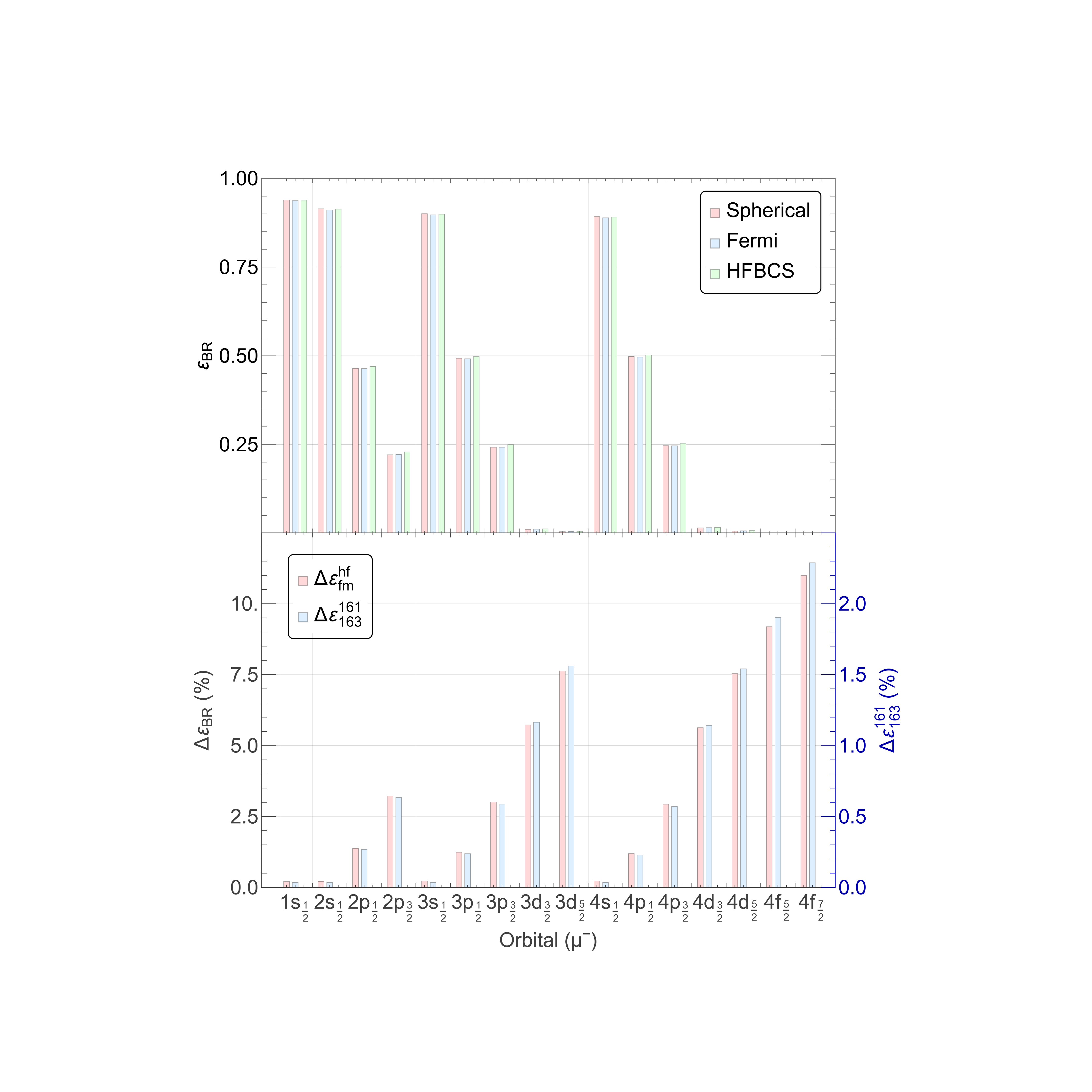}
\caption{(color online) Same as \cref{fig:a6} but for a muonic hydrogen-like ion.}  \label{fig:a7}
\end{figure}

\begin{figure}[ht!]
\centering
\includegraphics[width=.75\textwidth, trim={5.8cm 4.8cm 5.8cm 5cm},clip]{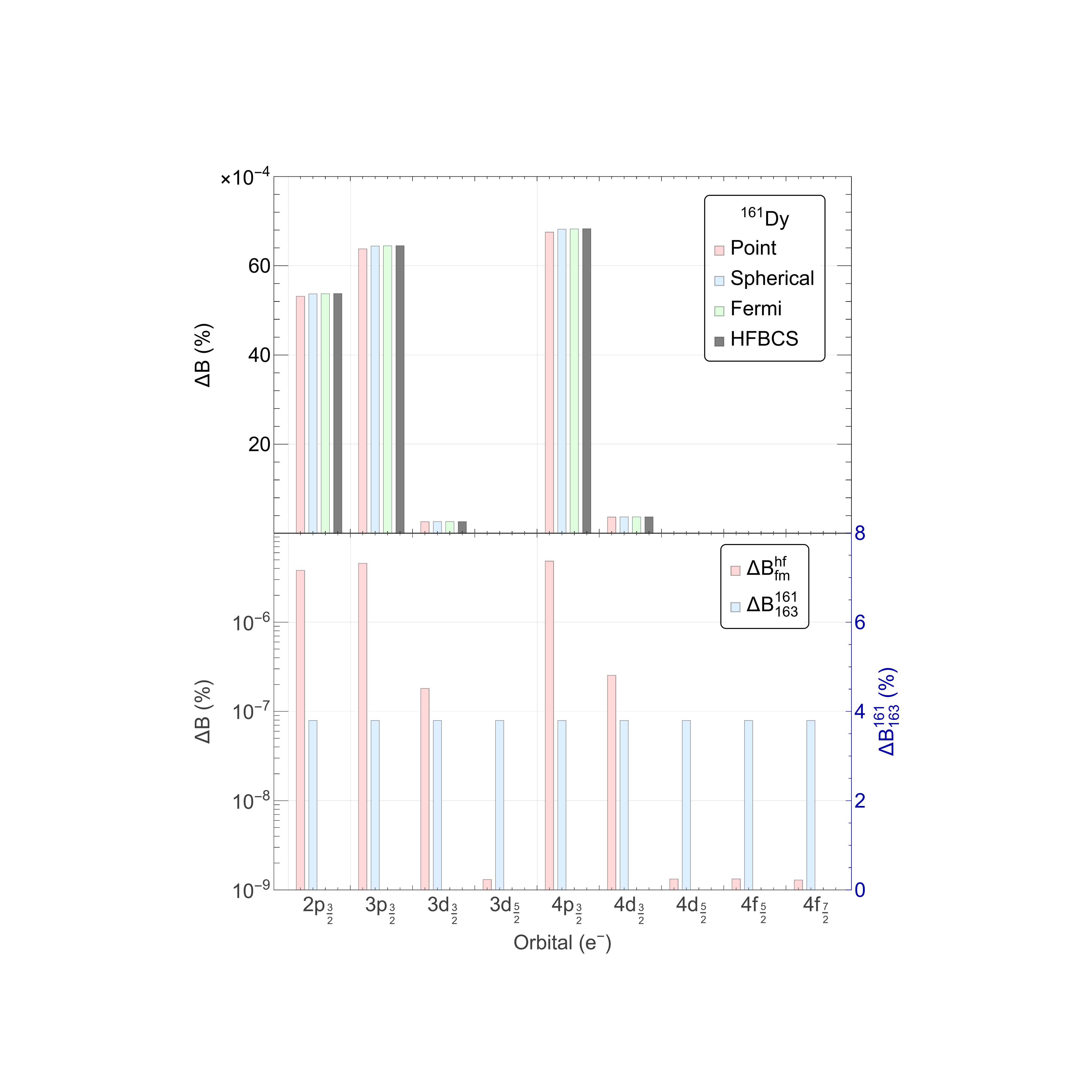}
\caption{(color online)  Quadrupole shifts for electronic $^{161}$Dy$^{65+}$. (Upper panel) ${\Delta B}^{m} \equiv \frac{b_{\mathrm{np}}^{\mathrm{pt}}-b_{\mathrm{wp}}^{m}}{b_{\mathrm{np}}^{\mathrm{pt}}}$ with $m\in\left\{\mathrm{pt}, \mathrm{sp}, \mathrm{fm}, \mathrm{hf}\right\}$ and subscript indices $\mathrm{np}$ and $\mathrm{wp}$ representing calculations with no penetration and with penetration, respectively. (Lower panel) Two quantities are depicted: ${\Delta B}\equiv \frac{b_{\mathrm{wp}}^{\mathrm{fm}}-b_{\mathrm{wp}}^{\mathrm{hf}}}{B_{\mathrm{wp}}^{\mathrm{fm}}}$ for $^{161}$Dy$^{65+}$ (in red and reading on the left axis) and ${\Delta B}_{161}^{163} \equiv \frac{b_{\mathrm{wp}}^{\mathrm{hf}}(161)-b_{\mathrm{wp}}^{\mathrm{hf}}(163)}{b_{\mathrm{wp}}^{\mathrm{hf}}(163)}$  (in blue and reading on the right axis). Note that the $s_{1/2}$ and $p_{1/2}$ contributions are zero.}  \label{fig:a14}
\end{figure}

\begin{figure}[ht!]
\centering
\includegraphics[width=.75\textwidth, trim={5.8cm 4.5cm 5.8cm 5cm},clip]{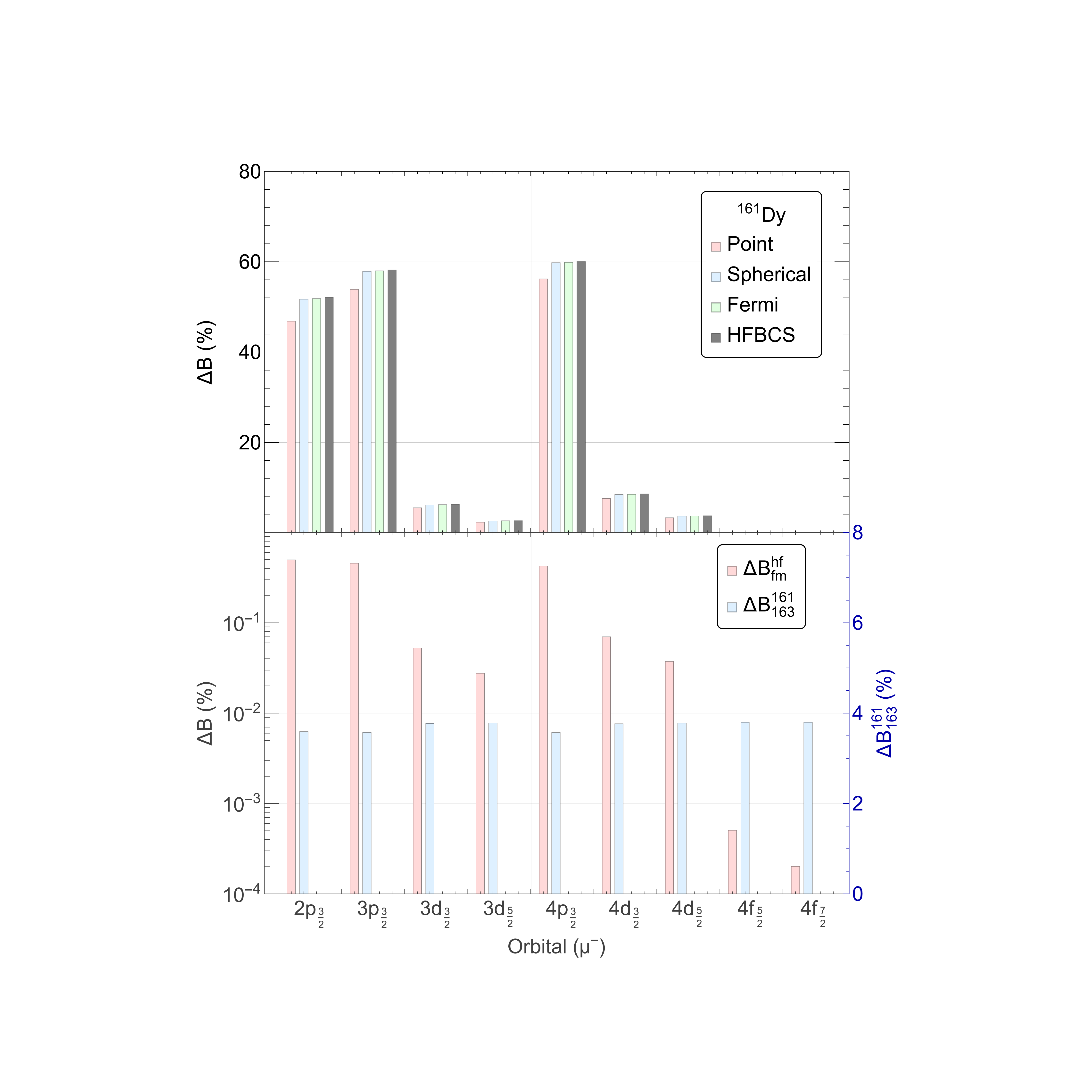}
\caption{(color online) Same as \cref{fig:a14} but for muonic $^{161}$Dy$^{65+}$.}  \label{fig:a15}
\end{figure}

\FloatBarrier

\bibliographystyle{apsrev}

% \bibliography{references}

\begin{thebibliography}{9}

\bibitem{ESSEN1955}L. Essen and J. V. L. Parry, Nature, {\bf 176}, 280 (1955).

\bibitem{PeikOkhapkin2015}
E.~Peik and Y.~N.~Okhapkin, C. R. Physique \textbf{16}, 516 (2015).

\bibitem{Karpeshin1998}F. F. Karpeshin, S. Wycech, I. M. Band, M. B. Trzhaskovskaya, M. Pfützner, and J. Żylicz, Phys. Rev. C {\bf 57}, 3085 (1998).

\bibitem{Tkalya2016}E. V. Tkalya and A. V. Nikolaev, Phys. Rev. C {\bf 94}, 014323 (2016).

\bibitem{Seiferle2019}
B.~Seiferle et al., Nature \textbf{573}, 243 (2019).

\bibitem{Campbell2011}
C. J. Campbell, A. G. Radnaev, and A. Kuzmich, Phys. Rev. Lett. {\bf 106}, 223001 (2011)

\bibitem{Morgan2025}
H. W. T. Morgan, H. B. Tran Tan, R. Elwell, A. N. Alexandrova, Eric R. Hudson, and Andrei Derevianko, Phys. Rev. Lett. {\bf 134}, 253801 (2025).

\bibitem{Meny2017}C. M{\'e}ny and P. Panissod, Modern Magnetic Resonance, Springer International Publishing (2017).

\bibitem{Thiele2014}
S. Thiele, F. Balestro, R. Ballou, S. Klyatskaya, M. Ruben and W. Wernsdorfer,
Science {\bf 344}, 1135 (2014).

\bibitem{Godfrin2017}
C. Godfrin, A. Ferhat, R. Ballou, S. Klyatskaya, M. Ruben, W. Wernsdorfer and F. Balestro, Phys. Rev. Lett. {\bf 119},
187702 (2017).

\bibitem{Yu2025}
X. Yu, B. Wilhelm, D. Holmes, A. Vaartjes, D. Schwienbacher, M. Nurizzo,
A. Kringhøj, M.R. van Blankenstein, A.M. Jakob, P. Gupta, F.E. Hudson,
K.M. Itoh, R.J. Murray, R. Blume-Kohout, T.D. Ladd, N. Anand,
A.S. Dzurak, B.C. Sanders, D.N. Jamieson and A. Morello, Nature Physics {\bf 21}, 362 (2025).

\bibitem{Pohl2010}
R. Pohl, A. Antognini, F. Nez, F.D. Amaro, F. Biraben, J.M.R. Cardoso,
D.S. Covita, A. Dax, S. Dhawan, L.M.P. Fernandes, A. Giesen, T. Graf,
T.W. Hänsch, P. Indelicato, L. Julien, C.-Y. Kao, P. Knowles,
E.-O. Le Bigot, Y.-W. Liu, J.A.M. Lopes, L. Ludhova, C.M.B. Monteiro,
F. Mulhauser, T. Nebel, P. Rabinowitz, J.M.F. dos Santos,
L.A. Schaller, K. Schuhmann, C. Schwob, D. Taqqu, J.F.C.A. Veloso
and F. Kottmann, Nature {\bf 466}, 213 (2010).

\bibitem{Antognini2013}
A. Antognini, F. Nez, K. Schuhmann, F.D. Amaro, F. Biraben, J.M.R. Cardoso,
D.S. Covita, A. Dax, S. Dhawan, M. Diepold, L.M.P. Fernandes, A. Giesen,
A.L. Gouvea, T. Graf, T.W. Hänsch, P. Indelicato, L. Julien, C.-Y. Kao,
P. Knowles, F. Kottmann, E.-O. Le Bigot, Y.-W. Liu, J.A.M. Lopes,
L. Ludhova, C.M.B. Monteiro, F. Mulhauser, T. Nebel, P. Rabinowitz,
J.M.F. dos Santos, L.A. Schaller, C. Schwob, D. Taqqu, J.F.C.A. Veloso,
J. Vogelsang and R. Pohl,
Science {\bf 339}, 417 (2013).

\bibitem{RosenthalBreit1932}
J.~E.~Rosenthal and G.~Breit,  Phys. Rev. \textbf{41}, 459 (1932).

\bibitem{BohrWeisskopf1950}
A.~Bohr and V.~F.~Weisskopf, Phys. Rev. \textbf{77}, 94 (1950).

\bibitem{Karpeshin2015}
F.~F.~Karpeshin and M.~B.~Trzhaskovskaya, Nucl. Phys. A \textbf{941}, 66 (2015).

\bibitem{Roberts2021}
B.M. Roberts and J.S.M. Ginges, Phys. Rev. A {\bf 104}, 022823 (2021).

\bibitem{Volotka2012}
A.V. Volotka, D.A. Glazov, O.V. Andreev, V.M. Shabaev, I.I. Tupitsyn and G. Plunien, Phys. Rev. Lett. {\bf 108}, 073001 (2012).

\bibitem{Bender2003}
M.~Bender, P.-H.~Heenen, and P.-G.~Reinhard, Rev. Mod. Phys. \textbf{75}, 121 (2003).

\bibitem{Moreno2018}E Moreno-Pineda, C Godfrin, F Balestro, W Wernsdorfer, M Ruben, Chemical Society Reviews {\bf 47}, 501 (2018).

\bibitem{Volotka2013}
A.~V.~Volotka, D.~A.~Glazov, G.~Plunien, and V.~M.~Shabaev, Ann. Phys. (Berlin) \textbf{525}, 636 (2013).

\bibitem{Boucard1998}
S. Boucard, PhD thesis, Université Pierre et Marie Curie – Paris VI (1998).

\bibitem{Boucard2000}
S. Boucard and P. Indelicato, Eur. Phys. J. D {\bf 8}, 59 (2000).

\bibitem{Jentschura2011}
U.D. Jentschura, Canadian Journal of Physics {\bf 89}, 109 (2011).

\bibitem{Oreshkina2022}
N.S. Oreshkina, Phys. Rev. Res. {\bf 4}, L042040 (2022).

\bibitem{Shabaev2001}
V.M. Shabaev, A.N. Artemyev, V.A. Yerokhin, O.M. Zherebtsov and G. Soff,
Phys. Rev. Lett. {\bf 86}, 3959 (2001).

\bibitem{Artemyev2001}
A.N. Artemyev, V.M. Shabaev, G. Plunien, G. Soff and V.A. Yerokhin, Phys. Rev. A {\bf 63}, 062504 (2001).

\bibitem{Persson2025}
J.R. Persson, arXiv:2310.16398 (2025).

\bibitem{PSISMS}
Paul Scherrer Institute (PSI), Swiss Muon Source (S$\mu$S), facility overview and beamlines (accessed Oct.\ 2025).

\bibitem{Friedrich2017} Harald Friedrich, Theoretical Atomic Physics, Springer (2017). 
\bibitem{Br55}K. A. Brueckner, Phys. Rev. {\bf 97}, 1353 (1955).
\bibitem{Ne70}J. W. Negele, Phys. Rev. C {\bf 1}, 1260 (1970).
\bibitem{Ne82}J. W. Negele, Rev. Mod. Phys. {\bf 54}, 913 (1982).
\bibitem{DB70}K. T. R. Davies, M. Baranger, Phys. Rev. C {\bf 1}, 1640 (1970).
\bibitem{CS72}X. Campi, D. W. Sprung, Nucl. Phys. A {\bf 194}, 401 (1972).
\bibitem{Go75}D. Gogny, Nuclear self-consistent fields, Trieste, 1975, eds. G. Ripka and M. Porneuf (North Holland, Amsterdam, 1975) p. 333.
\bibitem{DG80}J. Decharg\'{e} and D. Gogny, Phys. Rev. C {\bf 21}, 1568 (1980).
\bibitem{Mo70}S. A. Moszkowski, Phys. Rev. C {\bf 2}, 402 (1970).
\bibitem{Sk59}T. H. R. Skyrme, Nucl. Phys. {\bf 9}, 615 (1959).
\bibitem{VB72}D. Vautherin, D. M. Brink, Phys. Rev. C {\bf 5}, 626 (1972).
\bibitem{Va73}D. Vautherin, Phys. Rev. C {\bf 7}, 296 (1973).
\bibitem{GQ80}M. J. Giannoni, P. Quentin, Phys. Rev. {\bf 21}, 2076 (1980).
\bibitem{EBG75}J. Engel, D. Brink, K. Goeke, S. J. Krieger, D. Vautherin, Nucl. Phys. A {\bf 249}, 215 (1975).
\bibitem{HHB12}V. Hellemans, P. -H. Heenen, M. Bender, Phys. Rev. C {\bf 85}, 014326 (2012).
\bibitem{BMD15}L. Bonneau, N. Minkov, Dao Duy Duc, P. Quentin, J. Bartel, Phys. Rev. C {\bf 91}, 054307 (2015).
\bibitem{TAK14}Density Functional Theory in Quantum Chemistry, Takao Tsuneda. (2014). https://doi.org/10.1007/978-4-431-54825-6
\bibitem{KS65}W. Kohn and L. J. Sham, Phys. Rev.{\bf 140}, 1133 (1965).
\bibitem{BFN75}M. Beiner, H. Flocard, Nguyen van Giai and P.\ Quentin, Nucl.\ Phys.\ {\bf A283} (1975) 29.
\bibitem{BQB82}J. Bartel, P. Quentin, M. Brack, C. Guet and H.-B. H{\aa}kansson, Nucl. Phys. A {\bf 386}, 79 (1982).
\bibitem{CBH97}E. Chabanat, P. Bonche, P. Haensel, J. Meyer, R. Schaeffer, Nucl. Phys. A {\bf 627}, 710 (1997).
\bibitem{CBH98}E. Chabanat, P. Bonche, P. Haensel, J. Meyer, R. Schaeffer, Nucl. Phys. A {\bf 635}, 231 (1998).
\bibitem{LBB07}T. Lesinski, M. Bender, K. Bennaceur, T. Duguet, J. Meyer, Phys. Rev. C {\bf 76}, 014312 (2007).
\bibitem{BBD09}M. Bender, K. Bennaceur, T. Duguet, P.-H. Heenen, T. Lesinski, J. Meyer, Phys. Rev. C {\bf 80}, 064302 (2009).

\bibitem{GCP10}S. Goriely, N. Chamel, and J. M. Pearson, Phys. Rev. C {\bf 82}, 035804 (2010).
\bibitem{SDM13}S. Sadoudi, T. Duguet, J. Meyer, M. Bender, Phys. Rev. C {\bf 88}, 064326 (2013).
\bibitem{GCP13}S. Goriely, N. Chamel, J. M. Pearson, Phys. Rev. C {\bf 88}, 061302(R) (2013).
\bibitem{PC22}J. M. Pearson, N. Chamel, Phys. Rev. C {\bf 105}, 015803 (2022).
\bibitem{Sl51}J. C. Slater, Phys. Rev. {\bf 81}, 385 (1951).
%\bibitem{HiGi}S. Hilaire, M. Girod, "AMEDEE" database from  Hartree-Fock-Bogoliubov calculations with the D1S Gogny force (https://www-phynu.cea.fr)
\bibitem{DPP69}J. Damgaard, H. -C.\ Pauli, V. V. Paskhevich, and V. M. Strutinsky, Nucl. Phys. A {\bf 135}, 432 (1969).

\bibitem{FQKV}H. Flocard, P. Quentin, A. K. Kerman and D. Vautherin, Nucl. Phys. A{\bf 203}, 433 (1973).

\bibitem{CPG15}F. Chappert, N. Pillet, M. Girod, J. -F. Berger, Phys. Rev. C {\bf 91}, 034312 (2015).
\bibitem{GCV18}C. Gonzales-Boquera, M. Centelles, X. Vi\~nas, L. M. Robledo, Phys. Lett. B {\bf 779}, 195 (2018).
\bibitem{DFT84}J. Dobaczewski, H. Flocard and J. Treiner, Nucl. Phys. A {\bf 422}, 103 (1984).
\bibitem{Ring-Schuck}
P. Ring and P. Schuck, {\it The Nuclear Many-Body Problem} (Springer, 1980).
  
\bibitem{NRL19}Nurhafiza M. Nor, Nor-Anita Rezle, Kai-Wen Kelvin Lee, L. Bonneau, and P. Quentin, Phys. Rev. C {\bf 99}, 064306 (2019).

\bibitem{Herbut68} F. Herbut and M. Vuji$\rm\check{c}$i$\rm\check{c}$, Phys. Rev. {\bf 172}, 1031 (1968).

\bibitem{Pototzky10} K. J. Pototzky, J. Erler, P.-G. Reinhard, and V. O. Nesterenko, Eur. Phys. J. A {\bf 46}, 299 (2010).

\bibitem{Reich11_NDS112_Dy161} C. W. Reich and B. Singh, Nucl. Data
  Sheets {\bf 112}, 2497 (2011).

\bibitem{Reich10_NDS111_Dy163} C. W. Reich and B. Singh, Nucl. Data
  Sheets {\bf 111}, 1211 (2010).

\bibitem{Gambhir90} Y. K. Gambhir, P. Ring and A. Thimet,
  Ann. Phys. {\bf 198}, 132 (1990).

\bibitem{CODATA2022} P. J. Moher, D. B. Newell, B. N. Taylor, and
  E. Tiesinga, Rev. Mod. Phys. {\bf 97}, 025002 (2025).
  
\bibitem{St05}N. J. Stone, Atomic Data and Nuclear Data Tables {\bf 90} (2005) 75.
\bibitem{Palffy2010}A. P{\'a}lffy, Contemporary Physics, {\bf 51}, 471 (2010).

\bibitem{Stone16} N. J. Stone, At. Data Nucl. Data Tables {\bf
  111},1 (2016).

\bibitem{Reich12_NDS113_Tb159} C. W. Reich, Nucl. Data
  Sheets {\bf 113}, 157 (2012).
\bibitem{Singh24_NDS194_Ho165} B. Singh and J. Chen, Nucl. Data
  Sheets {\bf 194}, 460 (2024).
  \bibitem{Sc55}C. Schwartz, Phys. Rev. {\bf 97}, 380 (1955).

  \bibitem{Chemtob69} M. Chemtob, Nucl. Phys. A {\bf 123}, 449 (1969).
  
\bibitem{BM_V2}\AA. Bohr and B. R. Mottelson, "Nuclear Structure" Vol. 2, Benjamin, New York, 1975.
\bibitem{PBN68}O. Prior, F. Boehm, S. G. Nilsson, Nucl. Phys. A {\bf 110}, 257 (1968).

\bibitem{Sprung79} D. W. L. Sprung, S. G. Lie, M. Vallieres and
  P. Quentin, Nucl. Phys. A {\bf 326}, 37 (1979).

\bibitem{Strange2005}P. Strange, Relativistic Quantum Mechanics, Cambridge University Press, (2005).

\bibitem{Crespo1996}Jos{\'e} R. Crespo L{\`o}pez-Urrutia, P. Beiersdorfer, Daniel W. Savin, and Klaus Widmann, Phys. Rev. Lett. {\bf 77}, 826 (1996).

\bibitem{WR2007}W. R. Johnson, Atomic Structure Theory, Lectures on Atomic Physics, Springer (2007).

\bibitem{Rosenthal1932}J. E. Rosenthal and G. Breit, Phys. Rev. {\bf 41}, 459 (1932).
 
\bibitem{Bohr1950}A. Bohr and V. Weisskopf, Phys. Rev. {\bf 77}, 94 (1950).

\bibitem{Buttgenbach1984}S. B{\"u}ttgenbach, Hyperfine Interactions {\bf 20}, 1 (1984).

\bibitem{degroote2020}
R.P. de Groote and G. Neyens, Spins and Electromagnetic Moments of Nuclei, 
in Handbook of Nuclear Physics, 
ed. I. Tanihata, H. Toki and T. Kajino, 
Springer Nature, Singapore, (2020), pp. 1–36.

\bibitem{Vandeleur2025}
J. Vandeleur, G. Sanamyan, O.R. Smits, I.A. Valuev, N.S. Oreshkina and J.S.M. Ginges,
Phys. Rev. Lett. \textbf{134}, 093003 (2025).

\bibitem{Persson2023hyperfine}
J.~R.~Persson, arXiv:2304.11995, 2023.

\bibitem{Freeman1984}
A.~J.~Freeman, J.~V.~Mallow, J.~P.~Desclaux, and M.~Weinert,
Hyperfine Interact. \textbf{19}, 865 (1984).

\bibitem{PrezGalvn2007}
A.~Pérez~Galván, Y.~Zhao, L.~A.~Orozco, E.~Gómez, A.~D.~Lange, F.~Baumer, and G.~D.~Sprouse,
Phys. Lett. B \textbf{655}, 114 (2007).

\bibitem{Povel1973}
H.P. Povel, Nuclear Physics A {\bf 217}, 573 (1973).

\bibitem{Michel2017}
N. Michel, N.S. Oreshkina and C.H. Keitel, Phys. Rev. A {\bf 96}, 032510 (2017). 

\bibitem{Michel2019}
N. Michel and N.S. Oreshkina, Phys. Rev. A \textbf{99}, 042501 (2019).

\bibitem{VMK88}
D.A. Varshalovich, A.N. Moskalev and V.K. Khersonskii,
Quantum Theory of Angular Momentum, World Scientific, Singapore, (1988).
Da
\bibitem{Rowe70}
D.J. Rowe, Nuclear Collective Motion, Methuen, London, (1970).

\bibitem{BMII75}
A. Bohr and B.R. Mottelson, Nuclear Structure, Vol. II,
Benjamin, New York, (1975).

\bibitem{BMI69}
A. Bohr and B.R. Mottelson, Nuclear Structure, Vol. I,
Benjamin, New York, (1969). 

%\bibitem{ENSDF}https://www.nndc.bnl.gov/ensdf/

%\bibitem{PC_WI}https://en.wikipedia.org/wiki/Probability\_current.
%\bibitem{Che69}M. Chemtob, Nucl. Phys. A {\bf 123}, 449 (1969).

%\bibitem{Adler1956}K. Adler et al., Rev. Mod. Phys. {\bf 28}, 432 (1956).

%\bibitem{Fitzpatrick2014}R. Fitzpatrick (2014), http://farside.ph.utexas.edu/teaching/jk1/lectures/node31.html
%\bibitem{Jone_table1988}N. J. Stone, Table of Nuclear Magnetic Dipole and Electric Quadrupole Moments, Oxford Physics (1988).

%\bibitem{Finkbeiner1993}M. Finkbeiner, B. Fricke, Phys. Lett. A {\bf 176}, 113 (1993).

%\bibitem{fricke2004}
%G. Fricke and H. Schopper (eds.), Numerical Data and Functional Relationships in Science and Technology, 
%Vol. 20, Group I: Nuclear and Particle Physics, Nuclear Charge Radii, 
%Springer, Berlin Heidelberg, (2004).

%\bibitem{AngeliMarinova2013}
%I.~Angeli and K.~P.~Marinova, At. Data Nucl. Data Tables \textbf{99}, 69 (2013).

%\bibitem{Stone2005}
%N.~J.~Stone, At. Data Nucl. Data Tables \textbf{90}, 75 (2005).

%\bibitem{GaitaArino2019}
%A.~Gaita-Ari\~{n}o, F.~Luis, S.~Hill, and E.~Coronado, Nat. Chem. \textbf{11}, 301 (2019).

%\bibitem{Indelicato2013}
%P.~Indelicato, Phys. Rev. A \textbf{87}, 022501 (2013).

%\bibitem{wernsdorfer2019}
%W. Wernsdorfer and M. Ruben, \textit{Advanced Materials} {\bf 31}, 1806687 (2019).

%\bibitem{morenopineda2018}
%E. Moreno-Pineda, C. Godfrin, F. Balestro, W. Wernsdorfer and M. Ruben, 
%Chemical Society Reviews {\bf 47}, 501 (2018).

\end{thebibliography}

\end{document}